\documentclass[11pt,twoside,sort&compress]{elsarticle}

\usepackage{lineno,hyperref}

\journal{Progress in Nuclear and Particle Physics}

\usepackage{natbib}

\usepackage{setspace}

\usepackage{amsmath,amssymb,} 
\usepackage{graphics} 
\usepackage[utf8]{inputenc}
\usepackage[left=1in,right=1in,top=1in,bottom=1in]{geometry}
\usepackage{wrapfig}

\newcommand{\aap}{{Astron.\ Astrophys. }}
\newcommand{\apj}{{Astrophys.\ J. }}
\newcommand{\apjl}{{Astrophys.\ J.\ Lett. }}
\newcommand{\apjs}{{Astrophys.\ J.\ Suppl. }}
\newcommand{\apss}{{Astrophys.\ Space Sci. }}
\newcommand{\araa}{{Ann.\ Rev.\  Astron.\ Astrophys. }}

\newcommand{\mnras}{{Mon.\ Not.\ Roy.\ Astron.\ Soc. }}
\newcommand{\nat}{{Nature }}
\newcommand{\sovast}{{Sov. Astron. }}

\renewcommand{\vec}[1]{\boldsymbol{#1}}

\def\lsim{\raise0.3ex\hbox{$\;<$\kern-0.75em\raise-1.1ex\hbox{$\sim\;$}}}
\def\gsim{\raise0.3ex\hbox{$\;>$\kern-0.75em\raise-1.1ex\hbox{$\sim\;$}}}
\def\eps{\varepsilon}
\def\theta{\vartheta}
\def\R{{\cal R}}
\def\d{{\rm d}}
\def\be{\begin{equation}}
\def\ee{\end{equation}}
\def\ba{\begin{align}}
\def\ea{\end{align}}

\def\Xmax{$X_{\max}$ }
\def\X2{RMS($X_{\max})$}

\begin{document}

\begin{frontmatter}

  \title{Cosmic Ray Models}

\author{M.~Kachelrie{\ss}}
\address{Institutt for fysikk, NTNU, Trondheim, Norway}

\author{D.~V.~Semikoz}
\address{APC, Universit\'e Paris Diderot, CNRS/IN2P3, CEA/IRFU,\\
Observatoire de Paris, Sorbonne Paris Cit\'e, France 
}

\begin{abstract}
  We review progress in high-energy cosmic ray physics focusing on recent
  experimental results and models developed for their interpretation.
  Emphasis is put on the propagation of charged cosmic rays, covering the
  whole range from
  $\sim (20-50)$\,GV, i.e.\ the rigidity when solar modulations can be
  neglected, up to the highest energies observed. We discuss models
  aiming to explain the anomalies in Galactic cosmic rays, the knee, and the
  transition from Galactic to extragalactic cosmic rays.
\end{abstract}


\end{frontmatter}

\tableofcontents

\section{Introduction}
\label{introduction}

\paragraph{Cosmic ray measurements}

The presence of an ionizing radiation at the Earth's surface was already
recognized by Coulomb in 1785~\cite{Coulomb:1785}. More than a century later,
Hess showed conclusively that the ionisation rate increases with altitude,
suggesting that it has a cosmic origin~\cite{Hess:1912srp}. By the 1930s, the
observations of the geomagnetic latitude effect by Clay~\cite{Clay27} and
coincidence measurements using two Geiger-M\"uller counters by Bothe and
Kohlh\"orster~\cite{Bothe29} demonstrated that this ionizing radiation
consists mainly of charged particles, coined later ``cosmic rays''.
In the 1940s, measurements using cloud chambers and photographic plates
carried by balloons into the stratosphere showed that cosmic rays (CRs)
consist mainly of relativistic protons, with an admixture of heavier
nuclei~\cite{Schein:1941}.

The existence of extensive air showers triggered by high-energy CRs was
established by Kohlh\"orster, Auger, and their collaborators in the
1930s~\cite{1938NW.....26Q.576K,Auger:1938}.
After the second world war, large detector arrays were installed to measure
these extensive air showers, establishing a power law
${\rm d}N/{\rm d}E\propto 1/E^\alpha$
for the  energy spectrum of CRs with $\alpha\simeq 2.7$. At the energy
$E\simeq 4$\,PeV, a hardening of  the spectral index to $\alpha\simeq 3.1$
dubbed the CR knee was discovered by Kulikov  and Khristiansen
in the data of the MSU experiment in 1958~\cite{1959JETP...35....8K}.  
In the following years, the M.I.T.\ group deployed at the Volcano Ranch an
array of scintillation counters covering an area of 12\,km$^2$ which
recorded in 1962 an air shower with energy around
$10^{20}$\,eV~\cite{Linsley:1963km}.
At present, the two largest arrays observing CRs are the Pierre Auger
Observatory (PAO) located in Argentina covering an area of 3000\,km$^2$
and the Telescope Array (TA) in the USA covering 900\,km$^2$. Both 
are hybrid experiments combining surface detectors
to measure air showers on the ground and fluorescence detectors which can
follow the longitudinal development of the showers in the atmosphere.

A summary of CR intensity measurements is shown in the left panel of Fig.~1.
The (particle) intensity $I(E)$ is defined as the number $N$
of particles with energy $E$ crossing a unit area per unit time and
unit solid angle and is thus connected to the
(differential) number density of CRs with velocity $v$ as  
$n(E)=\frac{4\pi}{v} I(E)$. If the intensity $I(E)$ is isotropic, the
flux $F(E)$ through a planar detector is simply $F(E)=\pi I(E)$.
In such figures, the particle intensity $I(E)$ is often multiplied by a power
$\alpha$ of the energy $E$ such that $E^\alpha I(E)$ becomes approximately
flat, making thereby
structures in $I(E)$ more visible. In the flux of CR nuclei, which is the
dominating contribution to the total CR flux, additional to the CR knee
another break at $\simeq3\times 10^{18}$\,eV called the ankle and a cut-off like
feature around $10^{20}$\,eV are
visible. Below $\simeq 20$\,GeV, the CR spectrum is suppressed because
the magnetic field embedded within the Solar wind plasma
prevents that charged low-energy particles 
enter the Solar system. The second most-prominent species in the CR flux
are electrons which flux is reduced by a factor of order~100 relative to
the one of nuclei. The
fluxes of their antiparticles, antiprotons and positrons, are of comparable
magnitude and suppressed by two orders of magnitude relative to electrons.

In the left panel of Fig.~1,
the intensities $I_j(E)\propto {\rm d}N_j/{\rm d}E$ are multiplied with
$E^2$ which implies that the area $\int{\rm d}E\, E I_j(E)$ is proportional
to the energy density contained in particles of  type $j$.
Thus the energy carried by neutrinos (indicated by the magenta band) and
by the extragalactic gamma-ray background (EGRB)  (green) is of the same
order. A sizeable part of both the diffuse neutrino and gamma-ray flux could
be produced by extragalactic protons, if a large fraction of these protons
interacts in their sources.

Figure~1 contains also for the energies of the knee and ankle the number of
particles crossing a detector of a given size per year. Since the maximal area
of a balloon or satellite experiment is of the order of a few square meter,
the energy $10^{14}$\,eV marks the end of direct detection experiments.
These experiments have typically the ability to measure the charge  of
individual CRs and thus the fluxes of individual CR nuclei are
relatively well-known up to this energy. 
At higher energies, the CR flux drops to a level which prohibits to collect
them with high enough statistics using detectors of few m$^2$ size. However,
at these energies, the extensive showers of secondary particles initiated
by CR primaries interacting in the atmosphere start to reach the ground.
Detecting Cherenkov and fluorescence light of such showers in the atmosphere,
as well as the secondary particles on the ground allows one to reconstruct
the energy  and arrival direction of the primary CR rather precisely.
The determination of the primary mass has been, however, a challenging
problem for these indirect measurements, although considerable progress has
been made in the last 15~years.

\paragraph{Astrophysics of cosmic rays}
 
In 1934, Baade and Zwicky suggested presciently that CRs
draw their energy from supernovae explosions~\cite{1934PNAS...20..259B}.
Hiltner~\cite{1949Natur.163..283H} and Hall~\cite{1949Sci...109..166H}
discovered in 1949 an ubiquitous magnetic
field in the Milky Way through the polarisation of star light.
In the same year, Fermi \cite{ Fermi:1949ee} proposed his theory of CR
acceleration by moving ``magnetic clouds'',  explaining for the first
time how a power-law like energy spectrum can arise through the
combined action of acceleration and losses. One might view this year
as the birth date of the ``astrophysics of cosmic rays'', i.e.\
the  research field studying the acceleration and propagation of
cosmic rays. Five years later, Morrison, Olbert, and Rossi suggested that
the path length of CRs in the Milky Way should be small relative to their
interaction lengths, leading to the application of realistic diffusion models
to the propagation of Galactic CRs~\cite{Morrison:1954}. This approach
was worked out then in detail in the classic book of Ginzburg and
Syrovatskii~\cite{1969ocr..book.....G}.

Fermi's idea of second-order acceleration was developed further into the
theory of diffusive shock acceleration around 1977~\cite{1977ICRC...11..132A,Blandford:1978ky,1977DoSSR.234.1306K,1978MNRAS.182..147B}. In this theory,
the energy gain per cycle is linear in the  shock velocity, while it is
quadratic in the cloud velocity in Fermi's original model. Consequently,
diffusive shock acceleration leads to much larger maximal energies for the
same acceleration time.  Therefore it is today 
considered as the leading explanation for the acceleration of CRs in
a large variety of astrophysical environments, ranging from
shocks in the Solar corona, pulsar winds, and supernova remnants up to active
galactic nuclei and gamma-ray bursts. A crucial prediction of diffusive shock
acceleration is the slope of the energy spectra
${\rm d}N/{\rm d}E\propto 1/E^\beta$
of   accelerated particles~\cite{Drury:1983zz}:
In the test-particle approximation, nonrelativistic
shocks with Mach number ${\cal M}\equiv v_{\rm sh}/c_{\rm s}$ lead to energy
spectra with $\beta\simeq 2+1/4{\cal M}^2$. For supersonic shocks, the Mach
number  satisfies ${\cal M}= v_{\rm sh}/c_{\rm s} \gg 1$,
with $v_{\rm sh}$ denoting
the shock and $c_{\rm s}$ the sound speed, respectively. Thus shock acceleration
leads to energy spectra with $\beta\simeq 2$, which contain the same
amount of energy per decade up to a maximal energy $E_{\max}$. 

The maximal energy $E_{\max}$ achieved in an accelerator is determined by its
finite size and age. The most important theoretical limit to the maximal CR
energy arises from the condition that the Larmor radius 
\begin{equation}
\label{eq:larmor}
 R_{\rm L} = \frac{cp_\perp}{ZeB} =\frac{\R}{B} \simeq 1.08\, {\rm pc} \;
           \frac{\R}{\rm PV} \: \frac{\mu\rm G}{B}  
\end{equation}
of a particle with charge $Ze$ in a homogeneous magnetic field $B$
must be smaller by at least $v_{\rm sh}/c$ than
the dimensions of the acceleration region~\cite{1984ARA&A..22..425H}.
Here, $B$ denotes the local field strength of the magnetic field and $\R$
the rigidity $\R=cp_\perp/Ze$ of a CR with charge $Ze$ and momentum $p_\perp$
perpendicular to the direction of the magnetic field.  The factor  $v_{\rm sh}/c$
may be absent, if the magnetic field is quasi-parallel to the shock surface,
as argued by Jokipii~\cite{ 1987ApJ...313..842J}.This
criterion leads to the ``Hillas plot'' in the right panel of Fig.~1
which shows the maximal energy achievable in various types of CR sources
as function of their typical sizes and magnetic field strengths.
Sources able to accelerate protons to $10^{20}$\,eV should lie above the red
line, while the constraint for iron is relaxed to the green line. It is
immediately clear that only few source types might be able to accelerate CRs
to the highest observed energies. 
Two additional constraints on potential CR sources apply to their age
and their compactness: If the source is too compact, the strong magnetic
field leads to too large energy losses, reducing the allowed area for
$10^{20}$\,eV proton  sources to the light-grey area shown in the
Hillas plot. In the opposite case of a very extended source with a weak
magnetic field, the acceleration time may exceed the age of the source.
In the specific case of  CRs accelerated by SN shocks, the maximum energy
was estimated by Lagage and Cesarsky as 10\,TeV, assuming that the
magnetic field  is pependicular to the shock and its strength close to the
shock equals the ambient magnetic
field, $B\simeq 3\mu$G~\cite{Lagage:1983zz}. This result would exclude  shock
acceleration in SNRs as source of Galactic CRs up to the knee.

Going beyond the test-particle approximation for shock acceleration
leads to two
modifications: First, the pressure of CRs modifies the shock profile, and
as a result the CR spectra deviate from a simple power law and become concave.
Second, and more importantly, the escape current of CRs leads
to an efficient magnetic field amplification via the Bell
instability~\cite{2001MNRAS.321..433B,2004MNRAS.353..550B},
increasing thereby the possible maximal energies of CRs. The theoretical
suggestion of magnetic field amplification is supported by various
observations: For instance, the analysis of the morphology of
X-ray emission close to the outer shocks of SN1006~\cite{Berezhko:2003ej}
and Cas~A~\cite{Vink:2002yx,Bamba:2004zr} imply that strong magnetic fields
on the order of $100\mu$G are needed to explain rapid synchrotron losses by
high-energy electrons. The presence of stronger magnetic fields close to
shocks re-opens the possibility that supernovae accelerate CRs up to the
knee and beyond.

\begin{figure}
  \centering
  \includegraphics[width=0.49\columnwidth]{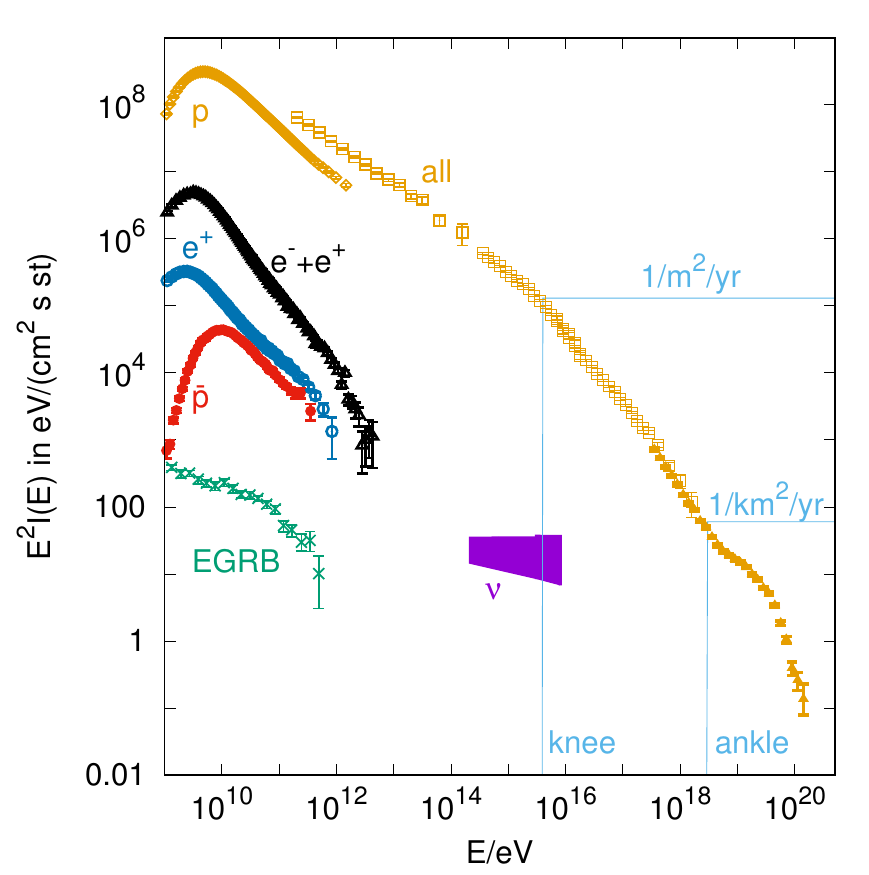}
  \includegraphics[width=0.49\columnwidth]{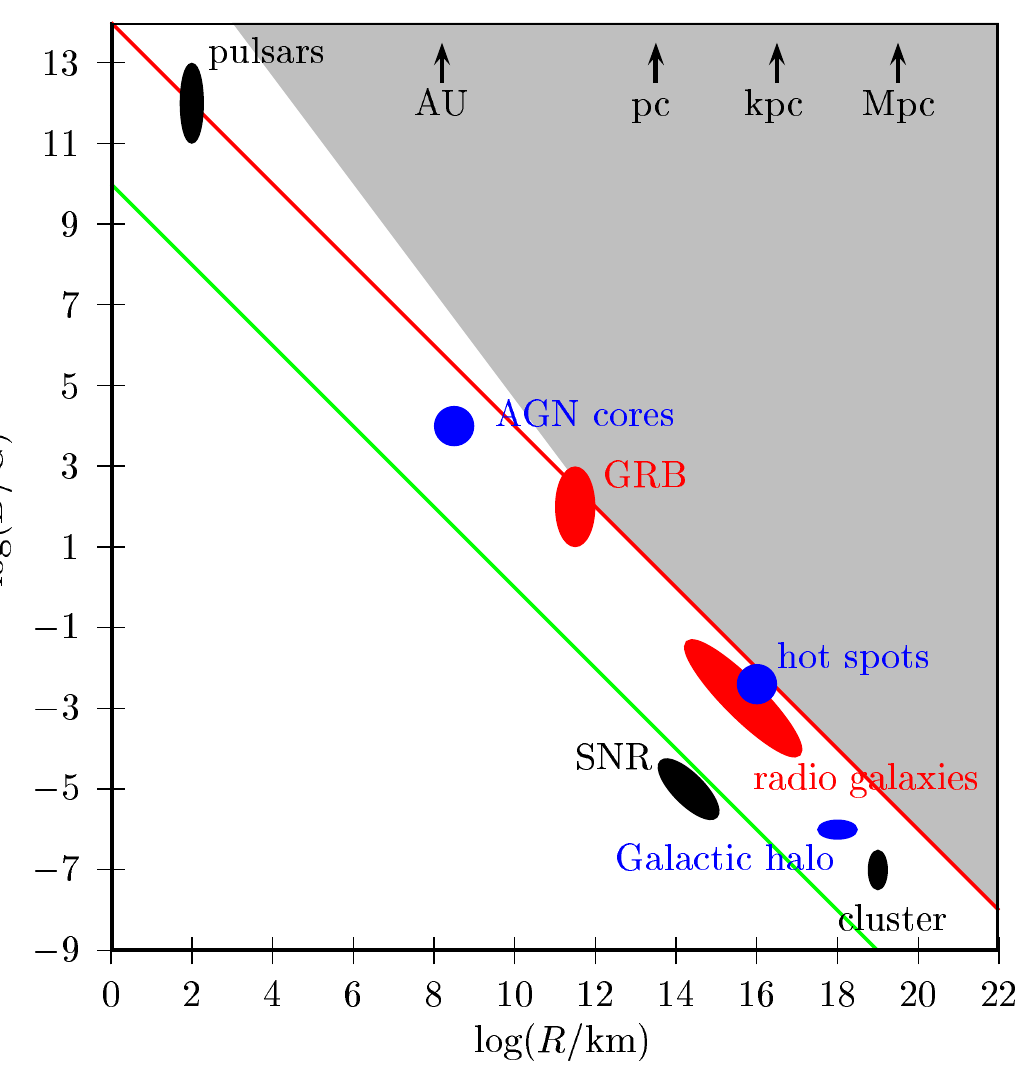}
\caption{\label{Hillas}
{\em Left panel:\/}
Summary of intensity measurements $E^2I(E)$ of CR nuclei and protons (orange),
electrons plus positrons (black), positrons (blue),
antiprotons (red) , neutrinos (magenta)  and diffuse photons (green). 
{\em Right panel:\/}
Magnetic field strength $B$ versus size $R$ of various suggested  CR sources;
adapted from Refs.~\protect\cite{Kachelriess:2008ze,Ptitsyna:2008zs}.
}
\end{figure}

Cosmic rays interacting with gas or background photons produce neutral
and charged pions whose decays in turn lead to secondary high-energy
photons and neutrinos. The combined study of potential CR sources
using charged CRs together with photons and neutrinos has developed
into the field of multi-messenger astronomy~\cite{Spurio:2018knn}.
Moreover, there is a close
relationship with searches for gravitational waves: Many suggested CR
sources tap their energy from the gravitational collapse of a compact
object, which leads to the emission of gravitational waves.
Vice versa it is expected that the merger of a binary system
involving one or two neutron stars leads to the acceleration of high-energy
particles, as it was observed for the first time in the case of  
GW170817~\cite{GBM:2017lvd}:
This event was observed extensively in the  optical, x-ray and gamma-ray
part of the electromagnetic spectrum, with a spectrum characteristic for a
short gamma-ray burst. In follow-up searches by the IceCube and ANTARES
neutrino observatories and the Pierre Auger Observatory, no neutrinos consistent
with this event were found.

\paragraph{Notation}

Many quantities in CR physics can be approximated
by broken power laws. In general, we will denote the power law for the
observed intensity as $I(\R)\propto \R^{-\alpha}$, for the
injection spectrum as $Q(\R)\propto \R^{-\beta}$, for the diffusion
coefficient as $D(\R)\propto \R^{\delta}$ and the power spectrum of the
turbulent magnetic field as ${\cal P}(k)\propto k^{-\gamma}$.
Depending on the context, we will prefer different energy variables:
Both the acceleration and diffusion in magnetic fields of CR nuclei
depends only on their rigidity $\R$, which favours this variable for the
discussion of CR propagation. The energy per nucleon $E/A$ is conserved in
spallation reactions and therefore convenient to use when CR spallation
plays a major role. Direct detection experiments present their data often
in terms of the kinetic energy $E_{\rm kin}=E-Am_p$. Finally, at the highest
energies the mass number $A$ cannot be determined reliably and one uses
therefore the total CR energy $E$.

\paragraph{Emphasis and structure}

We discuss CRs with rigidity above $\R\simeq 20$\,GV up to the highest
energies observed. { As a comparison of, e.g.,  electron spectra at different
times in the Solar cycle shows, the differences above this rigidity are
negligible relative to experimental uncertainties.}
Our choice for the lower limit allows us therefore to neglect
the effect of solar modulations.  We concentrate on the propagation of CRs
and the production of secondaries, omitting details of the acceleration
process in the sources. Instead we concentrate in this review on models
aiming to explain recent experimental results on the observed CR fluxes:
In the energy range below
the knee, we discuss mainly models which were suggested as solution
to the rise in the positron fraction, the breaks and the non-universality
of the CR nuclei spectra. In the case of extragalactic CRs, measurements
of the CR dipole and the mass composition favour a low transition energy 
between Galactic and extragalactic CRs and a mixed composition. Thus we
concentrate on  models able to explain the ankle as a feature of the
extragalactic CR spectrum.

For more general overviews and the topics neglected we recommend the
following resources: The textbooks~\cite{Ginzburg:1990sk,Schlickeiser:2002pg}
give a comprehensive introduction into the astrophysics of CRs. They are
nicely supplemented by the textbook~\cite{Gaisser:2016uoy}, which contains an
up-to-date discussion of observations and an introduction to the development of
extensive air showers.
The effect of solar modulations on low-energy CRs is thoroughly discussed in
Ref.~\cite{Potgieter:2013pdj}. 
{ Diffusive shock acceleration is reviewed in the classic
work~\cite{Drury:1983zz}, while more recent developments are covered, e.g.,
in Refs.~\cite{Marcowith:2016vzl,Bykov:2018wrt}.}
The standard diffusion approach to the propagation of Galactic CRs has been
described in detail in the textbooks~\cite{Ginzburg:1990sk,Schlickeiser:2002pg};
a discussion of the numerical approach used e.g.\ in GALPROP and its
main results is given in Ref.~\cite{Strong:2007nh}.
Gamma-ray studies using Cherenkov telescopes and satellite detectors
like Fermi-LAT which have revealed important informations on
CR sources are reviewed in Refs.~\cite{DeAngelis:2018lra,vanEldik:2015qla,Madejski:2016oqg}.
Recent reviews which provide additional details on ultrahigh energy cosmic
rays, in particular their mass composition and source candidates, are
Refs.~\cite{Mollerach:2017idb,Anchordoqui:2018qom}.  Some historical
background can be found in Refs.~\cite{Dorman:2014nka,Dorman:2014mka,DeAngelis:2010pj,Kampert:2012qen}.

In a scenario  alternative to the acceleration of CRs in astrophysical
sources, CRs are produced in decays or annihilations of relic particles.
While this possibility is of importance for the search for particle physics
beyond the standard model, it
can give only a subdominant contribution to the observed CR flux:
The main predictions of this scenario---a flat energy spectrum ($\propto
1/E^{1.9}$) of the decay products, a large photon fraction,  equal matter
and antimatter fluxes, and the (almost) absence of
nuclei~\cite{Kachelriess:2008bk}---constrain
this contribution e.g.\ in the energy range $10^{18}$--$10^{19}$\,eV
to be less than 0.1\% of the total CR flux~\cite{Aab:2016agp}.
In the PeV range, it has been suggested that the diffuse neutrino flux 
observed by IceCube is generated by decays of dark matter
particles~\cite{Bhattacharya:2019ucd}.
Below TeV energies, the search for dark matter in the form of
weakly interacting massive particles is a very active field, { and possible
connections to some anomalies in CR physics are reviewed, e.g.,
in Ref.~\cite{Gaskins:2016cha}. }

We start Section~\ref{below_knee} with a short review of our knowledge of
the regular and turbulent component of the Galactic magnetic field, followed
by a description of our local environment, the Local Bubble. The standard
approach to CR propagation in the Galaxy based on diffusion models,
the necessary inputs and the basic results are described next. Then we
discuss evidence that CRs diffuse strongly anisotropically, and the resulting
impact on the number of CR sources contributing to the locally observed flux.
After a review of the main experimental results and the observational
anomalies which have appeared during the last 10~years, we discuss
models suggested to explain these anomalies.

Section~\ref{model_knee} is devoted to the knee in the CR spectrum.
We start with a discussion of observations of the
knee in the all-particle spectrum and of knee structures in
the spectra of nuclear groups. Then we present models which explain the knee,
either by the maximal energy of different source populations or by
propagation effects.

In Section~\ref{transition}, we discuss ultra-high energy cosmic rays (UHECR).
After a discussion of measurements of the energy spectrum, we present 
results on the mass composition of UHECRs.
Then we discuss secondary gamma-rays and neutrinos produced by UHECRs,
before we review recent results on anisotropies.
Next we combine the experimental evidence
presented to argue that the transition between Galactic and extragalactic
CRs happens at a relatively low energy, and agrees with a spectral feature
called the second knee around $E\approx 5\times 10^{17}$\,eV.
We then discuss UHECR models which are able
to explain the presented data on the spectrum, composition and the
transition energy.
Finally, we summarize and conclude in section~\ref{conclusions}.

\section{Cosmic rays below the knee}
\label{below_knee}

\begin{figure}
  \centering
  \includegraphics[width=0.55\columnwidth]{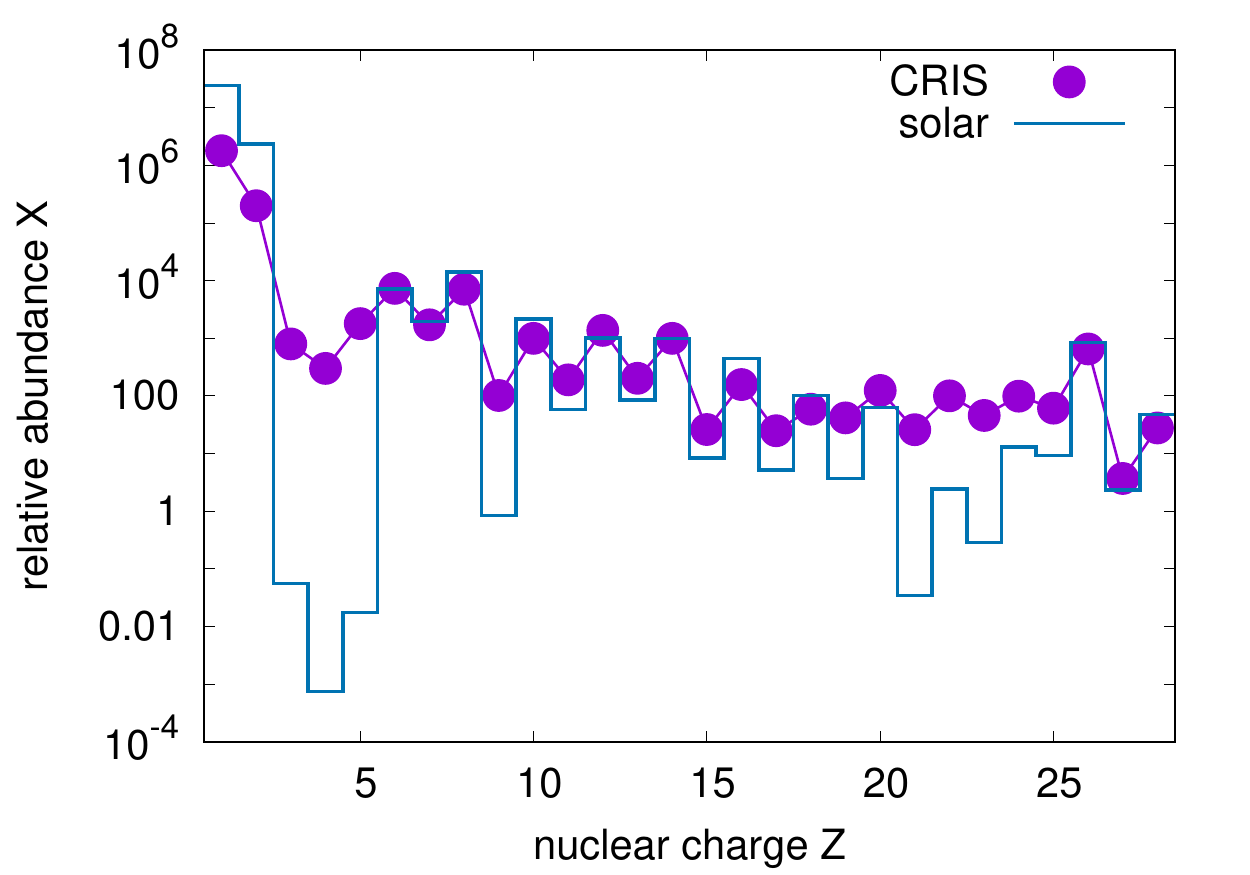}
\caption{\label{abund}
  Relative abundances $X$ normalised to $X({\rm Si})=1000$ for the proto-sun
  (solar system abundances) from Ref.~\protect\cite{2003ApJ...591.1220L} shown as boxes
  versus the abundances in CRs measured by BESS~\protect\cite{2002ApJ...564..244W} and
  CRIS~\cite{deNolfo:2006qj,2009ApJ...698.1666G}.}
\end{figure}

Cosmic rays are measured locally, with the two Voyager satellites as the most
distant experiments from Earth as the only exceptions. While these two
satellites have started
to probe the conditions outside the heliosphere, photons and neutrino
observations provide in addition directional information about CRs and the
physical processes taking place along the observed line-of-sight. Connecting
these and the local measurements to the physics occurring in CR sources
requires an accurate modelling of CR transport. Since the turbulent component
of the Galactic magnetic field (GMF) scatters efficiently CRs, they perform
a random walk and escape only slowly from the Galaxy. Before
they escape, they cross many times the Galactic disk producing secondary
CRs in hadronic interactions with gas. { Additionally, CRs accumulate
a grammage in their source region which typically has an increased density
compared to the average density in the Galactic disk.}
As a result of these interactions,
the abundance of elements which are produced rarely in big-bang and stellar
nucleosynthesis like the lithium-beryllium-boron group or titanium is strongly
enhanced in CRs, cf.~with
Fig.~\ref{abund}. The production of these secondaries provides an important
handle to constrain the parameters of a given propagation and source model.
Therefore we review first the properties of the GMF which determines
the propagation of Galactic CRs,
before we discuss the diffusion approach which has become the standard
method to study Galactic CR propagation. Then we describe several
observational anomalies, i.e.\ deviations from the ``naive'' expectations in
the simplest diffusion picture, which have appeared as the precision of new
experiments increased in the last decade. Finally, we critically discuss in
this section some of the solutions proposed as explanations for these
anomalies.

\subsection{Galactic magnetic field and our local environment}
\label{GMF}

\paragraph{Galactic magnetic field}

The observed distribution of CR arrival directions is highly isotropic,
with deviations from isotropy of only few parts in $10^{3}$ up to PeV
energies. Since Galactic CR sources
are strongly concentrated in the Galactic disc, an efficient mechanism for
the isotropisation of the CR momenta exists. Agents of this isotropisation
are turbulent magnetic fields, since CRs scatter efficiently
with field modes whose wavelength matches their Larmor radius $ R_{\rm L}$.

The magnetic field $\vec B$ can be divided into a regular component which is
ordered on large scales and a turbulent component. The turbulent magnetic field 
satisfies $\left\langle\vec B(\vec x)\right\rangle=\mathbf{0}$ and
$\left\langle\vec B(\vec x)^2\right\rangle \equiv B_{\rm rms}^2$, where
$\langle\cdot\rangle$ denotes the ensemble average. Decomposing
the turbulent field in Fourier modes with wave-vector $\vec k$, it can be
characterised by its power spectrum  $\mathcal{P}(\vec k)$ which
determines the magnetic energy density per mode $\vec k$.
Depending on its effect on observables, the turbulent field can be
split further into an anisotropic and an isotropic  component.

Turbulent  magnetic  fields  in  the interstellar medium (ISM) are produced
and shaped by two dominant  processes: the tangling and the  compression of
the mean field by mass flows as, e.g., stellar winds and  supernova shocks,
and the action of the fluctuation  dynamo~\cite{Brandenburg:2013vya}.
In the first case, the energy in the turbulent field is injected at
large scales $L_{\max}$ of order 1--100\,pc.
Then the energy cascades to smaller scales until it dissipates at the
damping scale  $L_{\min}$, which could be as low as an astronomical unit. 

Turbulent fields are often modelled as Gaussian random fields, in which case
all the information on the  magnetic field is encapsulated in the two-point
correlation function $\left\langle\vec B(\vec x)\vec B(\vec x')\right\rangle$.
The correlation length $L_{\rm c}$ of the field determines the scale
$k_{\rm c}=2\pi/L_{\rm c}$ above which approximately half of the energy density
of the turbulent field resides. In the case of a Gaussian random field with an
isotropic power-law spectrum $\mathcal{P}(k)\propto k^{-\gamma}$, it is equal to
\begin{equation}
 L_{\rm c} = \frac{L_{\max}}{2} \: \frac{\gamma - 1}{\gamma} \:
            \frac{1-(L_{\min}/L_{\max})^{\gamma}}{1-(L_{\min}/L_{\max})^{\gamma-1}}
\simeq
\frac{L_{\max}}{2} \: \frac{\gamma - 1}{\gamma} ,
\label{CorrelationLength}
\end{equation}
where the approximation is valid for $L_{\max}\gg L_{\min}$ and $\gamma>1$.
With $\gamma = 5/3,\,3/2$ and 1 respectively for Kolmogorov~\cite{1941DoSSR..30..301K}, Iroshnikov-Kraichnan~\cite{1964SvA.....7..566I,1965PhFl....8.1385K}
and Bohm turbulence, it follows $L_{\rm c} = L_{\max}/5$,
$L_{\rm c} = L_{\max}/6$ and $L_{\rm c} =L_{\min}$.
%
The turbulent fields obtained in MHD simulations are not Gaussian random
fields but intermittent: { Magnetic fields generated via the induction equation
by the random motion of the plasma reflect the non-Gaussian correlations
in the plasma velocity. For instance, Ref.~\cite{Shukurov:2017jxf} finds
that magnetic structures occupy a smaller proportion of the volume as
the magnetic Reynolds number---which controls the relative effects of
advection and diffusion---increases. It has been argued that this effect
reduces on
average the deflections of CRs scattering on magnetic irregularities and
enhances anisotropies in the CR propagation~\cite{Shukurov:2017jxf}.
How important the consequences of non-Gaussianity on the
propagation of CRs in the Milky Way are is, however, largely unexplored.
For a proper assessment, self-consistent  MHD simulations including
the CR fluid would be required.}

Observations of fluctuations in the thermal electron density $n_e$ show that
the corresponding power spectrum $\mathcal{P}(k)$ follows over twelve decades
$\mathcal{P}(k)\propto k^{-5/3}$~\cite{Armstrong:1995zc}, i.e.\ the slope
agrees with the one predicted by Kolmogorov~\cite{1941DoSSR..30..301K}.
Since electrons may not be purely passive tracers for the magnetic field, 
direct measurements of the magnetic turbulence are however necessary.
The only direct measurements of the magnetic field in the local ISM
have been performed by a magnetometer on board of Voyager~1: These
measurements are consistent with Kolmogorov turbulence and, extrapolated to
larger scales, a maximal scale of
$L_{\max}\simeq 10$\,pc~\cite{2015ApJ...804L..31B}.
Alternatively, one can combine different probes along a line-of-sight
to break the degeneracy between various parameters. For instance,
combining Faraday rotation measurements,
${\rm RM}\propto \int {\rm d}\vec{s}\cdot \vec{B}  n_e$,
with emission  measurements, ${\rm EM}\propto \int {\rm d}s\, n_e^2$, one can
infer the power spectrum of the turbulent magnetic field and the electron
density separately. On scales $L\lsim 4$\,pc, the power-spectrum of
magnetic turbulence is consistent with three-dimensional Kolmogorov
turbulence~\cite{1996ApJ...458..194M,Haverkorn:2008tb}. Its correlation length
grows from $\sim 1$\,pc in the spiral arms and $\sim 10$\,pc in the interarm
regions of the  Galactic disk to $\sim 100$\,pc in the halo according
to Refs.~\cite{Haverkorn:2008tb,Iacobelli:2013fqa}. The strength of
the Kolmogorov part of the turbulent field was estimated as
$B_{\rm rms}\simeq 0.6\mu$G~\cite{1996ApJ...458..194M} with a maximal scale
of $L_{\max}\simeq 4$\,pc. Above this scale, the slope of the turbulence
decreases, being close to $\alpha\simeq 2/3$ between
$4\,{\rm pc}<L<(70-100)$\,pc. Presumably, this break in the power spectrum
indicates a change from isotropic to anisotropic turbulence, when the field
components along the regular field are enhanced. For the propagation of CRs,
such a field acts locally like a part of the ordered field.

The magnetic field of the Milky Way is concentrated in the Galactic disk,
and its regular component approximately follows the spiral arms. These arms
are commonly modelled as logarithmic spirals. While a magnetic field
reversal on kpc scales has been unambiguously detected from pulsar RMs,
observations have not determined yet whether this reversal is a global one 
following the spiral arms, or whether it is a local feature.
The GMF  has an out-of-plane component, presumably similar to
the X-shaped halo fields detected in nearby edge-on spiral galaxies.
The strength of the coherent magnetic field as derived from pulsar RMs is
$\simeq 2\mu$G, generally consistent with values obtained modelling
synchrotron radiation. The total magnetic field strength is $\simeq 6\mu$G
at the Solar radius, increasing to $\simeq 10\mu$G at a Galacto-centric
radius of 3\,kpc.  Both the strength and the scale height of the halo field 
are rather uncertain, with estimates for the strength varying between
$\simeq 2\mu$G and $\simeq 12\mu$G and for the scale height between
$\simeq 1-5$\,kpc from pulsar RMs and $\simeq 5-6$\,kpc from synchrotron
emissivities; for reviews of the GMF see, e.g.,
Refs.~\cite{Beck:2000dc,Haverkorn:2014jka,Boulanger:2018zrk}.

Current models of the GMF like those of
Refs.~\cite{Pshirkov:2011um,Jansson:2012pc,Jaffe:2013yi} reproduce the
Faraday RM data and synchrotron emission maps to which they were fitted,
although the GMF morphology differs substantially between these models.
This implies, e.g., that it is at present not possible to correct the
deflection of a
UHECR by the GMF in a model-independent way. In contrast, CR propagation at
lower energies depends mainly on global features of the GMF, as, e.g., the
average escape time from the Galactic disk, and the differences between these
models play a smaller role.

Despite the variations between current GMF models and the relative large
uncertainties in the estimates of their parameters, we can draw a few
important conclusions: First, a significant contribution to the total field
strength is the anisotropic turbulent (also called ``ordered random'' or
``striated'') field whose fluctuations have typical length scales larger
than a pc. Thus for the propagation of CRs with energies below the knee,
this field acts locally like a part of the regular field. Second, the strength
of the Kolmogorov part of the turbulent field at smaller scales---which is
relevant for the scattering and isotropisation of CRs at these energies---is
smaller than the regular field. As a consequence, CR
propagate preferentially along the ordered field, and therefore CR
transport is anisotropic.
Moreover, the escape of CRs from the Galactic disk depends strongly on
the presence of an out-of-plane component of the regular magnetic field.

While it is natural to connect the isotropic Kolmogorov component to the
hydrodynamic turbulence in the ionized gas, the anisotropic component could
be generated by shock waves compressing a previously isotropic random
field, or by Galactic shear motions of the gas.
Another cause for an anisotropy in the turbulent magnetic field might be
that the regular field modifies the cascading of magnetic modes
from large to small scales. An example for such a model is Goldreich-Sridhar 
turbulence~\cite{Goldreich:1994zz}, where the power spectrum in the
perpendicular direction is Kolmogorov-like, and that in the parallel
direction is $\propto k^{-2}$. While such a type of turbulence is more
isotropic on scales close to the outer scale $L_{\max}$, it becomes
increasingly anisotropic for large wave-vectors. Thus such a behaviour
is opposite to the one observed, where the anisotropic field dominates the
large length scales. If the turbulence is compressible, fast magnetosonic
waves may exist in addition. Because of their smaller degree of anisotropy,
these waves may increase CR scattering as argued, e.g., in
Ref.~\cite{Cho:2002qi}. Finally, we note that at sufficiently low energies
the magnetic turbulence generated by CRs in form of  Alfv\'en waves may
become the dominating contribution to the turbulent field.

\paragraph{Our local environment}

The Sun resides inside a bubble of hot, tenuous plasma called the Local Bubble.
Such superbubbles are created around OB associations, when the wind-blown
bubbles of the individual stars encounter each other as they expand and merge
to form a single superbubble~\cite{vanMarle:2012bs,Krause:2013qar}.
Once the massive O and B stars explode at the end of their fusion cycle as 
core-collapse supernovae, shock waves are injected into the ISM. These shocks
expand quickly until they reach the bubble wall where they are typically
stopped~\cite{vanMarle:2015oca}. Therefore they do not form visible
supernova remnants (SNR), but instead power the expansion of the superbubble
in the ISM.

The Local Bubble extends roughly 200\,pc in the
Galactic plane, and 600\,pc perpendicular to it, with
an inclination of about $20^\circ$~\cite{2003A&A...411..447L}.
Observations~\cite{2019arXiv190107692M,2019arXiv190105971L}
and simulations~\cite{Breitschwerdt:1999sv,Schulreich:2017dyt} show that
the bubble walls are fragmented and twisted. Moreover, outflows away from
the Galactic plane may open up the bubble~\cite{Breitschwerdt:1999sv}.
The Local Bubble abuts with the Loop~1 superbubble, and with another
bubble towards the direction of the Galactic center~\cite{Wolleben:2010se}.

The magnetic field inside the bubble wall is expected to be enhanced by
the shocks compressing the ISM. In Ref.~\cite{2006ApJ...640L..51A},
the Chandrasekhar-Fermi method was used to derive $B_\perp=(8^{+5}_{-3})\mu$G
for the strength of the magnetic field in the wall of the Local Bubble.
Since this analysis is only sensitive to the field component perpendicular
to the line-of-sight,  the derived value can be seen as a lower limit on the
strength of the magnetic field in the wall. A similar result was
obtained in Ref.~\cite{Mao:2010zr} which measured Faraday RMs towards the
Galactic south and north pole. These authors used their upper limit of
$|{\rm RM}|\sim 7$rad/m$^{-2}$ towards the bubble wall to derive the
magnetic field inside the wall. Modelling the Local Bubble
as a cylindrical shell with radius
85\,pc and a wall thickness of 4\,pc, they derived $B\simeq 9\mu$G
inside the bubble wall, consistent with
the value obtained in Ref.~\cite{2006ApJ...640L..51A}.
In Ref.~\cite{2019arXiv190107692M}, evidence for a systematically varying
field strength in the bubble wall was presented: Using the Chandrasekhar-Fermi
method, one set of values around $B_\perp\simeq 8\mu$G, and another one with
strengths about three times higher, and up to about $40\mu$G was determined.
Such a variation is not too surprising in view of the fragmented structure
of the bubble wall and may be caused by an additional compression of the wall
by interactions of the outflow in the Local Bubble and opposing flows
by surrounding OB associations. 

The strength and direction of the magnetic field inside the Local Bubble
are only poorly constrained by observations: For instance, the difference
between the flow directions of interstellar He and H has been explained by
a mismatch of the local magnetic field direction and the flow of the ISM.
Using Voyager data, Ref.~\cite{Opher:2007ce} deduced $B\simeq 2\mu$G and
an angle of $\alpha=30^\circ-60^\circ$ between the local magnetic field
direction and the flow of the ISM. In Ref.~\cite{Wood:2007br},
theoretical MHD models of the heliosphere were used to predict the Ly$\alpha$
absorption along various lines of sight for different configurations of
the local magnetic field. Comparing the predicted and observed  Ly$\alpha$
spectra, models with $B\simeq 1.25-2.5\mu$G and $\alpha=15^\circ-45^\circ$
were found to fit best the data. 
 While these observations constrain the very local magnetic field, the field
between the bubble wall and the heliosphere is more difficult to determine.
In Ref.~\cite{2018A&A...611L...5A}, Planck observations of dust polarised
emission were fitted to a toy model describing the geometry of the magnetic
field in the bubble. While dust polarised emission provides no information
about the strength of the field, the direction of the local magnetic field
was determined as $(l,b)=(70^\circ\pm 11^\circ,43^\circ\pm 8^\circ)$ in
the northern and  $(l,b)=(78^\circ\pm 8^\circ,-14^\circ\pm 18^\circ)$ in
the southern Galactic polar caps. The large difference between the two
directions indicates that the magnetic field in the Local Bubble is
highly distorted.

Note also that the argument raised in Ref.~\cite{1998ApJ...493..715S}
that a magnetic field of up to $7\mu$G is required in the Local Bubble to
counter-balance the thermal pressure exerted by the enclosing hot X-ray gas
is obsolete: First, the discovery of X-ray emission associated with charge
exchange between solar wind ions and heliospheric neutrals has reduced
the need of non-thermal pressure support of the LB~\cite{2004A&A...422..391L}. 
Second,  a contamination from X-ray emission from the walls of the
Local Bubble~\cite{2004A&A...422..391L}, and an improper assumption of
collisional ionization equilibrium~\cite{2001Ap&SS.276..163B} may invalidate
the pressure balance argument of Ref.~\cite{1998ApJ...493..715S}.

\subsection{Standard approach to Galactic cosmic ray propagation}
\label{standard_prop}
\subsubsection{Method and inputs}

In the standard approach to Galactic CR propagation, the $N$-particle phase 
space distribution of CRs is approximated as a macroscopic fluid. Then the
time evolution of individual CRs propagating under the influence of the Lorentz
force is replaced by a diffusion process. Such a replacement corresponds to
a coarse-graining of CR trajectories on length scales
$\sim L_{\rm c}^{\alpha-1} R_{\rm L}^{2-\alpha}$. 
For CRs with rigidity $\R=100$\,TV and Kolmogorov turbulence with a correlation
length of order 10\,pc, the length $L_{\rm c}^{2/3} R_{\rm L}^{1/3}$ corresponds to
pc scales.  Note that particles with the same rigidity 
moving under the influence of only the Lorentz force follow the same 
trajectories in phase space. As a consequence, both diffusion in space and
in momentum (i.e.\ (re-) acceleration) proceed independently of 
the mass number of CR nuclei. Therefore one expects the same CR spectra
for different CR primaries, if they are expressed as function of rigidity
and interactions can be neglected.

The (spatial) diffusion tensor 
\begin{equation}
D_{ij}(\vec x_0,\R)= \lim_{t\to\infty}
\frac{1}{2Nt}\sum_{a=1}^N (x_i^{(a)}-x_{i,0})(x_j^{(a)}-x_{j,0})
\label{D_iso}
\end{equation}
can be determined numerically following the trajectories $x_i^{(a)}(t)$ of
$N\gg 1$ CRs injected at $\vec x_0$ for a given magnetic field
configuration~\cite{Casse:2001be,Parizot:2004wh,Giacinti:2012ar}.
Since these calculations are computationally expensive, one usually employs
analytical approximations instead.
For instance, the connection between the diffusion coefficient $D_{\|}$
parallel to the ordered field and the power spectrum
${\cal P}(k)$ of the turbulent magnetic field can be derived  analytically
in the  approximation of pitch-angle scattering, if the ordered field dominates
at the scale considered~\cite{Ginzburg:1990sk}. In this case,
the slope of the power spectrum ${\cal P}(k)\propto k^{-\gamma}$ 
determines the rigidity dependence of the diffusion coefficient parallel to
the ordered field as  $D_{\|}(\R)\propto \R^\delta$ with $\delta=2-\gamma$.
In Ref.~\cite{Strong:2007nh}, the numerical value of  $D_{\|}(\R)$ was
estimated as $D_{\|}\simeq 2\times 10^{27}(\R/{\rm GV})^{1/3}$\,cm$^2$/s for
a turbulent magnetic field with $B_{\rm rms}=5\mu$G, $L_{\max}=100$\,pc  and
Kolmogorov turbulence. In practise, most studies of CR propagation employ
instead a scalar diffusion coefficient with
a prescribed functional form as, e.g., $D(\R)=D_0 (\R/\R_0)^\delta$, and
determine the normalisation constant $D_0$ from a fit to secondary-to-primary
ratios as B/C (as we will discuss in the Section~2.2.2).

The CR fluid is coupled to the ISM and can drive, e.g., Galactic
winds~\cite{1991A&A...245...79B,Recchia:2017coe}. Its
pressure is in rough equipartition with the magnetic and dynamical pressure
in the ISM~\cite{Beck:2000dc,2005ARA&A..43..337C}, 
suggesting  the dynamical  importance of CRs for the ISM. 
Moreover, CR streaming can lead to wave turbulence and thus to the generation
of turbulent magnetic field modes. As a result, CRs, the ISM and the GMF
form a coupled, non-linear system which should be modelled self-consistently.
For recent studies in this direction see e.g.\
Refs.~\cite{2016ApJ...816L..19G,Pfrommer:2016mvy,Thomas:2018xnj,Evoli:2018nmb}.
Considering CRs of sufficiently high energy, $E\gsim E_\ast$,
this coupling can be neglected because the
CR density drops fast with energy. The numerical value of $E_\ast$ is
however very uncertain.
For instance,
Ref.~\cite{Ptuskin:2005ax} argued that  $E_\ast$ is as low as
$E_\ast\sim 1$\,GeV, i.e.\ the energy where a break in the diffusion
coefficient both from observation of synchrotron radiation from
electrons~\cite{2008JGRA..11311106W} and from secondary-to-primary rations
was deduced~\cite{Ptuskin:2005ax}. In contrast, the specific model
proposed in Refs.~\cite{Blasi:2012yr,Aloisio:2013tda,Aloisio:2015rsa} which
will be discussed as an example for the coupling between CRs and self-generated
turbulence in the next subsection argues for a high value,
$E_\ast\sim 300$\,GeV.

Restricting ourselves to energies $E\gsim E_\ast$, CRs can be propagated using
prescribed magnetic fields and gas densities as background. Adding then also
interactions,
the resulting transport equation for the (differential) density
$n^{(a)}(\vec x,p,t)={\rm d}N^{(a)}(\vec x,p,t)/({\rm d}p{\rm d}V)=
4\pi p^2 f^{(a)}(\vec x,p,t)$ of CR particles of type $a$ is given by
\begin{equation}
\begin{split}
  \lefteqn{
 \frac{\partial n^{(a)}}{\partial t} - 
 \nabla_i \left[ D_{ij} \nabla_j -  u_i\right]   n^{(a)} -
 \frac{\partial}{\partial p} \left[ p^2 D^{(p)}\frac{\partial}{\partial p} p^{-2}   n^{(a)} \right] 
  =   - \frac{\partial}{\partial p} \left( \beta^{(a)} n^{(a)} \right) }
\\ & -  
\left( c n_{\rm gas}\sigma_{\rm inel}^{(a)} +  \Gamma^{(a)} \right) n^{(a)}
 +  Q^{(a)} + \sum_b \left[ 
   c n_{\rm gas}  \int_E^\infty dE' \;\frac{\d\sigma^{ba}(E',E)}{\d\!E} +
   \Gamma^{ba} \right] n^{(b)} \,.
\label{transport}
\end{split}
\end{equation}
While the first line of Eq.~(\ref{transport}) describes the continuous time
evolution of the particle density $n^{(a)}$ due to (spatial) diffusion,
advection, diffusion in momentum space and continuous energy losses, 
the second line accounts for gain and loss processes. Particles are lost in
inelastic reactions and, if they are unstable, in decays; they are injected
by CR sources and produced as secondaries in interactions and  decays of
particles of type $b$. Most quantities, like the diffusion tensor $D_{ij}$,
the advection velocity $\vec u$, the injection rate $Q$, the energy loss
rates $\beta$, and the gas density $n_{\rm gas}$ depend on space and/or time.
For instance, the spatially varying strength of the GMF will lead to changes
in the  diffusion tensor and the synchrotron losses. Similarly, the discrete
nature of CR sources implies that their injection rate $Q$ is both space and
time dependent.

In order to reduce the complexity of this coupled set of partial differential
equations, one separates the problem of CR acceleration in their sources
from their propagation. Thus one employs for $Q$ an ansatz for the spectrum
of CRs after escape from their sources. Interactions and energy
losses (except for electrons at highest energies) do not modify strongly the
spectrum of primary CRs in the source and one expects therefore that they
share the same rigidity spectrum. The ansatz for this spectrum is typically
chosen as a power law,
$Q^{(a)}(\R)=Q^{(a)}_0(\R/\R_0)^{-\alpha}\exp(-\R/\R_{\max})$,
with a rigidity-dependent maximal energy $E_{\max}=Ze\R_{\max}$ and an
exponential cutoff.
The results from diffusive shock acceleration in the test-particle limit
motivate  the range $\alpha\simeq 2.0-2.2$ for the slope of the injection
spectrum. However, both the back-reaction of CRs on the shock and their
energy-dependent escape from their sources affect the test-particle
picture and modify the CR energy spectra. A potential resurrection of the
power-law spectra with $\alpha\simeq 2.0-2.2$  normally employed are the results
from Ref.~\cite{Schure:2013yga}: There it was shown that steep acceleration
spectra with $\alpha>2$ are converted into an escape spectrum with
$\alpha\simeq 2$. Moreover, steep acceleration spectra are required for the
large maximal energies needed to explain the extension of the Galactic
CR spectrum beyond the knee.  Similar conclusions were obtained in
Refs.~\cite{Cardillo:2015zda,Tatischeff:2018cyp}.

The relative abundances $Q^{(a)}_0$ of nuclei in the injection spectra are
either chosen close to the Solar ones, or are determined from a fit of
the produced abundances after propagation to the observed ones. Additionally,
the spatial distribution of the sources has to be fixed. In most models, the
injection of Galactic CRs is correlated with supernovae (SN). Therefore,
the spatial dependence of $Q^{(a)}_0$ is normally modelled according to the
observed distributions of supernova remnants (SNRs) or
pulsars~\cite{Green:2015isa}. Often, one neglects the spiral
structure of the Galactic disc as well as the enhanced SN activity in the
Galactic bulge. The latter assumption is justified, because of the small
volume of this region, except if one studies especially the photon emission
from this region~\cite{Carlson:2016iis}. Similarly, the spiral structure is
averaged out in many observational quantities, except e.g.\  for high-energy
electrons which can travel only short distances.

Magnetic fields are not purely static but move with a typical velocity of
the Alfv\'en speed, $v_A=B/\sqrt{4\pi\rho_{\rm ion}}\simeq (10-30)$\,km/s
with $\rho_{\rm ion}$ denoting the mass density of ionised atoms, and
thus diffusion occurs not only
in space but also in momentum. This results in second-order Fermi acceleration
during the propagation of CRs. The connection between the spatial and momentum
diffusion coefficients is given by $D^{(x)} D^{(p)}=\eps p^2 v_A^2$, with
$\eps(\beta)\sim 0.1$~\cite{Ginzburg:1990sk,Drury:2016ubm}.
Both analytical and numerical calculations indicate
that a large fraction of the total  energy in CRs is delivered by this process
which is often called  ``reacceleration''~\cite{Drury:2016ubm}. However,
reacceleration on Alfv\'en waves modifies the CR spectrum only at mildly
relativistic energies, i.e.\ at energies below our interest.

Advection, i.e.\ the bulk motion of the ISM, competes with diffusion as an
efficient way to transport CRs away from the Galactic disk. Since advection
is energy independent, it is the dominating transport mechanism at
sufficiently low
energies. For an uniform advection velocity, flat secondary-to-primary ratios
at low energies result. The observed flattening of the B/C ratio at few GeV was
interpreted as evidence for advection first in Ref.~\cite{1979ApJ...229..747J}.
However,  advection is in case of Kolmogorov turbulence not necessary to
reproduce the B/C data, and
Refs.~\cite{2002ApJ...565..280M,2002A&A...394.1039M} obtained zero
advection velocity as their best-fit case.
In a more realistic description, the advection
velocity increases with the distance to the Galactic plane. Moreover, in
regions of enhanced star formation activity superbubbles are formed which
may open towards large Galactic latitudes. In these ``Galactic fountains'',
ISM and CRs may be effectively advected out of the disk~\cite{Breitschwerdt:1999sv,2016ApJ...816L..19G}. Note also that a $z$ dependent increase of the
advection velocity $u$ can
modify the energy dependence of the diffusion coefficient deduced.

Continuous energy loss processes are only important at low energies, 
$E\lsim 1$\,GeV and, for electrons at high energies. In the latter case, 
the energy loss of an electron due to synchrotron radiation is given in
the Thomson regime by
\begin{equation}
 \beta \equiv -\frac{\d E}{\d t} \equiv   b E^2 =  3.79\times 10^{-18}\, 
 \frac{\rm GeV}{\rm s} \left( \frac{B}{\mu\rm G}\right)^2
 \left( \frac{E}{\rm GeV} \right)^2 \,.
\end{equation}
The losses due to inverse Compton scattering on cosmic microwave background
photons and starlight can be included using 
$u/({\rm eV/cm^3})\simeq (6.3\,B/\mu{\rm G})^2$.
As result of the losses, the energy of an electron degrades 
with time as $E(t) = E_0/(1+ b E_0t)$, and its half-life is 
\begin{equation}
  \tau_{1/2} = 1/(b E_0) 
 = 8.35\times 10^9\,{\rm yr} \left( \frac{\rm \mu G}{B} \right)^2
             \frac{\rm GeV}{E}  \,.                   
\label{ellosses}
\end{equation}
Thus electrons with energy 100\,GeV should be injected less than 5\,Myr
ago. This implies in turn that the sources of high-energy electrons have to
be local. Note also the difference between continuously injecting and 
bursting sources of electrons: In the first case, energy losses lead to a
break with $\Delta\alpha=1$ in a power-law energy spectrum, while
the energy spectrum of a bursting source has a sharp cutoff.

Finally, the last main ingredient in the transport equation are the
particle interactions which require as input cross sections and decay rates
as well as the distribution of the target material. For the latter one
can use either the Galactic mass distribution derived from kinematical 
studies~\cite{2011MNRAS.414.2446M} or the H\,I distribution from radio 
observations~\cite{2016A&A...594A.116H}. The scattering and decays of CR
nuclei are  determined by strong interactions and nuclear effects and, at
present, it is therefore not possible to calculate them from first principles.
Conventionally, one calls spallation reactions those interactions where  
CR nuclei fragment on gas of the ISM, loosing one or more nucleons 
and producing thereby lighter nuclei. In particular, one can assume in
spallation reactions zero energy transfer between the nuclei, if one neglects
the Fermi motion of
nucleons inside a nucleus. Therefore, the energy per nucleon $E/A$ is
conserved in such reactions. In this approximation, only the total production
cross sections $\sigma^{a\to b}$ as function of energy are required as input
from experimental measurements. Still, an accurate modelling with errors
smaller than, e.g.,  3\% of the lithium flux requires the knowledge of
twelve cross sections with much improved  accuracy compared to
current knowledge~\cite{Genolini:2018ekk}. A pilot
run~\cite{Aduszkiewicz:2287004} to collect new data on nuclear fragmentation
has recently been performed by the NA61/SHINE collaboration and a comprehensive
measurement campaign is proposed to take place after the long shutdown 2 of
the CERN accelerator facilities~\cite{Aduszkiewicz:2309890}.
Since measurements cannot cover the required energy range for all relevant
reactions, one has to rely on parametrisations for the spallation cross
sections. Parametrisations like the one of Ref.~\cite{Tripathi:1996} 
for the total inelastic cross section $\sigma_{\rm inel}^{(a)}$  and of
Ref.~\cite{2003ApJS..144..153W} for the partial spallation cross sections
$\sigma^{a\to b}$ assume that these cross sections are constant  above few GeV/n.
Since inelastic strong interaction cross sections grow only logarithmically
at high energies, the energy range experimentally covered is too small to
determine this
growth given the typical experimental precision. Using instead Monte Carlo
simulations like QGSJET-II-04~\cite{Ostapchenko:2010vb,Ostapchenko:2013pia}
for the determination of, e.g., the total inelastic cross section
$\sigma_{\rm inel}$ of $^{12}$C, one
finds an increase from 200\,mbarn to 255\,mbarn moving from
10\,GeV/n to 1\,TeV/n. Theoretically, one expects that the fragmentation
cross sections $\sigma^{a\to b}$ grow with energy proportionally to
the total inelastic cross section $\sigma_{\rm inel}$.
Thus constraining the slope of the diffusion coefficient
$D(\R)\propto \R^{\delta}$ using constant fragmentation cross sections may
overestimate $\delta$ by 0.05, cf.\ with Eq.~(\ref{fluxBc}).

At energies $E/A\gsim {\rm few}$\,GeV, it is not possible to separate between
spallation reactions and interactions with additional particle production:
Practically all inelastic scatterings are a combination, where the primary
nucleus fragments partly into pieces $j$ with energy $\simeq E/A_j$,
while a fraction
of its energy is converted into additional mesons and baryons. For a
sufficiently steeply falling CR primary spectrum, the products of these hard
``sub-reactions'' can be neglected, keeping track only of those secondaries
one is interested in, such as antiprotons, positrons, photons or neutrinos.

\subsubsection{Basic results}
\label{standard_prop_results}
\paragraph{Cosmic ray fluxes}

In order to solve the set of transport equations, several approximations are 
usually made. First, one notes that the measured ratios of radioactive CR
isotopes like Be$^{10}$ with a lifetime of $\simeq 1.6$\,Myr to the stable
Be$^{9}$, and of secondary-to-primary ratios
like B/C indicate a residence time of CRs of order $\tau_{\rm esc}\simeq
{\rm few}\times 10^7{\rm yr}\left(\R/5\, {\rm GV}\right)^{-\beta}$ with
$\beta\simeq 1/3$~\cite{1998A&A...337..859P,Aguilar:2015ooa}. 
If one then assumes that the main CR sources are SNe injecting
$\simeq 10^{50}$\,erg every $\sim 30$\,yr in the form of CRs,
then the flux from some $10^{4}$ sources accumulates at low rigidities,
forming a ``sea'' of Galactic CRs.
Since many sources contribute, the discrete nature of the CR sources
can be neglected. This allows one to consider the stationary limit
of Eq.~(\ref{transport}) and to use a smooth, time-independent source
distribution $Q(\vec x)$. As a second approximation, one also neglects the
spatial dependence of the diffusion term, replacing the Galaxy by a cylinder
with uniform propagation properties for CRs. Finally,  one replaces often
the tensor $D_{ij}$ by a scalar diffusion coefficient $D$, assuming that
the turbulent field dominates relative to the regular field. Note that the
last two
approximations clearly contradict our knowledge of the GFM which
indicates a strong variation of the magnetic field strength, both as
function of galactocentric radius and distance to the Galactic plane, as
well as an anisotropic diffusion of CRs. These approximations imply that
the fit results derived in such diffusion models for,
e.g.\ the normalisation $D_0$ of the diffusion coefficent, can be seen
only as effective parameters.

\begin{figure}
  \centering
  \includegraphics[width=0.55\columnwidth]{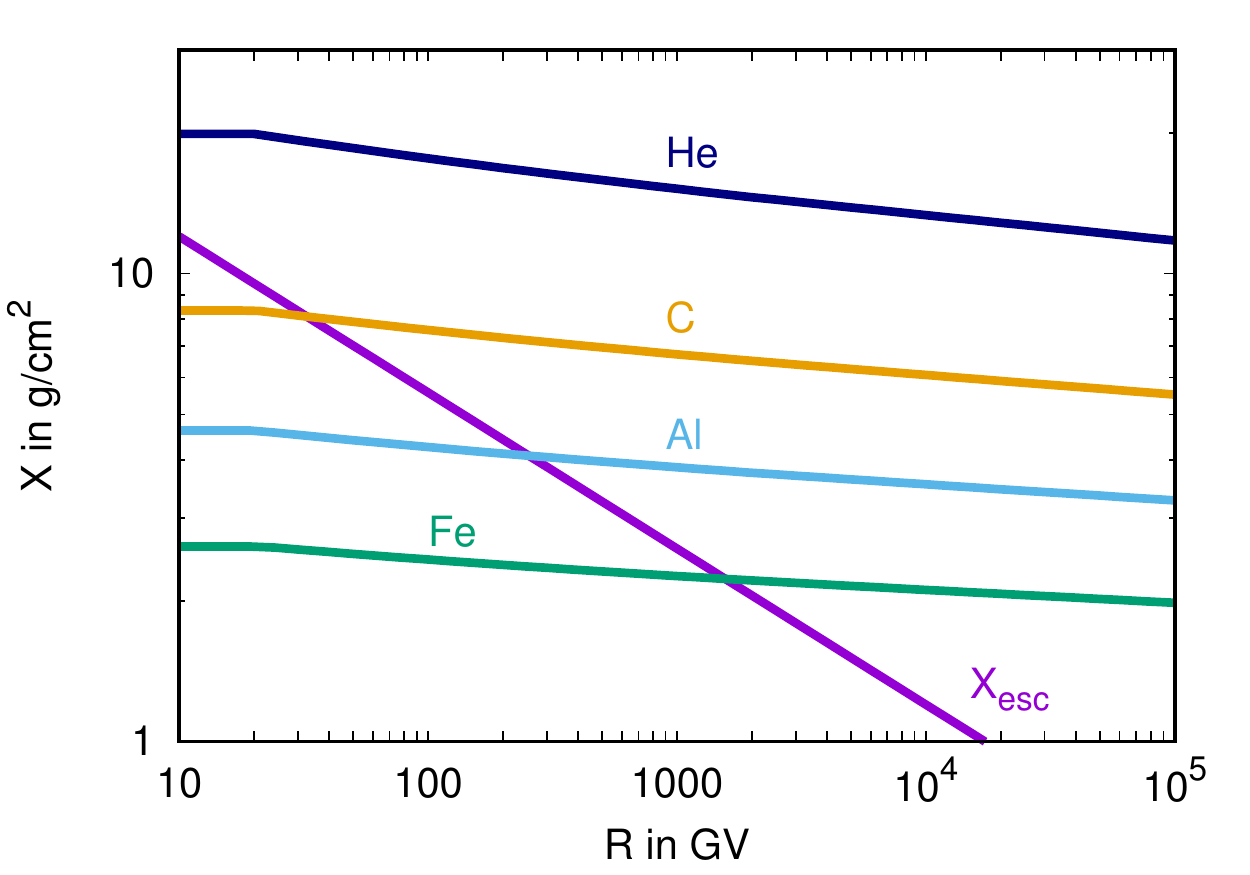}
  \caption{Interaction depths $X^{a}=m_p/\sigma_{\rm inel}^{(a)}$ calculated with
  QGSJET-II-04~\protect\cite{Ostapchenko:2010vb,Ostapchenko:2013pia} and the
  escape depth $X_{\rm esc}=X_0 (\R/\R_0)^{-1/3}$ as function of rigidity $\R$.
  \label{fig:Xint}}
\end{figure}

The basic behaviour of the solutions obtained solving numerically the
coupled transport equations~(\ref{transport}) using these appproximations
can be understood from a simple leaky-box model:
This model assumes that CRs inside a uniform confinement volume, e.g.\ a
CR halo with height $h$, have a constant escape probability per time which
is small, such that $\tau_{\rm esc}\gg h/c$. Neglecting all other effects
in Eq.~(\ref{transport}), these equations reduce to
\be
 \frac{\partial n^{(a)}}{\partial t} =
 \frac{n^{(a)}}{\tau_{\rm esc}} = D\Delta n^{(a)} \,,
\ee
and hence one can replace  the diffusion term by $n^{(a)}/\tau_{\rm esc}$. If we
consider now the steady-state solution, $\partial n^{(a)}/\partial t=0$, then
we obtain
\be
 \frac{n^{(a)}(E)}{\tau_{\rm esc}} = 
 Q^{(a)} - \left( c n_{\rm gas}\sigma_{\rm inel}^{(a)} 
   +\Gamma^{(a)}\right) n^{(a)}(E)
  +    c n_{\rm gas} \sum_b 
 \int_E^\infty\d\!E' \;\frac{\d\sigma_{ab}(E',E)}{\d\!E} \:
  n^{(b)}(E') \,.
\ee
For stable primary types like protons or $^{56}$Fe, the decay term
vanishes and production via fragmentation can be neglected. Introducing
$X_{\rm esc}=c\rho \tau_{\rm esc}$ and $X^{(a)}=m_p/\sigma^{(a)}_{\rm inel}$
as the amount of matter traversed by a relativistic particle
before escaping and interacting, respectively, it follows
\be \label{Eq:primary}
 n^{(a)} = \frac{Q^{(a)}\tau_{\rm esc}}{1+X_{\rm esc}/X^{(a)}}  \,.
\ee
The  interaction depths $X^{(a)}$ are compared
for some common nuclei in Fig.~\ref{fig:Xint} to the escape depth
$X_{\rm esc}=X_0 (\R/\R_0)^{-\delta}$, for which we assumed Kolmogorov
turbulence, $\delta=1/3$, and $X_0=12$\,g/cm$^2$ at $\R_0=10$\,GV.
For protons and helium, $X_{\rm p}>X_{\rm He}\gg X_{\rm esc}$
for all energies, and thus
\be
 n_{\rm p} = Q_{\rm p}\tau_{\rm esc}\propto Q_0 E^{-(\beta+\delta)} \,.
\ee
Hence the injection spectrum of protons and helium should be flatter than
the one observed: For  Kolmogorov turbulence, $\delta=1/3$, the observed
slope of the proton spectrum $\alpha=2.7$ requires the slope $2.4$ for
the injection spectrum.

For the other extreme case, iron, the interaction and  escape depths are
equal around $\R\simeq 2000$\,GV. Hence at lower rigidities, iron nuclei
are destroyed by interactions before they escape, $X_{\rm Fe}\ll X_{\rm esc}$,
and therefore the iron spectrum reflects the generation spectrum, 
$n_{\rm Fe}\propto Q_{\rm Fe}$. Starting from the energy where
$X_{\rm Fe}\approx X_{\rm esc}$, the iron spectrum should become steeper.
The observed iron spectrum is indeed flatter at low energies than, e.g., the
helium spectrum. However, the expected  steepening at $\R\simeq 2000$\,GV
is absent in the data.

The density of stable secondaries like boron follows with $\Gamma=Q=0$
and as 
\be
n^{(a)} = \frac{X_{\rm esc} \sum_b\sigma^{ba}/m_p \, n^{(k)}}
               {1+ X_{\rm esc}/X^{(a)}}.
\label{fluxBc}
\ee
Here, we have assumed that $E/A$ is used as variables which is approximately
conserved in fragmentation processes. 
Secondary-to-primary ratios like B/C are thus given by
\be \label{BCratio}
 \frac{n_{\rm B}}{n_{\rm C}} = \frac{X_{\rm esc} }{1+X_{\rm esc}/X^{(B)}}  
  \sum_{k>B}\frac{\sigma^{k\to B}}{m_p}\frac{n_k}{n_C}\propto \R^{-\delta} \,.
\ee
In the last step,  we have assumed  energy independent fragmentation cross
sections and $X_{\rm esc}/X^{(B)}\ll 1$. The latter condition is
satisfied at sufficiently high rigidities, $\R\gg 30$\,GV. As discussed
above, the first assumption leads to an overestimation of $\delta$ by 0.05.

Thus a measurement of a secondary-to-primary ratio like B/C at sufficiently
high energies allows one to determine the energy-dependence of the
diffusion coefficient, constraining thereby the power-spectrum of the
turbulent magnetic field modes. The most precise data on the B/C ratio
are those of the AMS-02 experiment~\cite{Aguilar:2016vqr}, which determine
in Eq.~(\ref{BCratio}) the slope as $\delta=0.333\pm 0.015$ using data above
65\,GV. Hence the data are consistent with a Kolmogorov power spectrum
of the turbulent magnetic field.

\begin{figure}
\center{
\includegraphics[width=0.75\columnwidth,angle=0]{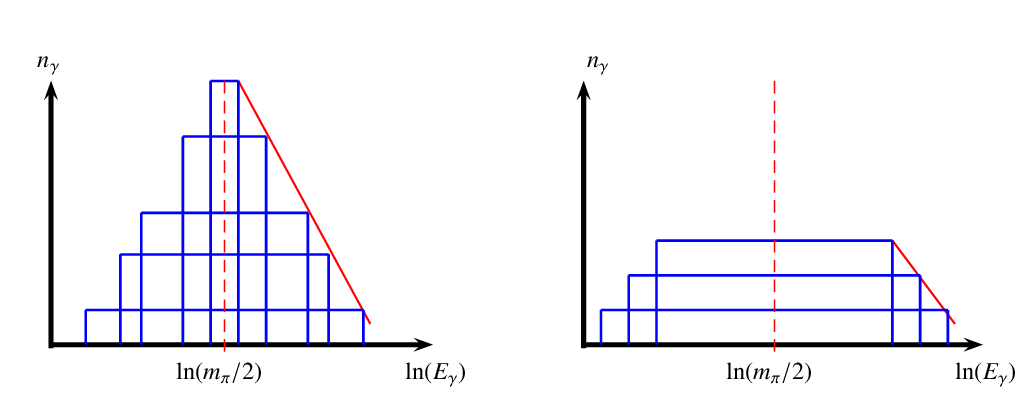}
}
\caption{The photon spectrum $n_\gamma={\rm d}N/{\rm d}E_\gamma$ produced in
  $\pi^0$ decays as function of $\ln(E_\gamma)$ for $Ap$ (left panel) and for
  $A\gamma$ (right panel) interaction: the total spectrum is the enveloppe
  (red line) of the boxes produced by pions with fixed energy.
  \label{box}}
\end{figure}

Let us next discuss a few basic properties of the secondary fluxes of
particles which are produced in hadronic interactions of CRs on gas, either
directly or via the decay of mesons. Photons are mainly produced by the decay
of neutral pions,  $\pi^0\to 2\gamma$, with a flat energy spectrum
${\rm d}n/{\rm d}E_\gamma=\;$const.\ for a given pion energy.
The maximal and minimal energy of these photons is
$E_{\min}^{\max}=\frac{m_{\pi}}{2} \sqrt{\frac{1\pm\beta}{1\mp\beta}}$.
The energy spectrum of the decay photons from pions with fixed energy ploted as
function of $\ln(E_\gamma)$ corresponds therefore to a symmetric box around
half of the pion mass, cf.\ the left panel of Fig.~\ref{box}. A signature of
photons from hadronic interactions is therefore the symmetry of the photon
spectra { as function of $\ln(E_\gamma)$} with respect to $m_{\pi}/2$.
The low threshold and the approximate Feynman scaling for forward
production spectra in hadronic interactions
implies then ${\rm d}N_\gamma/{\rm d}E\propto {\rm d}N_{\rm CR}/{\rm d}E$.
These arguments were used for the SNRs IC\;443 and W44 to argue for the
acceleration of CR protons in these sources~\cite{Ackermann:2013wqa}.
Note however that these observations do not cover the peak and the low-energy
side of the ``pion bump'' and depend therefore on the modelling of these
sources.

The intensity of secondaries is suppressed relative to the one of primaries
depending on the energy fraction $z=E_{s}/E_p$ transferred to
the secondaries and the slope $\alpha$ of the primary intensity. This
suppression can be accounted for in case of a power-law primary intensity,
$I_p(E)\propto E_p^{-\alpha}$, by defining  $Z$ factors as follows,
\begin{equation}
\label{Z_spec}
 Z_{s}(E_{s},\alpha) = \int_0^1 {\rm d} z \, z^{\alpha-1}\,
 \frac{{\rm d}\sigma_{s}(E_{s}/z,z)}{{\rm d} z}\,. 
\end{equation}
Here, ${\rm d}\sigma_{s}(E,z_{s})/{\rm d} z_{s}$ denotes the  inclusive spectrum
of secondaries. The intensity of secondaries of type $s$ is then
simply given by $I_s(E)=c \tau n Z_s(E,\alpha)I_p(E)$ with $\tau$ as the
time spent in the source region with target density $n$,
if the primary intensity can be described by a power-law.

As an important example, we give in Table~\ref{Ztab} the $Z$ factors for the
production of photons in $pp$ interactions for various values $\alpha$ and
photon energies $E_\gamma$ calculated with
QGSJet-II-04~\cite{Ostapchenko:2010vb,Ostapchenko:2013pia}. Moving from a flat
spectrum with $\alpha=2$ to a steep one with  $\alpha=3$, the $Z$ factor
decreases by a factor $\simeq 10$ and, thus, the secondary photon flux is
suppressed by the same factor, $I_\gamma(\alpha=3)\simeq I_\gamma(\alpha=2)/10$.
Similarly, the effect on the secondary yield  of heavier primary nuclei as
well as of the helium contribution in the gas depends strongly on the slope
of the CR fluxes. A simple and convenient way to account for the contribution
of heavier nuclei to the diffuse gamma-ray emission is provided by the nuclear
enhancement factor $\eps$. For the parametrisation of
Ref.~\cite{Honda:2004yz} for the CR flux, the numerical
values of the enhancement factor $\eps$ for the photon yield were determined
as 1.87, 1.98, 2.09 for
$E_\gamma=10, 100, 1000$\,GeV in Ref.~\cite{Kachelriess:2014mga}.
In general, however, production cross sections calculated for the  specific
primary and target nuclei should be used, which are provided e.g.\
by the parametrisation AAfrag~\cite{Kachelriess:2019}.

\begin{table}[b]
\center{
  \begin{tabular}{|c|cccccc|}
\hline
  $E_\gamma$ & $\alpha=2$ & $\alpha=2.2$ & $\alpha=2.4$ & $\alpha=2.6$ & $\alpha=2.8$ & $\alpha=3$
  \\ \hline
 10\,GeV &  5.45 & 3.06 & 1.84 & 1.17 & 0.771 & 0.529 \\
100\,GeV &  5.93 & 3.20 & 1.86 & 1.14 & 0.736 & 0.492 \\
  1\,TeV &  6.85 & 3.61 & 2.05 & 1.24 & 0.786 & 0.519 \\
\hline
\end{tabular}}
\caption{$Z_\gamma(E_{s},\alpha)$ factor in mbarn of the reaction $pp\to X\gamma$ for different values of $\alpha$ and $E_\gamma$.
\label{Ztab}}
\end{table}

Note also the special case of antiproton production where the threshold energy
is $E_{\rm th}=7m_p$. There are two main consequences of this high threshold:
First, the antiproton production cross section has a rather strong dependence
on the energy of the produced antiprotons up to $E_{\bar p}\simeq 100$\,GeV.
Second, the suppression of the antiproton production cross section 
below $E_{\bar p}\simeq 100$\,GeV is difficult to model in Monte
Carlo simulations of strong interactions, because it depends on poorly
constrained details in their hadronisation procedures, for a discussion
see Ref.~\cite{Kachelriess:2015wpa}. On the other side, parametrisations of
experimental data like those of Refs.~\cite{diMauro:2014zea,Kappl:2015bqa,Winkler:2017xor} have to be extrapolated outside the measured
kinematical range and rely often on physically poorly motivated assumptions.
Therefore, the theoretical uncertainty in the antiproton production cross
section below $E_{\bar p}\simeq 100$\,GeV is with 20\% relatively large.

\paragraph{Sources of CRs}

The possible sources of Galactic CRs are restricted by the energy
budget required to keep the energy contained in the escaping CRs
stationary. The local energy density of CRs can be determined from
Voyager data as $\rho_{\rm CR}\simeq 0.7$\,eV/cm$^3$~\cite{Drury:2016ubm}. 
If CRs are confined for the time $\tau$ inside the volume $V$ containing
the gas mass $M$, they cross the grammage $X=c\tau M/V$. Combining the
measured grammage,  $X\simeq 10$\,g/cm$^2$, and the CR luminosity
$L_{\rm CR}=\rho_{\rm CR}V/\tau$ leads to
\begin{equation}
 L_{\rm CR}=\rho_{\rm CR} \frac{cM}{X} \simeq 5\times 10^{40} {\rm erg/s}
\end{equation}
using $M=5\times 10^9 M_\odot$.

The classic source class suggested first by Baade and Zwicky
are SNe~\cite{1934PNAS...20..259B}:
In a successful core-collapse SN around $10\,M_\odot$ are ejected with
velocities $v\sim 5\times 10^8$\,cm/s. Assuming
$1/(30\,$yr) as SN rate in the Milky Way, the average output in
kinetic energy of Galactic SNe is $L_{\rm SN, kin} \sim 3\times 10^{42}$\,erg/s.
Hence, if the remnants of SNe can accelerate particles with an efficiency
$O(0.01)$, they could explain the bulk of Galactic CRs.
Note that an efficient magnetic field amplification which is required
such that SNe can accelerate CRs up to the knee and beyond implies that CRs
carry 20--30\% of the initial kinetic energy of the
SNe~\cite{2004MNRAS.353..550B,2001MNRAS.321..433B}.
Thus it is sufficient that a subset of all core-collapse SNe accelerates
CRs up to the end of the Galactic CR spectrum.

These energy considerations make SN explosions a very probable energy
source for Galactic CRs. However, core collapse SN explosions are not
randomly distributed in the Galaxy, since the majority of core-collapse
SN progenitors belong to OB associations, which are formed from the collapse
of a giant molecular cloud within a short time scale.
Therefore several tens of SNe occur within a few million years inside
a superbubble created by the strong winds of the massive stars
in the OB association.
The larger dimensions of superbubbles, the presence of turbulence stirred
by the stellar winds and of multiple shocks promote superbubbles to attractive
acceleration sites reaching PeV energies~\cite{Parizot:2004em}.
Evidence for CR acceleration in superbubbles comes from the 
the so-called superbubble model
for Li, Be and B production~\cite{Parizot:2000ts}.
This model proved capable of accounting for all the current observational
constraints pertaining to the nucleosynthesis and  evolution of light element:
In particular, it explains the observed very efficient production of Li, Be
and B in the early Galaxy, when C and O nuclei were still very rare.
Another consequence would be a substantial enrichment of CRs by
freshly synthesized nuclei, from SN ejecta and stellar winds.
This might offer a natural way to explain the large $^{22}$Ne/$^{20}$Ne
ratio observed in CRs~\cite{1996ApJ...466..457D} or the hardening
the CR spectra of heavier elements~\cite{2011ApJ...729L..13O,Ohira:2015ega}.

{ In a related scenario, most massive stars in an OB association explode
before their winds merged and the superbubble forms only later. If
the kinetic energy of such a SN is sufficiently large,
$E_{\rm kin}\gsim 10^{51}$\,erg, and the magnetic field strong, 
diffusive drift acceleration may be  operating additionally to diffusive shock
acceleration. As a result, massive stars exploding into their winds may  able
to accelerate CRs up to the ankle, for a detailed review of this option see
Ref.~\cite{Biermann:2018clk}. 
Observations of radio SNe in the starburst galaxy M82 are consistent with
the strong magnetic fields required in this scenario~\cite{Soderberg:2009ps}.
Moreover, they find large shock velocities, $v_{\rm sh}\simeq 0.1c$, suggesting
that these sources can accelerate CR protons beyond the knee.
}

Another interesting test of the SN hypothesis has been recently performed in
Ref.~\cite{Neronov:2017syf}: Using gamma-ray observation of the
Constellation~III region in the Large Magellanic Cloud, it was argued
that SNe are favoured compared to preceding stellar winds as a site of
CR acceleration. Moreover, it was shown that the energy injected in CRs
equals $(1.1_{-0.2}^{+0.5})\times 10^{50}$\,erg/supernova,  with  a power-law
spectrum and slope $2.09_{-0.07}^{+0.06}$. Thus both the energy and the
slope match well with the idea of shock acceleration in SNR.

A nova explosion produces similarly to a SN an expanding shell, however,
with a reduced energy of $\sim 10^{46}-10^{47}$\,erg. Since the frequency
of nova explosion is 100/yr in the Milky Way, the total energy input per
time by novae is similar or larger than the one of SNe. Thus it is possible
to associate nova with a steep CR component having a maximal rigidity of
200\,GV~\cite{Zatsepin:2006ci}.
In contrast, Gamma-Ray Bursts (GRB) are potential CR sources with a high
energy output, but a small rate $\lsim 10^{-4}/$yr. In
Ref.~\cite{Wick:2003ex}, it was argued that the Galactic CR flux above
the knee could be caused by a single Galactic GRB $\sim 500$\,pc away that
took place around 200,000\,yr ago. In
Refs.~\cite{1993ApJ...418..386L,Eichler:2016mut}, it was suggested that
even the observed UHECRs could be explained by a Galactic GRB.
Common to both works is that the propagation of CRs was treated in
a simplified diffusive approach which  is not justified any more at these
energies. Therefore, the conclusions of these works should be taken with
caution.

The Galactic center (GC) with its supermassive black hole is another potential
site of CR acceleration. While the GC at present is quiet, an active episode
in the past has been connected to the creation of the Fermi Bubbles.
For instance, Ref.~\cite{Miller:2016chr} estimated the average energy output
of the GC as $(1-7)\times 10^{42}$\,erg/s, what exceeds the energy output
of SNe.
In Ref.~\cite{Jaupart:2018eev}, it was shown that particles accelerated
during such active episodes around the GC can account for a
significant fraction of the locally observed CRs with energies up to knee,
if the diffusive halo is  large and the slope of the diffusion coefficient
is high, $\delta=0.5$. Since electrons and positrons lose energy fast as they
propagate, the GC can only contribute secondary $e^\pm$. Therefore
additional local sources of electrons and positrons have to contribute
the observed high-energy part of the lepton spectra in this scenario.

\paragraph{Challenges for the simple diffusion model}

The diffusion approach based on the approximations described above has
been sufficient to describe the bulk of experimental data obtained until
$\simeq 2005$. With the increased precision of newer experiments like the
CREAM balloon detector, the PAMELA and Fermi satellites or
AMS-02 on the International Space Station
several discrepancies have emerged. These observational anomalies
and ideas for their solutions will be described in the next two subsections.
Before that we will discuss a more conceptional challenge for the
approximations employed in the  standard diffusion
approach~\cite{Giacinti:2017dgt}.

\begin{figure}
\vskip-0.3cm
\center{
\includegraphics[width=0.45\columnwidth,angle=270]{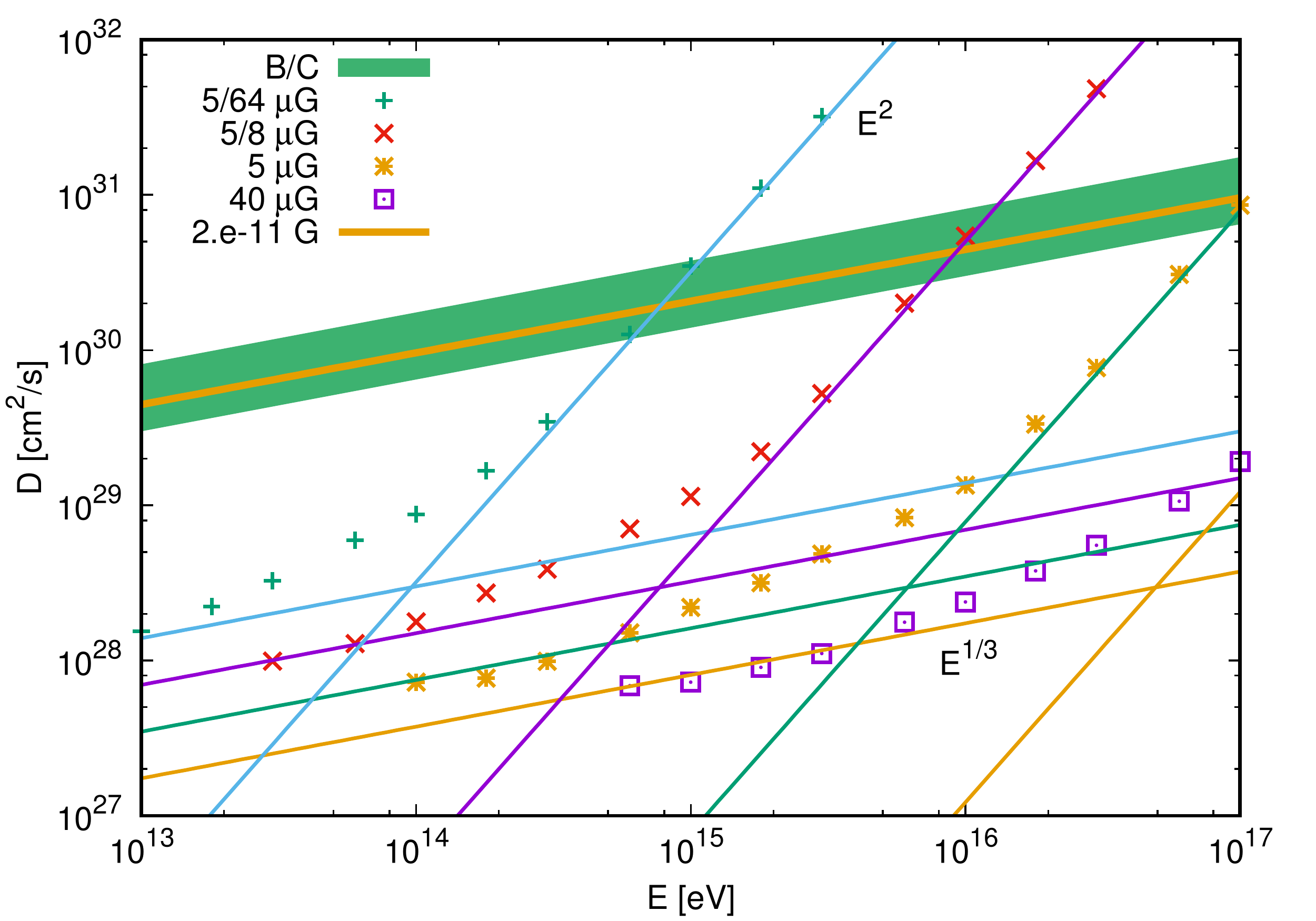}}
\vskip0.6cm
\caption{The CR diffusion coefficient in pure isotropic Kolmogorov turbulence 
  with $L_{\max}=25$\,pc and for four values of
  The asymptotic behaviours at 
low and high energies are shown with the solid lines. 
\label{iso1}}
\end{figure}

 In the diffusion picture, one can model the propagation of CRs as a 
random walk with an energy dependent effective step size. For a
pure isotropic random field, one expects therefore  as functional dependence
of the diffusion coefficient
\be  \label{diff}
  D = \frac{cL_0}{3} 
 \left[ (R_{\rm L}/L_0)^{2-\gamma} + (R_{\rm L}/L_0)^2 \right] ,
\ee
where the condition $R_{\rm L}(E_{\rm tr})=L_0$ determines the transition from 
small-angle scattering with $D(E)\propto E^2$ to
large-angle scattering with $D(E)\propto E^{2-\gamma}$. At even higher
energies, CRs enter the ballistic regime and the concept of a
diffusion coefficient becomes ill-defined.

The numerical value of $L_0$ should scale with the correlation length 
as $L_0\propto L_{\rm c}$, but the proportionality factor has to be
determined numerically. In Refs.~\cite{Parizot:2004wh,Giacinti:2017dgt},
it was found that $L_0 \simeq L_{\rm c}/(2\pi)$ provides a good 
fit to their numerical results. The presence of the factor $1/(2\pi)$
becomes evident recalling that we compare in Eq.~(\ref{diff}) the linear
length $L_0$ with the radius $R_{\rm L}$. Numerically, the transition energy
is given by
\begin{equation}
\label{eq:Ecr}
E_{\rm tr} =  2\times 10^{14}{\rm eV} \: (L_{\rm c}/{\rm pc}) \: (B/{\rm \mu G})\,.
\end{equation}
Note that, for $L_{\rm c}\sim {\rm few}$\,pc and $B\sim {\rm few}\,\mu$G,
the transition energy $E_{\rm tr}$ is in the knee region. Thus the change
in CR propagation at $E_{\rm tr}$ may be a possible reason for the spectral
break at the knee.

\begin{figure}
\centering
  \includegraphics[width=0.5\columnwidth]{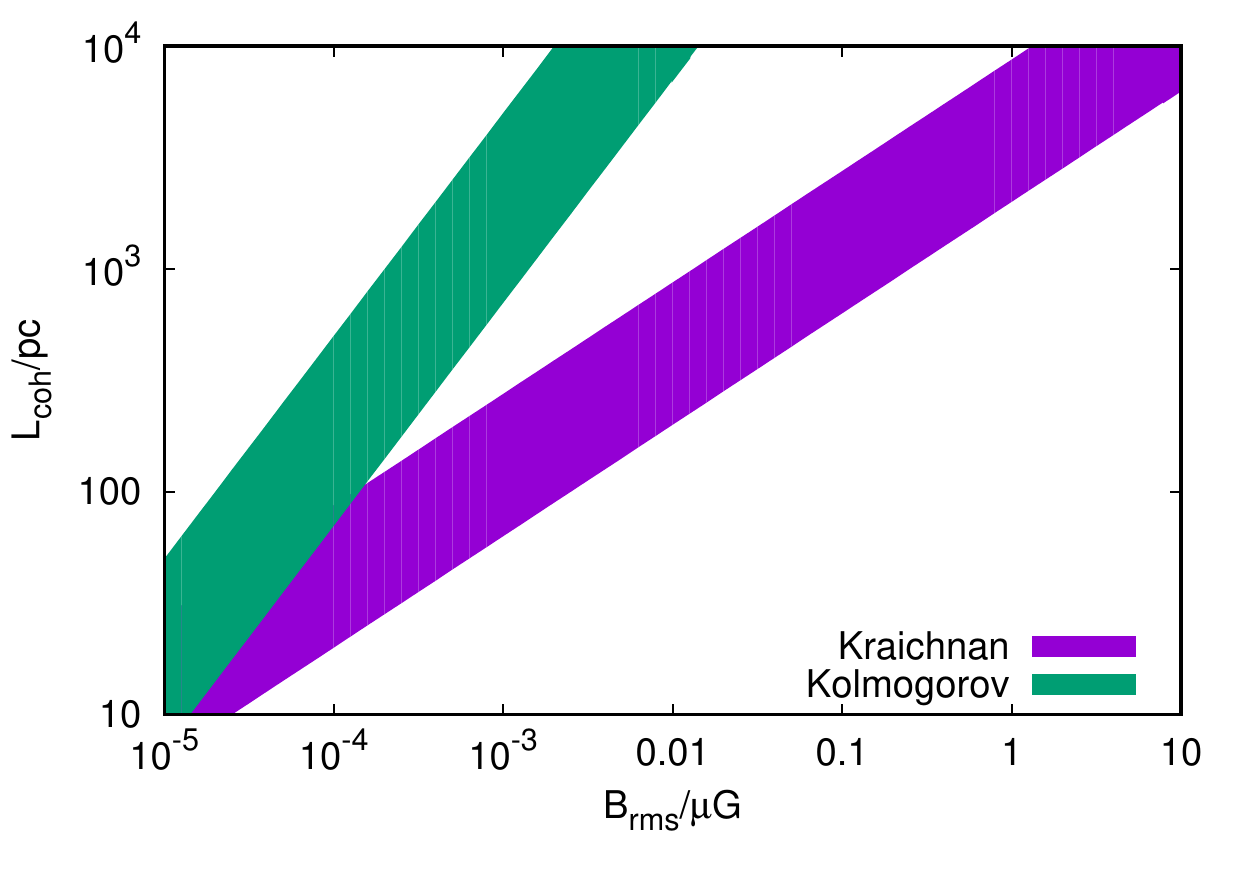}
  \includegraphics[width=0.49\columnwidth]{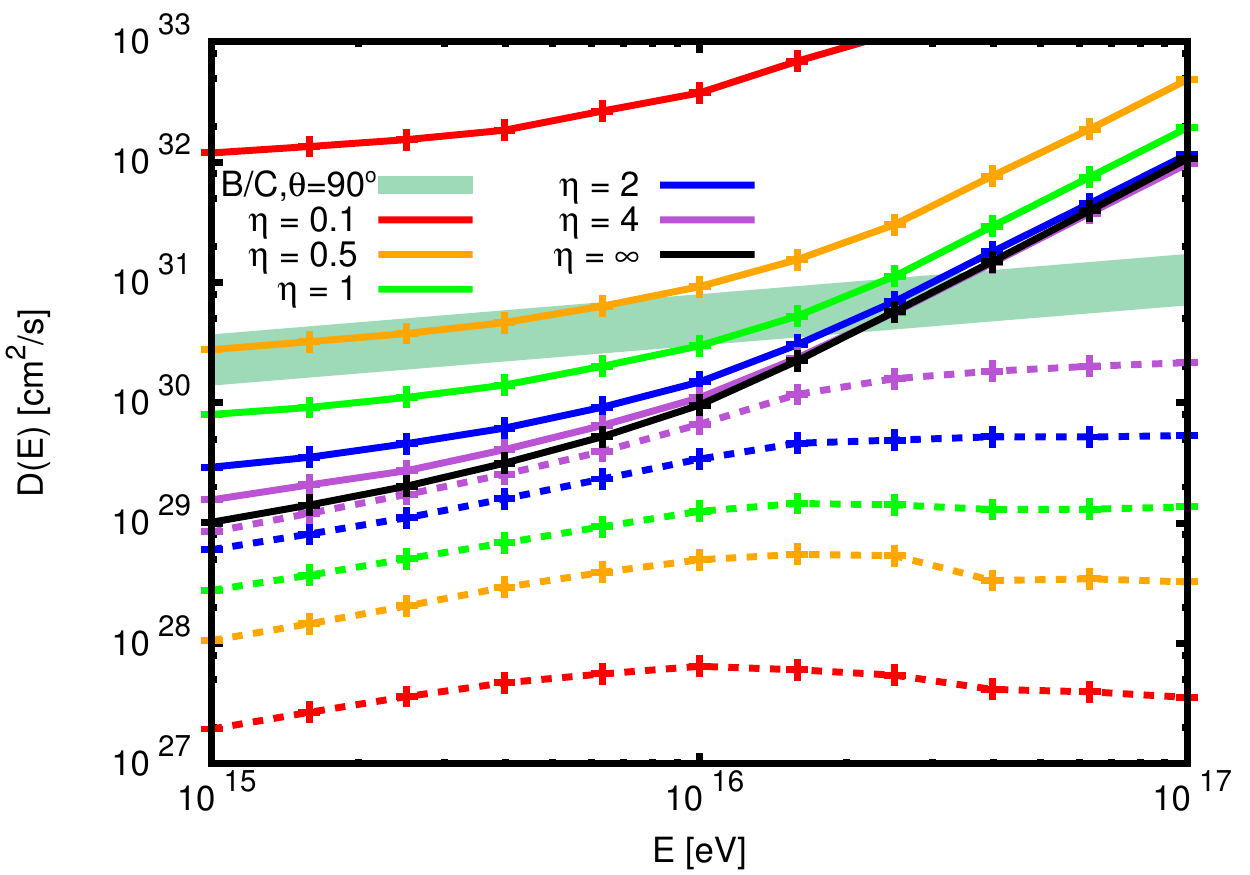}
\caption{{\em Left panel:\/}
  Allowed range of $B_{\rm rms}$ and $L_{\rm coh}$ compatible
with $D_0=(3-8)\times 10^{28}$cm$^2$/s at $E_0=10$\,GeV for
Kolmogorov and  Iroshnikov-Kraichnan turbulence.
{\em Right panel:\/}
Parallel and perpendicular diffusion coefficient for isotropic
turbulence with a regular field.
\label{iso2}}
\end{figure}

In Fig.~\ref{iso1}, we  show the diffusion coefficient calculated using
Eq.~(\ref{D_iso}) for a pure random field following Kolmogorov turbulence
with $L_{\max}=25$\,pc for various field strengths. The transition between
the  asymptotic low-energy ($D\propto E^{1/3}$, large-angle scattering) and 
high-energy ($D\propto E^{2}$, small-angle scattering) 
behaviour is clearly visible. However, for all used field strengths  
the diffusion coefficients are much smaller than those extracted using,
e.g., Galprop~\cite{Johannesson:2016rlh} or DRAGON~\cite{Evoli:2008dv}
from fits to secondary-to-primary ratios like B/C: Typical values found
are  in the range $(3-8)\times 10^{28}$cm$^2$/s
at 10\,GeV; their extrapolation to high energies is shown as green band
in the figure.
Requiring that the numerically determined diffusion coefficient
for pure isotropic turbulence lies in this band determined from the
B/C ratio, the possible range for the field strength and the correlation
length of the turbulent field shown in the left panel of Fig.~\ref{iso2}
follows: The weak field dependence of $D(E)\propto B_{\rm rms}^{-1/3}$ requires
a reduction of $B_{\rm rms}$ by a
factor $1/100^3=10^{-6}$ for Kolmogorov turbulence keeping $L_{\rm c}$ constant.
Keeping instead  $B_{\rm rms}\sim {\rm few}\times\mu$G, the correlation length
should be comparable to the size of the Galactic halo. Using instead
Iroshnikov-Kraichnan turbulence, even larger correlation lengths would be
required.
Therefore CR propagation has to be necessarily anisotropic, because otherwise
CRs overproduce secondary nuclei like boron for any reasonable values of the
strength and the correlation length of the turbulent field.
Such an anisotropy may appear if
the turbulent field at the considered scale does not dominate over the 
ordered component, or if the turbulent field itself is anisotropic.

Adding a uniform magnetic field along the $z$ direction to the isotropic
turbulent field, the diffusion tensor  becomes anisotropic with 
$D_{ij}={\rm diag}\{ D_\perp,D_\perp,D_\| \}$ and $D_\|>D_\perp$.
In the right panel of Fig.~\ref{iso2}, we show $D_\|$ (solid lines) and
$D_\perp$ (dashed lines) for different values of the  turbulence level
$\eta \equiv B_{\rm rms}/B_0$, where $B_0$ denotes the strength of the regular
field. The total magnetic field strength is chosen as 
$B_{\rm tot} = \sqrt{B_{\rm rms}^{2} + B_0^{2}} = 1\,\mu$G, and the outer scale of
the turbulence  is set to $L_{\max} = 100$\,pc. Decreasing 
$\eta$, the difference between $D_\|$ and $D_\perp$ increases, while keeping the 
order $D_\|>D_\infty>D_\perp$ intact, where $D_\infty(E)$ denotes the diffusion 
coefficient for pure isotropic turbulence.

\begin{figure}
\centering
  \includegraphics[width=0.49\columnwidth]{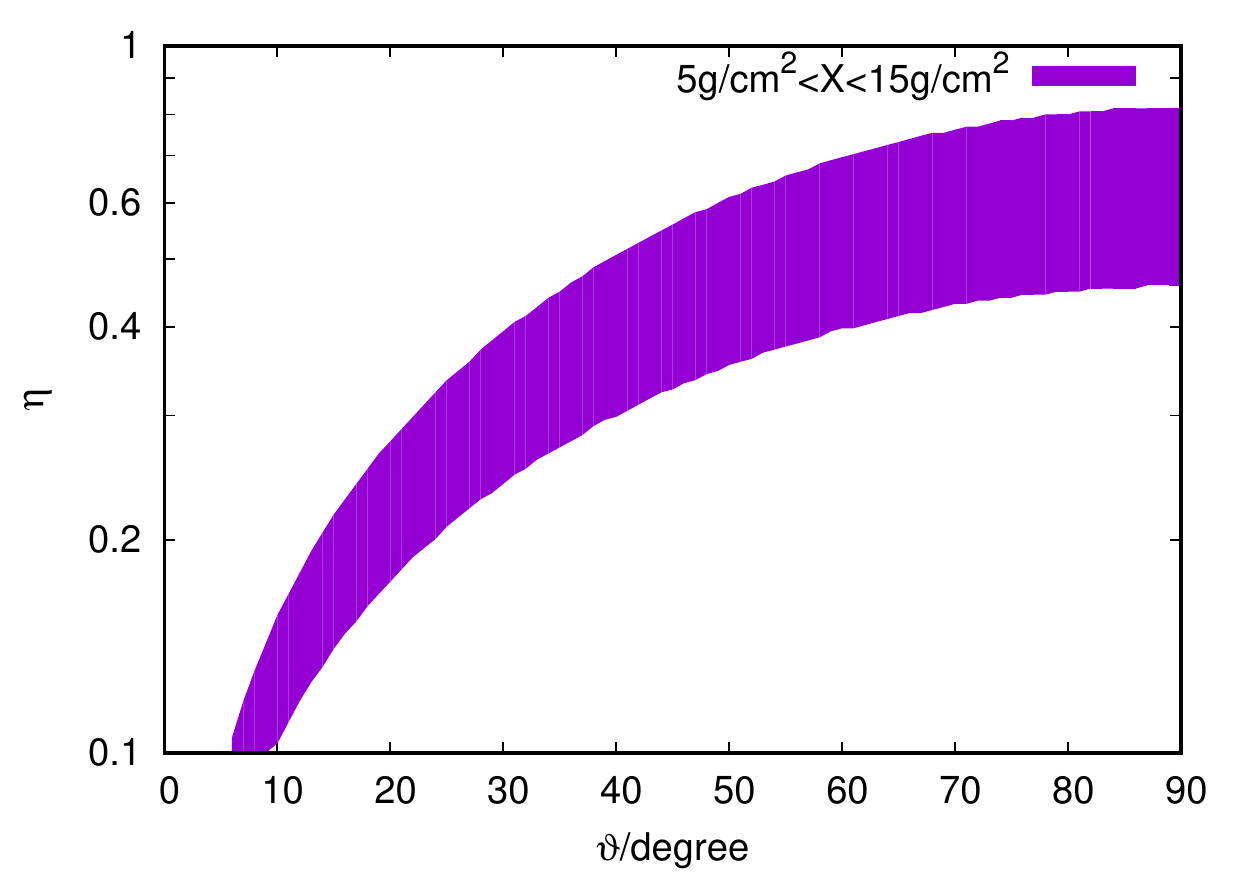}
  \includegraphics[width=0.49\columnwidth]{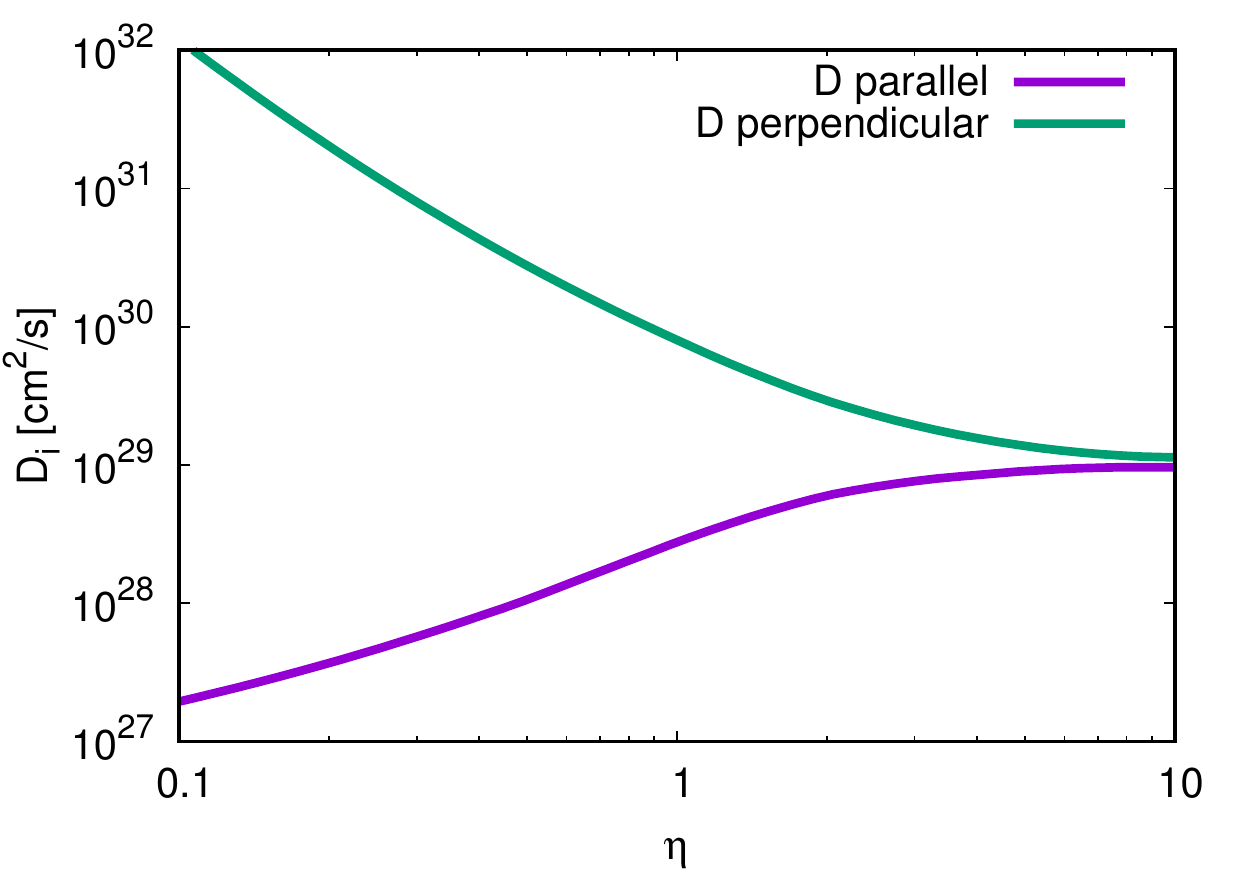}
\caption{{\em Left panel:\/}
Grammage $X$ crossed by CRs in a ``disc and halo''
  model, for the diffusion coefficients shown in the right panel of
  Fig.~\ref{iso2},
  as a function of the tilt angle $\theta$ between the regular magnetic field
  and the Galactic plane, and of the turbulence level $\eta$.
  {\em Right panel:\/} Fit of the diffusion coefficients $D_\|$ and $D_\perp$
  from the right panel of Fig.~\ref{iso2} at $E=10^{15}$\,eV as 
  function of $\eta$.
\label{fig:Xeta}}
\end{figure}

One can estimate the level of anisotropy required to obtain consistency
with the diffusion coefficient fitted to B/C considering the following
toy model: Let us assume a thin matter disc with density
$\rho/m_{\rm p}\simeq 1$/cm$^3$ and height $h=150$\,pc around the Galactic
plane, while CRs propagate inside a larger halo of height $H=5$\,kpc. The
regular magnetic field inside this disc and halo has a tilt angle $\theta$
with the Galactic plane, so that the component of the diffusion tensor
relevant for CR escape is given by
\be
D_z = D_\perp\cos^2\theta + D_\|\sin^2\theta \,.
\ee
Applying a simple leaky-box approach, the grammage follows as
$X = c \rho hH /D_z$.
Using now as allowed region for the grammage e.g.\ $5\leq X\leq 15$\,g/cm$^2$,
the permitted region in the $\theta$--$\eta$ plane shown in the left panel of
Fig.~\ref{fig:Xeta} follows. For not too large values of the tilt angle,
$\theta\lsim 30^\circ$, the regular field should strongly dominate,
$\eta\lsim 0.35$.

Note that the authors of Ref.~\cite{2014ApJ...785..129K} also argued
that CR diffusion has to be strongly anisotropic. They used the argument
that the CR flux from the young, nearby SNR Vela has to be suppressed
compared to the expectation for isotropic diffusion. Such a suppresssion
could be caused in the case of anisotropic diffusion by a large perpendicular
distance from the Sun to the
magnetic line through Vela. In models of the global GMF like the one of
Jansson-Farrar~\cite{Jansson:2012pc}, the Sun and Vela are however connected
by a magnetic field line. The reason for the suppression of the CR flux
from Vela may be instead the distortion of the global GMF in the Local
Bubble, as shown in Ref.~\cite{Bouyahiaoui:2018lew}.

As a result of the anisotropic CRs propagation, the diffusion 
coefficient perpendicular to the ordered field can be between two and three 
orders of magnitude smaller than the parallel one, $D_\bot\ll D_{||}$, as shown
in the right panel of Fig.~\ref{fig:Xeta}.
Then the  $z$ component of the regular magnetic field can drive CRs efficiently
out of the Galactic disk. For instance, the ``X-field'' in the  Jansson-Farrar
model~\cite{Jansson:2012rt} for the GMF leads to the correct CR escape time,
if one chooses as turbulence level $\eta\simeq 0.25$~\cite{Giacinti:2014xya}.
For this choice, the diffusion coefficients satisfy $D_\|\simeq 5 D_{\rm iso}$
and $D_\perp\simeq D_{\rm iso}/500$, where $D_{\rm iso}$ denotes 
the isotropic diffusion coefficient
$D_{\rm iso}$ satisfying the B/C constraints. In the regime, where the
CRs emitted by a single source fill a Gaussian with volume
$V(t)=\pi^{3/2} D_\perp D_\|^{1/2}t^{3/2}$, the CR density is increased
by a factor $500/\sqrt{5}\simeq 200$ in the case of anisotropic diffusion.
The smaller volume occupied by CRs from each single source leads to a smaller 
number of sources contributing substantially to the local flux, with only 
$\sim 10^2$ sources at $\R\sim 10$\,GV and about $\sim 10$~most recent SNe in 
the TeV range. This reduction of the effective number of sources 
may invalidate the assumption of continuous CR injection and a stationary
CR flux.

\paragraph{Impact of the Local Bubble}

Another challenge for the simple diffusion picture with a constant
diffusion coefficient throughout the Milky Way is the observation that
the Sun resides inside the Local Bubble. This implies, e.g., the question
how biased local measurements are.  Surprisingly, the impact of the Local
Bubble on the propagation of CRs has been largely neglected so far. An
exception is, e.g., Ref.~\cite{Donato:2001eq}, where the influence of a local
under-density on the propagation of radioactive isotopes was examined 
using a two-zone diffusion model.
However, this work assumed that no sources reside inside the Local
Bubble---in contrast to the picture that the bubble was created by recent
SN explosions. In Ref.~\cite{Moskalenko:2002yx}, it was suggested that the
locally measured spectra of primary nuclei contain at low energies a
component which was accelerated in the Local Bubble. The presence of this
additional component allowed the authors to explain both the antiproton
flux in the GeV range and B/C ratio  without using artificial breaks
in the diffusion coefficient and the primary injection spectrum.
In Ref.~\cite{2005ICRC....3..157S},
it was speculated that the knee may be caused by the fast
escape of CRs generated inside the bubble above 4\,PeV. The more recent
studies~\cite{An17,Andersen:2017yyg,Bouyahiaoui:2018lew}
suggest that the effects of the Local Bubble can be profound, changing
in particular strongly the contribution from recent nearby CR sources
as Vela to the locally observed CR flux.

\subsection{Observations and anomalies}
  \label{galactic_observations}

\subsubsection{Anisotropy of cosmic rays up to the knee}
  \label{sec_anisotropy}

\begin{figure}
  \centering
  \includegraphics[width=0.54\columnwidth,angle=0]{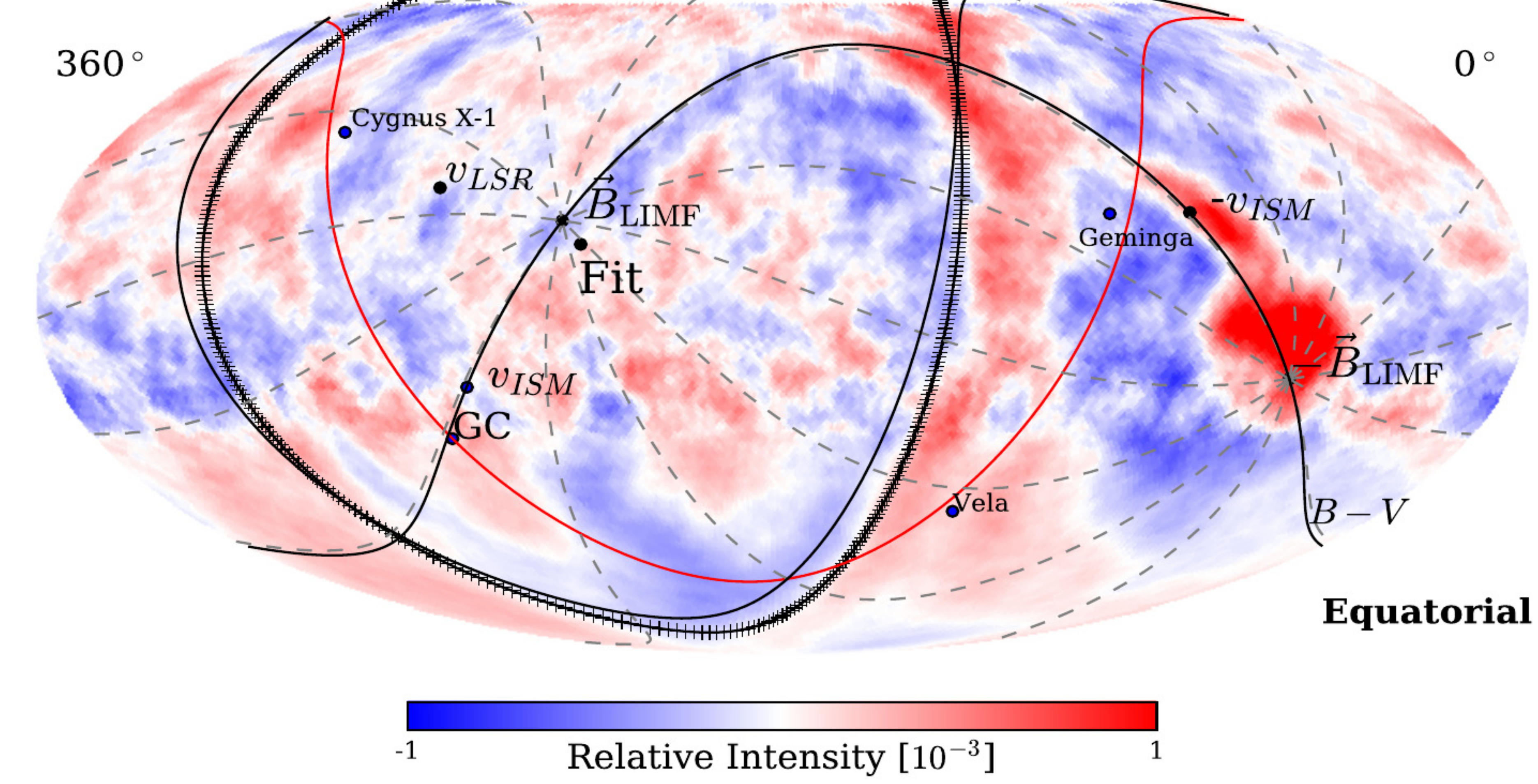}
  \hskip0.5cm
  \includegraphics[width=0.41\columnwidth,angle=0]{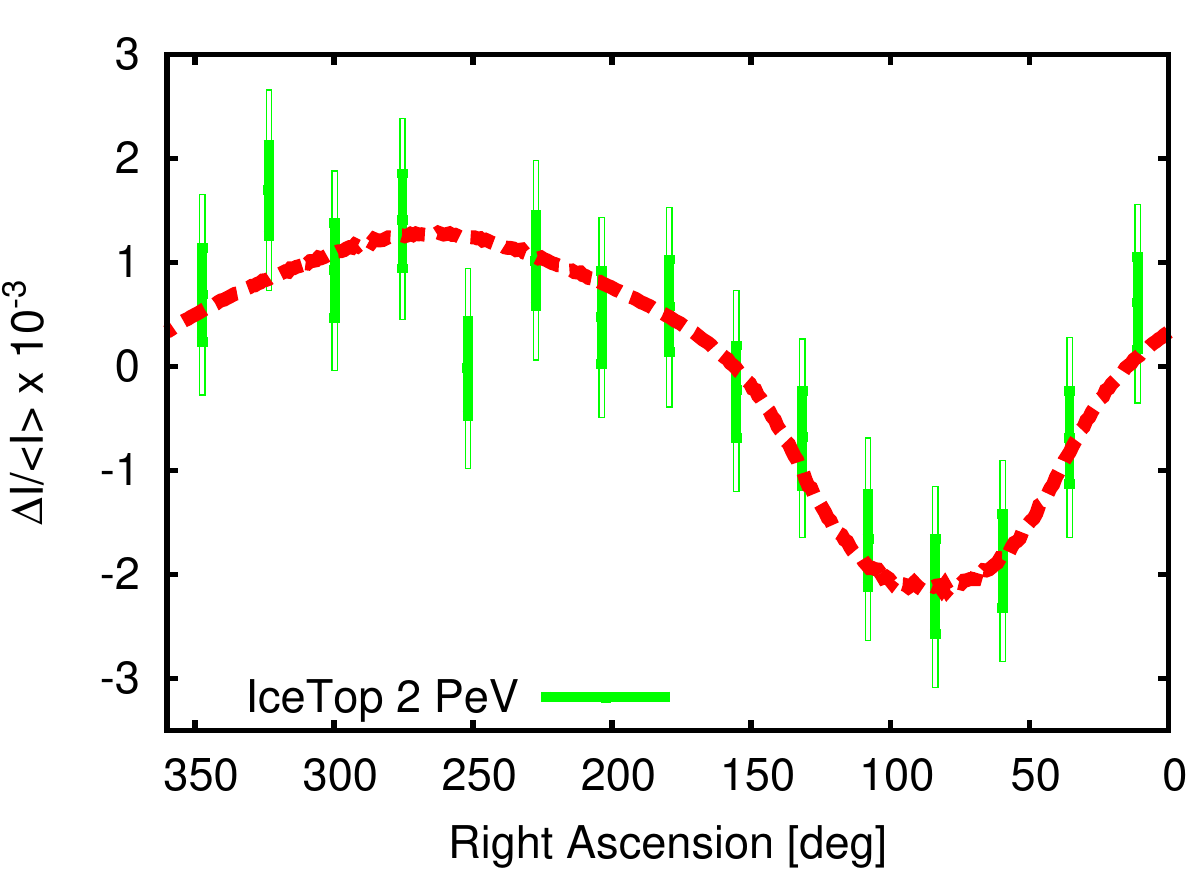}
  \caption{{\em Left panel:\/} All-sky map of the CR intensity combining HAWC
    and IceCube data at 10\,TeV median energy after subtracting multipoles with
    $\ell\leq 3$. The magnetic equator as the plane containing the magnetic
    field in the local ISM~\protect\cite{2016ApJ...818L..18Z} is shown
    as a black curve, the Galactic plane as a red curve and the positions of
    two nearby SNR, Geminga and Vela, are indicated too; adapted from
    Ref.~\protect\cite{Aartsen:2018ppz}.
    {\em Right panel:\/}
    The relative intensity projected on the equatorial plane as function
    of right ascension at 2\,PeV, adapted from Ref.~\protect\cite{Giacinti:2016tld}.
\label{anisotropy}}
\end{figure}

The observed intensity of CRs is characterised by a large degree of isotropy
up to the highest energies. This indicates that turbulent magnetic fields
inside the Milky Way are able to isotropize the Galactic part of the CR flux
at least up to knee.
A CR dipole anisotropy at the level of $10^{-3}$ was measured  above 100\,GeV
starting from the 1970s. However, only recently an almost all-sky coverage
was reached  at multi-TeV energies combining HAWC data from  the northern
and IceCube data from the souther hemisphere~\cite{Aartsen:2018ppz}.
In addition to the expected dipole anisotropy and other large-scale
anisotropies reflecting the non-uniformity of the CR source distribution,
higher multipoles have been detected which are visible by eye in the map shown
in the left panel of Fig.~\ref{anisotropy}. These small-scale anisotropies
are most likely connected to the local structure of the turbulent GMF and
the heliosphere; for a detailed discussion see
Refs.~\cite{Ahlers:2016rox,Deligny:2018blo}.
Additionally, the shape of the anisotropy shown in the right panel which
deviates from the cosine shape expected for a pure dipole contains useful
information about the type of magnetic field fluctuations on which CRs
scatter~\cite{Giacinti:2016tld}.
Here we consider only the dipole component of the anisotropy.

The magnitude of the dipole anisotropy  $\vec \delta$ of the CR intensity
$I= c/(4\pi)n$ is defined by
\begin{equation} \label{delta_diff0}
 \delta \equiv \frac{I_{\max}-I_{\min}}{I_{\max}+I_{\min}} .
\end{equation}
In most cases, CR experiments measure only the projection of the dipole
vector on the equatorial plane. This implies in particular that the measured
magnitude is smaller than the true one, except  $\vec \delta$ would be
contained in the equatorial plane. Moreover, only the phase, i.e.\ the
right ascension of the projected dipole vector, is experimentally determined.

\begin{figure}
  \hspace*{-.6cm}
  \includegraphics[width=0.35\columnwidth,angle=270]{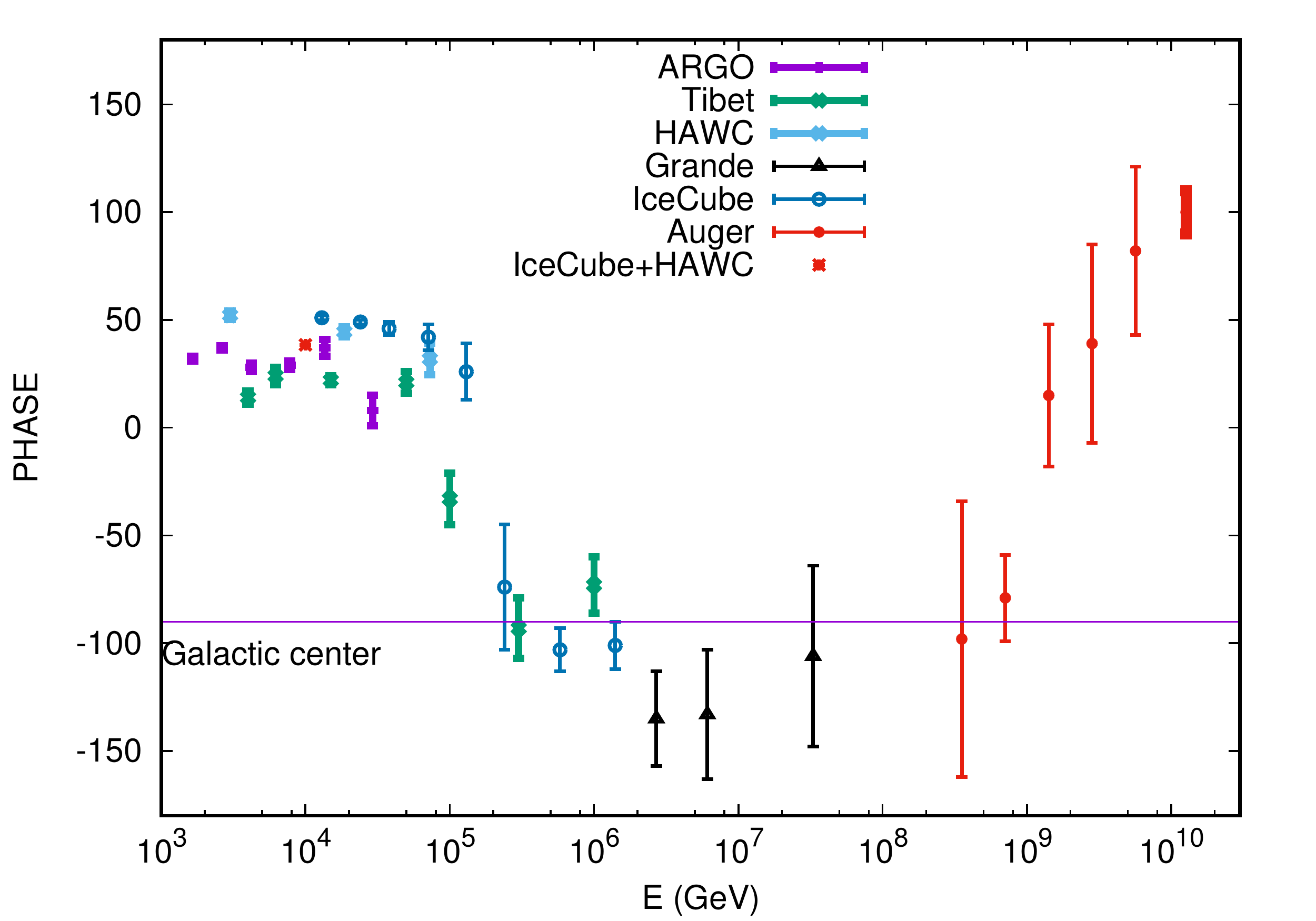}
  \includegraphics[width=0.35\columnwidth,angle=270]{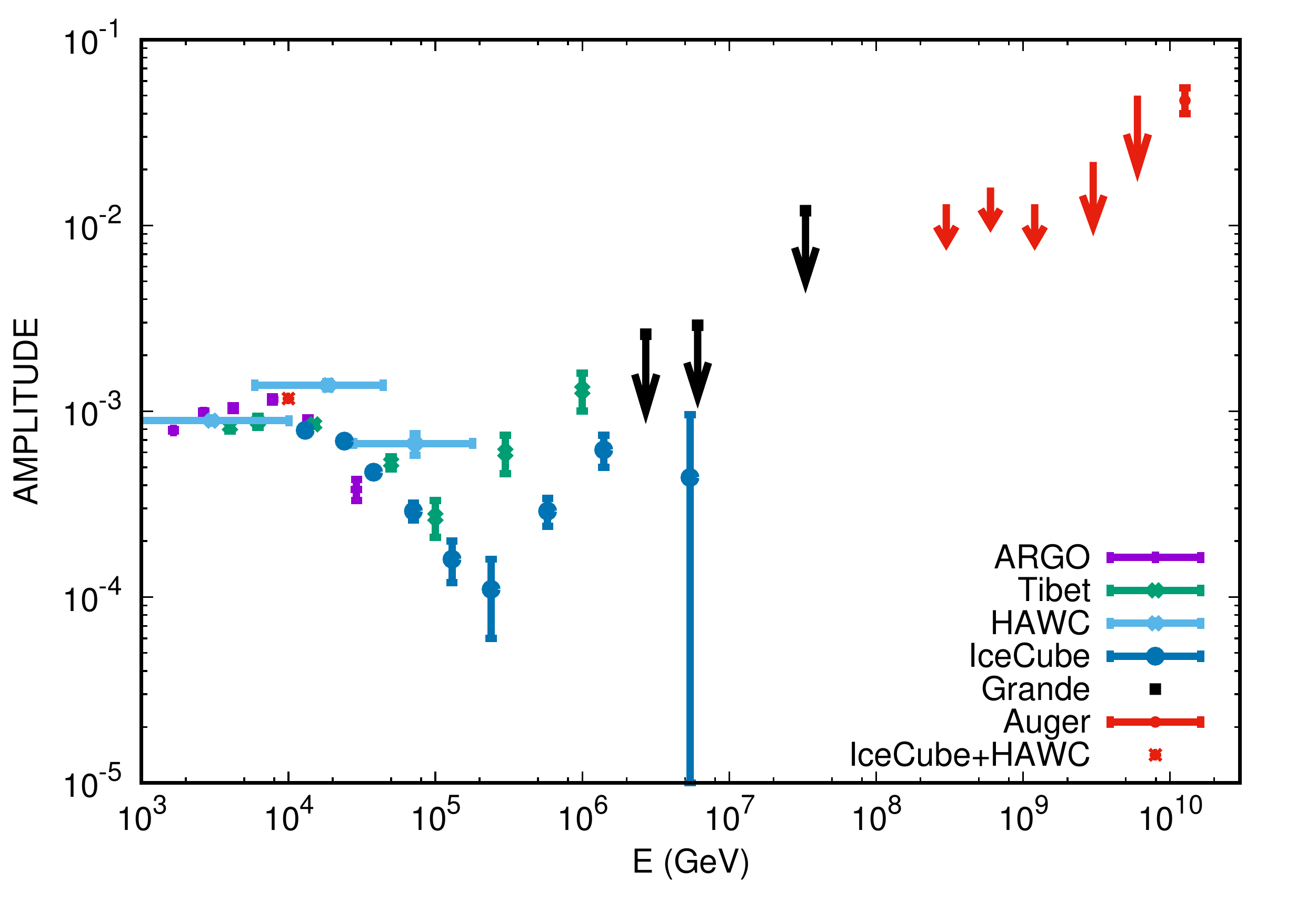}
  \vskip0.4cm
  \caption{The dipole component of the CR anisotropy as function of energy
    as measured by ARGO~\protect\cite{Bartoli:2015ysa}, Tibet~\protect\cite{Amenomori:2005dy,Amenomori:2017jbv}, HAWC~\protect\cite{Abeysekara:2018qho}, 
    IceCube~\protect\cite{Aartsen:2016ivj}, combining HAWC and IceCube~\protect\cite{TheHAWC:2017uyf},   KASCADE-Grande~\protect\cite{Chiavassa:2015jbg}
    and the PAO~\protect\cite{Aab:2018mmi}: the phase (left panel) and the amplitude
    (right panel) as function of energy. 
\label{anisotropy_dipole_data}}
\end{figure}

The (projected) dipole anisotropy $\delta$ measured by seven experiments is
shown in
Fig.~\ref{anisotropy_dipole_data} as function of energy. The phase
of the anisotropy shown in the left panel is close to constant up to
200\,TeV, flips then by approximately $180^\circ$ and stays again
constant up to 100\,PeV. At even higher energies, the phase changes smoothly.
The amplitude of the dipole anisotropy shown in the right panel 
changes rather smoothly at low energies, being first approximately constant
followed by a decrease in the range between 10--200\,TeV. This decrease stops
abruptly at 200\,TeV, i.e.\
at the same energy where the phase flips  by $180^\circ$. Above 10\,PeV,
only limits by KASCADE-Grande and PAO exist up to the energy
$E>8$\,EeV, where the dipole is again detected.

The behaviour of the dipole anisotropy $\delta$ as function of energy
up to 200\,TeV seems at first sight difficult to reconcile with  diffusive
CR propagation: 
In this picture, CRs are  most efficiently scattered by those 
turbulent field modes which wave-length equals their Larmor
radius. Therefore the scattering rate is energy dependent and determined by 
the fluctuation spectrum of the turbulent GMF. 
As result, both the diffusion coefficient and the CR anisotropy are expected 
to increase with energy with the same rate. Thus the decrease of the
CR anisotropy appears to be in contradiction to Kolmogorov-like diffusion
which is supported, e.g., by the AMS-02 result on the B/C
ratio~\cite{Aguilar:2016vqr}. Moreover, the abrupt change of the dipole phase
is difficult to explain in the simplest diffusion approach where the dipole
is aligned with the CR flux, $\vec j\propto \vec\nabla n$.  In this picture,
only small
variations of the dipole direction are expected, if the CR sea is smooth
and many sources contribute. Finally, it has been often stressed that the
observed amplitude of the dipole anisotropy is small compared to the
theoretical expectation~\cite{Hillas:2005cs,Blasi:2011fi}.
These discrepancies were dubbed the ``CR anisotropy problem'' by
Hillas~\cite{Hillas:2005cs}.

\subsubsection{Primary cosmic ray nuclei}
\label{sec_spec_breaks}

The energy spectra of primary CRs measured by direct detection experiments,
i.e.\ up to energies $E\simeq 10^{14}$\,eV, were until 2010 well described
by a featureless power law with slope $\alpha\simeq 2.7$.
The absence of structures indicates that a common acceleration mechanism in
CR sources is at work and that features connected to the age or maximal
energy of individual sources are averaged out, because a large number of
sources contributes to the locally observed CR flux. In other words, a
``sea'' of Galactic CRs exists at these energies, which is well mixed. As
a consequence, the energy spectra of primary cosmic rays should be also
universal, being, e.g., independent of the Galactic longitude. Moreover,
CR spectra expressed as function of rigidity should not depend on the type
of nuclei as long as interactions can be neglected. 
However, the idea of a perfectly mixed CR sea is an approximation, and
thus deviations from this picture should appear as soon the experimental
sensitivity is sufficiently improved.

\paragraph{Breaks in the rigidity spectrum}

\begin{figure}
  \centering
  \includegraphics[width=0.7\columnwidth,angle=0]{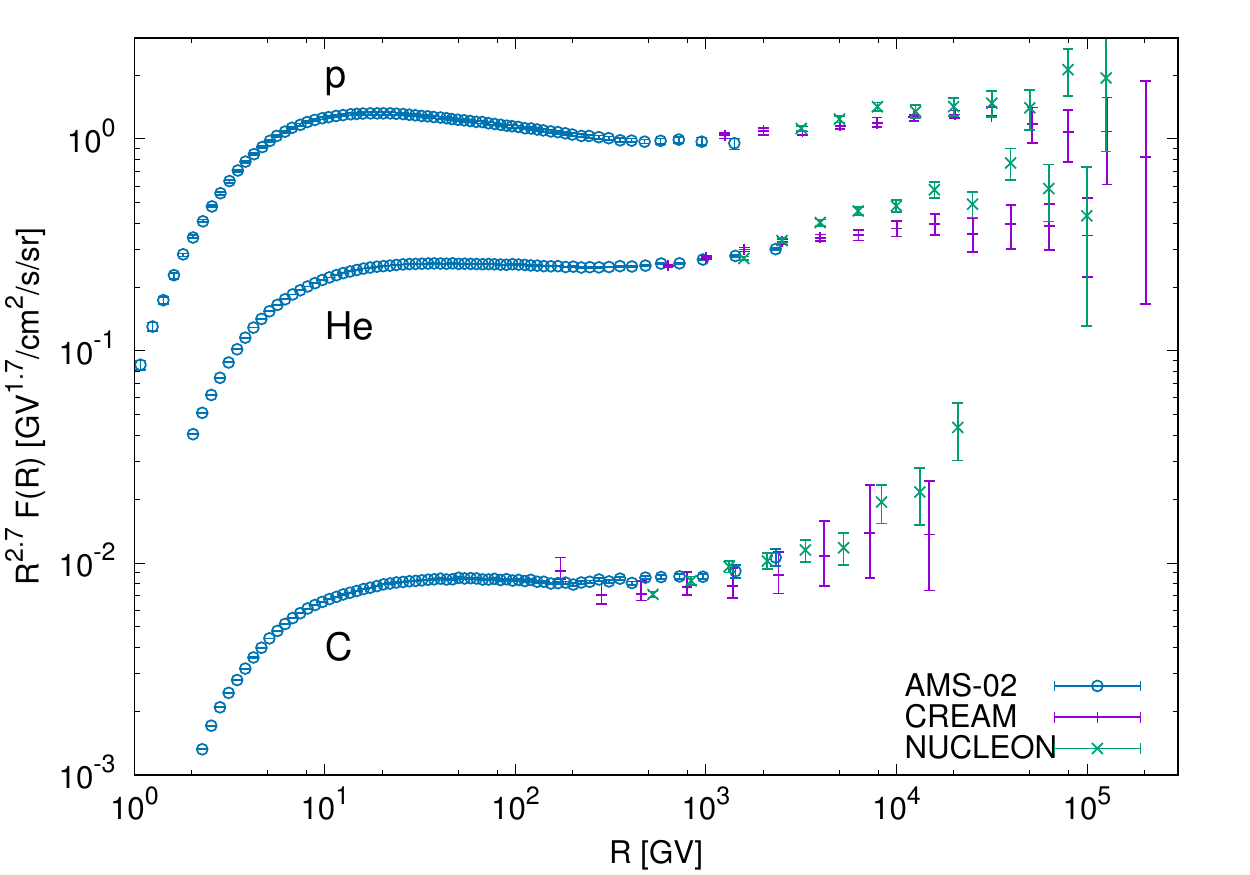}
  \caption{The spectrum of proton, helium and carbon as function of  rigidity, measured by the AMS-02~\protect\cite{Aguilar:2015ooa,Aguilar:2017hno}, CREAM~\protect\cite{Yoon:2017qjx} and NUCLEON experiments~\protect\cite{Gorbunov:2018stf,Atkin:2019xxt}.
\label{allparticle_speclow}}
\end{figure}

In 2010, the CREAM collaboration announced results from two balloon
flights for the spectra of CR nuclei with energies between 2.5\,TeV
and 250\,TeV~\cite{Ahn:2010gv}. Compared to the extrapolation of
data at lower energies, the proton and helium spectra measured were much
flatter, with the helium flux $4\sigma$ higher than expected. Similar
results were obtained for the fluxes of heavier nuclei, which were
all consistent with a break around 200\,GeV/n.
These results were later confirmed by the PAMELA~\cite{Adriani:2011cu},
Fermi-LAT~\cite{Ackermann:2014ula}
and AMS-02 experiments~\cite{Aguilar:2015ooa,Aguilar:2015ctt}. 
A summary of recent measurements is shown in
Fig.~\ref{allparticle_speclow} for the spectra of protons, helium and
carbon nuclei. The best-fit obtained by the  AMS-02 collaboration to the
proton flux between  45\,GV and 1.8\,TV using a broken power law has
its break at $336^{+68}_{-44}$\,GV where the slope changes from $\alpha=2.85$
to $\alpha=2.72$~\cite{Aguilar:2015ooa}. For helium, a fit of the flux
between  45\,GV and 3\,TV  gave a break at $245^{+35}_{-31}$\,GV, where the
slope changes from $\alpha=2.78$ to $\alpha=2.66$. 
At higher energies, there are indications for additional features in the
energy spectrum. The new space experiment NUCLEON measured a  knee-like
feature at $R=10$\,TV with $3\sigma$ significance in its first two years of
observations ~\cite{Atkin:2018wsp}. This result still needs confirmation with better statistics.
In Fig.~\ref{allparticle_speclow}, one can see, that the results of both the
CREAM and the NUCLEON experiments are consistent  with AMS-02 in their
common energy range, but their results are somewhat differ at higher energies.

\paragraph{Deviation from rigidity dependent power laws}
 \label{sec_spec_diff_power_law}

If the breaks in the CR nuclei spectra discussed above are caused by
acceleration or propagation effects, the rigidity spectra of different
nuclei should have the same shape and differ only in their normalisation. 
However, already the first CREAM results in 2010 provided strong evidence that
the spectra of proton and helium differ above the break. Recall also from
Fig.~\ref{fig:Xint} that  interactions do not influence the proton and
helium spectra even at the lowest energies. The findings of CREAM
were confirmed by the AMS-02 experiment which found a clear change in the
ratio of the proton and He fluxes~\cite{Aguilar:2015ctt}, cf.\ with
Fig.~\ref{allparticle_speclow}: Their results indicate that 
the spectral index $\gamma_{\rm p/He}$ of the
p/He flux ratio increases with rigidity up to 45\,GV and becomes then
constant, $\gamma_{\rm p/He}=-0.077\pm0.02$~\cite{Aguilar:2015ctt}.
As a result of the harder helium spectrum, the proton and He flux are
crossing over in the energy range 3--10\,TeV.

\paragraph{Gamma-ray observations}

Since the locally measured CR spectrum is up to $\sim 30$\,GV affected
by the Solar wind, indirect measurements of the CR flux using gamma-rays
are a valuable alternative. In this case, one uses the knowledge
of the differential hadronic production cross section of photons to infer
the shape of the primary CR flux. At very low energies, a
possible contribution of electron Bremsstrahlung or changes of CR
propagation inside dense molecular clouds has to be taken into account.

\begin{figure}
  \centering
  \includegraphics[width=0.49\columnwidth,angle=0]{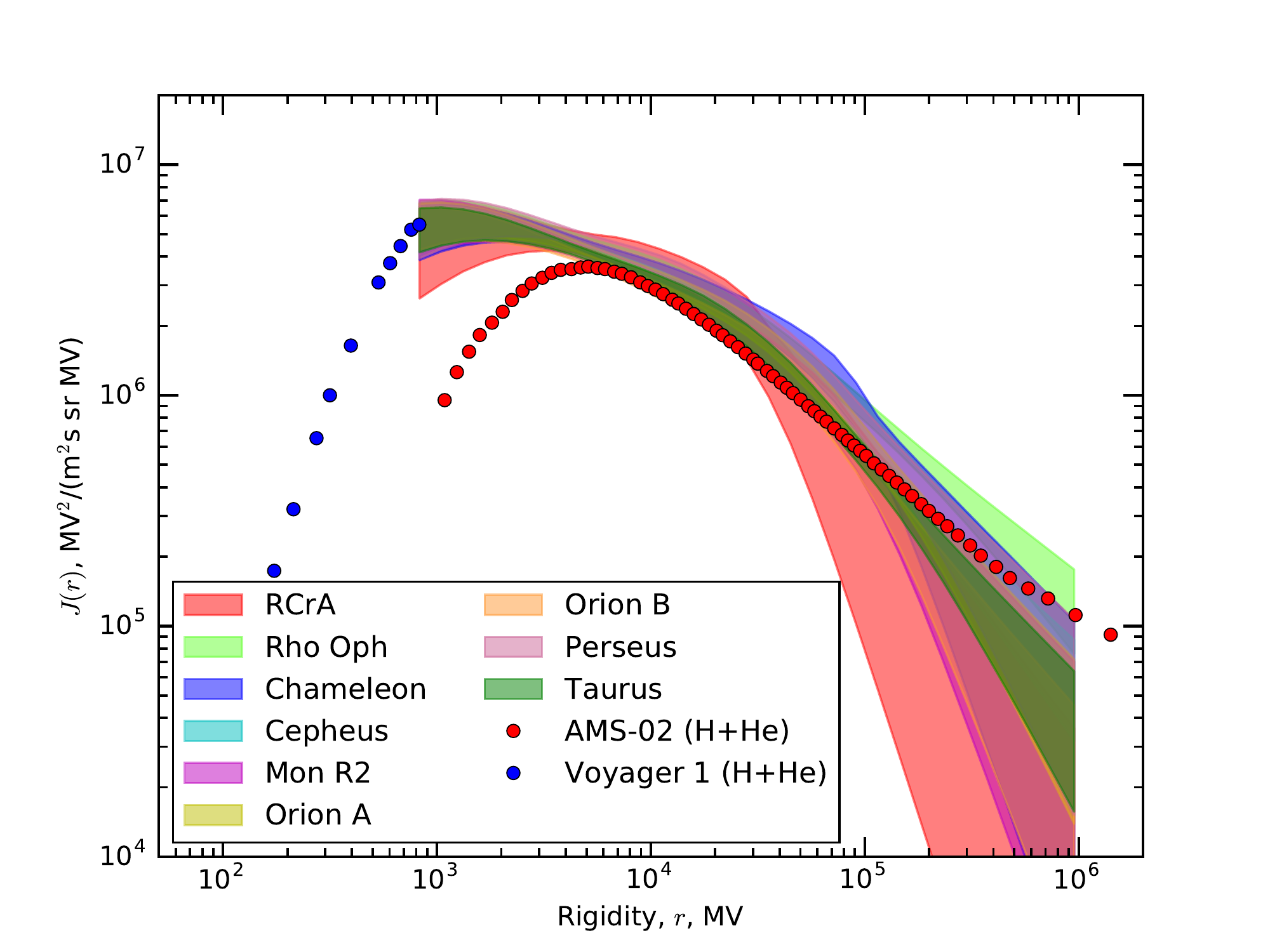}
  \includegraphics[width=0.49\columnwidth,angle=0]{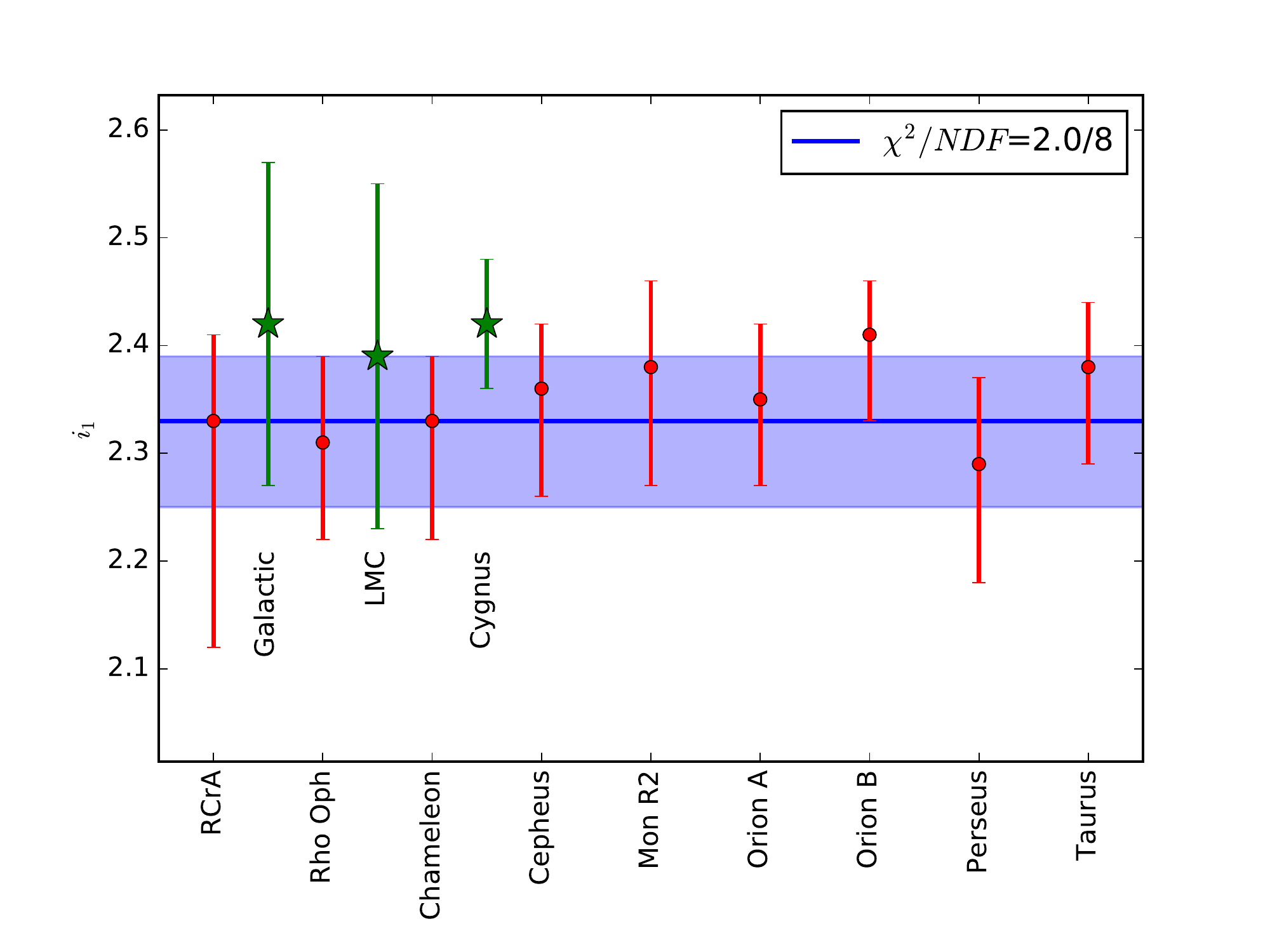}
  \caption{ Proton plus helium spectrum of individual molecular clouds as function of rigidity ({\em left\/}) and  the corresponding exponent of the power law
    below the break ({\em right\/});
    from  Ref.~\protect\cite{Neronov:2017lqd}. \label{clouds}}
\end{figure}

One possibility to perform such an indirect determination of the CR flux
is to use the gamma-ray flux measured by the Fermi-LAT
experiment from giant molecular clouds in the Gould belt, which is located
at the distance 200--500\,pc from us. Using the first years of Fermi data,
Ref.~\cite{Neronov:2011wi} concluded that the measured gamma-ray spectrum
can not be fitted with a single power law for the CR spectrum, 
but requires a break around  $E \simeq 9$\,GeV. The claim of such a break
was initially disputed~\cite{Kachelriess:2012fz,Dermer:2012bz}, but became
stronger with larger statistics.
The most recent analysis using almost 10~years of data had enough events
to study individual clouds separately~\cite{ Neronov:2017lqd}.
In  the left panel of Fig.~\ref{clouds}, the derived CR spectra
of individual clouds as function of rigidity are shown. The  spectra from all
molecular clouds are fitted with a broken power law and are
normalized to the Voyager~1 data at low energies. In the right panel, the
indices of the power law below the break determined for individual clouds
are compared to the one of the central part of the Milky Way, the
Large Magellanic Cloud, and the Cygnus region. One can see that the
molecular cloud spectra below the break 
are consistent with the spectral index $\alpha=2.3\pm 0.1$.
These values are also consistent with the spectrum from the central Galaxy
and from the Large Magellanic Cloud, but deviate from the locally observed
slope $\alpha\simeq 2.7$. 
In contrast to the spectrum below the break, both the position of the
break and the slope after the break differ in individual
clouds~\cite{Neronov:2017lqd}.

Gamma-ray observations can be also used to compare the CR spectrum
as function of Galactic longitude. In Ref.~\cite{Neronov:2015vua},
the spectral slope in the central Galaxy was determined as $\alpha\simeq 2.4$
compared to the steeper $\alpha\simeq 2.7$ found from local measurements.
In the more detailed study of Ref.~\cite{Yang:2016jda}, it was then shown
that the slope of the CR spectrum strongly depends on the Galactic longitude:
Considering only the Galactic plane, $|b|\leq 5^\circ$, the slope varies
between $\alpha\simeq 2.5$ towards the Galactic center and $\alpha\simeq 2.8$
in the outer Galaxy. Note, however, that these results were challenged
recently by Ref.~\cite{Aharonian:2018rob} which finds a rather homogeneous
CR sea below 100\,GeV.

\subsubsection{Primary electrons}
 \label{sec_electrons}

\begin{figure}
  \centering
  \includegraphics[width=0.45\columnwidth,angle=0]{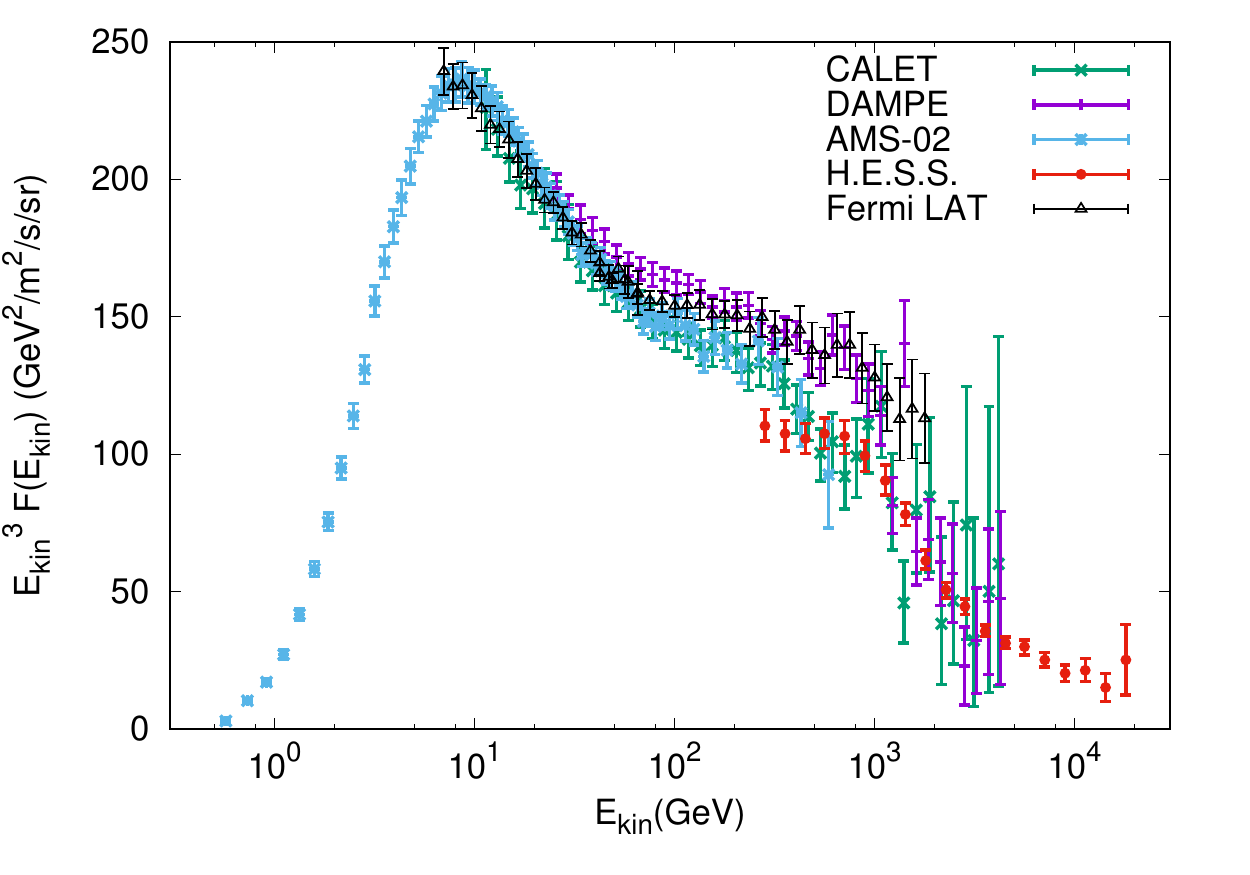}
  \includegraphics[width=0.45\columnwidth,angle=0]{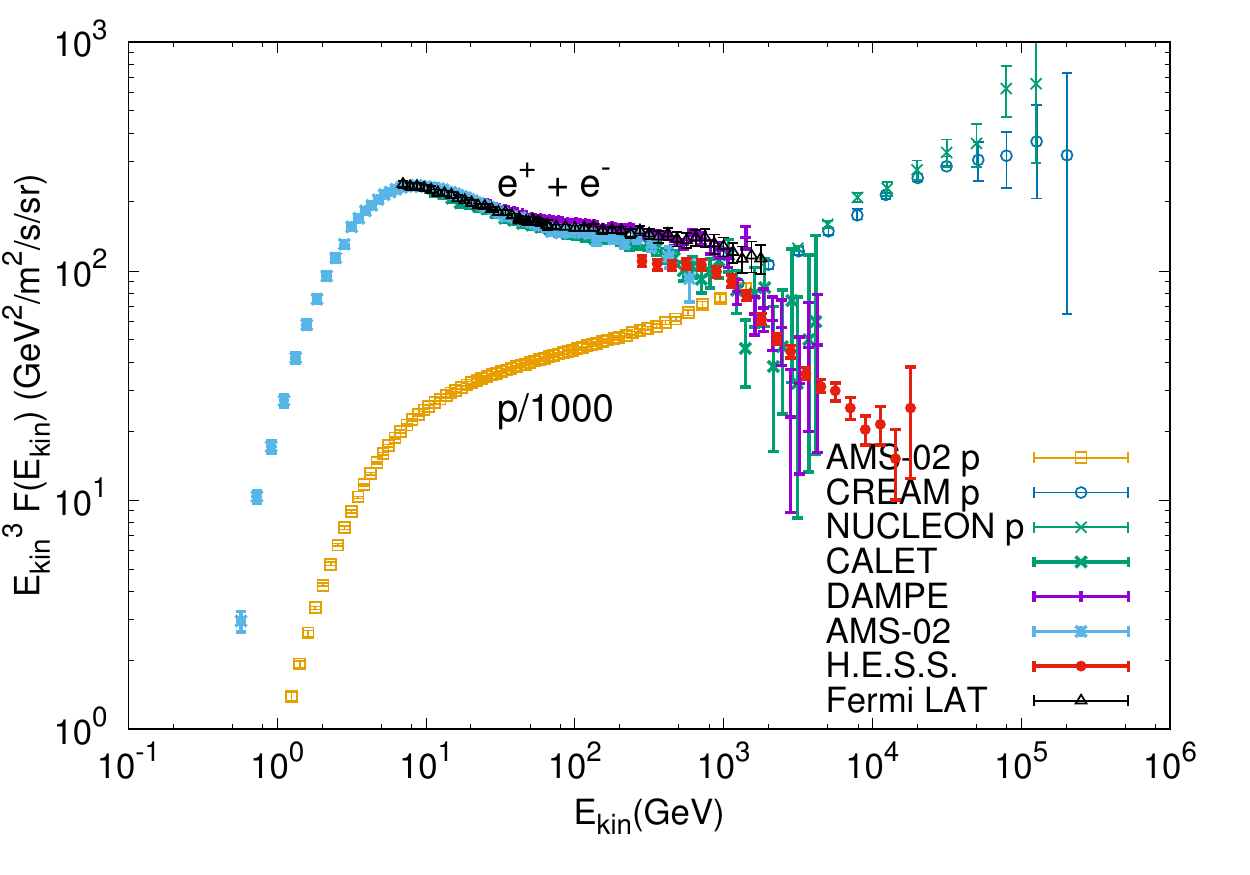}
    \caption{{\em Left panel:\/} Cosmic ray electron plus positron energy spectrum measured by AMS-02~\protect\cite{Aguilar:2014mma}, Fermi-LAT~\protect\cite{Abdollahi:2017nat}, H.E.S.S.~\protect\cite{Abdalla:2017brm}, DAMPE~\protect\cite{Ambrosi:2017wek}, and CALET~\protect\cite{Adriani:2018ktz}. {\em Right panel:\/} Comparison of the proton spectrum from Fig.~\ref{allparticle_speclow} (divided by 1000) and the electron plus positron spectrum. 
     \label{electron_spec}}
\end{figure}

The flux of CR electrons has been measured recently by several experiments.
While the magnetic spectrometers PAMELA and AMS-02 can separate electrons and
positrons, other experiments like H.E.S.S., Fermi-LAT, DAMPE, and CALET
perform calorimetric measurements of the sum of electrons and positrons.
In the left panel of Fig.~\ref{electron_spec}, we show the combined electron
plus positron flux $F(E)$ multiplied with $E^3$ measured by
AMS-02~\cite{Aguilar:2014mma},
Fermi-LAT~\cite{Abdollahi:2017nat}, H.E.S.S.~\cite{Abdalla:2017brm},
DAMPE~\cite{Ambrosi:2017wek}, and CALET~\cite{Adriani:2018ktz}.
The spectrum $E^3 F(E)$ has a peak at 10\,GeV and is affected by solar
modulations at energies $E\lsim 20$\,GeV, as one can see from the differences
between the data of PAMELA and AMS-02 which measured the electron flux at
different times, for more  details see Ref.~\cite{Lipari:2018usj}.
Between 20 and 50\,GeV, the spectrum hardens gradually and is then
up to $\simeq 500$\,GeV well described by a power law with spectral index
$\alpha_{\rm e^-}\simeq 3.17$. Finally, the combined electron plus positron
flux has a strong break at 1\,TeV, as it was first shown by data from 
H.E.S.S. The shape of this break and the suppression measured by the
different experiments vary, indicating underestimated systematic errors. 
From the measurement of the positron flux (discussed later and
shown in Fig.~\ref{positron_data}), one can estimate a flux ratio
of positron-to-electrons $\lsim 10\%$ at 1\,TeV.  Thus the break in the
combined electron plus positron flux at 1\,TeV is a break in
the electron flux. The data, in particular of H.E.S.S., above the
break are consistent with a new, steeper power-law. 

In the right panel of Fig.~\ref{electron_spec}, we compare the proton and
electron (plus positron) spectra. The two spectra have different slopes and
normalisations
of the power law. The difference in normalisation is caused by the different
injection mechanism into the acceleration process
for electrons and protons, while the steeper slope  of electrons should
be caused by their energy losses. 
In the diffusion picture, the energy losses of electrons should lead to two
cooling breaks in the observed flux of electrons~\cite{Ginzburg:1990sk}. A
first break, where the slope changes from $F(E)\propto Q(E)\propto E^{-\alpha}$
to $F(E)\propto E^{-(\alpha+1/2)}$ is expected at the energy when the average
path length of electrons becomes comparable to the height $H$ of the CR halo.
A second break with a steepening to $F(E)\propto E^{-(\alpha+1)}$ should occur,
when the average path length becomes comparable to the height $h$ of the
thin disk containing CR sources. Combining the distance
$d\simeq\sqrt{2D\tau_{1/2}}\simeq\sqrt{2D/(bE)}$ an electron
diffuses during its energy loss time (cf.\ with Eq.~(\ref{ellosses})),
we can estimate the energy of the first break: Assuming $H=5$\,kpc,
Kolmogorov diffusion with $D_0=5\times 10^{28}$cm$^2$/s at $E_0=10$\,GeV
and $b=1.4\times 10^{-16}({\rm GeV\,s})^{-1}$, it follows $E_{1/2}\sim 1$\,GeV.
Thus the break is hidden in the energy region where solar modulations
strongly modify the shape of the electron flux. At higher energies,
say above 30\,GeV, one expects thus the slope $\alpha\simeq 2.85+0.5=3.35$
for the electron flux. The second break with a steepening to
$\alpha\simeq 3.85$ is expected at 10--100\,TeV.
This simple picture is in clear contradiction to the energy dependence of
the measured flux. In particular, the gradual hardening of the spectrum
between 20 and 50\,GeV is difficult to explain as the contribution from
an uniform background of electron sources. If the hardening would be a
propagation effect, some imprint should be also seen  at the same energy range
in the proton spectrum which is shown in the right panel of Fig.~\ref{electron_spec}.
However, the  hardening in the proton spectrum happens at a higher energy and
is less pronounced. A possible explanation to both features is the
appearance of a new component in the CR flux: Such a new contribution
is expected to dominate the steeply falling electron spectrum at lower
energies than the proton spectrum.

\subsubsection{Secondary nuclei}
   \label{sec_secondary_nuclei}

\begin{figure}
  \centering
  \includegraphics[width=0.7\columnwidth,angle=0]{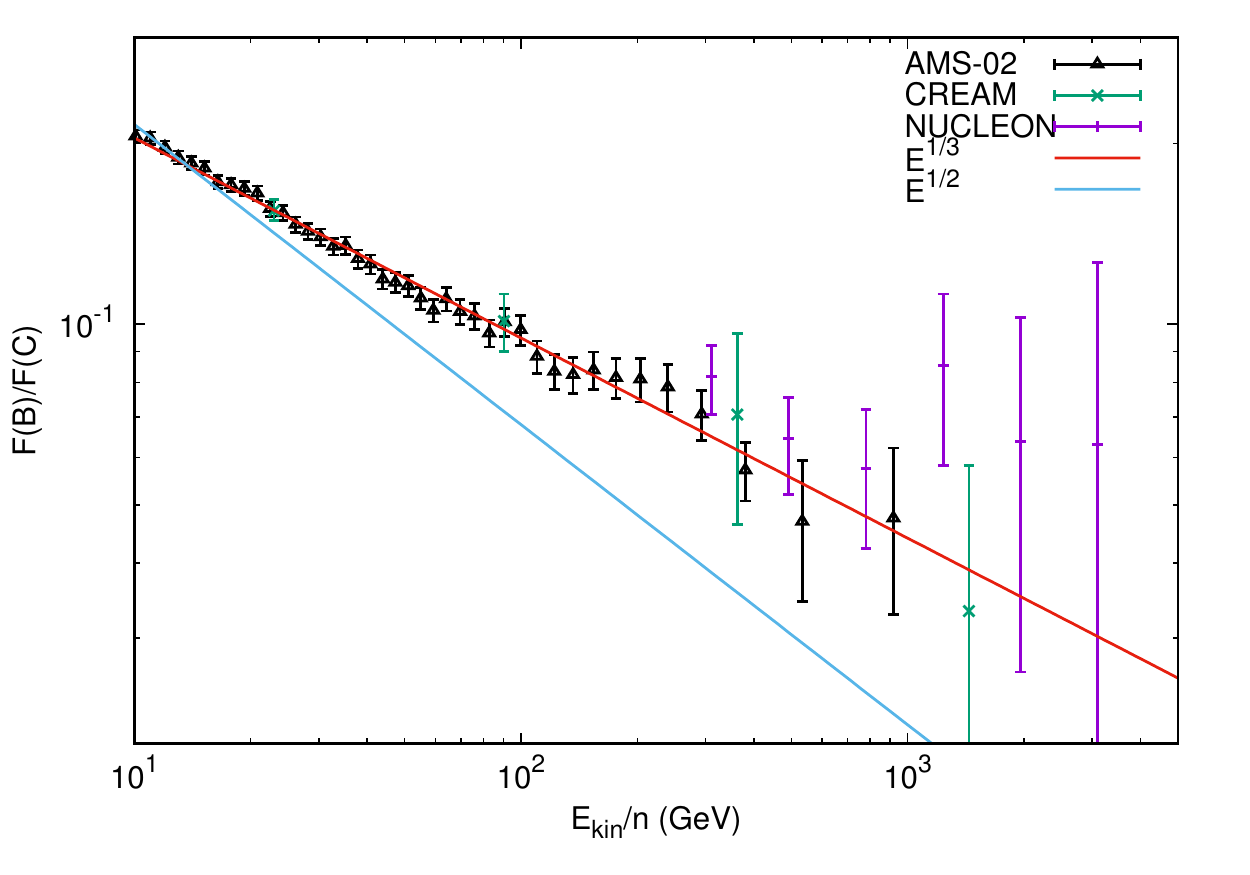}
    \caption{Boron-to-carbon ratio as function of kinetic energy per nucleon measured by the CREAM~\protect\cite{Ahn:2008my}, AMS-02~\protect\cite{Aguilar:2016vqr},  and NUCLEON~\protect\cite{2018arXiv180909665G}  experiments. The power law predicted by Kolmogorov turbulence ($1/E^{1/3}$) and  Iroshnikov-Kraichnan  turbulence ($1/E^{1/2}$) at high energies are also shown.  
     \label{BtoC_data}}
\end{figure}

In the diffusion picture, the fluxes of light primary CRs given by
Eq.~(\ref{Eq:primary}) are connected by 
$n_p = Q_p\tau_{\rm esc}\propto Q_0 E^{-(\alpha+\delta)}$ to the
slopes $\alpha$ and $\delta $ of the injection spectrum and the diffusion
coefficient, respectively. Combining the locally observed  $1/E^{2.7}$ spectrum
with the prediction from diffusive shock acceleration, $\alpha=2.0-2.2$,
suggests therefore $\delta=0.5$ corresponding to  Iroshnikov-Kraichnan
turbulence.
On the other hand, $\delta=0.5$ leads to a fast increase of the CR dipole
anisotropy, in contradiction to the behavior of $\delta$ as function of
energy shown in Fig.~\ref{anisotropy_dipole_data}.

The value of $\delta$ can be determined from secondary-to-primary ratios
like B/C which scale as $ \R^{-\delta}$, if $X_{\rm esc}/X^{(B)}\ll 1$, cf.\ with
Eq.~(\ref{BCratio}). From Fig.~\ref{fig:Xint}, we see that the latter
condition is satisfied at $\R\gg 30$\,GV.
In Fig.~\ref{BtoC_data}, we show the B/C ratio as function of the
kinetic energy per nucleon  measured by the AMS-02~\cite{Aguilar:2016vqr},
CREAM~\cite{Ahn:2008my} and NUCLEON~\cite{2018arXiv180909665G}  experiments. 
One can see that the data are consistent with the slope $\delta=1/3$
predicted by Kolmogorov turbulence, while the decrease of the B/C ratio
predicted by  Iroshnikov-Kraichnan turbulence is too strong.
The data above $E_{\rm kin}/n\gsim 500$\,GeV hint for a flattening of the
B/C ratio at high energies, but the large errorbars prevent any firm
conclusion.

Taken at face value, the observation that the boron-to-carbon ratio follows
Kolmogorov turbulence has important consequences: It implies that either
the acceleration spectrum of Galactic CR sources is softer than expected,
being close to $1/E^{2.4}$. Such soft spectra would compound the problem
of reaching sufficiently high maximal energies in Galactic CR sources.
Or the simple diffusion picture which was used in our argument has to be
modified. For instance, models which invoke strong advection can reproduce
the B/C data using  Iroshnikov-Kraichnan turbulence. 
Alternatively, the locally observed CR spectra may deviate from the global
average. Such a deviation may be expected if the number of CR sources
is relatively small.

\subsubsection{Radioactive isotopes}
   \label{sec_radiactive_nuclei}

\begin{figure}
  \centering
  \includegraphics[width=0.4\columnwidth,angle=270]{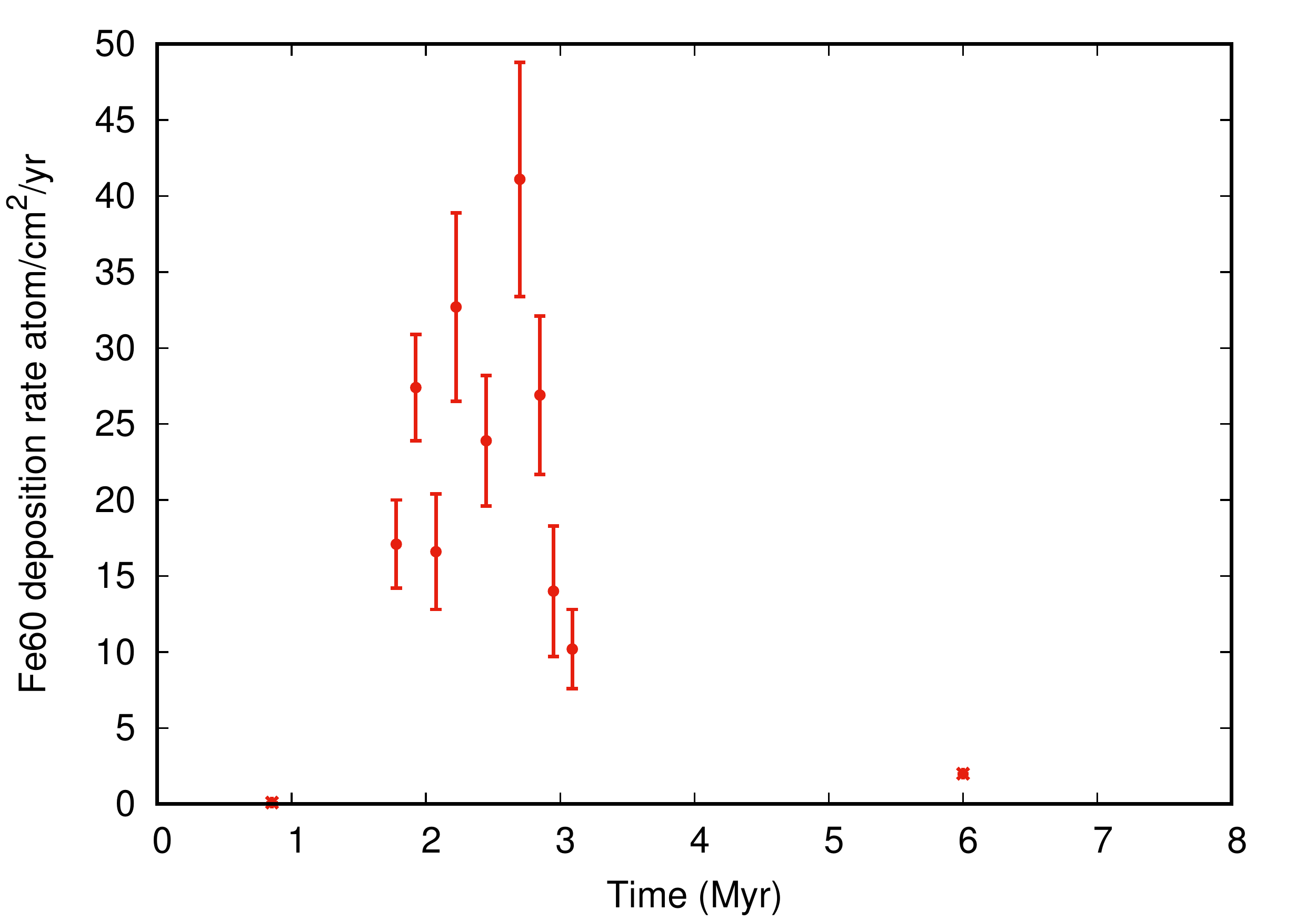}
  \vskip0.6cm
    \caption{Deposition rate of $^{60}$Fe  in the ocean crust as function of
      time from Ref.~\protect\cite{2016Natur.532...69W}.
     \label{sediments}}
\end{figure}

The average time $\tau_{\rm esc}$ CRs spent in the Milky Way before escape can
be deduced from the suppression, due to radioactive decay, of the flux
of unstable nuclei that have a lifetime comparable with their residence time
in the Galaxy.
Comparing the fluxes of two isotopes of the same chemical element, one
stable and the other unstable, allows one to measure this suppression, what in
turn can be used to estimate $\tau_{\rm esc}$.
Beryllium is a very rare element in ordinary matter, and essentially all
beryllium nuclei in CRs  are secondaries  formed by the fragmentation of
heavier nuclei, cf.\ with Fig.~\ref{abund}. It has two stable isotopes
( $^{9}$Be, and $^{7}$Be, if it is fully ionized) and one
unstable ($^{10}$Be) with the half-life
$\tau= (1.51 \pm 0.04)$\,Myr. The CRIS collaboration used a leaky-box model
to convert their measurements  of the beryllium ratio
$^{10}{\rm Be}/(^{7}{\rm Be}+^{9}{\rm Be})$
in three different bins of kinetic energy around 100\,MeV/nucleon into the
estimate $\tau_{\rm esc}= (15.0 \pm 1.6)$\,Myr for the
escape time~\cite{2001ApJ...563..768Y}.

In 1996, it was suggested that rare, long-lived radioactive isotopes can be
used as probes of nearby SN explosions~\cite{Ellis:1995qb}. Such isotopes as
$^{60}$Fe with a half-lifetime of $2.6$\,Myr~\cite{Rugel:2009zz} can be
searched for in sediments on the bottom of oceans. The age of the ocean
crust can be determined with $^{10}$Be dating. The first successful measurement
was carried out in 1999, finding an enhanced $^{60}$Fe concentration in a
ferromanganese crust in the Southern Pacific~\cite{Knie:1999zz}. This
observation was confirmed using marine sediments at other
locations~\cite{Benitez:2002jt,Fitoussi:2007ef,2016Natur.532...69W} and the
Moon~\cite{2016PhRvL.116o1104F}.
In Fig.~ \ref{sediments}, we show results for the abundance of $^{60}$Fe in
the Earth crust from Ref.~\cite{2016Natur.532...69W}. The peak in the
$^{60}$Fe abundance points to a SN event in the vicinity of the Earth
$\simeq (2-3)$\,Myr ago. The study of Ref.~\cite{2016Natur.532...73B}
suggested that more than one SN contributed to this peak.
Modelling the transport of dust grains containing $^{60}$Fe,
Ref.~\cite{Fry:2014yqa} suggested a distance of order 100\,pc to the SNe. 
For a review and discussion of other radioactive elements than
$^{60}$Fe see Refs.~\cite{Feige:2012be,Fields:2019jtj}.

\subsubsection{Secondary positrons and antiprotons}
   \label{sec_secondary_positrons}
 
\def\Rpos{R_{e^+/e^-}}
\def\Rpbar{R_{\bar p/p}}

Positrons and antiprotons are produced as secondaries by CR protons and nuclei
in the interactions with interstellar gas. While positrons also can be produced
as primaries in pulsars through $e^+e^-$ pair production, astrophysical sources
of primary antiprotons do not exist, offering an opportunity to search for
new phenomena and exotic physics. Similarly to the case of  secondary nuclei,
the decrease of the escape time $\tau_{\rm esc}=\tau_0 (E/E_0)^{-\delta}$
with energy should soften the slope of the positrons and antiprotons
secondaries relative the primary spectrum by $\delta=1/3$ in the case of
Kolmogorov turbulence. For positrons, energy losses due to
synchroton radiation and Thompson scattering on background photons result
in an additional softening, as discussed for primary electrons. Thus the
expectation in the standard diffusion picture is that both the
positron-to-electron ratio $\Rpos$ and the antiproton-proton ratio $\Rpbar$
should decrease with energy.

\begin{figure}
  \centering
  \includegraphics[width=0.49\columnwidth,angle=0]{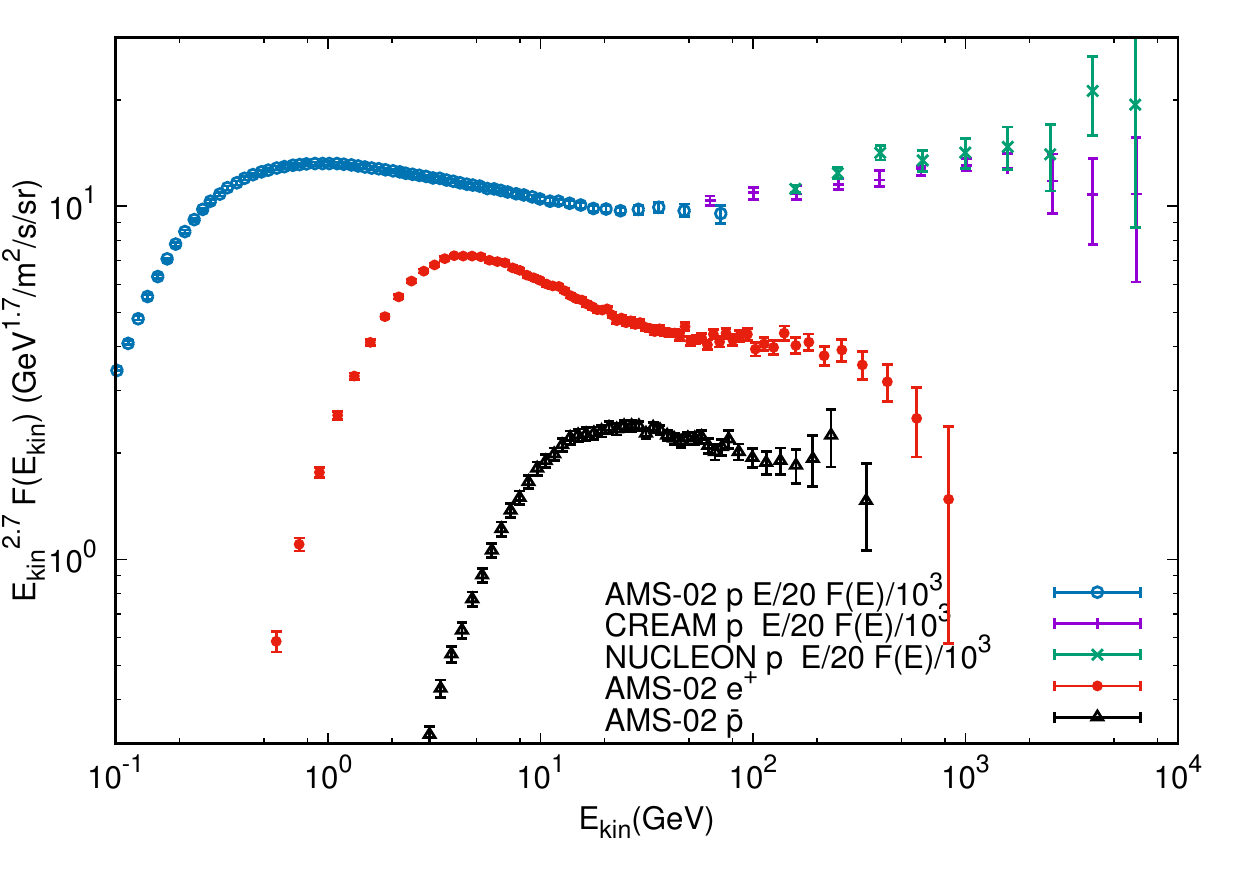}
  \includegraphics[width=0.49\columnwidth,angle=0]{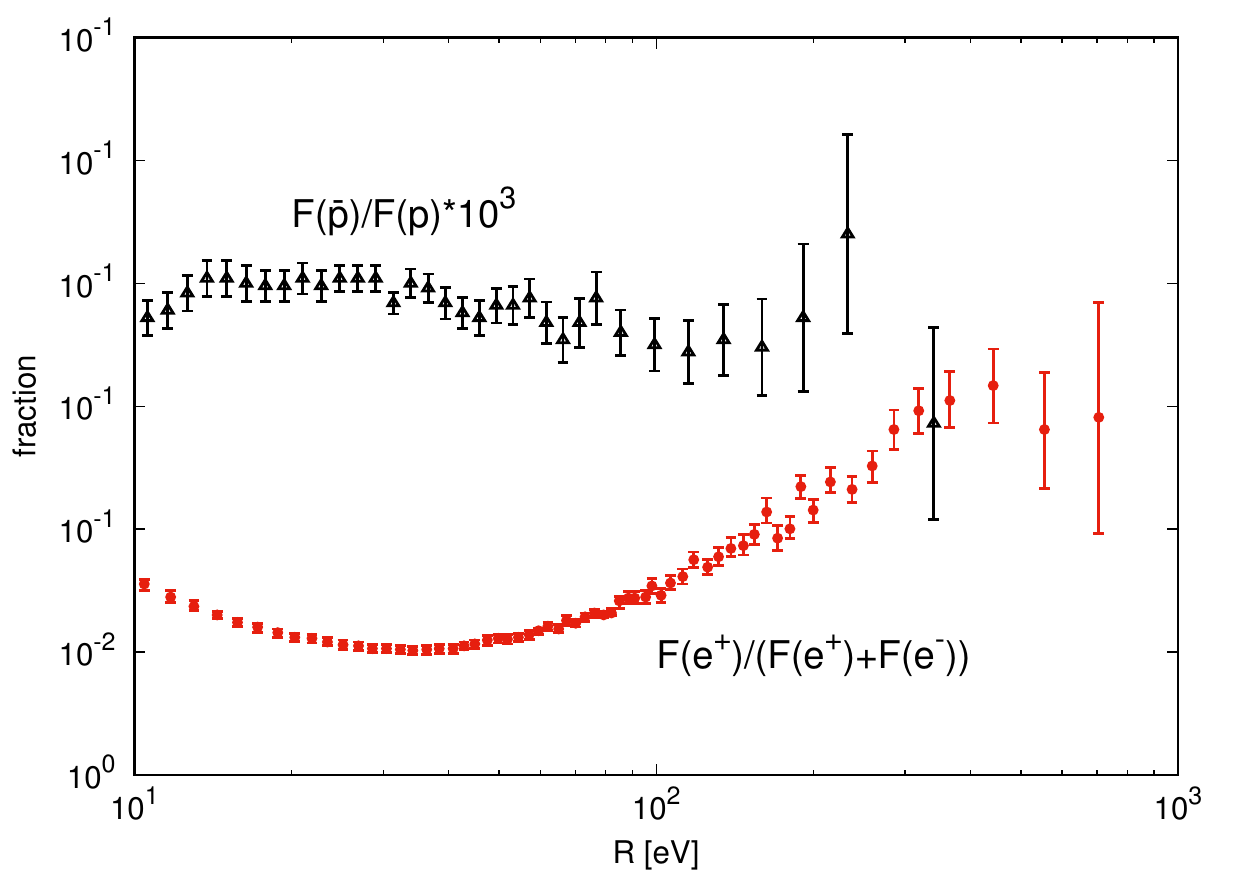}
  \caption{
      {\em Left panel:\/} Positron and antiproton spectra from the AMS-02
      experiment~\protect\cite{Aguilar:2019owu,Aguilar:2016kjl}. For comparison, we show the  proton spectrum from Fig.~\ref{allparticle_speclow}, divided by $10^3$
      and rescaled in energy  $E/20$.
      {\em Right panel:\/} The positron-to-electron ratio $\Rpos$ from
      Ref.~\protect\cite{Weng:2018} and the antiproton-to-proton ratio $\Rpbar$ from
      Ref.~\protect\cite{Aguilar:2016kjl}.
    \label{positron_data}}
\end{figure}

In Fig.~\ref{positron_data}, we present in the left panel the AMS-02 positron
flux from Ref.~\cite{Aguilar:2019owu} as function of energy with red errorbars.
Above $\simeq 30$\,GeV, when the effect of solar modulations can be neglected,
an ankle-like feature around 50\,GeV and a break at 300\,GeV are visible.
This break in the positron spectrum happens at lower energy, and is much
softer than the one in the electron spectrum. In the right panel, we show
the corresponding positron-to-electron ratio $\Rpos$, which clearly
deviates from the expectation for a pure secondary production.
Additionally, we show in the left panel with black errorbars the antiproton
flux measured by AMS-02~\cite{Aguilar:2016kjl}.  As it was noted in
Refs.~\cite{Kachelriess:2015oua,Aguilar:2016kjl,Lipari:2016vqk},
a positron-to-antiproton flux ratio close to 2 is consistent with
expectations that both are produced in hadronic interactions.
To illustrate this fact, we also show the proton
flux measured by AMS-02~\cite{Aguilar:2019owu} and CREAM~\cite{Yoon:2017qjx},
changing the normalisation by a factor $10^{-3}$. Additionally, we  rescaled
the energy of the proton flux by a factor 20 down, taking thereby into
account that the energy transferred to positrons is around 5\%.
One can see, that below 50\,GeV, the positron flux is much steeper than the
proton flux, while above this energy up to the break around 300\,GeV,
the slopes of the two fluxes are similar.  The antiproton flux above 50\,GeV
repeats the slope of the proton flux, but the relatively large errorbars
at high energies prevent any  definite conclusion on its high energy behaviour.

While the positron-to-electron ratio $\Rpos$ shown in the right panel of
Fig.~\ref{positron_data} is clearly incompatible with pure secondary
production of positrons, the larger errorbars both in theoretical
predictions and the experimental flux determination make the situation
for antiprotons less conclusive: For instance, Ref.~\cite{Giesen:2015ufa}
found no unambiguous evidence for a significant excess with respect to
expectations. This conclusion requires however, that all uncertainties
in the cross section and propagation model are by chance correlated,
increasing thereby the prediction. Similarly, Ref.~\cite{Boschini:2017fxq}.
found an antiproton spectrum which is only slightly lower than the data combining
a GALPROP diffusion model including advection and reacceleration for the
interstellar and HelMod for the heliospheric propagation. Thus a
conclusive statement that antiproton measurements
require an additional source of antiprotons requires a reduction of the
current errors, both in theoretical predictions and experimental measurements.

\subsection{Models}
 \label{models_below_knee}

We now review the main types of models suggested as explanation for
the anomalies discussed in the previous section. First we discuss possible
explanations of the observed behaviour of the CR anisotropy as function
of energy, restricting ourselves to the the dipole part of the anisotropy
and energies up to the transition to extragalactic CRs,
$E\lsim {\rm few}\times 10^{17}$\,eV. Then we discuss models addressing
the spectral features observed in the CR energy spectra, starting from
models aiming to explain the anomalies in the secondary fluxes. After that,
we review models for the breaks in the CR nuclei spectra, and finally
models which try to to explain several features simultaneously.

\subsubsection{Explaining the anisotropy}
 \label{models_below_knee_anisotropy}

In the diffusion approximation,  Fick's law is valid and the net CR current
$\vec j(E)$ is determined by the gradient of the differential CR number
density  $n(E)=\d N/(\d E \d V)$ and the diffusion tensor $D_{ij}(E)$ as
$j_i = -D_{ij}\nabla_j n$. Then the dipole vector $\vec \delta$ of the CR
intensity $I= c/(4\pi)n$ follows as
\begin{equation} \label{delta_diff}
 \delta_i 
          =  \frac{3}{c} \frac{j_i}{n} 
          = - \frac{3D_{ij}}{c}\frac{\nabla_j n}{n} \,.
\end{equation}
If the ordered magnetic field $\vec B$ dominates, the tensor structure
of the diffusion tensor simplifies,
\be
D_{ij}= D_\| e_ie_j + D_\perp(\delta_{ij}-  e_ie_j) + D_A \eps_{ijk} e_k
     \simeq  D_\| e_ie_j ,
\ee
except for observers which are nearly perpendicular to the
magnetic field line through the source.  
Here, $\vec e$ is a unit vector in the direction of the ordered magnetic
field, while the diagonal elements of the diffusion tensor describe diffusion
along ($D_{\parallel }$) and perpendicular ($D_{\perp }$) to the ordered field.
The off-diagonal antisymmetric component ($D_{\rm A}$) appears only, if 
drift terms for CRs are included.
Thus in the limit of a strong ordered field,  $D_{ij}\simeq  D_\| e_ie_j$,
the CR gradient is projected onto the magnetic field
direction~\cite{1990ApJ...361..162J}. Therefore anisotropic diffusion
predicts that the dipole anisotropy should
align with the local ordered magnetic field instead of pointing to the
source~\cite{1990ApJ...361..162J,Savchenko:2015dha}. Note that the ordered
magnetic field corresponds to the sum of the regular
magnetic field and the sum of turbulent field modes with wavelengths larger
than the Larmor radius at the corresponding CR energy. This implies that
CRs with $R_{\rm L}\ll L_{\rm c}$ propagate anisotropically even in the absence
of a regular magnetic field, since the field modes with
$k/(2\pi)\gg R_{\rm L}$ act locally as an ordered
field~\cite{Giacinti:2012ar}.

In the case of a (three-dimensional) Gaussian CR density $n$, the
formula~(\ref{delta_diff}) can be evaluated analytically. The result
$\delta=3R/(2cT)$ for a single source with age $T$ and distance
$R$ is independent of the regular and turbulent magnetic field.
In Ref.~\cite{Savchenko:2015dha}, it was shown that the CR density of
a single source is quasi-Gaussian, if CRs propagate over length scales
$l\gg L_{\rm coh}$. This implies in particular that there is no
``mis-alignment effect'' which can reduce the absolute value of the
dipole anisotropy: For instance, an observer perpendicular to the
magnetic field line through the source will pick up $D_\perp \nabla_\perp n$
in Eq.~(\ref{delta_diff}). But since $D_\perp$ will be cancelled taking the
derivative,  its small value has no influence on the anisotropy.
Numerically,  the dipole anisotropy $\delta$ of a
source  contributing the fraction $f_i$ to the total observed CR
flux is thus
\be \label{single}
 \delta_i = f_i \,\frac{3R}{2cT} \simeq
  5.0 \times 10^{-4} \:f_i\,\left(\frac{R}{200\,{\rm pc}}\right)
 \left(\frac{T}{2\,{\rm Myr}}\right)^{-1} \,.
\ee

The plateau in the range 2--200\,TeV visible in the experimental data for the
dipole anisotropy shown in Fig.~\ref{anisotropy_dipole_data} is
naturally explained by the energy-independent contribution to the dipole
anisotropy of a single source. This is supported by the fact that the dipole
phase remains approximately constant in this range too, before it flips
by $\sim 180^\circ$.  Such a flip is naturally explained by  the projection
effect on the magnetic field line, if above 200\,TeV another source, which
is located in the opposite hemisphere, dominates the CR dipole anisotropy.

\begin{figure}
\centering
\includegraphics[width=0.45\columnwidth,angle=270]{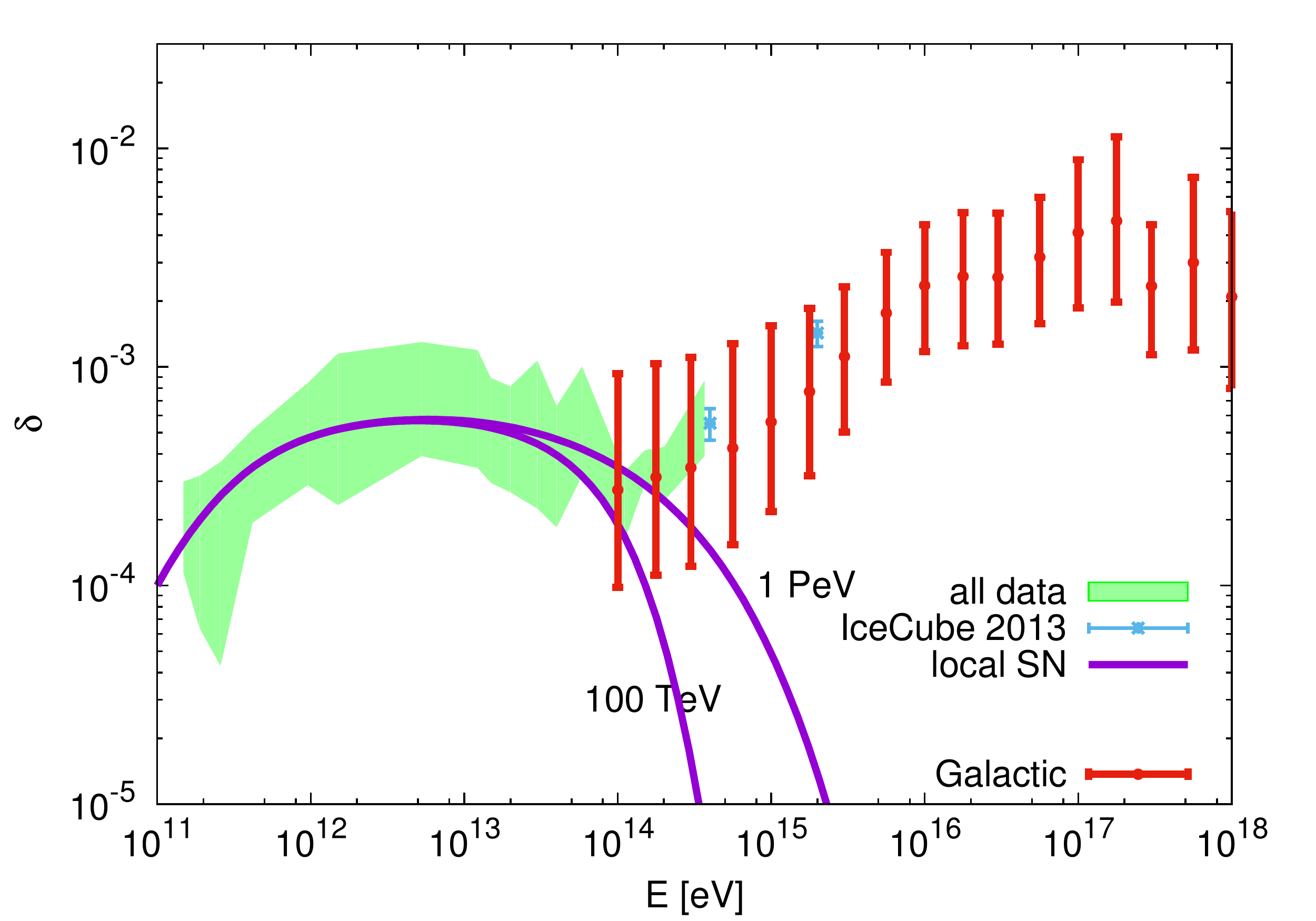}
\caption{
  Lower and upper limit (green band) 
on the dipole  anisotropy and data from IceCube (blue errorbars) compared 
to the contribution from the local  source (magenta, for two values of
$E_{\max}$) 
and from all Galactic SNe (red) calculated in the escape model of Ref.~\protect\cite{Giacinti:2014xya,Giacinti:2015hva}.
\label{aniso_single}}
\end{figure}

More specifically, it was suggested in Ref.~\cite{Savchenko:2015dha}
that a 2--3\,Myr old source at the distance 200--300\,pc dominates
the dipole anisotropy in the range 2--200\,TeV. 
The contribution of this local source is shown in Fig.~\ref{aniso_single}
by two magenta lines for
two different high-energy cutoffs: In one case, it was assumed that the
source can accelerate up to 100\,TeV, in the other that it is a PeVatron.
In both cases, the CR flux was calculated following the trajectories of
individual CRs, as discussed in Refs.~\cite{Giacinti:2014xya,Giacinti:2015hva,Kachelriess:2015oua,Kachelriess:2017yzq}.
Additionally, the total anisotropy beyond $10^{14}$\,eV of all Galactic SNe
is shown by red error-bars which was calculated in the escape model which uses
the same magnetic field configuration as the one used for the loal
source~\cite{Giacinti:2014xya,Giacinti:2015hva}.

A characteristic feature of this proposal is that a relatively old source
dominates the observed CR flux. This is only possible in the case of
anisotropic diffusion, and requires additionally that the perpendicular
distance $d_\perp$ of the Sun to the magnetic field line connecting it to
the source is not too large. Even for small $d_\perp$, the CR flux from the
single source is suppressed at low-energies, because of the slower
perpendicular diffusion.
In Ref.~\cite{Kachelriess:2017yzq}, the value $d_\perp\simeq 70$\,pc was
estimated requiring that the low-energy break in the source spectrum
explains the breaks in the energy spectra of CR nuclei. For this choice of
$d_\perp$,
the flux of the local source is suppressed below $\simeq 1$\,TeV (cf.\ with
Fig.~2 of Ref.~\cite{Kachelriess:2017yzq}), leading to a decreasing $f_i$
and the transition to the standard $\delta\propto E^{1/3}$ behavior below
this energy.

Another choice for the age of the source was suggested in Refs.~\cite{2013APh....50...33S,Ahlers:2016njd}.
Here, Vela with the age around 11,000\,yr and distance 300\,pc was
proposed as the single source responsible for the observed plateau in the
dipole anisotropy. In this case, the
contribution of Vela to the dipole amplitude has to be suppressed by a factor
$\simeq 200$.
Three mechanisms for such a suppression may be operating:  First, if
the regular magnetic field and the CR gradient are not parallel,
the projection effect in $D_{ij}\nabla_j n$ can reduce the
dipole~\cite{Mertsch:2014cua,Ahlers:2016njd}.
Second, the measured CR dipole is a projection into the equatorial
plane and is thus reduced compared to the true one.
Finally, the CR flux contributed by Vela may be small. 
Calculating the CR fluxes from nearby young sources using the standard
isotropic diffusion coefficient and taking into account these effects,
Ref.~\cite{Ahlers:2016njd} argued that Vela leads to the correct level of
anisotropy.
There are however two caveats in this conclusion: First, we recall that
the projection effect in $D_{ij}\nabla_j n$ does not change the dipole
amplitude, if the CR density is quasi-Gaussian~\cite{Savchenko:2015dha}.
Second we note that Ref.~\cite{Ahlers:2016njd} calculated the CR fluxes
from individual sources for isotropic diffusion.
This requires that the regular magnetic field is weak enough, $\eta\gsim 5$,
so that $D_\perp\simeq D_\|\simeq D_{\rm iso}$.
Otherwise, a calculation of the CR flux following the lines of
Refs.~\cite{Savchenko:2015dha,Kachelriess:2015oua,Kachelriess:2017yzq} would
be required.

Finally, we note that none of these works took into account that the Sun
resides inside the Local Bubble: As we discussed in section~2.1, the
strength and the structure of the magnetic field in the bubble wall and
the bubble interior is changed relative to the surrounding. Thus both
the magnitude and the direction of the dipole amplitude may be changed
compared to the predicition of Eq.~(\ref{single}).

\subsubsection{Explaining secondaries}
 \label{models_below_knee_secondaries}

\begin{figure}
\centering
\includegraphics[width=0.7\columnwidth,angle=0]{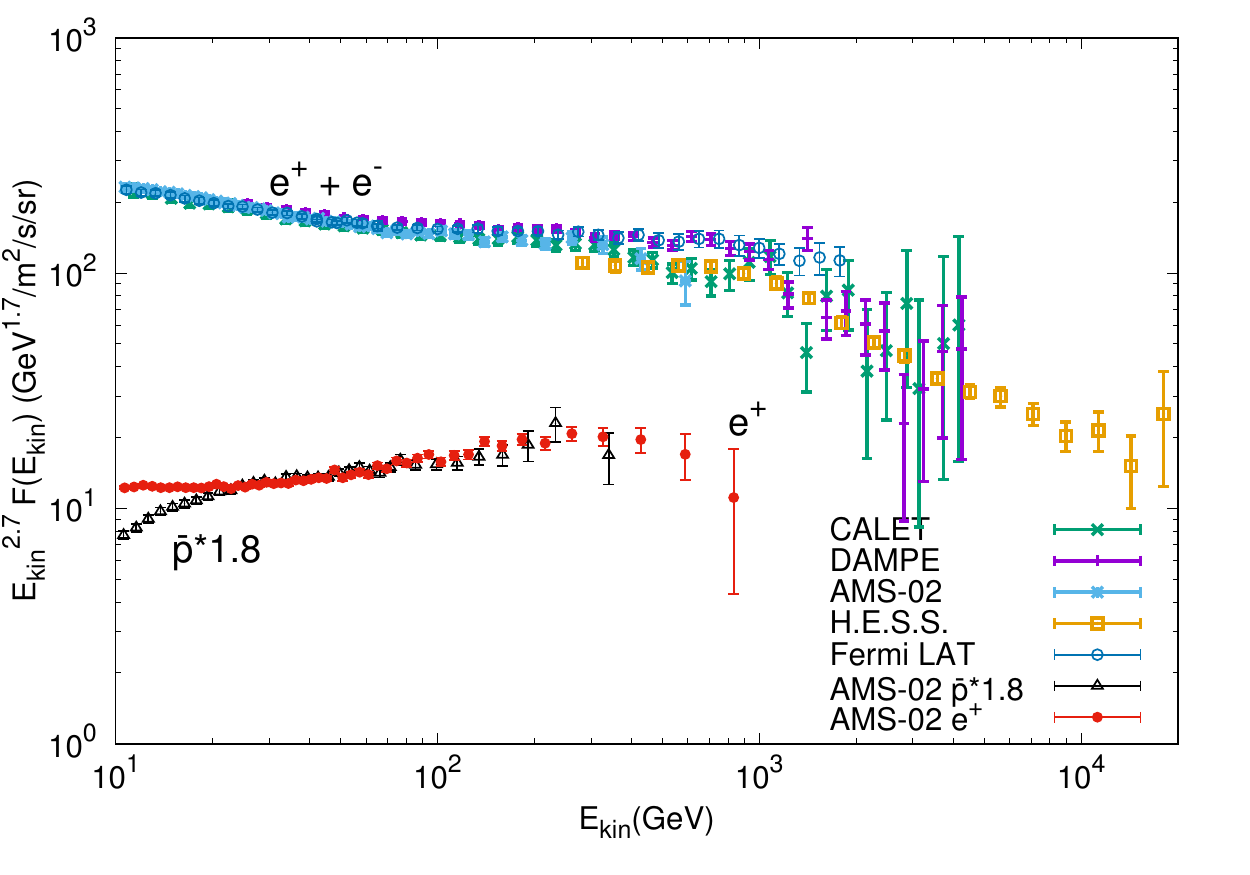}
\caption{Positron and antiproton fluxes measured by AMS-02~\cite{Aguilar:2019owu,Aguilar:2016kjl} compared to the
  electron plus positron flux. The antiproton flux is multiplied by the factor 1.8 expected for hadronic secondary production. 
  The experimental data for the electron plus positron fluxes are the same as in
  Fig.~\ref{electron_spec}.}
\label{secratios}
\end{figure}

Before reviewing the different classes of explanations for the
excess of secondaries compared to the expectation in the standard diffusion
picture, we want to stress a peculiarity of the observed  flux ratios:
The positron and antiproton fluxes above 100\,GeV repeat the 
spectral shape of the proton flux, see  Fig.~\ref{positron_data}.
In contrast, the electron flux is considerably steeper than the positron and
antiproton fluxes, as seen in  Fig.~\ref{secratios}. From Fig.~\ref{secratios},
we can observe that the break of the positron flux at 300\,GeV does not
coincide with the break in the electron plus positron flux  at 1\,TeV.
Moreover, the positron contribution to the combined electron plus
positron flux is small. Therefore the slopes and breaks in the primary
electron flux are unrelated to those in the positron flux.

The nearly scale invariance of hadronic 
interactions implies that the secondary fluxes produced in interactions of 
CRs on interstellar gas have a shape similar to the primary flux, if the 
grammage CRs cross is energy independent. Such an energy independence of
the grammage is achieved, e.g., if a relatively young source contributes
significantly  to the observed local CR flux. 
Moreover, the ratio $R(E)=F_{e^+}(E)/F_{\bar p}(E)$ of the positron
and antiproton fluxes is then mainly predicted by the properties of hadronic 
interactions,  being proportional to the ratio of their $Z$ factors,  
$R=F_{e^+}/F_{\bar p}\propto Z_{e^+}/Z_{\bar p} \simeq 1.8$~\cite{Kachelriess:2015oua,Lipari:2016vqk}. 
Such models predict also a corresponding increase of secondary nuclei like boron, which starts however at much higher rigidities.

The authors of Ref.~\cite{Blum:2017iwq} studied the secondary production
of CRs avoiding the use of a specific model as far as possible. Assuming 
universal CR spectra, they derived the energy dependent grammage crossed
by CRs from the B/C ratio measured by AMS-02. Then they used the grammage
to derive the resulting antiproton flux as well as an upper bound on  the
positron flux. Their analysis supports a secondary origin of the
positron and antiproton fluxes.

\paragraph{Pulsars}

The most straightforward explanation for the observed rise of the
positron-electron ratio $\Rpos$ assumes that primary sources of positrons
exist. Pulsars are natural candidates for such primary positron sources,
since electromagnetic pair cascades in their magnetospheres lead
to a large positron fraction, $\Rpos\simeq 1$. { In contrast, the antiproton 
flux would stay in this model on the level predicted in the standard
difffusion picture, if the multiplicity of electron-positron pairs is
large, as it is usually assumed for a cold MHD wind~\cite{Kirk:2007tn}.}
Since the energy losses of electrons increase fast with energy, the 
high-energy part of the $e^++e^-$ spectrum should be dominated by local 
sources, as pointed out already 30~years
ago~\cite{1987ICRC....2...92H,1989ApJ...342..807B}.
The expected electron and positron fluxes from pulsars, as well as the
resulting anisotopy, have been studied in detail using the isotropic
diffusion approximation~\cite{Hooper:2008kg,Grasso:2009ma}. From the known
pulsars, Geminga and PSR~B0656+14 (which was thaught to be associated with
the Monogem SNR) are the most promising candidates: They are located within
250--300\,pc from the Sun and relatively young, $T=370$\,kyr and $T=110$\,kyr,
respectively. 
For a pair-conversion efficiency around $\eps_{e^+e^-}\sim 40\%$
and an injection slope $\beta\sim 1.5$--1.7, these two pulsars 
were found to give a reasonable fit to the PAMELA positron data
and the total electron+positron flux measured by
Fermi-LAT~\cite{Hooper:2008kg,Grasso:2009ma}. The resulting anisotropy
of the  electron+positron flux peaks at $\sim 1$\,TeV with $\delta\sim 1$\%.

Recent HAWC observations of extended TeV gamma-ray emission from these two
pulsars confirmed that they are
local sources of accelerated electrons~\cite{Abeysekara:2017old}. The spatial
extension of the emission profile can be used to constrain the properties
of the magnetic turbulent field in the region close to these sources.
While the HAWC data are compatible with both isotropic Kolmogorov and
Iroshnikov-Kraichnan turbulence, the absence of an asymmetry in the emission
profile bounds the correlation
length as $L_{\rm c}\lsim 5$\,pc~\cite{Lopez-Coto:2017pbk}. For the best-fit
value of $B_{\rm rms}\simeq 3\mu$G for the strength of the turbulent field,
the resulting diffusion coefficient at 100\,TeV is
$D\lsim 5\times 10^{27}$cm$^2$/s.
This value is compatible with those shown in Fig.~\ref{iso1}, but much lower
than the one required to reproduce, e.g., the observed B/C ratio.
As a consequence of this slow diffusion, the HAWC collaboration concluded
that the contribution of Geminga and PSR B0656+14 to the positron excess
is negligble~\cite{Abeysekara:2017old}.

As we argued in Sec.~2.1, such a low value of the  diffusion coefficient is
connected to the absence of a strong regular field. Thus the hydrodynamic
turbulence close to the source
might have entangled the regular field, increasing the value of the turbulence
level $\eta$. Alternatively, instabilities driven by escaping CRs could have
increased $\eta$~\cite{DAngelo:2015cfw,Nava:2016szf}.
The possibility of strong magnetic field amplification is however
disfavoured by the small value of $B_{\rm rms}\simeq 3\mu$G and the
relatively small energy density of CRs compared to the ISM.
In either case, the region with a reduced diffusion coefficient should
be restricted to the close surrounding of the source. Thus two-zone
diffusion models have been used to reevaluate the positron contribution of
nearby pulsars, see e.g.\
Refs.~\cite{Hooper:2017gtd,Fang:2018qco,Profumo:2018fmz}.
Using a standard diffusion coefficient outside the sources, the
positron flux is strongly increased relative to the analysis of
Ref.~\cite{Abeysekara:2017old} and can fit well the positron data.
However, the presence of positrons with energies in the 10--1000\,GeV range
in the sources can be tested directly looking for GeV photons using Fermi-LAT.
No GeV photon halo around Geminga and PSR~B0656+14 has been found in
the search performed in Ref.~\cite{Shao-Qiang:2018zla},
and the derived upper limits were used
to constrain their contribution to the observed positron flux as
$\lsim 15\%$. The similar analysis~\cite{DiMauro:2019yvh} detected a weak
GeV halo around Geminga and set an upper limit  of $20\%$ to its contribution
to the observed positron flux. These limits disfavour young pulsars as
explanation for the positron excess.
The question if the ensemble of old pulsars can explain the positron excess
was studied in Ref.~\cite{Cholis:2018izy}. Scanning over a wide parameter
space in pulsar properties and diffusion models, they identified cases
in which the observed positron data can be reproduced. Successful
models using Kolmogorov diffusion predict a positron fraction 
rising to $\Rpos\simeq 0.4$ at $\simeq 300$\,GeV which flattens at
higher energy---a behaviour which is in tension with the latest AMS-02
data.

Finally, we mention three alternative pulsar scenarios: In the first
one, the contribution from the population of all millisecond pulsars
was studied~\cite{Venter:2015gga}. While  pair cascades from the
magnetospheres of isolated millisecond pulsars cut off around a few
tens of GeV and thus cannot contribute to the high-energy rise of $\Rpos$,
electron-positrons are accelerated up to tens of TeV in the strong
intra-binary shocks of black widow and redback binary systems. In particular,
Ref.~\cite{Venter:2015gga}  argued that 
the contribution to the positron flux by black widows and redbacks may
reach levels of a few tens of percent at tens of TeV, depending on model
parameters. 
A second alternative pulsar scenario was suggested recently in
Ref.~\cite{Lopez-Coto:2018ksn}: Using the low diffusion coefficient
deduced from HAWC observations, an undecteted pulsar with
age $\lsim 300.000$\,yr inside the Local Bubble, i.e.\ at a distance
of 30--80\,pc, was proposed as explaination for the positron excess. 
This model predicts a rasing positron fraction $\Rpos$ reaching 25\%
at 1\,TeV, which is disfavoured by the indication of a break at 300\,GeV
in the positron spectrum measured by AMS--02.
A third alternative pulsar scenario are pulsar wind nebulae (PWN) with
bow shocks~\cite{Blasi:2010de,Bykov:2017xpo}. These shocks form, if PWNe
move relative to the ambient ISM
with supersonic speeds. Particles accelerated at the termination surface of
the pulsar wind may undergo reacceleration in the converging flow 
formed by the outflow from the wind termination shock and the
inflow from the bow shock, leading to a very hard energy spectra,
$\beta\simeq 1.2$. For a conversion efficiency of $\eta\simeq 20\%$ of the
pulsar spin-down energy into the acceleration of pairs and steepening
of the spectrum at $E= 500$\,GeV to $\beta\simeq 2.3$, the measured
positron fraction can be well reproduced.

\paragraph{Reacceleration in SNRs}     

\begin{figure}
\centering
\includegraphics[width=0.49\textwidth]{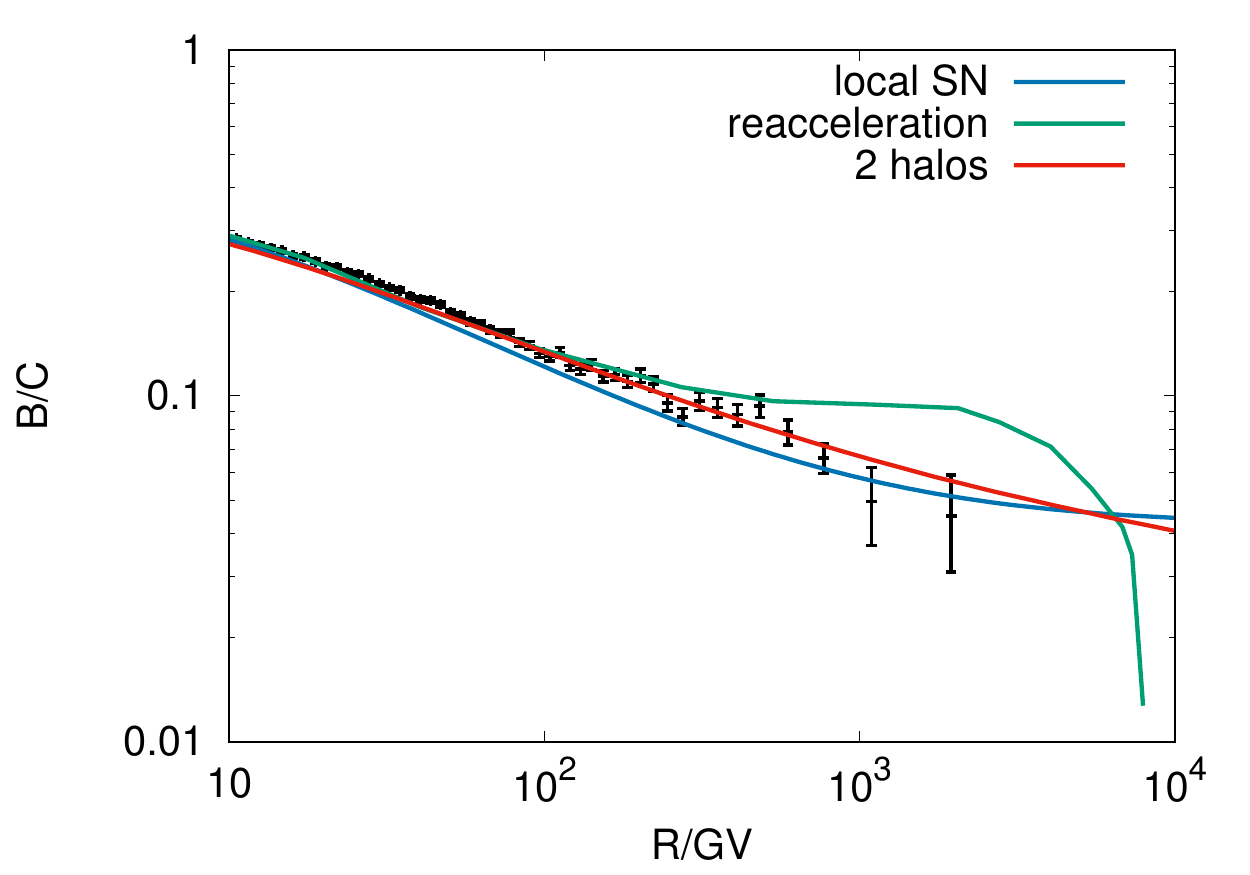}
\includegraphics[width=0.49\textwidth]{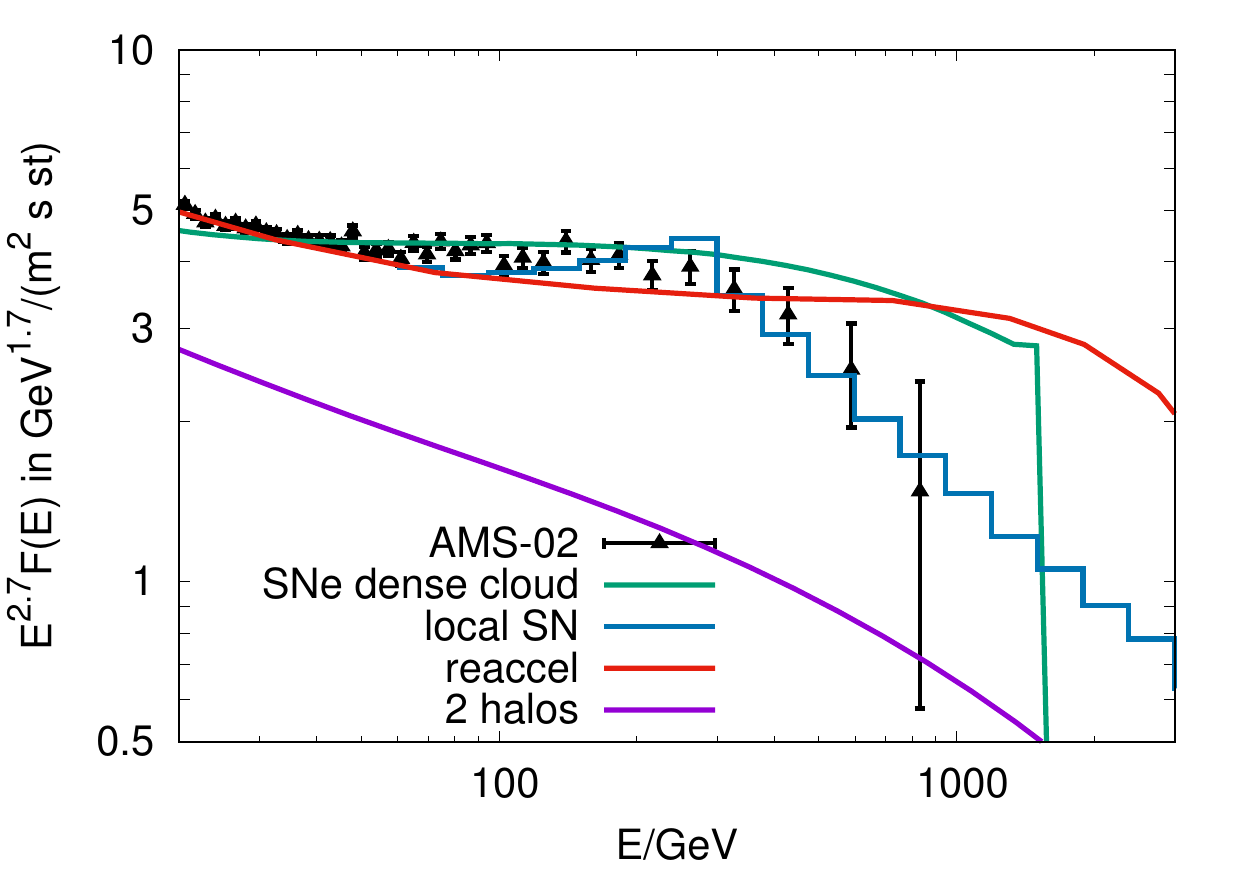}
\caption{{\em Left panel:\/}
  Boron-to-carbon ratio as function of rigidity predicted in the
  reacceleration model~\cite{Mertsch:2014poa} for $R_{\max}=3\times 10^3$\,GV
  in the two-halo model from Ref.~\protect\cite{Tomassetti:2015mha},
  and from
  a local SN~\protect\cite{Kachelriess:2017yzq}.
  {\em Right panel:\/} 
  The positron flux in the reacceleration model from Ref.~\protect\cite{Mertsch:2014poa}
  for $R_{\max}=3\times 10^3$\,GV,
  from SNe in a dense cloud~\protect\cite{Kohri:2015mga},
  in the two-halo model from Ref.~\protect\cite{Tomassetti:2015mha},
  and from
  a local SN~\protect\cite{Kachelriess:2017yzq}.
\label{models}}
\end{figure}

Positrons produced by hadronic interactions in the vicinity of a SNR shock
participate
in the acceleration process and have therefore a flatter energy spectrum than
primary electrons. In Ref.~\cite{Blasi:2009hv}, it was suggested that 
the total spectrum of all positrons produced in a SNR is therefore flatter
and that thereby the rise of the positron-electron ratio can be explained.
Later, this scenario was applied to antiprotons~\cite{Blasi:2009bd} and
secondary nuclei like boron or titanium nuclei~\cite{Mertsch:2009ph}.
All these works solved the diffusion equation in the stationary limit. Since
therefore the number of produced secondaries is infinite, their production
rates have to be normalised by hand: While the rate of positrons produced
in the shock vicinity and thus taking part in the acceleration process is
proportional to the time $t_{\max}$ acceleration is effective,
the rate of positrons generated downstream is proportional to
$D(E_{\max})/u^2_1$. Here, $D(E_{\max})$ is the diffusion
coefficient at the maximal acceleration energy $E_{\max}$ and $u_1$ the
advection velocity upstream. In general, these parameters are connected by
the relation $D(E_{\max})/u^2_1\sim t_{\max}/20$~\cite{Lagage:1983zz}, and 
can therefore not be chosen arbitrarily. If however these parameters are 
treated as independent quantities, the
relative size of the two components can be changed. Choosing them such
that the reaccelerated component is enhanced, a raising secondary-to-primary
ratio can be generated. Reference~\cite{Blasi:2009hv} argued that such a choice
of parameters is motivated as an effective way to include the time evolution
of the SNR. The boron-to-carbon ratio predicted in these
reacceleration models for $R_{\max}=3$\,TV is shown in the left panel
of Fig.~\ref{models}~\cite{Mertsch:2014poa}. For larger values of 
$R_{\max}$, the rise of B/C becomes more pronounced and evolves into a
bump, while it disappears for $R_{\max}\lsim 1$\,TV. In the right panel
of Fig.~\ref{models}, the positron flux predicted for $R_{\max}=3$\,TV is
shown.

In Refs.~\cite{Kachelriess:2011qv,Kachelriess:2012ag}, time dependent 
Monte Carlo simulations of the acceleration process and secondary
production were performed. It was found that at any time, the
reaccelerated and the downstream contributions to the secondary spectra
add up approximately to a standard $1/E^2$ spectrum. This result suggests
that the time dependence of parameters like the shock velocity or the
magnetic field strength do not lead to an enhancement of secondary fluxes
and, therefore,  that reacceleration of secondaries in SNRs cannot explain
the observed rising positron fraction. This conclusion is supported by the
non-observation of a rise in other secondary ratios, as e.g.\ in B/C.

\paragraph{Supernovae in dense clouds}

The authors of Refs.~\cite{Fujita:2009wk} suggested that a
recent cluster of SN explosions happened in a nearby dense gas cloud.
In this scenario, the cloud has been ionized by the ultraviolet radiation
of OB stars. Then it is argued that particle acceleration continues
in the radiative phase of the evolving SNRs, leading to a hard CR 
spectrum with slope $\alpha<2$~\cite{Yamazaki:2006uf}. Protons interacting
with the dense gas inside the cloud produce secondaries until the cloud is
destroyed. Then both the primary CRs and the produced secondaries are released.

In the subsequent work~\cite{Kohri:2015mga}, the same authors adapted the
parameters of their model using now as slope of the CR spectrum $\beta= 2.15$
and as injected energy in CRs $E_{\rm CR}=3\times 10^{50}$\,erg. Moreover,
they assumed a gas cloud with density  $n\sim 50$/cm$^3$ at the distance
100--200\,pc.  Particles were
released $5\times 10^{5}$\,yr ago, and diffuse afterwards isotropically with
$\delta=0.4-0.6$. For these values, the resulting flux of secondaries can
explain the rising positron fraction and leads also to an increase of
the antiproton flux~\cite{Kohri:2015mga}. The results of this model
are compared in Fig.~\ref{models} to the positron 
flux measured by AMS-02. The model predicts a rather sharp cutoff in the
positron flux, since the positron
production stopped $5\times 10^{5}$\,yr ago. Therefore synchrotron cooling
leads to a cutoff at $E\sim $\,TeV in the positron spectrum. In contrast,
the AMS-02 data indicate a break  at $\sim 300$\,GeV,
suggesting the continuous injection of positrons from one (or several) older
sources.

\paragraph{Short CR confinement time}

In Ref.~\cite{Lipari:2016vqk}, it was stressed that the ratio
$R=F_{e^+}/F_{\bar p}\propto Z_{e^+}/Z_{\bar p} \simeq 1.8$
suggests that hadronic  secondary production is the main source
of these antiparticles. In a detailed study of how the softening
feature induced by the energy losses emerges in the electron and
positron spectra, two possibilities and their consequences were
outlined~\cite{Lipari:2018usj}: In the traditional view discussed in
Section~2.3.3, the softening break in the electron and the positron
spectrum appears at the energy $\sim 1$\,GeV. The alternative option
advocated in Ref.~\cite{Lipari:2018usj} identifies the break in the electron
flux at $\simeq 1$\,TeV seen by the H.E.S.S.\ collaboration as the softening
caused by the energy losses of electrons. This implies a much
smaller confinement time of CRs, around 1\,Myr at a few GeV.
Since the  confinement time is smaller than in the standard scenario,
the diffusion coefficient is accordingly increased. As a result, the number
of CR sources contributing to the locally observed flux is larger
than usually.
Another important prediction of this scenario are equal propagation
properties of protons, electrons and their antiparticles below 1\,TeV.
In this scenario, the indications for a break in the positron spectrum
at 300\,GeV have to be therefore either wrong---or this break is caused
not by energy losses, but is instead connected to spectral
features of the positron sources.

\subsubsection{Explaining the breaks and non-universal rigidity spectra}
\label{models_below_knee_breaks}

We review next models which explain the break at 300\,GV in the CR nuclei
spectra, the rigidity dependent p/He ratio and/or the longitude dependence
of the CR spectrum.

\paragraph{Radial dependence of the turbulence}

We discussed in section~\ref{sec_spec_breaks} that gamma-ray
observations show evidence for a slope of the CR spectrum
varying with Galactic longitude.
In Refs.~\cite{Gaggero:2014xla,Gaggero:2015xza},
a gradual increase in the slope $\delta=\delta(R)$  of the CR diffusion
coefficient with increasing distance $R$ from the
Galactic centre was proposed. As a consequence, the CR flux and thus the
resulting gamma-ray spectra are harder in the inner Galaxy.
Combined with strong advection at
$R<6.5$\,kpc, the model could fit the CR proton flux, B/C and the diffuse
gamma fluxes choosing $\delta(R)=aR+b$ with $a=0.035$\,kpc$^{-1}$ and
$b=0.21$. This model did not address the origin of the breaks in the
CR spectra or the positron excess. It is however an example how
Iroshnikov-Kraichnan diffusion, $\delta(R_\odot)=0.5$, can be made
compatible with the observed B/C
ratio. Note also that a change in the slope $\delta$  of the CR diffusion
coefficient implies a corresponding change in the power spectrum
${\cal P}(k)\propto k^{-\gamma}$ of the turbulent magnetic field, which are
connected by $\gamma=2-\delta$. While it is natural that the normalisation
of $D$ and ${\cal P}(k)$ changes towards the Galactic center, it is
surprising that the type of turbulence depends on the Galactocentric
distance $R$.

\paragraph{Two-halo models}

In Ref.~\cite{Tomassetti:2015mha}, it was suggested to divide the Galactic
CR halo into two zones in which the rigidity dependence of the diffusion
coefficient differ. In a version of this model updated
after the release of the B/C data by AMS-02~\cite{Tomassetti:2017umm},
CRs diffuse  slower than usually assumed, $D_{\rm i}(\R)\propto \R^{0.18}$,
in an inner halo around the Galactic disk with the vertical extension
$|z|<L_{\rm i}\simeq 0.5$\,kpc. In contrast, CRs diffuse faster than usually
in the outer halo, $D_{\rm o}(\R)\propto \R^{0.73}$, which has the extension
$L_{\rm i}<|z|<L_{\rm o}$ with $L_{\rm o}\simeq 4$\,kpc. 
The observed energy spectrum of primaries, which is usually given by
$F(E)\propto Q(E) L/D(E)$, is then modified to  
$$
F(E)\propto \frac{Q(E)}{D_{\rm i}(E)/L_{\rm i}+D_{\rm o}(E)/L_{\rm o}}.
$$
As a result, there are breaks in the slopes of primaries as well as
in the secondary-to-primary
ratios at high energies. This model describes successfully the hardening of the 
proton flux, but requires a different slope of the proton injection spectrum
than that of other nuclei. Another prediction of this model is the decrease
of the B/C ratio as $\propto \R^{-0.15}$ at high energies, cf.\ with
Fig.~\ref{models}. Moreover, it was speculated that the weak energy
dependence  $D_{\rm i}(\R)\propto \R^{0.15}$ may explain the plateau in the
CR dipole anisotropy.   While the positron flux
is enhanced relative to standard diffusion models, the flux shown in
the right panel of Fig.~\ref{models} falls short of
the observed data. Therefore this model requires that pulsars generate
the observed high-energy positron flux. Moreover, the antiproton flux
is in this model in tension with the AMS-02 data, even if an additional
antiproton component from interactions in CR sources is
included~\cite{Tomassetti:2017izg}.

\paragraph{Self-generated turbulence}

The authors of Refs.~\cite{Blasi:2012yr,Aloisio:2013tda,Aloisio:2015rsa}
connected the breaks in the
CR spectrum with a change in the energy dependence of the diffusion
coefficient. In particular, they associated the break at 300\,GV
with the transition from self-generated CR turbulence to external
turbulence injected by SN explosions and stellar winds associated with
OB associations.

In such a model, the transport equation~(\ref{transport}) of CRs is coupled
to an evolution equation of plasma waves which describes the generation and
damping of these modes,
\begin{equation}
\frac{\partial}{\partial k} \left[ D_{kk}\frac{\partial W}{\partial k} \right]
- \Gamma_{\rm CR} W = Q_w(k)  ,
 \label{modes}
\end{equation}
Here, $ Q_w(k)$ is the generation rate of magnetic turbulence with wave-number
$k$ by SNe and stellar winds, while $W(k)$ is the energy density of the magnetic
turbulence normalised such that $B_{\rm rms}^2=B_{\rm tot}^2 \int dk\, W(k)$. 
The turbulent cascade is determined by $D_{kk}=c_k|v_A|k^{7/2}W^{1/2}$ with
the constant $c_k\sim 0.05$~\cite{1995ApJ...452..912M}.  
Finally, $\Gamma_{\rm CR}$ is the rate by which CRs  amplify 
the magnetic turbulence. In particular, a CR gradient along the $z$ direction
will lead to a streaming instability with growth rate~\cite{1975MNRAS.173..255S}
\begin{equation}
  \Gamma_{\rm CR}  = \frac{4\pi}{3}  \frac{v_A}{kW(k)B_{\rm tot}^2}
  \left[ p^2 n(p) \right]_{p_{\rm res}} ,
 \label{streaming}
\end{equation}
with $p_{\rm res}$ denoting the CR momentum resonant with the turbulent
field mode $k$.
The streaming instability can lead to self-confinement of CRs, if its
growth rate exceeds the rate of wave damping. Since less than 1\% of the
mass in the ISM is ionized, neutral-ion damping was considered first as
the main damping mechanism. But despite the large mass fraction, partly
ionized regions occupy only a small fraction of the volume of the ISM.
Therefore ion-neutral friction prevents self-confinement of CRs even at
low energies only inside cold clouds.
In a fully ionized gas, nonlinear Landau damping where thermal ions
Landau resonate with two turbulent wave modes is operative.
Additionally, inhomogeneities in the ordered field lead to wave damping.
In particular, the long-wavelength modes of the ``standard'' turbulence
injected by  SNe and stellar winds leads to a shearing of Alfv\'en waves,
an effect which is however not well understood. As a result, the
energy $E_\ast$ below which self-generated turbulence dominates
can be only estimated, with $E_\ast \sim 100$\,GeV as a typical
value~\cite{2013PhPl...20e5501Z}.  Note also that the variation in the CR
density and the damping rate between, e.g., the Galactic disk and halo,
should lead to a corresponding change in the value of $E_\ast$.

\begin{figure}
\centering
\includegraphics[width=0.47\textwidth]{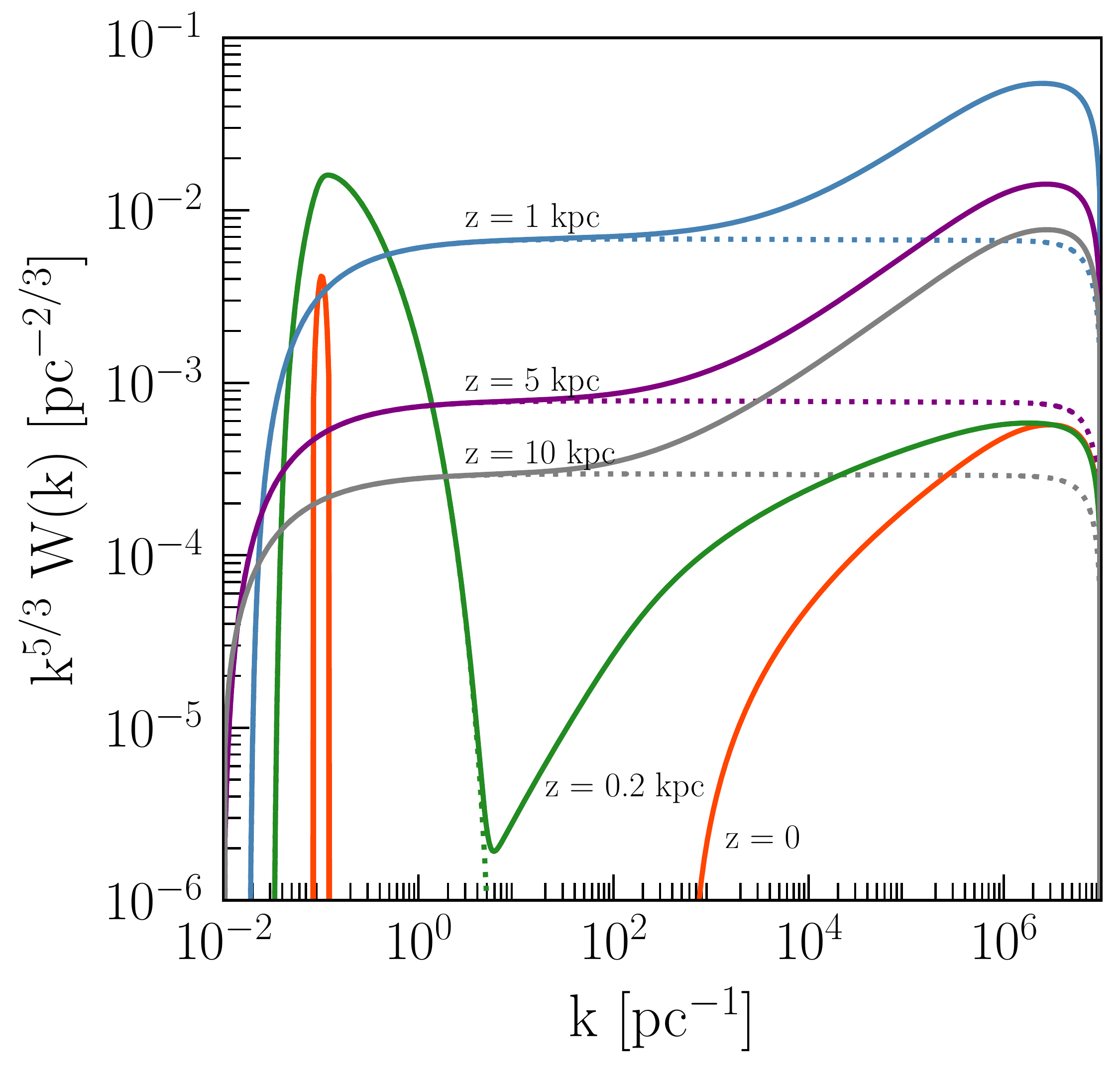}
\hskip0.2cm
\includegraphics[width=0.47\textwidth]{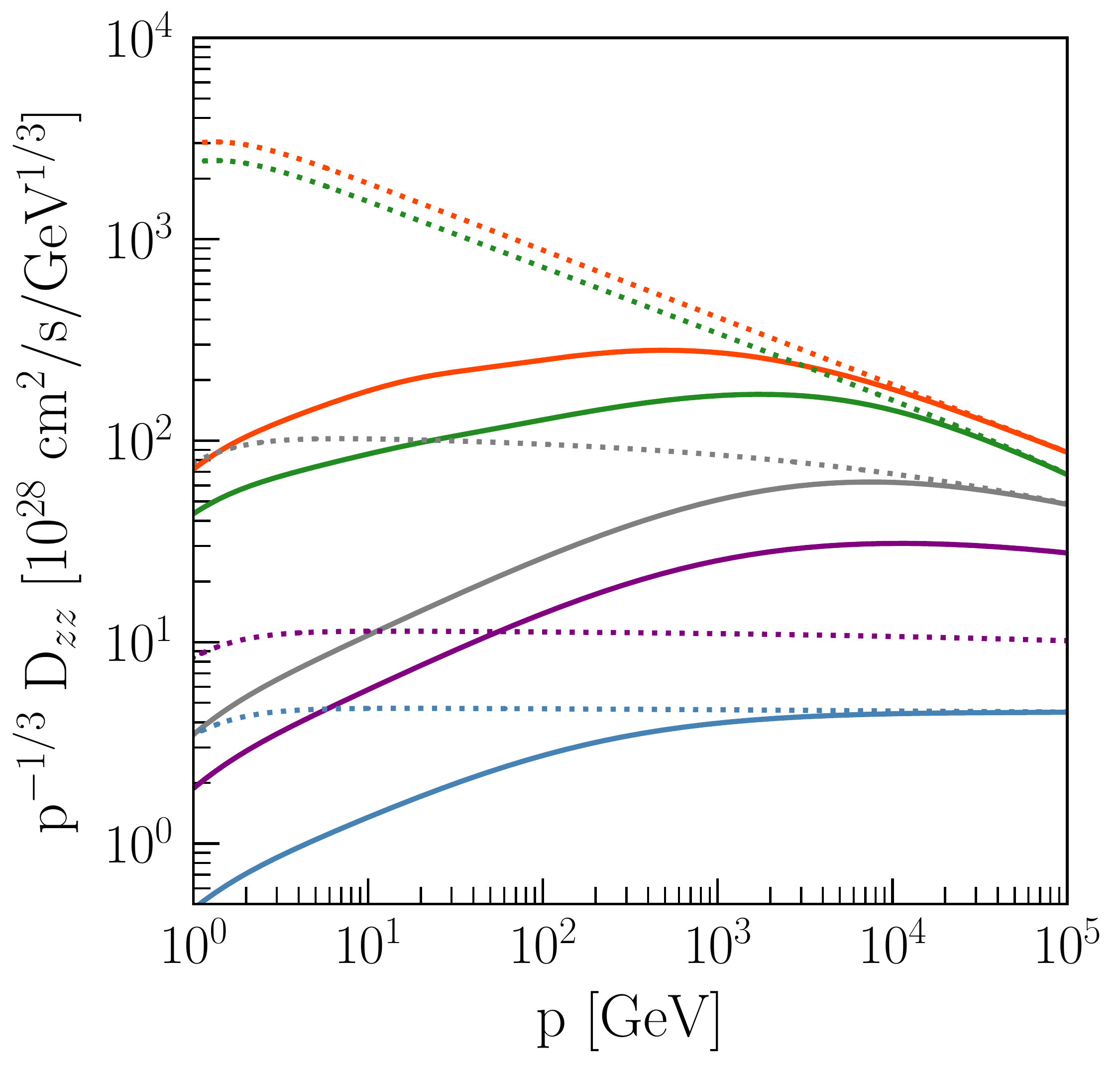}
\caption{{\em Left panel:\/} Wave spectrum as a function of $k$ with (solid lines) and without (dashed lines) the contribution of self-generated waves at different $z$. 
  {\em Right panel:\/} Diffusion coefficient as a function of momentum for different values of $z$; figures adapted from Ref.~\protect\cite{Evoli:2018nmb}.
\label{Halo}}
\end{figure}

In the model of Refs.~\cite{Blasi:2012yr,Aloisio:2013tda,Aloisio:2015rsa},
the turbulent cascade introduces a scale $z_0\sim v_AL_{\max}^2/D_{kk}$
below which the turbulence is mainly self-generated. Effectively, this
distance defines the boundary of the CR halo where CRs diffuse in
Kolmogorov-like turbulence and an outer halo where CRs scatter on
self-generated turbulence. Thus this model may be seen as physically motivated
realisation of a two-halo model, although with quite different parameters.
It reproduces the spectral break at rigidity $\simeq 300$\,GV, but does not
address the origin of the harder proton spectrum.

In Ref.~\cite{Evoli:2018nmb}, the coupled transport equation~(\ref{transport}) 
for CR protons and the evolution equation~(\ref{modes}) of plasma waves were
generalised to include a dependence on the vertical distance $z$ to the
Galactic disk and the advection of turbulence. In this way, a Galactic halo
of CRs 
and self-generated magnetic turbulence was generated in a self-consistent way 
without introducing by hand a boundary for the CR halo. 
 In the left panel of Fig.~\ref{Halo}, the resulting
wave spectrum at different distances $z$ to the Galactic disk is shown
with (dashed lines) and without (solid lines) self-generated turbulence.
The resulting diffusion coefficients shown in the right panel of
Fig.~\ref{Halo} deviate for energies $\lsim 1$\,TeV from the values
for the Kolmogorov background turbulence. This implies that on scales smaller
than $10^{-3}$\,pc, the slope of the power spectrum of the turbulent magnetic
field should deviate from $\gamma=5/3$. The proton spectrum obtained in
this model
reproduces well the experimental data from Voyager-1 up to the CREAM energy
range.

\paragraph{Two-population models}

This class of models explains the break at $\R\simeq 200$\,GV as the
transition  between a source type dominating the low-rigidity and
another type dominating the high-rigidity part of the spectrum. For instance,
the authors of Ref.~\cite{Zatsepin:2006ci} proposed a three component
model fitting the CR spectra between 10\,GeV and 100\,PeV. They suggested
that the lowest energy component with $\R_{\max}=200$\,GV is generated
by novae, the intermediate component up to the knee with $\R_{\max}=4$\,PV
by isolated SNe, while the third component above the knee is generated
by multiple SNe occurring inside of superbubbles.

In Refs.~\cite{Thoudam:2011aa,Thoudam:2013iia}, the hardening was connected to
the transition between the flux from many background sources at low-energies 
to a component dominated by local sources at higher energies. 
They summed up the contributions from 10~SNRs with distance $<1$\,kpc
and age less than 1\,Myr assuming isotropic diffusion with $\delta=0.54$ and
$D_0=1.6\times 10^{28}$cm/s$^2$ at 3\,GV, while the background was fitted
to a static diffusion model. At energies $E\gsim 1$\,TeV, the total
flux is dominated by the contribution fom Vela, while at lower energies
CRs from Vela have not reached yet the Earth. It is this transition which
leads to the break in the CR nuclei spectra.
A similar analysis was performed in Ref.~\cite{Bernard:2012pia}, which
treated remote and local sources in a consistent way, and found results
in agreement with those of Ref.~\cite{Thoudam:2013iia}.

\subsubsection{Explaining several features}
\label{models_below_knee_global}

On the first sight, the excess of secondaries as positrons and antiprotons
and the spectral features in the CR nuclei fluxes seem to be disconnected.
Most models presented up to now were therefore tailored to explain only one of
these anomalies. Typically, models aimed to explain the breaks in the nuclei
spectra were suggested to be combined with pulsars as sources of
positrons. There are three reasons why a unified solution to all
theses anomalies is desirable: First, the pulsar explanation for
the positron excess does not address the antiproton excess.
Second, the absence of a strong GeV photon halo around Geminga and
PSR~B0656+14 disfavours these pulsars as  explanation for the positron excess.
Last but not least, Occam's razor favour models explaining these
anomalies by a single model.

An example for such a model is Ref.~\cite{Tomassetti:2015cva} which tried to
explain the observed excess of positrons and antiprotons as well as the
hardening of the rigidity spectra of CR nuclei at the same time. They
considered the suggestion that secondaries are reaccelerated in old SNRs
to explain the rising positron fraction. At the same time, the non-observation
of a corresponding rise in B/C limits the maximal rigidity to which such
sources can accelerate to $\R_{\max}\simeq 1$\,TeV. Thus the old SNRs
responsible for the observed positrons and antiprotons cannot explain
at the same time the CR flux up to the knee. Assuming then a second population
of distant SNRs with stronger shocks and magnetic fields leading to a flatter
acceleration spectrum results then in a hardening of the CR fluxes at the
energy, when this new source population dominates the total flux. This model
predicts a decrease of B/C as in conventional diffusion model,
${\rm B/C}\propto \R^{-\delta}$, with ${\rm B/C}\simeq 0.01$ at
$\R=10^5$\,GV.

A closely related model is the one of  Ref.~\cite{Yang:2018nhs}. It
combined the idea of a two-population model 
with the suggestion that the grammage $X_s$ accumulated by CRs close to the
sources might be larger than traditionally supposed. A physical motivation
for this larger grammage might be the confinement of CRs in self-generated
turbulence~\cite{DAngelo:2015cfw}. For an energy-independent escape of
CRs from sources, $X_s\simeq 1.5$\,g/cm$^2$ and an acceleration spectrum
with $\alpha=1.9$ for the high-energy population the positron and
antiproton data as well as the B/C ratio could be reproduced.

{
The production of secondaries as positrons and antiprotons and the spectral
features in the CR nuclei fluxes were discussed also within the model of
massive stars exploding into their winds~\cite{Biermann:2018clk}. A good
fit to the AMS-02 positron data is obtained assuming that they to originate
from triplet pair production on background photons close to the source.
However, the fit in Ref.~\cite{Biermann:2018clk} is based on a monochromatic
energy spectrum for the background photons; employing instead a realistic
energy spectrum would impact the resulting positron spectrum. Since the
CR primary spectrum consists of two components with the slope
$E^{-2}$ and $E^{-7/3}$, respectively, which can be combined with two
different spectra of magnetic turbulence, the exact shape and the position
of possible breaks in the secondary spectra of antiprotons and nuclei like
lithium are difficult to predict. 
}

\paragraph{Local source and anisotropic diffusion}

\begin{figure}
\centering
\includegraphics[width=0.7\columnwidth,angle=0]{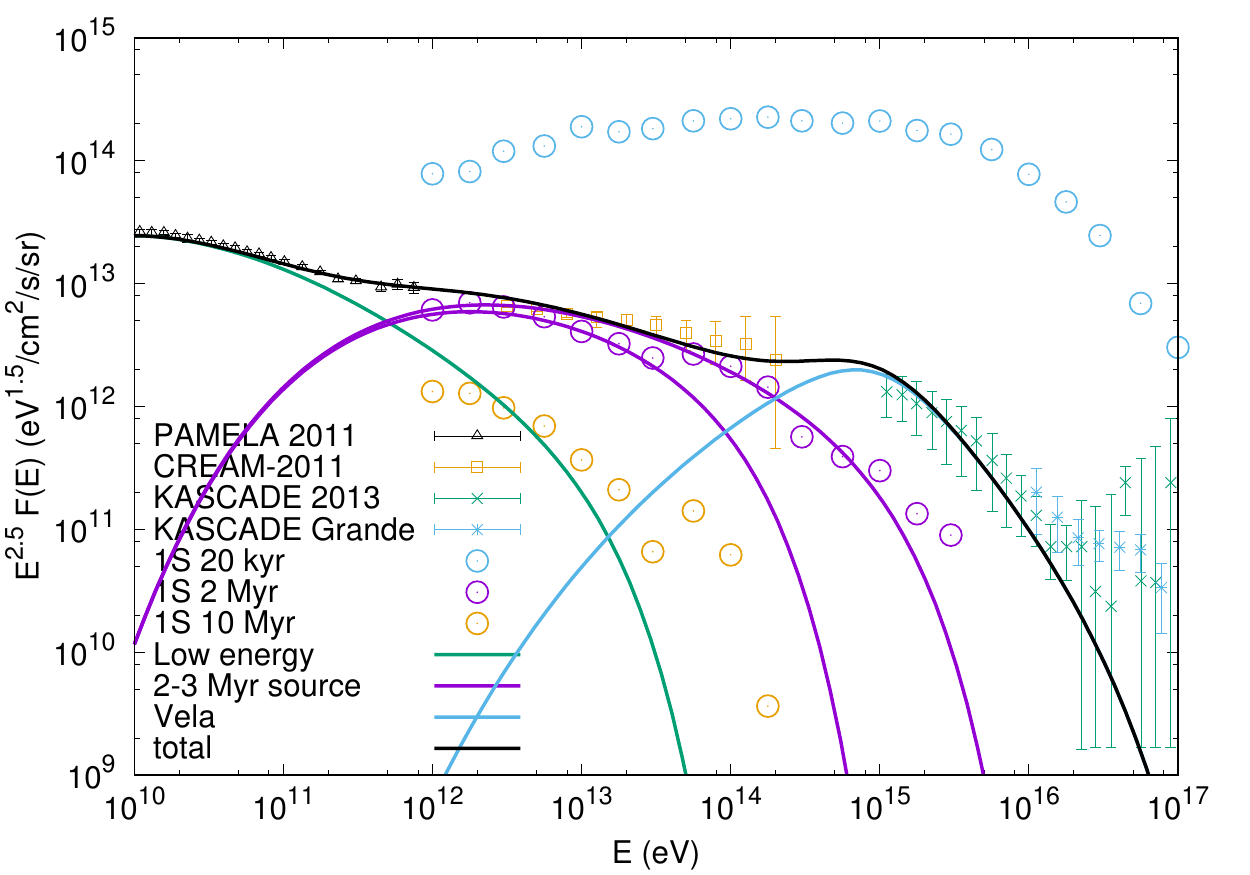}
\vskip0.5cm
\caption{Contribution from the average of all Galactic (blue line and
  green erorbars) sources and a local source (magenta lines, for $E_{\max}=10^{14}$
  and $10^{15}$\,eV) to the CR proton intensity together with experimental data
  (orange and blue errorbars). 
  Additionally, the flux from the local source is shown by dots for three
  source ages without energy cutoff.
\label{Adata}}
\end{figure}

We have already discussed that a source with age  $T=2$--3\,Myr and distance
$R=300$\, pc can explain the observed plateau in the dipole anisotropy, if
its contribution to the total CR flux in this energy range is of order of one. 
Now we review how the model of a local source combined with anisotropic
diffusion can address successfully also the spectral anomalies in the CR flux. 
For the age $T=2$--3\,Myr, the CR density below 100\,TeV is still in the
quasi-Gaussian regime~\cite{Kachelriess:2015oua}, when particle escape
can be neglected. This implies an approximately energy-independent grammage
crossed by CRs, explaining thereby the same slope of the secondary and
proton spectra.

In Fig.~\ref{Adata}, the proton flux of the local SN is
shown with magenta lines for two values of the maximal acceleration  energy,
$E_{\max}=10^{14}$ and $10^{15}$\,eV, respectively. The contribution to the
proton flux of the background of older SNe at low energies is assumed to
have the slope $\alpha=2.5$ (green line), while the contribution at
higher energies (blue line) was calculated in Ref.~\cite{Giacinti:2015hva}.
Summing up these three contributions reproduces the observed Galactic flux of
protons, represented by the data of PAMELA~\cite{Adriani:2011cu},
CREAM~\cite{Yoon:2011cr}, KASCADE and KASCADE-Grande~\cite{Apel:2011mi}.
Additionally to the flux at 2\,Myr, the flux from the local source at
20\,kyr and 10\,Myr is shown. The time dependence of the observed spectrum
from a single source of both the amplitude and the shape of the spectrum
can be clearly seen. At early times, 20\,kyr after explosion, the spectrum
is close to the injected $1/E^{2.2}$ slope. 
The amplitude of the flux at this early time is much higher than the one
observed, because it was assumed that the Earth is close to the magnetic
field line going through the source. Since CRs propagate preferentially
along field lines, the CR flux is enhanced compared to the isotropic case.
At late times, 10\,Myr, the flux is below the observed one at all energies.
Note that the cutoff at high energies in the observed flux is caused by the
maximal acceleration energy of the source. In contrast, the cutoff at low
energies is introduced by the perpendicular distance of the Earth to
the magnetic line going through the source. Since the diffusion in
perpendicular direction is slow, low energy CRs did not have time to reach
the Earth~\cite{Kachelriess:2017yzq}.

\begin{figure}
  \centering
\includegraphics[width=0.5\columnwidth,angle=0]{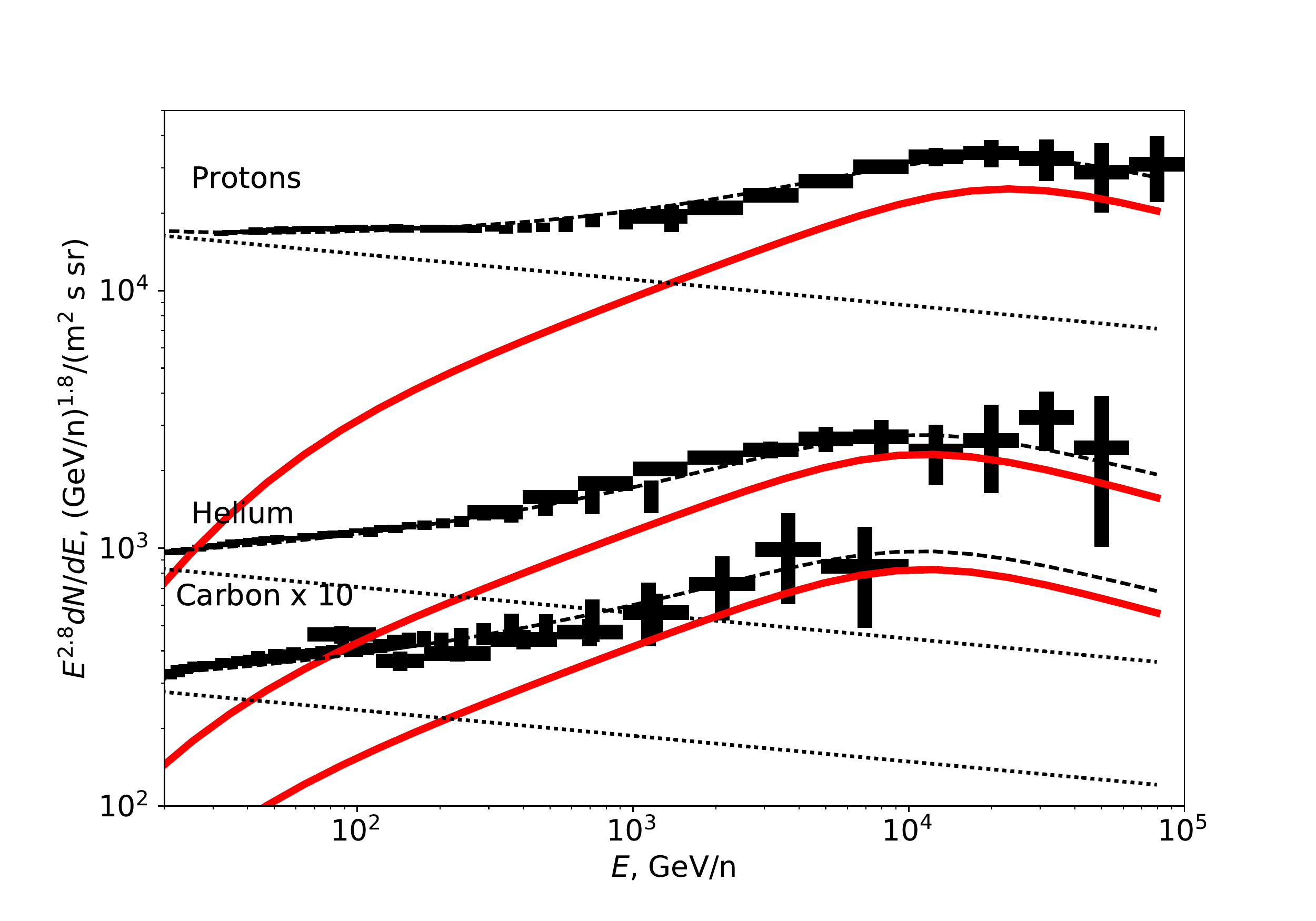}
\includegraphics[width=0.45\columnwidth]{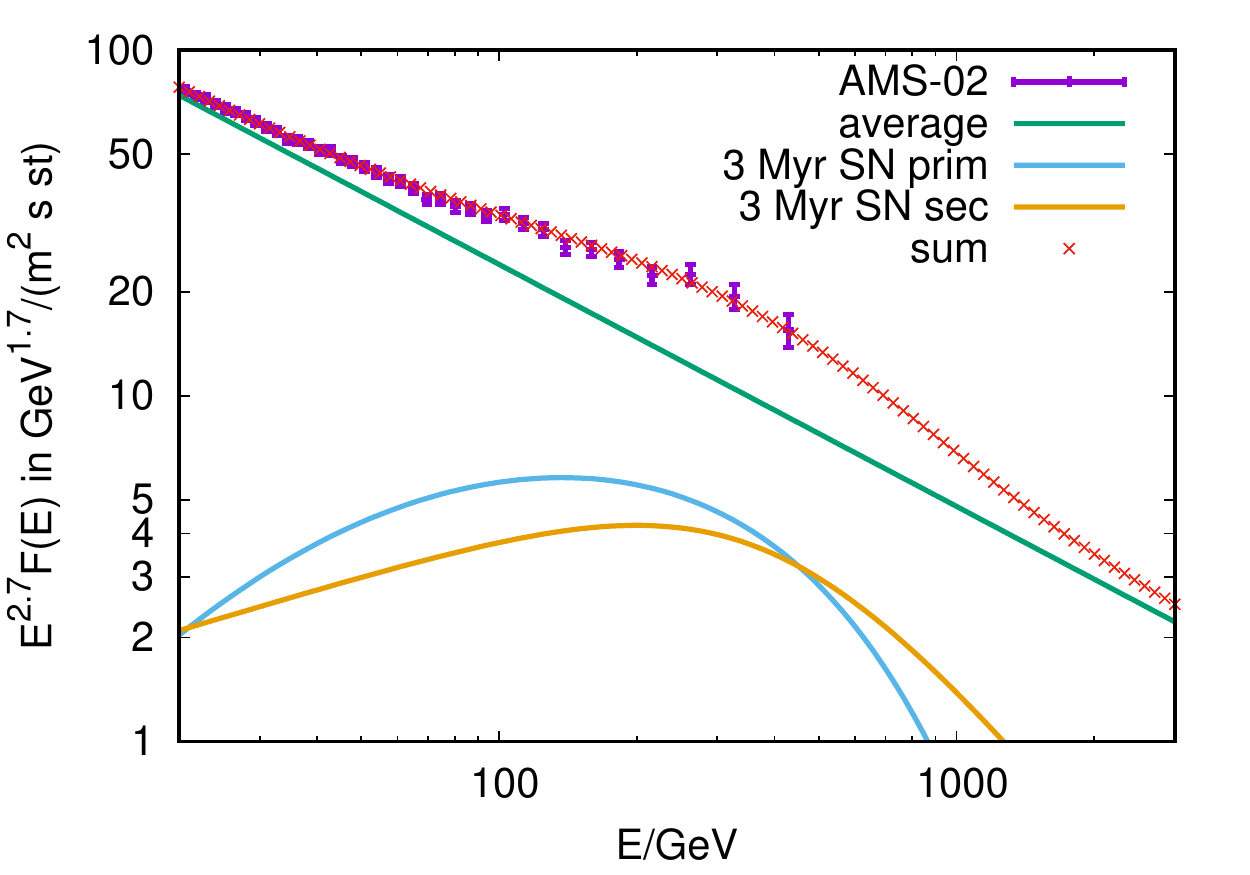}
\caption{{\em Left panel:\/} The flux of CR protons, helium and carbon
  measured by AMS-02 and CREAM-III as function of energy/nucleon shown
  together with  a  two-component model consisting of the average CR spectrum
  (dotted lines)  and the local source contribution (solid red  lines),
  from Ref.~\protect\cite{Kachelriess:2017yzq}.
  {\em Right panel:\/} Electron flux measured by AMS-02 compared to the 
  contributions for the local source (blue and orange lines) and average
  (green line) components. 
\label{fig:single} }
\end{figure}

In the left panel of Fig.~\ref{fig:single}, the proton, helium and carbon
nuclei fluxes as function of energy per nucleon are shown together with
data from the AMS-02 and CREAM experiments~\cite{Kachelriess:2017yzq}. 
Each spectrum consists of two contributions: 
The low-energy component $F^{(1)}(\R)$ represents the average CR flux in 
the local interstellar medium, while the second component $F^{(2)}(\R)$
representing the CR contribution from the local SN  dominates at high energies.
The small differences in the abundance of nuclei in the two components---which
is expected if the SN exploded in a star-forming region---is responsible for
the deviations from the universality of the total flux of CR nuclei.
The boron-to-carbon ratio as function of rigidity predicted in the
local SN model is shown in the left panel of Fig~\ref{models}. The
contribution of the
local SN leads to a flattening of the B/C ratio around $\R\simeq 1$\,TV.

In the right panel of Fig.~\ref{fig:single}, the contribution of the local SN to
the electron flux is shown. Neglecting energy losses, the contribution to
the electron spectrum from the local source has the same functional
form as the proton spectrum. Normalising it by  choosing for the
electron/proton ratio $K_{\rm ep}$ at injection $K_{\rm ep}=4\times 10^{-3}$
and adding an exponential cutoff due to energy losses gives the flux of
primary electrons shown as blue line. Additionally, electrons are
produced continuously by the CR protons from the local source. This
contribution has a break at $\simeq 300$\,GeV and is shown by the orange
line. Adding finally the contribution from the background of distant sources,
the observed spectrum is reproduced.
The positron flux predicted in the local SN model shown in the right
panel of Fig.~\ref{models} agrees also well with the data. Thus in this
model, all the observational anomalies we have discussed can be understood
as the imprint of a 2--3\,Myr old SN on the local CR flux.

\section{Cosmic rays around the knee}
\label{model_knee}

\subsection{Observations}
 \label{knee_observations}

\begin{figure}
  \centering
  \includegraphics[width=0.7\columnwidth,angle=0]{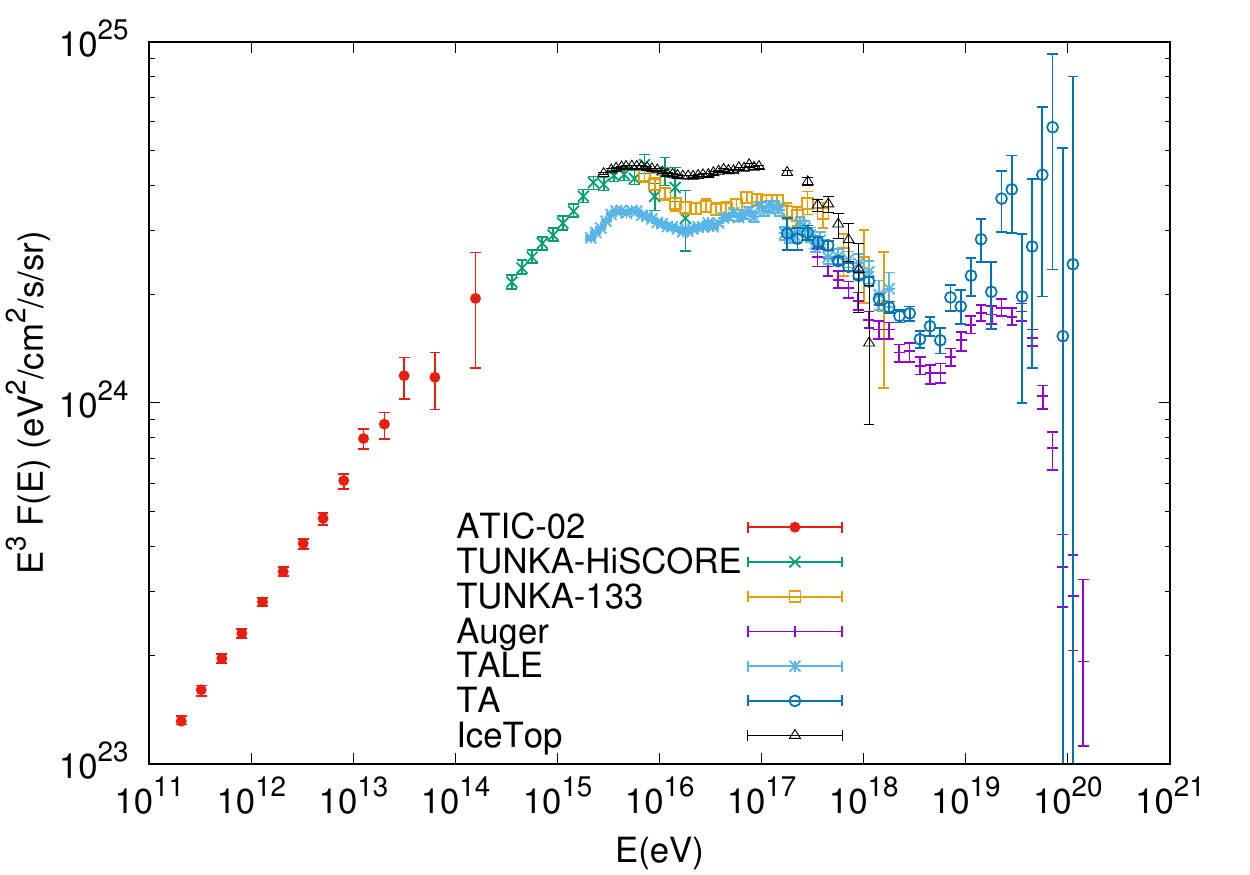}
  \caption{The all-particle CR spectrum multiplied with $E^{3}$ as function
  of energy together with experimental data from ATIC-2~\protect\cite{ATIC_spec},
  TUNKA~\protect\cite{TUNKA_spec}, TUNKA-HiSCORE~\protect\cite{TUNKA_HS_spec}, IceTop~\protect\cite{IceTop_spec}, TALE~\protect\cite{Abbasi:2018xsn}, the
  PAO~\protect\cite{PAO_spec}, and the TA~\protect\cite{TA_spec}. All data are consistent after a 10\% energy shift, i.e.\ within their experimental
  uncertainty.
\label{allparticle_spec}}
\end{figure}

The all-particle energy spectrum extending from $10^{11}$\,eV up to the
highest energies as measured by a few recent experiments is shown in
Fig.~\ref{allparticle_spec}. The CR intensity measured directly
in satellite experiments like ATIC~\cite{ATIC_spec} extends smoothly
into the range of indirect measurements via ground arrays and
Cherenkov or fluorescence detectors like TUNKA~\cite{TUNKA_spec},
TUNKA-HiSCORE~\cite{TUNKA_HS_spec},  IceTop ~\cite{IceTop_spec}, TALE ~\cite{Abbasi:2018xsn}, the PAO~\cite{PAO_spec} and the
TA~\cite{TA_spec}. The spectrum can be approximated by a multiply broken
power-law,
with a first break at 200\,GeV as discussed in the previous section.
At the energy $E_{\rm k}\simeq 4$\,PeV, a second pronounced change
dubbed the CR knee occurs where the spectral index changes from
$\beta\sim 2.7$ below to $\beta\sim 3.1$ above the knee.   
The second knee corresponds to a change in the spectral slope
of the all-particle energy spectrum at $\simeq 5\times 10^{17}$\,eV where
the slope softens by $\Delta\beta\simeq 0.2$. This softening was only
discovered in 1992 by the Akeno experiment~\cite{Nagano:1991jz}.

The elemental composition of the CR flux below $E\sim 10^{14}$\,eV is
relatively well determined by direct measurements using satellite and
balloon experiments. At higher energies, the low CR flux prevents direct
measurements. Moreover, large fluctuations in the development of
extensive air showers make the determination of the mass number of 
individual CR primaries very difficult.  Therefore experiments present their
results summing up various elements into spectra of few groups of elements,
or investigate only the mean mass number $A$ of the CR flux. A frequently
used quantity to characterize the composition is the mean logarithmic
mass, defined as $\langle\ln A\rangle=\sum_i f_i \ln A_i $ with $f_i$ as the
relative fraction of nuclei with mass number $A_i$. Experimentally,
$\langle\ln A\rangle$ can be deduced applying three main methods: Firstly,
the quantity is connected to the ratio of the number $N_i$ of electrons
and muons measured at the ground level as
$\langle\ln A\rangle \simeq C  +15\log(N_e/N_\mu)$. The
coefficient $C$ has to be determined with the
help of Monte Carlo simulations for hadronic interactions and is
therefore model dependent. Secondly,
$\langle\ln A\rangle$  is proportional to the observed depth $X$ of the shower
maximum in the atmosphere, which depends through the relation 
$X_{\max}^A=X_{\max}^p - D_{10}\ln A$ on the mass number $A$. Here, $X_{\max}^p$
and $X_{\max}^A$ denote the depth of the shower maximum
initiated by a proton and by a nucleus with mass number $A$, respectively,
while the elongation rate $D_{10}={\rm d}X_{\max}/{\rm d}\log(E)$ varies
with energy between 50 and 70\,g/cm$^2$. Hence, the difference in the
observed depth of the shower maxima between an iron and a proton
induced shower is $\simeq 100$\,g/cm$^2$, what should be compared to
the uncertainties from interaction models ($\Delta X\sim 25$g/cm$^2$)
and the systematic error ($\Delta X\sim 10$g/cm$^2$) in the
PAO experiment.
Finally, the fluctuations of an iron induced shower are smaller than that
of a proton induced shower. Thus the shape, or in first approximation,
the width of the observed $X_{\max}$ distribution can be used to derive the
composition. For a more detailed review of experimental methods see, e.g.,
the review~\cite{Bluemer:2009zf,Mollerach:2017idb}.

\begin{figure}
  \center{\includegraphics[width=0.7\columnwidth,angle=0]{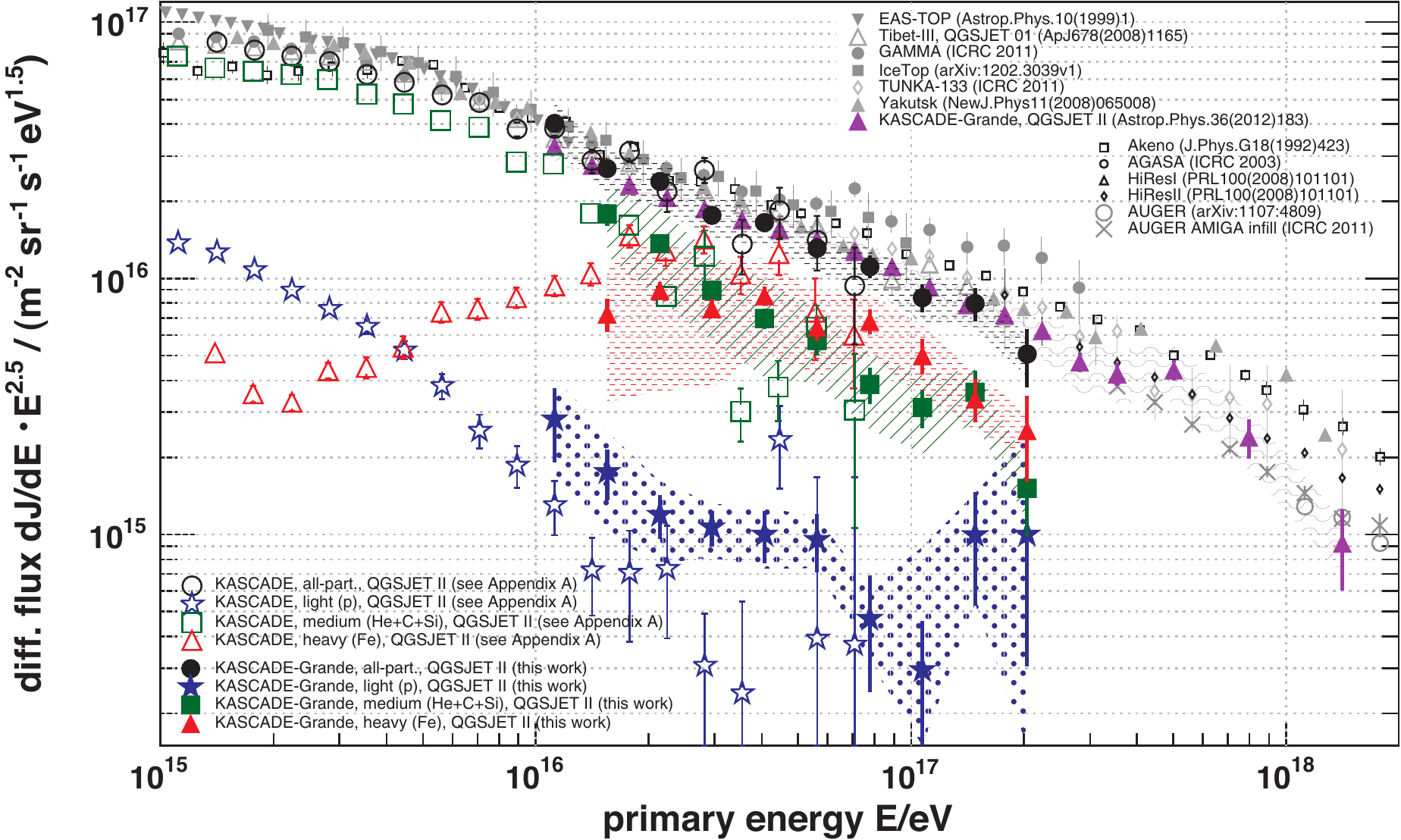}}
    \vskip0.6cm
    \caption{The CR spectrum for few elemental groups deduced from KASCADE and
      KASCADE-Grande date using QGSJET-II, from Ref.~\protect\cite{Apel:2013uni}.
\label{knee_kasc}}
\end{figure}

While there is a general agreement about the position of the knee in the
total CR spectrum at $E_{\rm k} \simeq 4$\,PeV,
and that the composition becomes increasingly heavier in
the energy range between the knee  and $10^{17}$\,eV~\cite{Aglietta:2004np,Antoni:2005wq,IceCube:2012vv,Apel:2013uni}, there exist yet substantial
uncertainties concerning the partial contributions of different mass groups
to the primary CR  composition. In particular, the question at which energy
the suppression in the light components starts remains unclear.
The results from KASCADE and KASCADE-Grande which as first experiments
presented energy spectra for individual groups of elements, suggested
a proton knee around 4\,PeV. The spectra for the heavier elements were
consistent with a rigidity-dependent knee following  $E_k^{(Z)}=ZE_p$,
cf.\ with Fig.~\ref{knee_kasc}. Since these results were obtained
by measuring the number of electrons and muons on the ground,
a rather large dependence on the used Monte Carlo simulation resulted. In
particular, the relative fraction of light elements changed considerably
going, e.g.,  from QGSJET-II-02 to Sybill. 
Moreover, the results of KASCADE and KASCADE-Grande are hampered by large
fluctuations, since the experiment was close to sea-level.

\begin{figure}
  \includegraphics[width=0.49\columnwidth,angle=0]{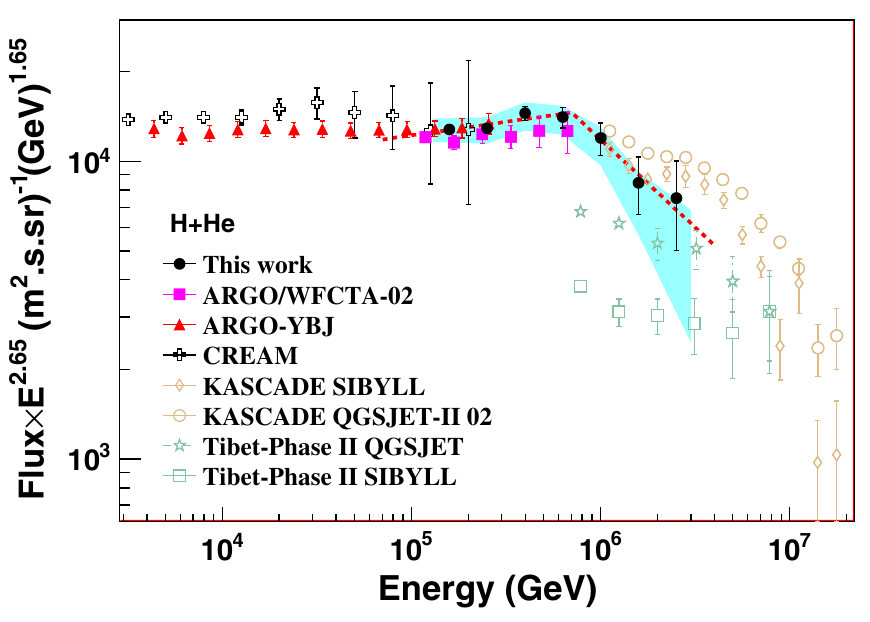}
  \hskip0.3cm
  \includegraphics[width=0.49\columnwidth]{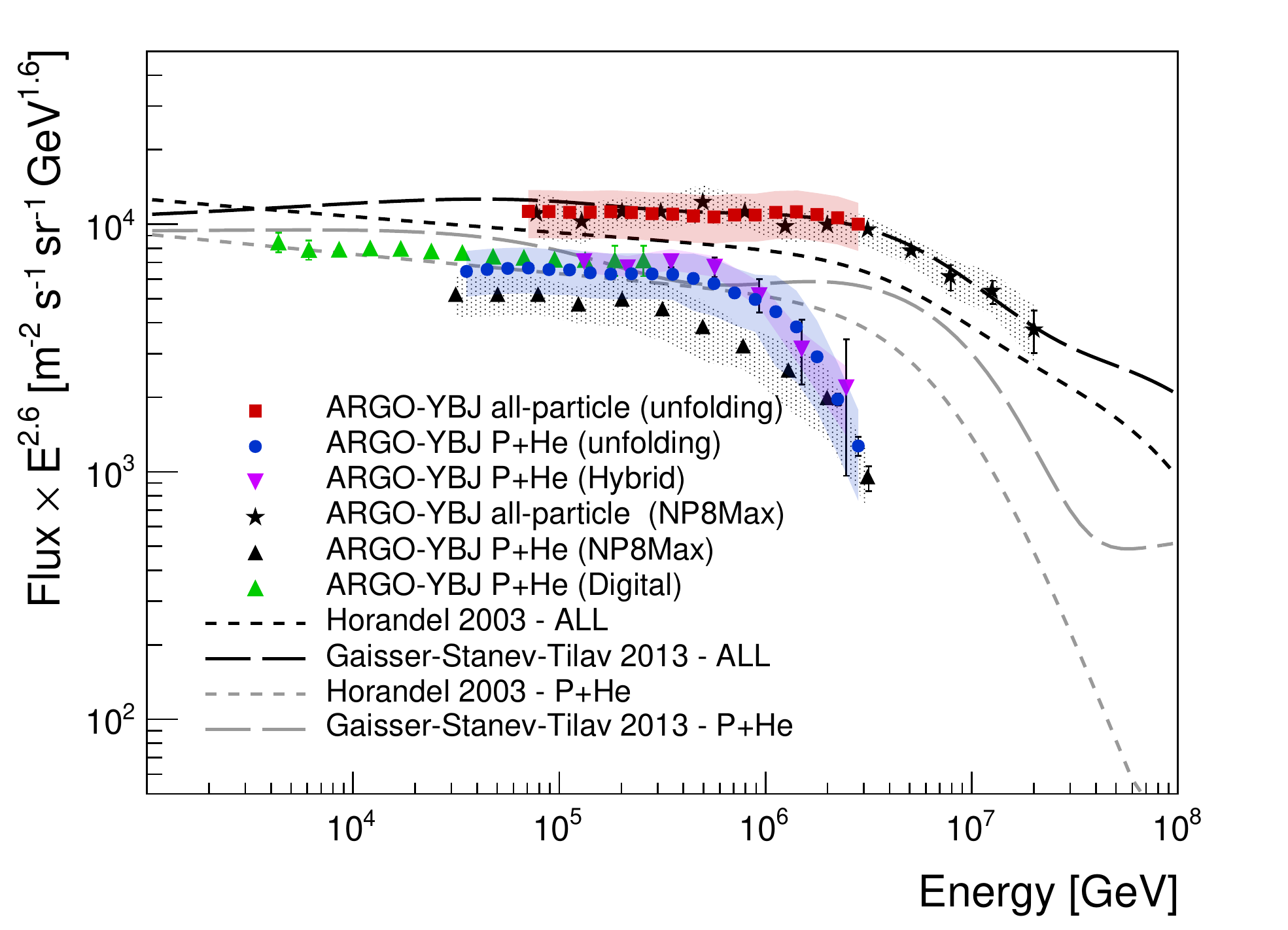}
\caption{{\em Left panel:\/}  H+He spectrum obtained by the hybrid experiment with
  ARGO-YBJ and the imaging Cherenkov telescope. 
  The p+He spectra by
CREAM~\protect\cite{Yoon:2011cr}, ARGO-YBJ~\protect\cite{Bartoli:2015fhw} and the hybrid
experiment~\protect\cite{Bartoli:2014eua} below the knee,
the spectra by Tibet AS~\protect\cite{Amenomori:2011zza} and KASCADE~\protect\cite{Antoni:2005wq,Apel:2013uni} above the knee are
shown for comparison.
  {\em Right panel:} The all-particle and proton plus helium energy spectra
  measured by ARGO-YBJ by using different experimental techniques;
  from Ref.~\protect\cite{Montini:2016osn}.
\label{fig:kneeARGO}}
\end{figure}

Conflicting results for the knee in the flux of light elements were
obtained by experiments using air shower arrays at high altitudes.
Earlier results from CASA-MIA, BASJE-MAS, and Tibet AS$\gamma$
suggested that the suppression of the proton flux starts at 500\,TeV, i.e.\
earlier than the knee in the total spectrum. Similar results have been
obtained by the ARGO-YBJ experiment. This experiment combined
results from its ground array having a large coverage with a wide field of
view  Cherenkov telescope. The left panel of Fig.~\ref{fig:kneeARGO} shows
the knee-like structure in the combined p+He flux  around 700\,TeV
obtained in these hybrid measurements~\cite{Bartoli:2014eua}, while the
right panel compares the p+He flux with the all-particle spectrum.
Note that for experiments at high altitudes like ARGO-YBJ on one hand the
fluctuations in the electron number at ground are reduced, but on the
other hand the shower maximum is close to the ground level. Moreover,
fluctuations in the muon number are large, because most charged pions
have no time to decay.

Finally, we want to stress the importance of determining the knee energy in
the flux of the light elements. The combined p+He flux is dominated
by the He component. Thus the ARGO-YBJ results would imply for the energies
of a rigidity-dependent knee $E_{\rm p}\simeq 0.53$\,PeV and
$E_{\rm Fe}\simeq 20$\,PeV. Associating these relatively low values with
the maximal energy of typical Galactic CR sources reduces---at a first
glance---considerably the pressure on acceleration models. However, if
the source spectra
would have  an exponential cut-off at the rigidity $\R\simeq 0.53$\,PV, 
then the Galactic CR spectrum would end well below 0.1\,EeV.
This would require either an additional Galactic component, or an
extremely early transition to extragalactic CRs. In the first case, a new
type of Galactic CR sources able to accelerate to such energies is needed,
and the acceleration problem reappears. Requiring instead the presence of
extragalactic CRs at such low energies is
problematic, because they may be hidden by magnetic horizons.
If on the other hand the proton knee is at $E_{\rm p}\simeq 4$\,PeV,
typical Galactic CR sources should be PeVatrons and an additional
Galactic component at high-energies may be avoided.
Moreover, knowing the energy of the knee may help to exclude
some models: For instance,  an energy of the proton knee as low as
$E_{\rm p}\simeq 0.5$\,PeV is  difficult to achieve in models which explain
the knee by propagation effects, as it would require too small correlation
lengths of the turbulent magnetic field.

\subsection{Models}
 \label{knee_models}

Explanations for the origin of the knee fall in three main categories. 
In the first one, it was speculated that a sudden change in hadronic 
interactions above 2\,TeV in the center-of-mass system leads to the observed 
change of the CR spectral slope at the knee.  Alternatively, the CR flux
may be suppressed above the knee because of the opening of additional energy 
loss channels. These possibilities were excluded by measurements at the
Large Hadron Collider (LHC) which have not revealed any strong deviation from
the expected characteristics of hadron production at multi-TeV 
energies~\cite{dEnterria:2011twh}. In the second class of models, the knee
is connected to properties in the energy spectrum of the Galactic
CR sources. For instance, the knee might correspond to the maximal
rigidity to which CRs can be accelerated by the 
population of Galactic CR sources dominating the CR flux below 
PeV~\cite{Stanev:1993tx,Kobayakawa:2000nq,Hillas:2005cs}.
Alternatively, the knee may be caused by a break in the source CR energy
spectrum at this rigidity~\cite{Drury:2003fd,Cardillo:2015zda}.
A variant of this model is the suggestion that the spectrum below the
knee is dominated by a single, nearby source and that the knee correspond to
the maximal energy of this specific source.
Finally, the knee could be connected to a change in the propagation of Galactic CRs.
Such a change might be induced by a transition from pitch angle scattering to
Hall diffusion or drift along the regular magnetic field~\cite{1993A&A...268..726P,Candia:2002we,Candia:2003dk}.
Another possibility is the case suggested in Ref.~\cite{Giacinti:2014xya,Giacinti:2015hva} that the knee energy corresponds to the rigidity at which the 
CR Larmor radius $R_{\rm L}$ is of the order of the correlation length $L_{\rm c}$
of the turbulent magnetic field in the Galactic disk. In both cases, as a
result a transition 
from large-angle to small-angle scattering or Hall diffusion is expected,
as visible in Fig.~\ref{iso1}. Therefore
the energy dependence of the confinement time changes which in turn induces 
a steepening of the CR 
spectrum~\cite{1971CoASP...3..155S,1993A&A...268..726P,Candia:2002we,Candia:2003dk,Giacinti:2014xya,Giacinti:2015hva}.
All these models except the first (excluded) one
lead to a rigidity-dependent sequence of knees at $ZE_{\rm k}$, a behaviour
first suggested by Peters~\cite{Peters61,Zatsepin62}. Measurements of the
nuclear composition of the CR flux can therefore not distinguish between
them. In contrast to models in category 2, those of category 
3 predict both the position of the knee and the rigidity dependent 
suppression of the different CR components for a given model of the 
Galactic magnetic field~\cite{Giacinti:2014xya,Giacinti:2015hva}.

\paragraph{Maximal energy of source populations}

\begin{figure}
\centering
  \includegraphics[width=0.45\columnwidth,angle=0]{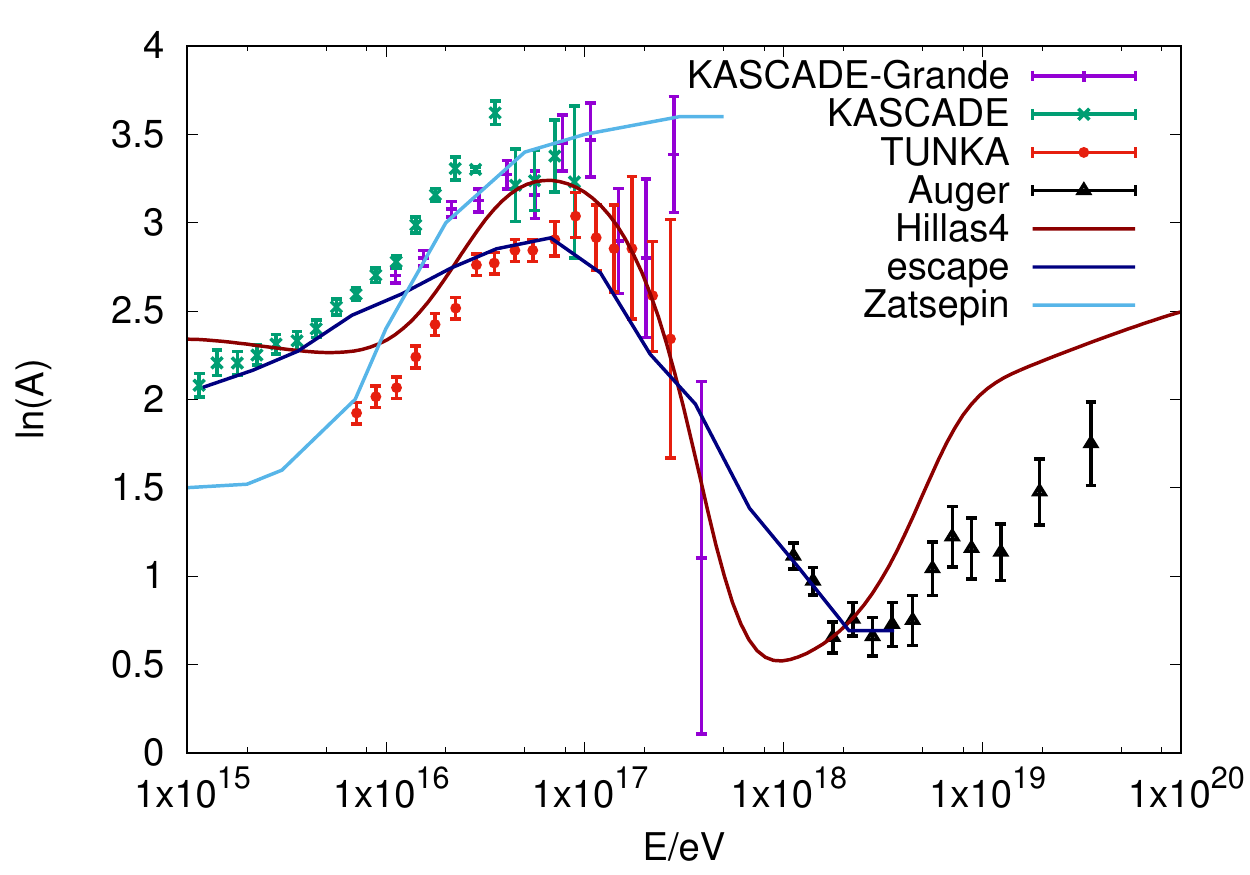}
  \caption{Predictions for $\ln(A)$ in the Zatsepin et al.~\cite{Zatsepin:2006ci}, in the Hillas~$4^\ast$~\cite{Gaisser:2013bla} and in
  the escape model~\cite{Giacinti:2014xya,Giacinti:2015hva}
  compared to experimental data.
\label{knee_lnA}}
\end{figure}

Two specific examples for the models inside class~2 are the ones proposed
by Hillas~\cite{Hillas:2005cs} and by Zatsepin et al.~\cite{Zatsepin:2006ci}.
Such models require at least two  populations of
Galactic CR, one dominating the CR spectrum below and one above the knee.
A natural explanation associates these two populations with two different
types of SN progenitors: Cosmic rays below the knee might be accelerated, 
e.g., in SN explosions of isolated stars with  masses  $M=8-15M_\odot$,
while CRs with higher energy are generated by  massive OB or
Wolf-Rayet stars in superbubbles~\cite{Zatsepin:2006ci}. A characteristic
feature of this model is a fast change of the mass composition
around $10^{16}$\,eV, as it is visible in its prediction for $\ln(A)$
shown in Fig.~\ref{knee_lnA}.

\begin{figure}
\centering
  \includegraphics[width=0.45\columnwidth,angle=0]{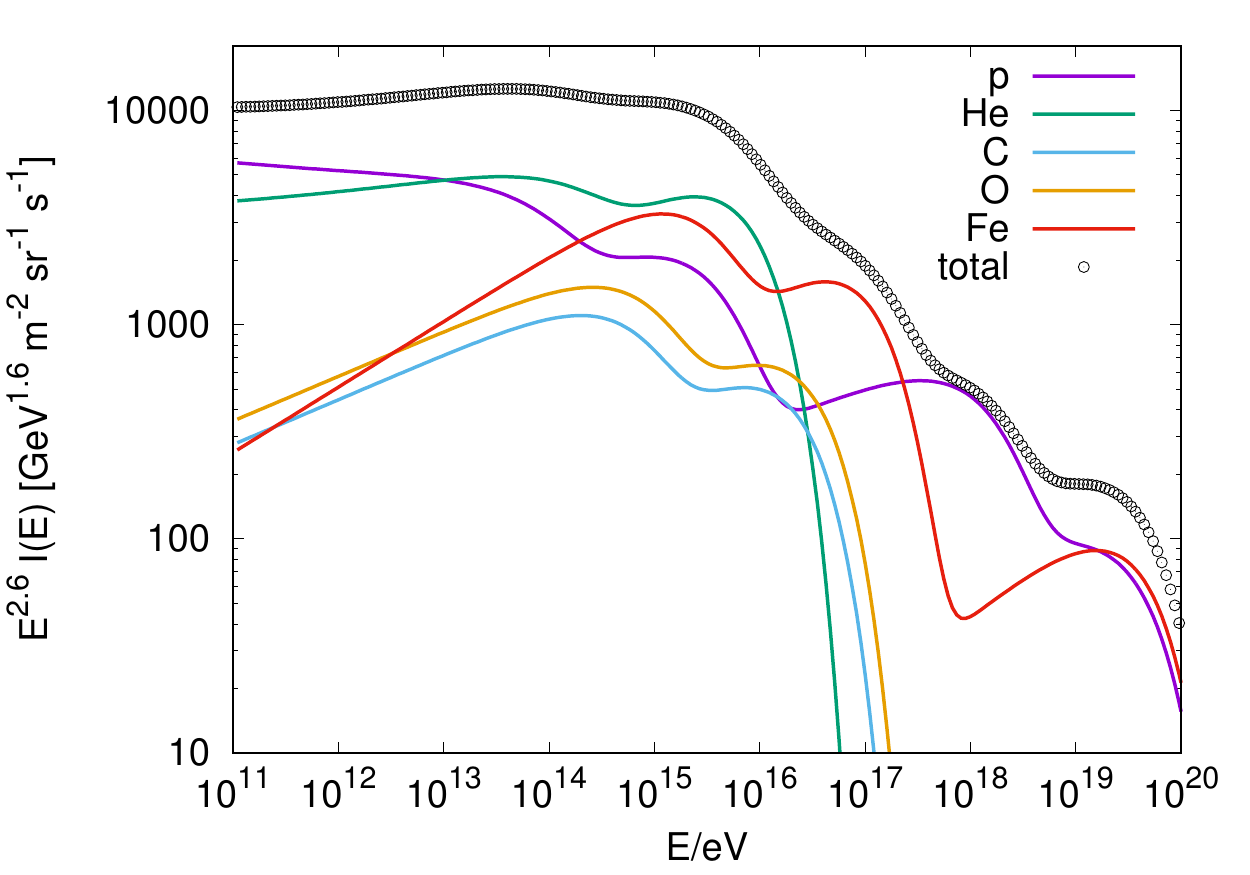}
  \includegraphics[width=0.45\columnwidth,angle=0]{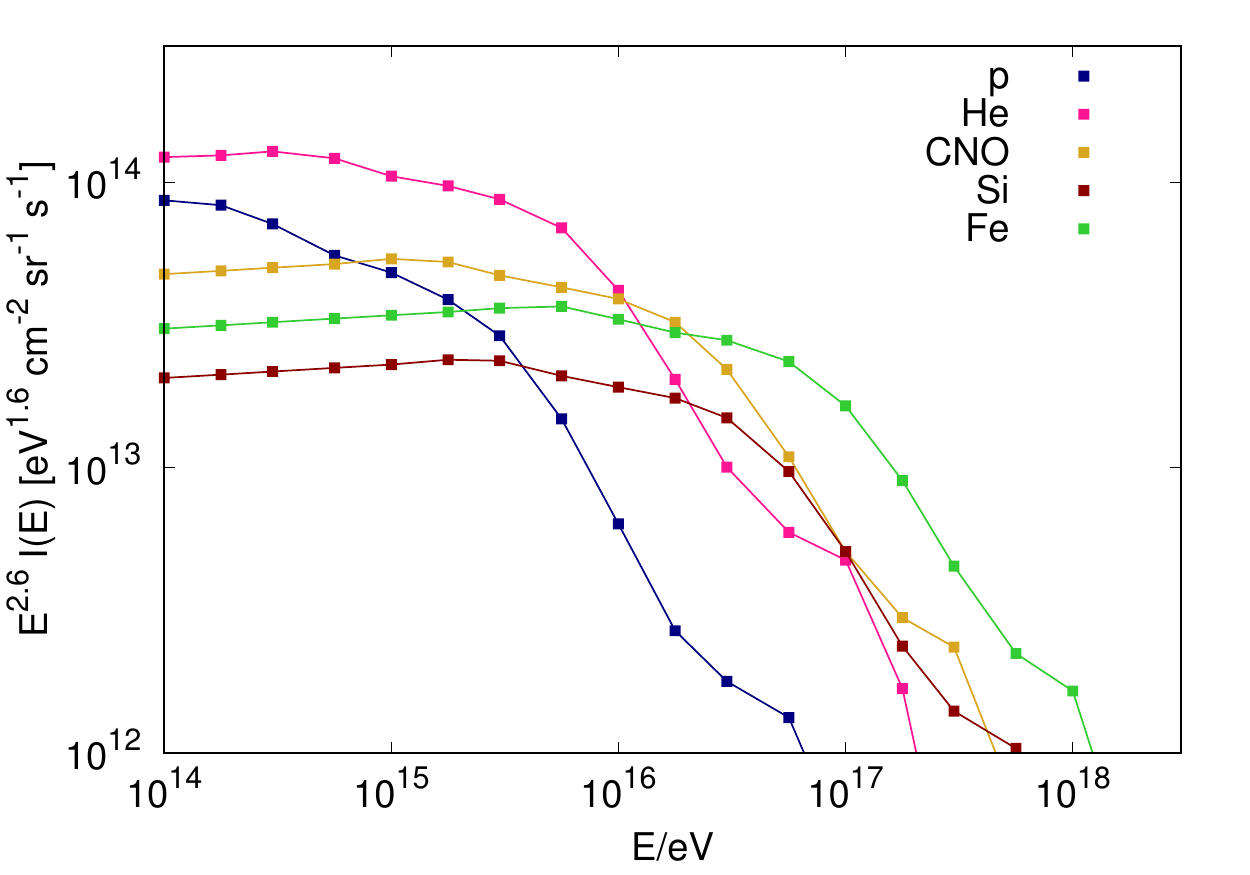}
\caption{All-particle CR spectrum  $E^{2.6}I(E)$ and individual 
  contributions of five elemental groups as a function of the energy.
  {\em Left panel\/} for the Hillas model from
  Ref.~\protect\cite{Gaisser:2013bla}; {\em right panel\/} for the ``escape model''
  from Ref.~\protect\cite{Giacinti:2014xya,Giacinti:2015hva}.
\label{knee_spectra}}
\end{figure}

Another often used model is an updated variation of the Hillas model
presented in Ref.~\cite{Gaisser:2013bla}. The model takes into account
the hardening of the CR spectrum above 200\,GV and uses as cutoff for
the first population 120\,TV. The second population accelerates CRs
up to the knee, while two additional populations of extragalactic
CRs dominate the CR flux above ${\rm few}\times 10^{17}$\,eV. 
The all-particle spectrum
and the individual contributions of five groups of nuclear elements
in this model are shown in the left panel of Fig.~\ref{knee_spectra}
for the case called $4^\ast$ with four populations. The third population
in this model  contains only proton and iron, and, in order to improve
the predicted $\ln(A)$ value, a fourth population containing only protons
was added. In Fig.~\ref{knee_lnA}, we compare the prediction for this case to
experimental data of $\ln(A)$ and a good agreement is visible. However,
such a two-component mixture of protons and irons is in contradiction to
the narrow width \X2 of  observed air showers in the atmosphere, as we
will discuss in Section~ \ref{UHECR_composition}.

In the model of Ref.~\cite{Thoudam:2016syr}, two alternatives as explanation
for a second component of Galactic CRs above the knee were investigated. 
In the first one, CRs are re-accelerated at the termination shock of a
Galactic wind, while in the second one SN explosions of Wolf-Rayet
stars lead to a CR component with an exponential cutoff of protons at
$E_{\max}\simeq 1.5\times 10^{17}$\,eV. The latter model leads to a large
contribution of intermediate elements like helium and carbon in
the CR spectrum. As we will discuss in the next section, the prediction
of a large helium and CNO fraction in the energy range around $10^{17}$ and
$10^{18}$\,eV in the  Wolf-Rayet model is in line with composition measurements
of, e.g., the PAO. However, such a light or intermediate component has to
be extragalactic, since otherwise the limits on the dipole amplitude
will be violated~\cite{Giacinti:2011ww}.

\paragraph{Break in the  source energy spectrum}

The knee may be caused alternatively by a break in the injection
spectrum $Q(\R)$ of Galactic CR sources~\cite{Drury:2003fd,Cardillo:2015zda}.
Typically, one models the end of the CR injection spectrum $Q(\R)$ 
by an exponential cutoff, induced either by the finite life-time or size
of the accelerator. However, the shape of the cutoff depends strongly on
the escape conditions, as it was shown, e.g., in Ref.~\cite{Protheroe:2004rt}
using Monte Carlo toy models. Since the injection efficiency
and the shock velocity, and thus the maximal acceleration energy, are
time-dependent, a break could be also caused by the transition from the
free-expansion to the Sedov-Taylor phase of the SNR.
In particular, a steepening of the injection spectrum at $\R_{\rm br}=2$\,PV
by $\Delta\beta=0.9$ was found in  Ref.~\cite{Drury:2003fd}:
Including strong field amplification as suggested
by Bell and Lucek~\cite{2001MNRAS.321..433B,2004MNRAS.353..550B}
into a toy acceleration model, these authors found a
break in the energy spectrum of accelerated protons, coinciding for typical
values SNR parameters with the knee region. The strength $\Delta\beta$
of the steepening depends among others on the injection history, and in
a typical test particle ansatz $\Delta\beta=0.9$ was found.
Similar results were obtained by Ref.~\cite{Cardillo:2015zda}.
While the steepening in the spectrum of a single source is thus too hard,
a superposition of sources with varying break energies may lead to
a break in the total flux compatible with observations.
We conclude therefore that the often presented conclusion that a second Galactic
population of CR sources above the knee is required is based on the
assumption of a hard, exponential cutoff. In contrast, a single class of
sources may explain the CR below and above the knee, if the high-energy tail
of these sources is sufficiently strong.

{
A steepening of the  source energy spectrum was found also in the model
of massive stars exploding into their
winds~\cite{Biermann:1993wy,Biermann:2018clk}. These supernova produce
two CR components: One shaped by only diffusive shock acceleration
with a flat energy spectrum and a cutoff close to the knee, and
second component accelerated additionally by drift acceleration
and a steeper spectrun close $E^{-3.2}$. Thus the predicted spectral slope
above the knee agrees well with the observed one.
}

\paragraph{Single source}

In Ref.~\cite{Erlykin:1997bs,Erlykin:2000jm}, Erlykin and Wolfendale
suggested that the CR flux in the knee region is dominated by a single
young  nearby source. They argued that the sharpness of the knee requires
the dominance of a single source, since the unavoidable spread in the
properties of, e.g., SN progenitors and environments would lead to a
corresponding spread in their maximal energies, thereby smoothing out the
break. In an up-dated version of this model, they suggested Vela, a
SNR with the age $T \simeq 11000$\,yr and distance $R=270$\,pc, as a
possible source candidate~\cite{Erlykin:2015hha}.

\begin{figure}
  \includegraphics[width=0.5\columnwidth,angle=0]{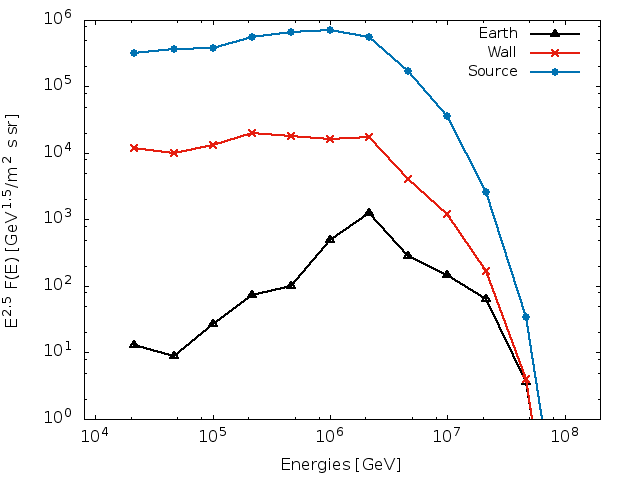}
  \hskip0.3cm
  \includegraphics[width=0.5\columnwidth]{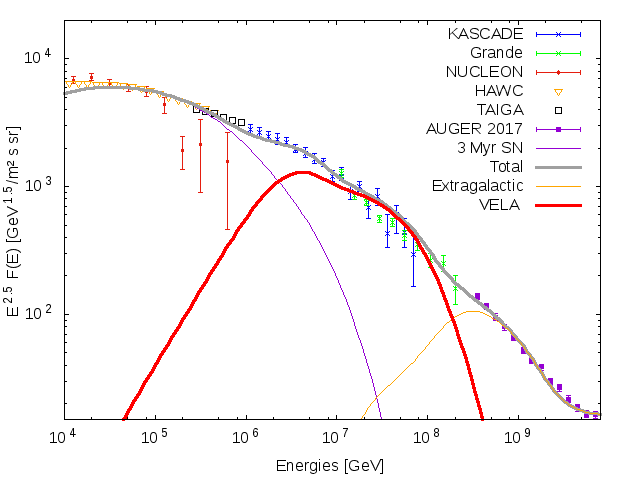}
  \caption{{\em Left panel:\/} The CR flux of protons around Vela, in the wall
      and at Earth as function of energy $E$.
      {\em Right panel:} The all-particle flux at Earth from a 2--3\.Myr old
      SN (purple) and Vela (red) together with experimental data. Adapted from
      \protect\cite{Bouyahiaoui:2018lew}.
\label{fig:Vela}}
\end{figure}

Vela is connected with the Solar system by a magnetic field line in models
of the global Galactic magnetic field as, e.g., the Jansson--Farrar
model~\cite{Jansson:2012rt}. In the case of anisotropic diffusion, CRs
propagate preferentially along the magnetic field lines and thus the
locally observed CR flux from Vela would be strongly enhanced: The
resulting flux shown in Fig.~\ref{Adata} by the blue circles for
$T=10^4$\,yr overshoots the locally measured one by three orders of
magnitude. Such an excess is avoided, if one takes into account that the
Sun is located inside the Local Bubble: In Ref.~\cite{Bouyahiaoui:2018lew},
the Local Bubble was modelled as a cylinder with base radius $R=100$\,pc,
a bubble wall of thickness to $w=3$\,pc and $B_{\rm sh}=12\mu G$,
and $B_{\rm out} = 1\mu G$ outside the bubble.
The Sun was assumed to be at the centre of the Local Bubble, while Vela was set
at the distance 270\,pc from the Sun along the magnetic field.
In the left panel of Fig.~\ref{fig:Vela}, the resulting CR flux of protons
close to
Vela, in the bubble wall and at Earth is shown.  While at high energies,
when the CR Larmor radius is large compared to the thickness of the bubble
wall, the bubble wall is transparent, protons start to be trapped in
the wall around $E\sim 1$\,PeV. At lower energies, the flux inside the
bubble is increasingly suppressed.
In the right panel of Fig.~\ref{fig:Vela}, we show the resulting all-particle
flux  from Vela together
with the  flux from a 2--3\,Myr old SN in the model of
Refs.~\cite{Kachelriess:2015oua,Kachelriess:2017yzq}. The combined flux of
these two sources covers the energy range from 200\,GeV up to the
extragalactic transition region, fitting well the experimental data.

\paragraph{Propagation}

The possibility that the knee is caused by an increased leakage of
CR from the Galaxy was suggested already by Syrovatsky in
1971~\cite{1971CoASP...3..155S}. Assuming for the Galactic magnetic field
a strength $B\simeq {\rm few}\;\mu$G and a correlation length
$L_{\rm c}\simeq 100$\,pc, the transition from the diffusive to the
small-angle scattering regime is expected at ${\rm few}\times 10^{16}$\,eV.
Thus for such parameters, the energy range of the knee can be studied
in the diffusion approach. In Refs.~\cite{1993A&A...268..726P,Candia:2002we,Candia:2003dk}, it was assumed that the knee corresponds to a transition between the
dominance of pitch angle scattering to Hall diffusion or drift along the
regular field. Since the diffusion coefficients derived for conventional
pitch angle scattering and hall diffusion have different energy slopes,
the cross over between the two diffusion scenarios will lead to
a corresponding change in the energy spectrum of Galactic CRs.
While Ref.~\cite{1993A&A...268..726P} used a simplistic spatial dependence
for the diffusion coefficient, \cite{Candia:2002we,Candia:2003dk} employed
a more realistic model for the GMF. The diffusion coefficients used in the
latter work were extracted from the numerical results obtained in
\cite{Casse:2001be}. The mean $\ln(A)$ predicted in this model for the
energy range $10^{15}-10^{17}$\,eV agrees well with observations.

The escape model developed in Refs.~\cite{Giacinti:2014xya,Giacinti:2015hva}
is an alternative approach which connects the knee also with a change in the
propagation of Galactic CRs. In contrast to previous works, the escape of CRs from
our Galaxy was studied in these works calculating trajectories of individual
CRs in models of the regular and turbulent Galactic magnetic field like
the models of Pshirkov et al. or of
Jansson-Farrar~\cite{Pshirkov:2011um,Jansson:2012pc}. For a 
correlation length $L_{\rm c} \simeq (2-5)$\,pc of the turbulent field and a 
weak turbulent magnetic field, a knee-like structure at 
$E/Z={\rm few}\times 10^{15}$\,eV was found,
which is sufficiently strong to explain the proton knee observed by KASCADE.
The resulting intensity of four other elemental groups are shown in the
right panel of Fig.~\ref{knee_spectra}. They
are consistent with the energy spectra of CR nuclei determined by KASCADE 
and KASCADE-Grande. Moreover, the strength of the turbulent GMF component
was such that at low energies the B/C data could be successfully
reproduced.

\section{Transition from Galactic to extragalactic cosmic rays}
\label{transition}

\subsection{Observations}

We restrict our discussion to the basic experimental results,
concentrating mainly on those measurements which help to understand
the transition between Galactic and extragalactic CRs discussed in the
next subsections. We consider the energy range above $10^{17}$\,eV which
corresponds roughly to the lower energy cut in most experimental analyses
performed at the PAO and TA. Moreover, we will later argue that the
transition between Galactic and extragalactic CRs takes place around
$5\times 10^{17}$\,eV. Choosing  $10^{17}$\,eV as the lower end of the
energy range considered guaranties thus that all measurements relevant
for the transition are included. To be definite, we call all particles
above  $10^{17}$\,eV ultra-high energy cosmic rays (UHECR).

\begin{figure}
  \centering
  \includegraphics[width=0.45\columnwidth]{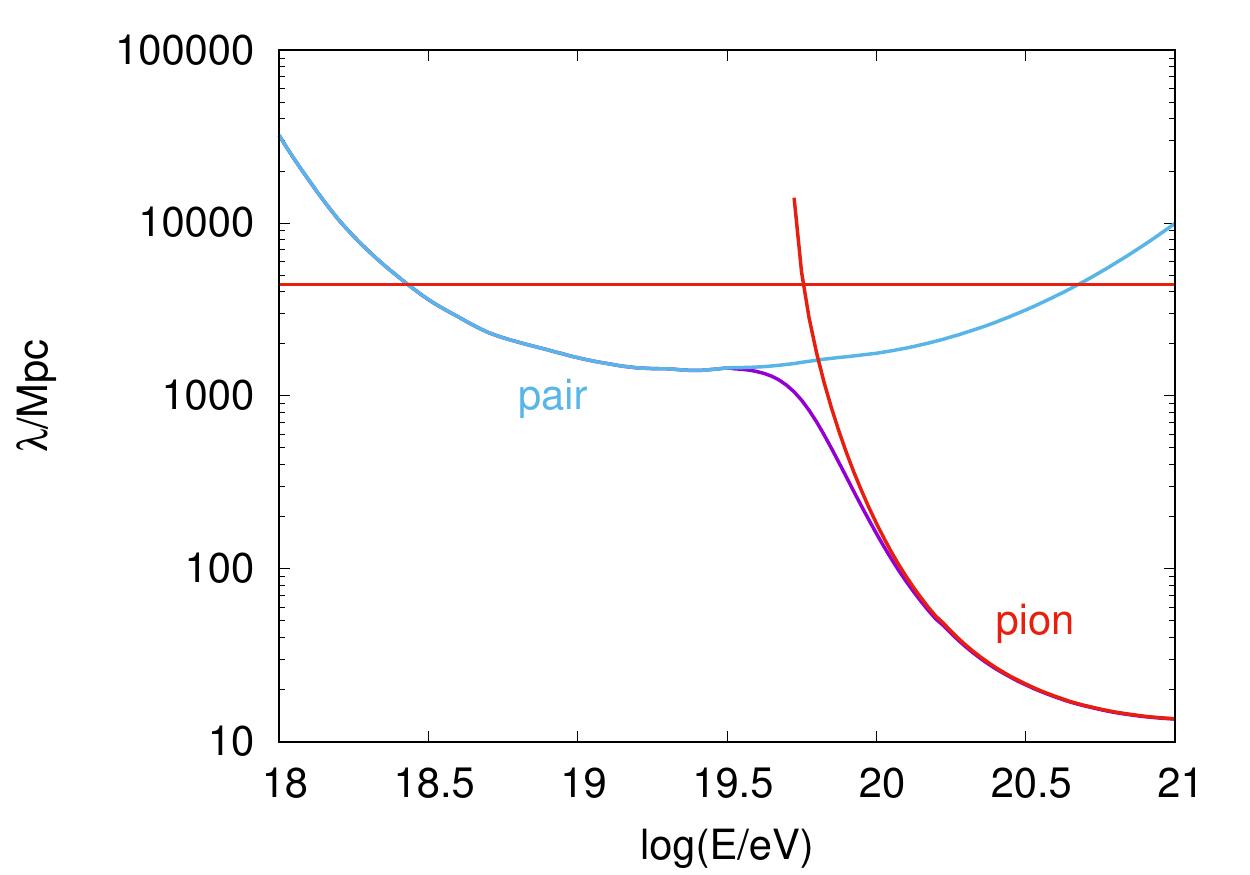}
  \includegraphics[width=0.45\columnwidth]{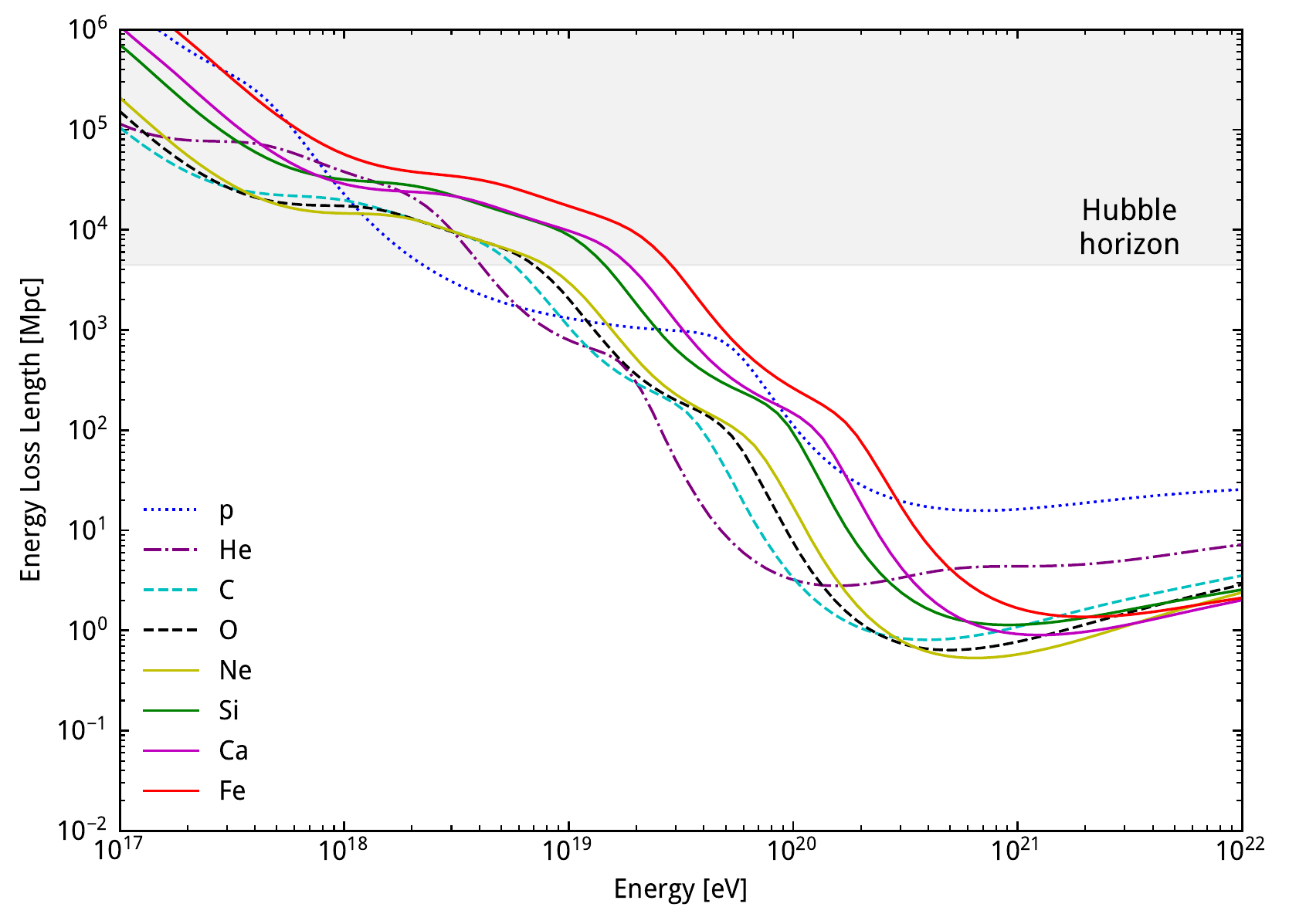}
  \caption{{\em Left panel:\/}
    The energy-loss horizon $\lambda$ of UHECR protons as function of energy.
    {\em Right panel:\/} Energy-loss horizons of different UHECR nuclei derived
    from CRPropa~\protect\cite{Batista:2016yrx,Sigl:2017wfx} as function of energy.}
\label{fig:UHECR_losses}
\end{figure}

\subsubsection{Energy spectrum}
 \label{UHECR_spectrum}

Interactions of UHE protons with  the cosmic microwave background (CMB)
leave their imprint on the UHECR  energy spectrum in the form of the
Greisen-Zatsepin-Kuzmin (GZK) cutoff~\cite{Greisen:1966jv,Zatsepin:1966jv},
a bump and a
dip~\cite{Hill:1983mk,Berezinsky:1988wi,Stanev:2000fb,Berezinsky:2002nc}. 
The GZK cutoff is a very pronounced steepening of the proton spectrum at the
energy  $E_{\rm GZK} \simeq(4 - 5)\times 10^{19}$\,eV, caused by photo-pion
production due to interactions of UHE protons with CMB photons. This effect
was predicted one year
after the discovery of the CMB in 1966 by Greisen and independently by
Zatsepin and Kuzmin. It implies that at the highest energies only local
sources within $\simeq 100$\,Mpc can contribute to the observed UHECR flux.

In the left panel of Fig.~\ref{fig:UHECR_losses}, the energy-loss horizon
$$
 \lambda_{\rm hor}(E,z=0)=
 \left( \frac{1}{E}\frac{{\rm d}E}{c{\rm d}t} \right)^{-1}
$$
for protons as function of energy is shown. The losses are caused by three
processes which dominate in different energy ranges: Pion production at
the highest energies, $e^\pm$ pair production  in the intermediate energy
range  $3\times 10^{18} {\rm eV }\lsim E\lsim 5\times 10^{19}$\,eV,
and redshift losses due to the expansion of the Universe. Note that the
energy losses at the redshift $z>0$ can obtained
by a simple rescaling from the present ones,
$\lambda_{\rm hor}(E,z)^{-1}=(1+z)^3\lambda_{\rm hor}^{-1}((1+z)E,z=0)$,
since they are caused by CMB photons.
The GZK cutoff is caused by the strong increase of the pion production
rate through the resonant process $p+\gamma_{\rm CMB}\to\Delta^+\to p+\pi^0$, 
when the peak of the Planck distribution of CMB photons is above
threshold and participates in this reaction. While the suppression is
very pronounced, the shape of this steepening is strongly model-dependent
and difficult to distinguish from, e.g., a cutoff due to the maximal
acceleration energy in the UHECR sources. Interacting with the CMB,
protons lose energy and accumulate in the
form of a bump at the energy $E_{\rm b}<E_{\rm GZK}$. These bumps
are clearly seen in the spectra calculated for single sources,
but disappear in the diffuse spectrum, because bumps from sources
at different distance are located at different energies.  
The dip visible in $\lambda_{\rm hor}$ is  produced by
$p+\gamma_{\rm CMB}\to p+e^++e^-$ interactions, which leads to a
corresponding spectral feature in flux of UHECR protons.

In the right panel of Fig.~\ref{fig:UHECR_losses}, the energy loss distances
of different UHECR nuclei as function of energy are shown.  The main energy
loss processes of nuclei are the photo-disintegration on CMB photons at
high energies and on infra-red photons at lower energies.  Only iron nuclei
can travel distances comparable to those of protons, around 100\,Mpc,
at the highest energies $E\gsim 10^{20}$\,eV. Thus for any
composition, one expects a cutoff in the UHECR energy spectrum.
While the existence of this cutoff was long time debated, it was
experimentally confirmed by the HiRes experiment in 2007, 41~years 
after its prediction~\cite{Abbasi:2007sv}. It is still unclear
if the cutoff is caused by the maximal energy of sources or by the GZK effect.
In the case of UHECR protons, a second dip at $6\times 10^{19}$\,eV was
suggested as model-independent probe for the GZK
effect~\cite{Berezinsky:2006mk}. However, an observation of this narrow
dip would require a much larger statistics than presently available, in
particular if the proton flux is only subdominant.

\begin{figure}
  \centering
  \includegraphics[width=1.\columnwidth]{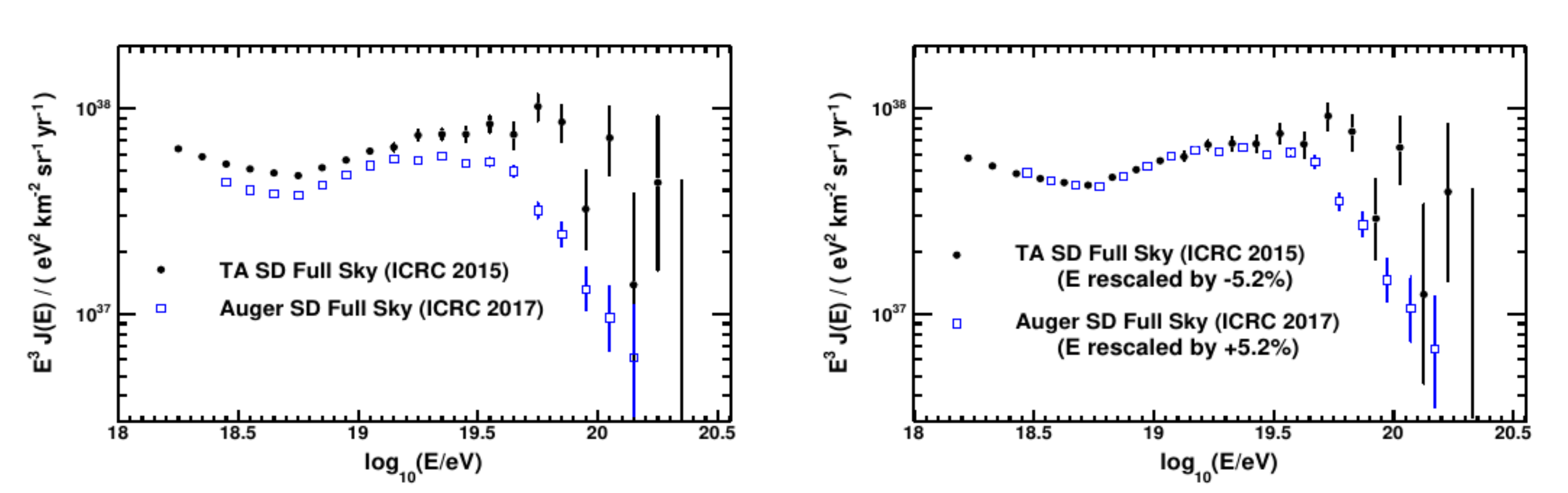}
\caption{UHECR spectrum measured by the PAO (Auger) and TA experiments: 
  {\em Left panel\/} before and {\em right panel\/}
  after the shift in the energy scale; from Ref.~\protect\cite{Ivanov:2017juh}. }
\label{fig:UHECR_spectrum}
\end{figure}

In Fig.~\ref{fig:UHECR_spectrum}, we show measurements of the UHECR flux
from the two most recent experiments, the PAO and the TA. The uncertainty in
the absolute energy scale of
these experiments is around 10\%, leading to large shifts in a plot of
$E^3F(E)$. The two experiments can be cross-calibrated using that the
UHECR flux up to $10^{19.2}$\,eV is very isotropic: Applying a relative
shift of the energy scale of the two experiments by 10.4\%, their all-particle
fluxes shown in the right panel of Fig.~\ref{fig:UHECR_spectrum} agree well
up to $4\times 10^{19}$\,eV.
At higher energies, the deviations increase, with the cutoff in the TA
spectrum shifted to higher energies. Apart from unaccounted systematic
effects, the different fields-of-view of the two experiments can explain
these differences: Above $4\times 10^{19}$\,eV, the mean free path of
UHECRs drops, such that differences in the large-scale structure (LSS)
are not averaged out any
longer. Since the number of UHECRs contributing to the flux decreases,
differences in the maximal energies of the most important sources in
the northern or southern sky can become important. Alternatively,
when the cutoff is caused by the GZK effect, differences in the distance
to the dominating sources in two hemispheres may cause the change in
the flux observed by PAO and TA.
Note that the energy $E_{1/2}\simeq 5\times 10^{19}$\,eV where the integral
flux drops for a pure proton composition by a factor two relative to the flux
expected without pion production~\cite{Berezinsky:2002nc} deviates
significantly from the value determined from the PAO data,
$E_{1/2}\simeq (2.3\pm 0.4) \times 10^{19}$\,eV. Such a low value of $E_{1/2}$
may hint towards an intermediate mass composition at the highest energies.

\subsubsection{Composition}
 \label{UHECR_composition}

\begin{figure}[tb]
  \centering
  \includegraphics[width=0.52\linewidth]{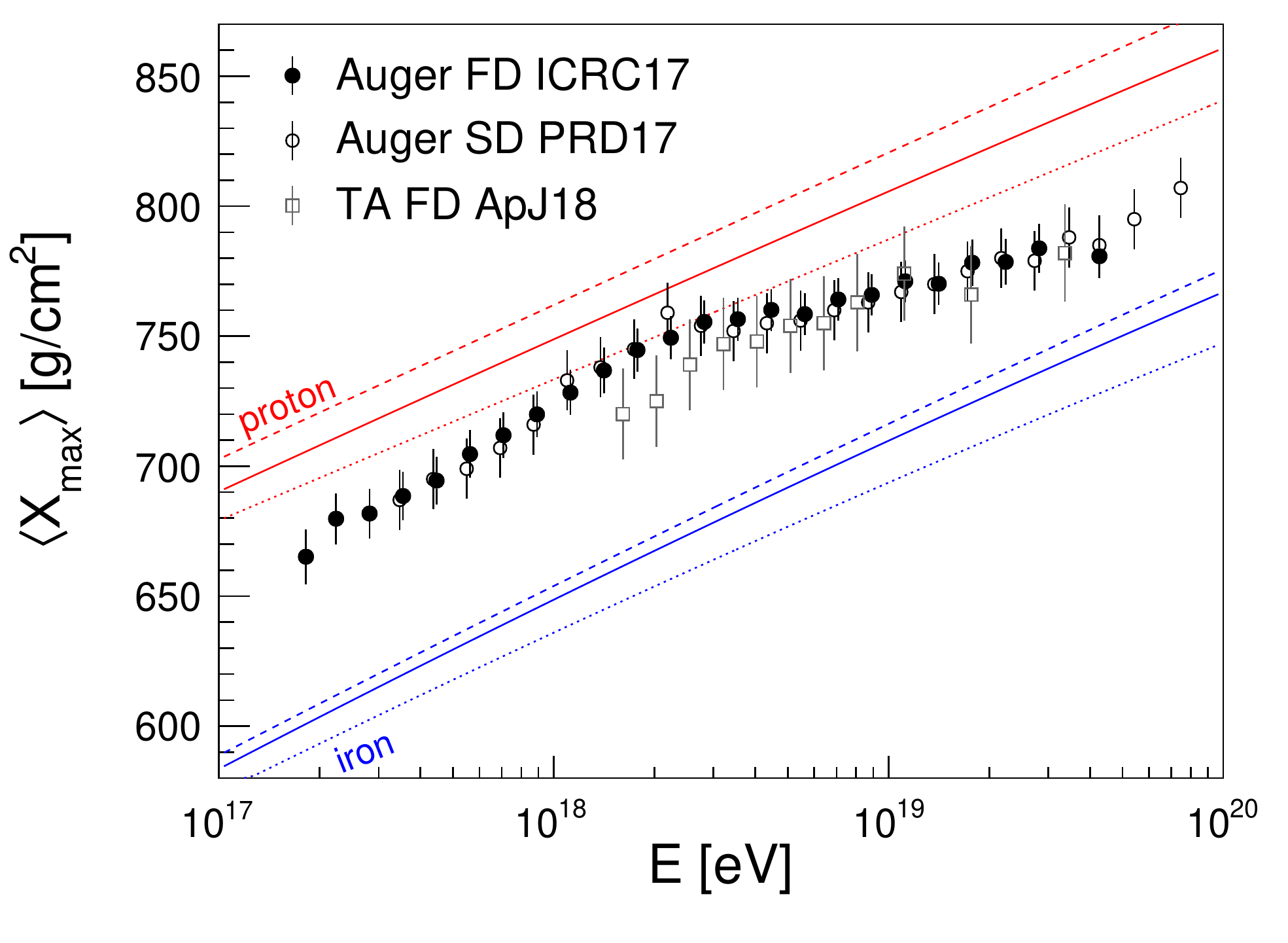}\hspace*{-0.4cm}
  \includegraphics[width=0.52\linewidth]{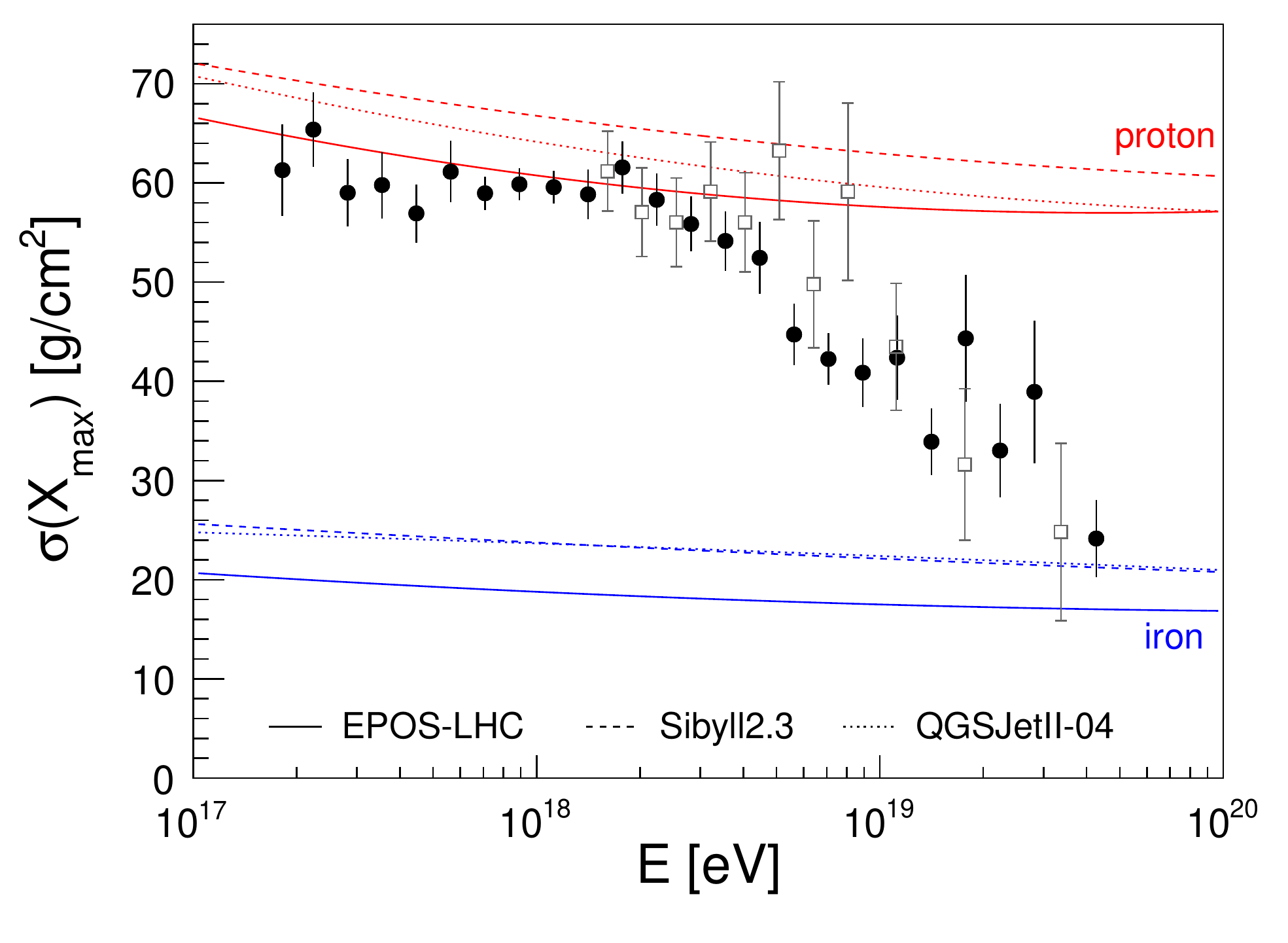}
  \caption{Measurements~\cite{Aab:2017cgk, Bellido:2017cgf, Abbasi:2018nun} of
    the mean ({\it left\/}) and standard
    deviation ({\it right  panel\/}) of the distribution of shower maximum as a
    function of energy.  The
         energy evolution of the mean and standard deviation of \Xmax
         obtained from simulations~\cite{Bergmann:2006yz} of proton-
         and iron-initiated air showers are shown as red and blue
         lines, respectively; from Ref.~\protect\cite{AlvesBatista:2019tlv}.
         \label{fig:UHECR_composition}    }
\end{figure}

The mass composition of UHECRs can be inferred from  the atmospheric depth
\Xmax where the number of particles in an air shower reaches its maximum.
The fluorescence technique has been used both by PAO and TA to determine \Xmax,
but only the former publishes its results in a form such that they can be
directly compared to model predictions.
In the left panel of Fig.~\ref{fig:UHECR_composition}, we show the \Xmax
values obtained by the PAO using fluorescence detectors (FD) as filled dots.
Additionally,
results from the surface array (SD) are shown as open dots. Its higher duty
cycle allows one to extend the energy range, while the energy calibration using
FD data avoids the use of hadronic interaction models. From the evolution of
\Xmax with energy, one can conclude that the composition becomes
lighter between $10^{17.2}$ and $10^{18.33}$\,eV, qualitatively in
agreement with the expectation for a transition from Galactic to
extragalactic CRs in this energy region. Above $10^{18.33}$\,eV,
this trend is reversed and the composition becomes heavier.
The data from the Telescope Array for \Xmax shown by squares
are approximately corrected for detector effects by
shifting the mean by +5\,g/cm$^2$~\cite{Yushkov:2018}. 
Moreover, the TA data points were shifted down by 10.4\% in energy to
match the energy scale of PAO~\cite{Ivanov:2017juh}. After accounting for these
corrections, the \Xmax data from the two experiments are in good
agreement.
In the right panel of Fig.~\ref{fig:UHECR_composition}, we show the
width \X2 of the \Xmax distributions. Again, the \X2 distribution from
TA has to be corrected for the detector resolution by subtracting as
\Xmax resolution  15\,g/cm$^2$~\cite{Abbasi:2018nun} in quadrature.
A wide distribution as obtained at low energies can be caused either by
a light or a mixed composition. At higher energies, the distribution becomes
more narrow, pointing to a purer and heavier composition.

\begin{figure}
  \centering
  \includegraphics[width=0.7\columnwidth]{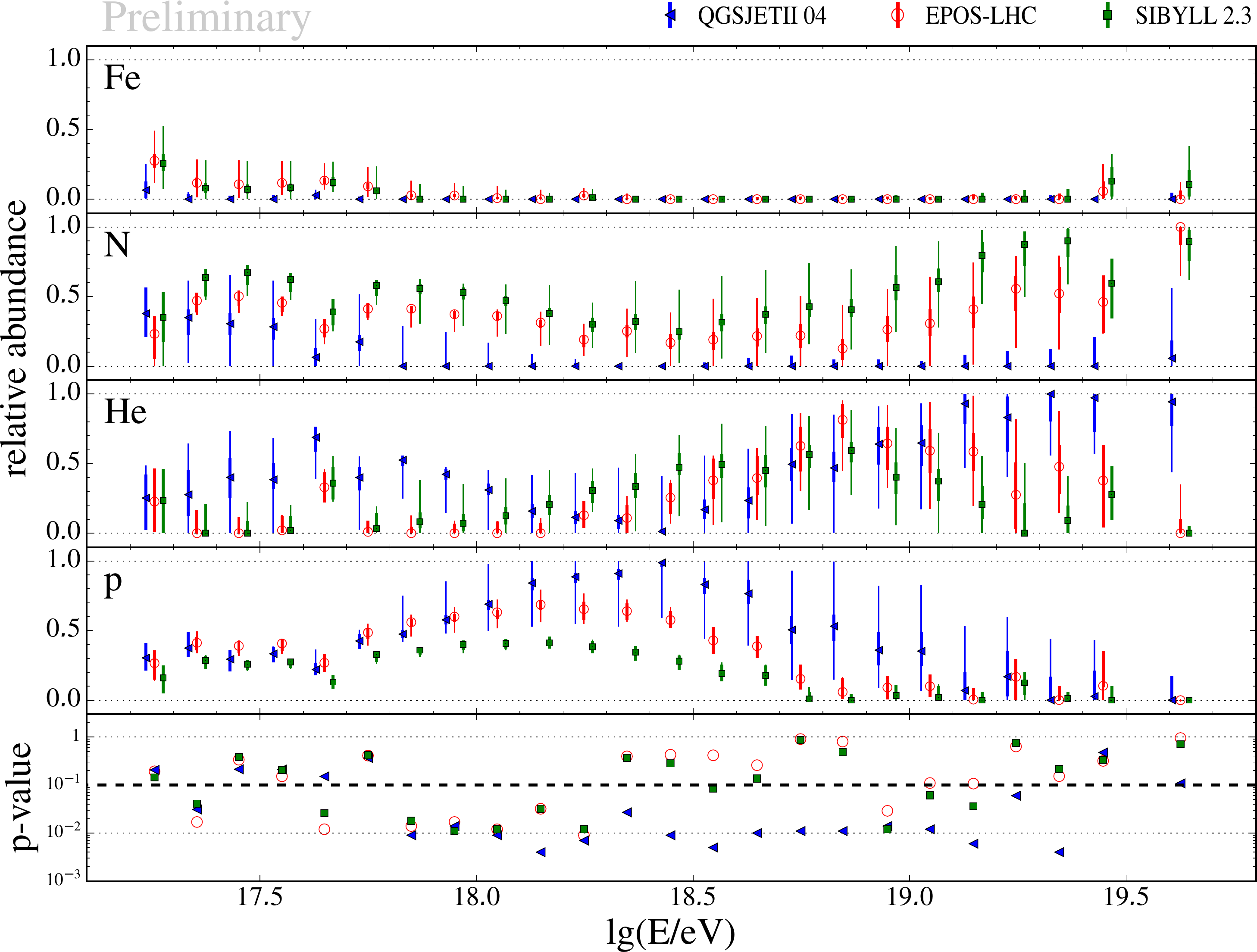}
  \caption{Fraction of the four elements in the UHECR flux;
    from Ref.~\protect\cite{unger:2017cr}. }
\label{fig:auger_composition2}
\end{figure}

Additionally, in both panels of
Fig.~\ref{fig:UHECR_composition}, the predictions from three simulations
for hadronic interactions are shown. All three models,
QGSJET-II-04~\cite{Ostapchenko:2010vb,Ostapchenko:2013pia},
EPOS-LHC~\cite{Pierog:2013ria}, and
SIBYLL 2.3c~\cite{Riehn:2017mfm}, were tuned
to LHC data. The residual difference in the predictions of these models
can be used as a rough estimate for the uncertainties in the
theoretical predictions. Using these simulations, one can compare
the predicted \Xmax distributions for a mixture of CR nuclei to the
observed  distribution and fit the relative fraction of the  CR nuclei.
The result of such a fit for a mixture of proton, helium, nitrogen and
iron nuclei is shown
in Fig.~\ref {fig:auger_composition2}.
Above $10^{18}$\,eV, the dominant component in the UHECR flux changes
successively from protons, to helium and nitrogen, a behaviour
suggestive for the presence of a Peters' cycle. At the lowest energies,
there is evidence for a non-zero iron fraction which drops then to zero.

The TA exposure is a factor 8 smaller than the one of PAO,
leading to rather large statistical uncertainties in particular
in analyses using the FD. Within theses errors, the data agree
with the one from the PAO; for a discussion of the good overall
compatibility of the \Xmax measurements from the PAO and TA see
Ref.~\cite{deSouza:2017wgx}. As a note of caution, we mention
that Monte Carlo simulations of strong interactions as those of
Refs.~\cite{Ostapchenko:2010vb,Ostapchenko:2013pia,Pierog:2013ria,Riehn:2017mfm}
used to infer the mass composition cannot reproduce all details of
the experimental shower measurements. In particular, data on the
muon component in EAS show rather strong deviations~\cite{Dembinski:2019uta},
indicating that the details of a composition analysis like the one shown
in  Fig.~\ref{fig:auger_composition2} have to be interpreted with care.

\subsubsection{Photons and neutrinos as secondaries}
\label{sec4second}

High-energy cosmic rays can interact with gas or photons in their
sources, and with photons from the extragalactic background light (EBL)
during propagation. Any process involving 
hadronization leads mainly to the production of pions, and isospin symmetry
fixes then the ratio of charged to neutral pions produced.
The production of neutrinos is thus intimately tied to the one of 
photons, and both depend in turn on the flux of primary CRs.
Therefore the observation of these CR secondaries can provide important
information on extragalactic CRs.

We discussed already in section~\ref{standard_prop_results} the basic
properties of the secondary photon and neutrino fluxes produced in hadronic
interactions on nuclei. The main difference of secondary production on
background photons is the higher threshold energy $E_{\rm th}$ and the
resulting suppression of the secondary flux at  $E\leq E_{\rm th}$,
$$
{\rm d}N_s/{\rm d}E \sim
\begin{cases} E^{-1}  & \text{for $E<E_{\rm th}$}, \\
              {\rm d}N_{\rm CR}/{\rm d}E & \text{for $E>E_{\rm th}$} .
\end{cases}
$$
For instance, in p$\gamma$ interactions the threshold energy is
$E_{\rm th}\gsim m_\pi m_p/\eps_\gamma$ with
$\eps_\gamma$ as the typical energy of the background photons.
Cosmogenic neutrinos are
mostly produced in interactions on EBL photons with energy
$\eps_\gamma\lsim 10$\,eV.
Taking into account that $\langle E_\nu\rangle=E_p/20$,  this implies that
the flux of cosmogenic neutrinos is suppressed below
$E\approx 2\times 10^{17}$\,eV. If neutrinos are produced by p$\gamma$
interactions in the source, e.g.\
on radiation from an accretion disk with $\eps_\gamma\lsim 1$\,eV, one
expects as threshold $E_{\rm th}\approx 2\times 10^{18}$\,eV.
In contrast, $pp$ interactions lead to a neutrino flux without threshold.

\begin{figure}
  \centering
  \includegraphics[width=0.65\columnwidth,angle=0]{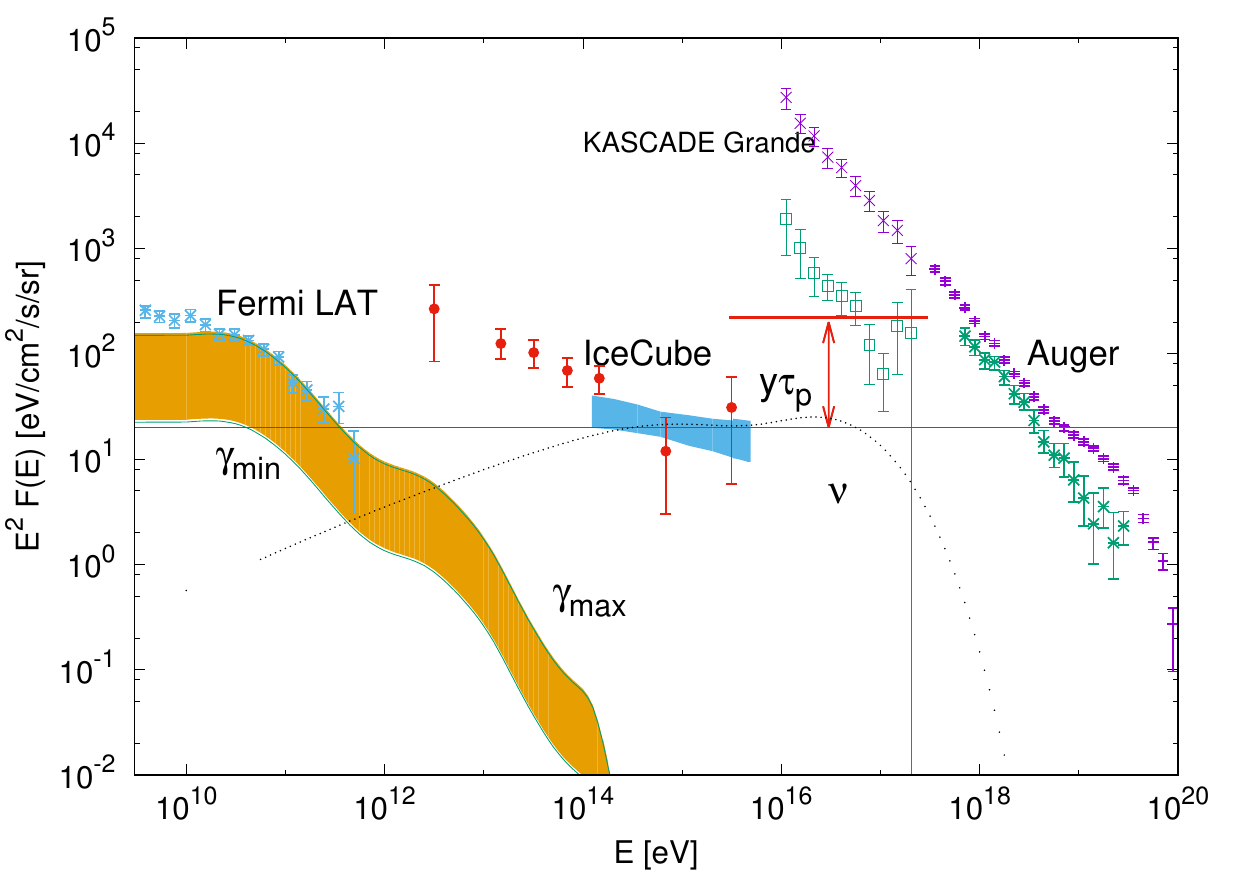}
  \caption{Expected secondary photon and neutrino fluxes from UHECR sources
    compared to the EGRB measured by Fermi-LAT, diffuse neutrino flux by
    IceCube and the all-particle and proton fluxes by KASCADE Grande and
    PAO. The neutrino flux example shown as thin dotted line is from
    Ref.~\protect\cite{Kachelriess:2017tvs}. 
\label{fig:secondary}}
\end{figure}

\paragraph{Diffuse extragalactic gamma-ray background and diffuse neutrino flux}

The Universe is opaque to the propagation of gamma-rays with energies in the
TeV region and above. Such photons are absorbed by pair production on
the EBL. As a result, the extragalactic photon flux at energies
$E\gsim 1$\,TeV is strongly attenuated. High-energy photons are however not
really absorbed but initiate electromagnetic cascades, via the  processes
$\gamma+\gamma_b \to e^++e^-$ and $e^\pm +\gamma_b \to e^\pm + \gamma$~\cite{strong74,Berezinsky:1975zz}.
The cascade develops very fast until it reaches the pair creation threshold.
Thus the Universe acts as a calorimeter for electromagnetic radiation,
accumulating it in the MeV-TeV range as an extragalactic gamma-ray background
(EGRB). The observed EGRB limits therefore all processes that inject
electromagnetic energy. 

The idea to use the EGRB to bound the neutrino flux was first suggested in
Ref.~\cite{Berezinsky:1975zz}. 
In Refs. ~\cite{Kalashev:2007sn,Berezinsky:2010xa},
measurements of the EGRB by EGRET and later Fermi-LAT were used to constrain
strongly evolving UHECR models.
In the meanwhile, the measurement of the EGRB was extended to higher
energies~\cite{Ackermann:2014usa} and, as a result, the limits on the allowed
cascade radiation $\omega_{\rm casc}$
and on the cosmogenic neutrino flux dropped by a factor~3 to
$\omega_{\rm casc}\leq 2 \times 10^{-7}$\,eV/cm$^3$. 

The main contribution to the point-source flux in Fermi-LAT is given  at high
energies by blazars. In Ref.~\cite{Neronov:2011kg}, the contribution of
unresolved sources to the ERGB was studied and it was found that the high
energy part of this background is dominated by unresolved BL~Lacs.
The same conclusion was reached later in the more detailed study of
Ref.~\cite{DiMauro:2013zfa}. Finally, in Ref.~\cite{TheFermi-LAT:2015ykq}
the Fermi collaboration  concluded that up to 86\% of the EGRB comes from
unresolved blazars. Taking these results at face value, the room for any
additional injection of photons is very limited. It is therefore desirable
that the same source class explains both UHECRs and the observed neutrino
flux by IceCube.

In Fig.~\ref{fig:secondary}, we present the expectation for the secondary
photon and neutrino fluxes produced by UHECR sources. The EGRB derived by
the Fermi collaboration in Ref.~\cite{Ackermann:2014usa} shown by blue errorbars
limits the secondary photon fluxes from UHECR sources which has a universal
shape indicated by the orange band~\cite{Ginzburg:1990sk,Kachelriess:2011bi}.
The total UHECR flux is shown by magenta error-bars, while the green error-bars
give the flux of UHECR protons derived from the PAO and KASCADE-Grande
composition measurements~\cite{unger:2017cr,Apel:2013uni}.
The blue band represents the 8~years astrophysical muon neutrino flux multiplied
by three and the red points the 4~years cascade neutrinos measured by
IceCube~\cite{Aartsen:2016xlq,Aartsen:2017mau}. The horizontal thin red line
shows the average neutrino flux level in the case of
an $1/E^2$ flux, which is of the order of 10\% of the EGRB. In the case of
an $1/E^{2.15}$ neutrino flux, the accompanying photons saturate the gamma-ray
bound, as it was first stressed in Ref.~\cite{Murase:2013rfa}.
The thick red line shows the expected level of the proton flux required 
to produce the diffuse neutrino flux. The ratio $y\tau_p$ of the proton
and neutrino fluxes is determined by the corresponding proton interaction
probability  $\tau_p$  and the spectrally weighted average energy
tranfer $\langle y\rangle$ to neutrinos. For a spectral slope close
to two,  $y\sim 0.2$, implying that a large fraction of protons has to
interact inside their sources.

A diffuse flux of surprisingly large magnitude was discovered by IceCube~\cite{Aartsen:2016xlq}: 
While its high-energy part is consistent with $\alpha\approx 2.1$
and a normalisation close to the cascade bound, at lower energies a softer
component appears. The sources of this soft component have to be either
extragalactic and hidden, or Galactic but close to isotropically distributed.
Examples of extragalactic hidden sources are failed type IIn
SNe~\cite{Murase:2010cu,Katz:2011zx,Zirakashvili:2015mua}, while
a close to isotropic Galactic flux could be produced by decays of
PeV dark matter particles, a large CR halo around the Milky
Way~\cite{Taylor:2014hya,Blasi:2019obb}, or extended local neutrino sources
as the wall of the Local Bubble~\cite{Andersen:2017yyg}.
The prime way to distinguish between these two types of solutions is
the detection of the accompanying photon flux, which extends in the Galactic
case up to PeV energies~\cite{Ahlers:2013xia,Neronov:2018ibl}. Such a
detection can be achieved, e.g., by the CARPET
experiment~\cite{Dzhappuev:2017vfy}.

{ Since the majority of neutrino sources is rather weak, 
an identification of neutrino sources via neutrinos multiplets has failed
so far. This constraints the combination of source density and luminosity,
favoring low luminosity sources~\cite{Murase:2016gly}.  However, the first
source  for which there is evidence of a correlation with IceCube neutrino 
event~170922A is the blazar TXS 0506+056, which is located at relatively
high redshift $z=0.3$~\cite{IceCube:2018dnn}. There are two possible
explanations: If blazars are the only dominant neutrino sources in IceCube, 
TXS 0506+056 at $z=0.3$ combined with the absence of high-energy neutrino
multiplets~\cite{Murase:2016gly} favors a subclass of bright BL Lacs and
low luminosity FSRQ as neutrino source~\cite{Neronov:2018wuo}.
 
Alternatively, one can assume that blazars like TXS~0506+056 give a
subdominant contribution, while a new class of weak sources gives the major
contribution to the extragalactic IceCube signal. Because 86\% of the EGRB
originates from unresolved blazars, this is possible only for sources which
give a minor contribution to EGRB, i.e. which have a flux $\propto 1/E^2$,
cf.\ with the horizontal red line in  Fig.~\ref{fig:secondary}. Such sources
should be abundant to obey the multiplet constraint of
Ref.~\cite{Murase:2016gly}, as for example star-burst galaxies 
or ordinary low luminosity AGNs.  To avoid the additional contribution of
UHECR sources to the EGRB, it is still important that such sources are major
contributors to UHECR, as in the generic case of the minimal model of
Ref.~\cite{Kachelriess:2017tvs}. 
}

\subsubsection{Anisotropies}
\label{transition_anisotropy}

The possibility to observe in the extragalactic CR flux anisotropies connected
to the large-scale structure (LSS) of CR sources, or even to identify CR
sources, depends on the number density $n_s$ of CR sources and the strength
$B_{\rm exgal}$ of the extragalactic magnetic field (EGMF). If the energy-loss
horizon $\lambda_{\rm hor}$ of CRs is large 
compared to the scale of inhomogeneities in their source distribution,
then the flux of extragalactic UHECRs is isotropic in the rest frame
of the CMB.
In the same energy range, peculiar velocities average out on cosmological 
scales and the UHECR flux is thus isotropic at leading order.
The movement of the Solar system with velocity  $u=(368\pm 2)$\,km/s with
respect to the CMB results in a cosmological Compton-Getting  effect with
dipole amplitude
\begin{equation}
  \delta_{\rm CG}\equiv \frac{I_{\rm max}-I_{\rm min}}{I_{\rm max}+I_{\rm min}}
  =\left(2-\frac{\d\ln I}{\d\ln E}\right)\,u \,.
\end{equation}
Taking into account the observed spectrum $I(E)\propto E^{-2.7}$ of
cosmic rays above the ankle, the numerical value of the dipole amplitude
follows as $\delta_{\rm CG}=(2+2.7)\,u\simeq 0.6\%$~\cite{Kachelriess:2006aq}.

Moving to higher energies, the free mean path of CRs decreases and anisotropies
connected to  the LSS of CR sources should become more prominent
and replace the cosmological Compton-Getting effect.
Finally, at sufficiently high energies deflections in magnetic fields may
become negligible, and a small enough number of bright point sources
results in  small-scale clusters of arrival directions around or near the true
source positions. Depending on the actual values of $n_s$ and $B_{\rm exgal}$,
two out of the three cases described may not be relevant: The
cosmological Compton-Getting effect may be only realized at such low
energies that the extragalactic CR flux is subdominant. Or sufficiently
small deflections of CRs may require unrealistic high energies, where the
statistics of UHECR events is too low. We start therefore this subsection
with a discussion of the expected deflections of CRs in the  Galactic and
extragalactic magnetic fields.

\paragraph{Deflections of UHECRs}

Propagating the distance $d$, a CR with charge $Ze$ and perpendicular
momentum $p_\perp$ is deflected in a regular magnetic field by the angle
\begin{equation}
\theta \simeq 0.52^\circ Z
\left(\frac{p_{\perp}}{10^{20}\,{\rm eV}}\right)^{-1}
\left(\frac{d}{{\rm kpc}}\right)
\left(\frac{B}{10^{-6}{\rm G}}\right) \,.
\label{eq:theta}
\end{equation}
This formula also holds in a turbulent field for distances $d\ll L_{\rm c}$.
In the opposite limit,  $d\gg L_{\rm c}$, the CR propagation resembles a random
walk. If the average deflection angle per correlation length $L_{\rm c}$ is
small, $R_{\rm L}\gg L_{\rm c}$, the variance of the deflection angle after the
distance $d$ is given by
\begin{equation}
\label{eq:thetarms}
 \theta_{\rm rms} \equiv \langle \theta^2\rangle^{1/2}
 \simeq  \frac{(2dL_{\rm c}/9)^{1/2}}{R_{\rm L}} = 25^\circ  Z
 \left( \frac{10^{19}{\rm eV}}{E}\right)
 \left(\frac{d}{100\,{\rm Mpc}}\right)^{1/2}
 \left(\frac{L_{\rm c}}{1\,{\rm Mpc}}\right)^{1/2}
 \left(\frac{B}{10^{-9}{\rm G}}\right) \,,
\end{equation}
where the numerical prefactor $2/9$ was determined in Ref.~\cite{Waxman:1996zn}.
The increased path-length compared to straight-line propagation leads
to the time-delay 
\begin{equation}
\label{eq:Deltat}
 \Delta t \simeq  \frac{d\theta_{\rm rms}^2}{4} = 1.5\times 10^3 \, {\rm yr}\, Z^2
 \left( \frac{10^{20}{\rm eV}}{E}\right)^2
 \left(\frac{d}{10\,{\rm Mpc}}\right)
 \left(\frac{L_{\rm c}}{1\,{\rm Mpc}}\right)
 \left(\frac{B}{10^{-9}{\rm G}}\right)^2 
\end{equation}
of charged CRs relative to photons~\cite{MiraldaEscude:1996kf}.

Note that the increase in the path-length of CRs propagating in magnetic
fields can result in the formation of a magnetic
horizon~\cite{Parizot:2004wh,Berezinsky:2005fa}:
The maximal distance a CR can travel is in the diffusion picture given by
\be
 r_{\rm hor}^2 = \int_0^{t_0} \d t \:D(E(t)) 
               = \int_{E_0}^{E} \frac{\d\!E'}{\beta}\: D(E'(t)) \,,
\ee
where $t_0$ is the age of the  Universe. If we consider CRs
with energy below
$E_0\lsim 10^{18}$eV, the energy losses are mainly due to the expansion
of the Universe, $\beta=\d\!E/\d t=-H$. For an estimate, we can use a
``quasi-static'' Universe, $H(t)=H_0$ and $t_0H_0=1$. Then
$E(t)=E_0\exp(-H_0t)$ and
\be
 r_{\rm hor}^2 = \frac{cL_{\rm c}}{6H_0} \: \left(\frac{E}{E_{\rm tr}}\right)^2 
                (\exp(2)-1) \,,
\ee
using $D(E)=cL_{\rm c}/3 \:(E/E_{\rm tr})^2$ valid for $R_{\rm L}\gsim L_{\rm c}$
with $E_{\rm tr}$ defined in Eq.~(\ref{eq:Ecr}).
If we assume that a magnetic field with correlation length $L_{\rm c}\sim\,$Mpc
and strength $B\sim 
0.1\,$nG exists in a significant fraction of the Universe, then the size
of the magnetic horizon at $E=10^{18}\:$eV is $r_{\rm hor}\sim 100\,$Mpc.
Hence, similar to the GZK suppression above $6\times 10^{19}\,$eV, we
see a smaller and smaller fraction of the Universe for lower and lower 
energies. As a consequence, the spectrum of extragalactic CRs
visible to us steepens at  $E<10^{18}\,$eV and the extragalactic component
becomes sub-dominant.

From Eqs.~(\ref{eq:theta}) and~(\ref{eq:thetarms}), we can estimate the
deflection of CR protons with energy $E=5\times 10^{19}$\,eV by the GMF.
Excluding paths skimming
the Galactic plane, we can set $d=500$\,pc and $B=3\mu$G both for the
regular and the turbulent field, resulting in a deflection
$\theta\sim 1^\circ$ by the regular and $\theta_{\rm RMS}\sim 0.2^\circ$ by the
turbulent  Galactic magnetic field, respectively, what agrees with the
results of the numerical
calculations of Ref.~\cite{Giacinti:2011uj}. Thus the deflection
is comparable to the angular resolution of UHECR experiments, and is dominated
by the deflection in the regular field. In contrast, iron nuclei would
be deflected by the GMF around $30^\circ$ at this energy,
making astronomy with heavy nuclei close to impossible. 
While the deflections of CR protons by the GMF should not prevent
UHECR astronomy, the selection of proton events requires however an
estimate of the mass number of the primary CRs on an event-by-event basis.

The impact of the EGMF on the deflections of UHECRs  cannot be
reliably estimated, since no convincing theory for its origin and its
amplification mechanism exists. The seed fields of the EGMF could be created
in the early
universe, e.g. during phase transitions, and then amplified by MHD
processes. Alternatively, an early population of starburst galaxies or
AGN could have generated the seeds of the EGMFs at redshift between five
and six, before galaxy clusters formed as gravitationally bound
systems. In both cases, a large fraction of the universe may be filled
with seed fields for EGMFs. Another possibility is that the ejecta of
AGN magnetized the intra-cluster medium only at low redshifts, and
that thus the EGMF is confined within galaxy clusters and groups. 
Other mechanisms have been suggested and hence no unique model with
unique predictions for the EGMF exists.
Simulations of the LSS  including magnetic fields differ both in the input
physics (seed fields, amplification mechanism), the numerical algorithms and
the extraction of the results~\cite{Sigl:2004yk,Dolag:2004kp,Bruggen:2005ti}.
Therefore, it should be not too surprising that
their results disagree strongly:
In the simulation performed in Ref.~\cite{Sigl:2004yk}, a significant
fraction of all UHECRs suffers deflections comparable or larger than
given by Eq.~(\ref{eq:thetarms}), while the simulation performed
in~\cite{Dolag:2004kp} favour considerably smaller values. If the
results of the latter simulation are closer to reality, deflections
in extragalactic magnetic fields may be negligible at least for
protons even at relatively low energies such as $4\times
10^{19}$\,eV. Similar conclusions were obtained in Ref.~\cite{Hackstein:2016pwa}
which studied the propagation of UHECR protons in extragalactic magnetic 
simulated with the cosmological ENZO code.
Currently, the best upper limits in the strength of EGMFs are
1.7\,nG for correlation lengths close to the Jeans scale, and
0.65\,nG  for correlation lengths close or larger than the observable
Universe~\cite{Pshirkov:2015tua}.

\paragraph{Dipole anisotropy}

The (projected) dipole anisotropy $\delta$ measured by several experiments has
been shown in Fig.~\ref{anisotropy_dipole_data} as function of energy. Below
$10^{17}$\,eV, the phase of the anisotropy shown in the left panel is
approximately constant, except for a flip at $\simeq 200$\,TeV. By contrast,
at energies above $10^{17}$\,eV, the phase changes smoothly towards
100\,degrees in right ascension (R.A.), i.e.\ it points roughly towards
the Galactic anticentre.
This clearly suggests that the extragalactic CR flux starts at $10^{17}$\,eV
to become sizeable. Considering next the strength of the dipole, shown in
the right panel of Fig.~\ref{anisotropy_dipole_data}, we note that above
$10^{16}$\,eV only upper limits have existed until recently. Only in 2017, the
PAO could detect the dipole performing a one-dimensional harmonic analysis
in R.A. splitting the events in two energy bins, 4--8\,EeV and
$>8$\,EeV~\cite{Aab:2017tyv,Aab:2018mmi}. While the amplitude of the first
harmonic in the low-energy energy bin was consistent with isotropy, the
amplitude $6.5^{+1.3}_{0.9}\%$ in the second energy $>8$\,EeV deviates more than
5\,$\sigma$ from isotropy.
Thus the observed amplitude of the dipole anisotropy is a factor~10
larger than the one predicted for the extragalactic Compton-Getting
effect~\cite{Kachelriess:2006aq}.
Higher-order harmonics like the quadrupole moment
are consistent with isotropy. Therefore a combined harmonic analysis in
R.A. and azimuth could be performed, and thereby the dipole vector could
be reconstructed, with coordinates ${\rm R.A.}=100^\circ\pm 10^\circ$ and
${\rm dec}=-24^\circ\pm 13^\circ$.
The estimated direction may be connected to an
overdensity in the local galaxy distribution, seen e.g.\ in the 2MRS
catalogue~\cite{Erdogdu:2005wi}. Thus the behaviour of the dipole direction
suggests that the transition from Galactic to extragalactic CRs
starts at $10^{17}$\,eV.

\begin{figure}
  \centering
  \includegraphics[width=0.65\columnwidth]{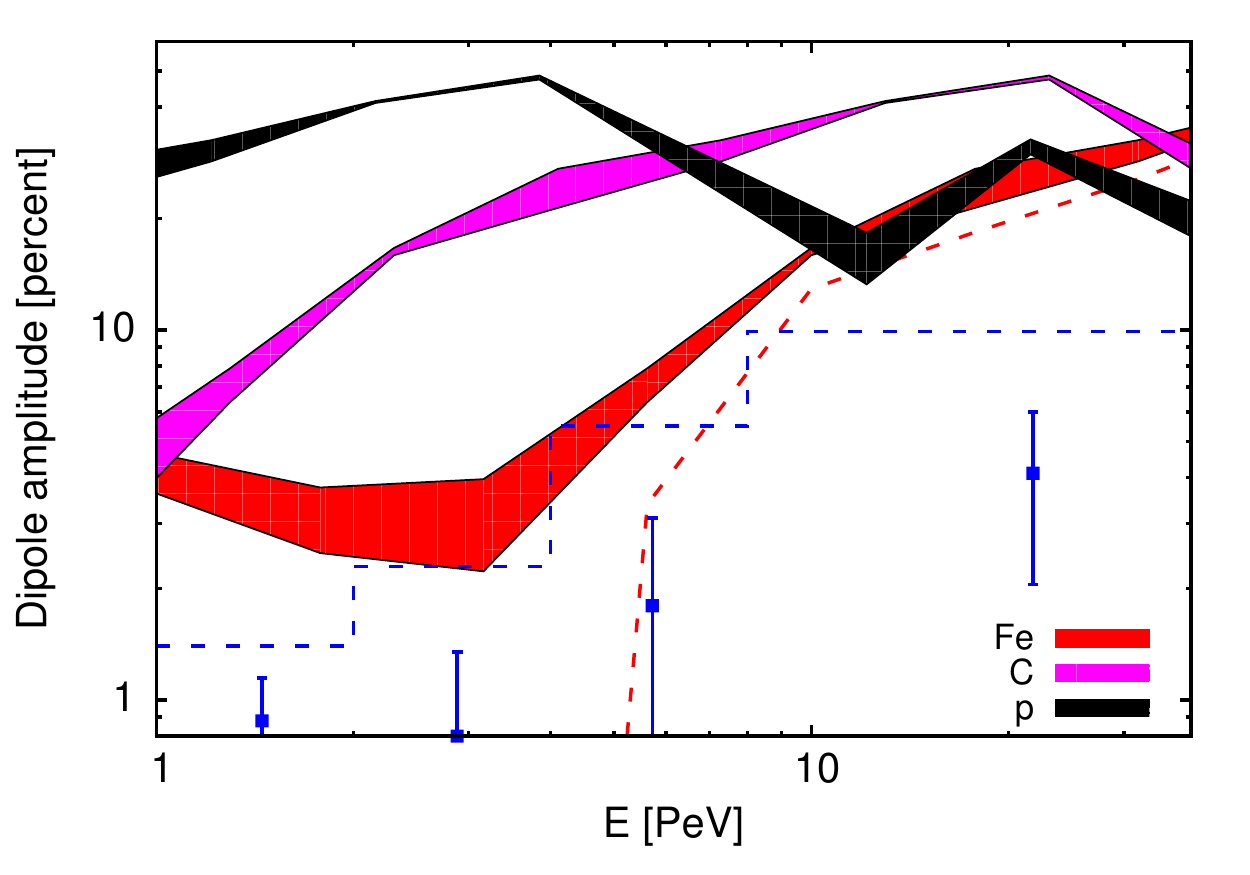}
  \caption{Predicted dipole from Galactic sources of nuclei as function of
  energy. The band corresponds to a variation of the vertical scale height
  $z_0$ of the turbulent GMF between 2 and 8\,kpc;
  adapted from Ref.~\protect\cite{Giacinti:2011ww}.}
\label{fig:Gal_dipole}
\end{figure}

This conclusion can be strengthened considering the amplitude of the
dipole. In Ref.~\cite{Giacinti:2011ww} it was shown
that a light (intermediate) Galactic CR flux leads to a dipole of order $20\%$
(10\%), cf.\ with Fig.~\ref{fig:Gal_dipole}. 
This overshoots clearly both the
limits below 8\,EeV and the observed value at  $>8$\,EeV. Thus the dominant
light-intermediate contribution to the CR flux measured by the PAO above
$3\times 10^{17}$\,eV  has to be extragalactic. Similar results were
obtained later in Ref.~\cite{Abreu:2012ybu}. In both
Refs.~\cite{Giacinti:2011ww,Abreu:2012ybu}, the trajectories of individual
CRs were followed solving the Lorentz equation. In contrast, claims like
the one in Ref.~\cite{Kumar:2013jaa} that Galactic protons are  consistent
with the observed dipole amplitude are typically based on the diffusion
or a simplified random walk approach which are not justified at these
energies.

\paragraph{Medium scale anisotropies}

\begin{figure}
  \centering
  \includegraphics[width=0.65\columnwidth]{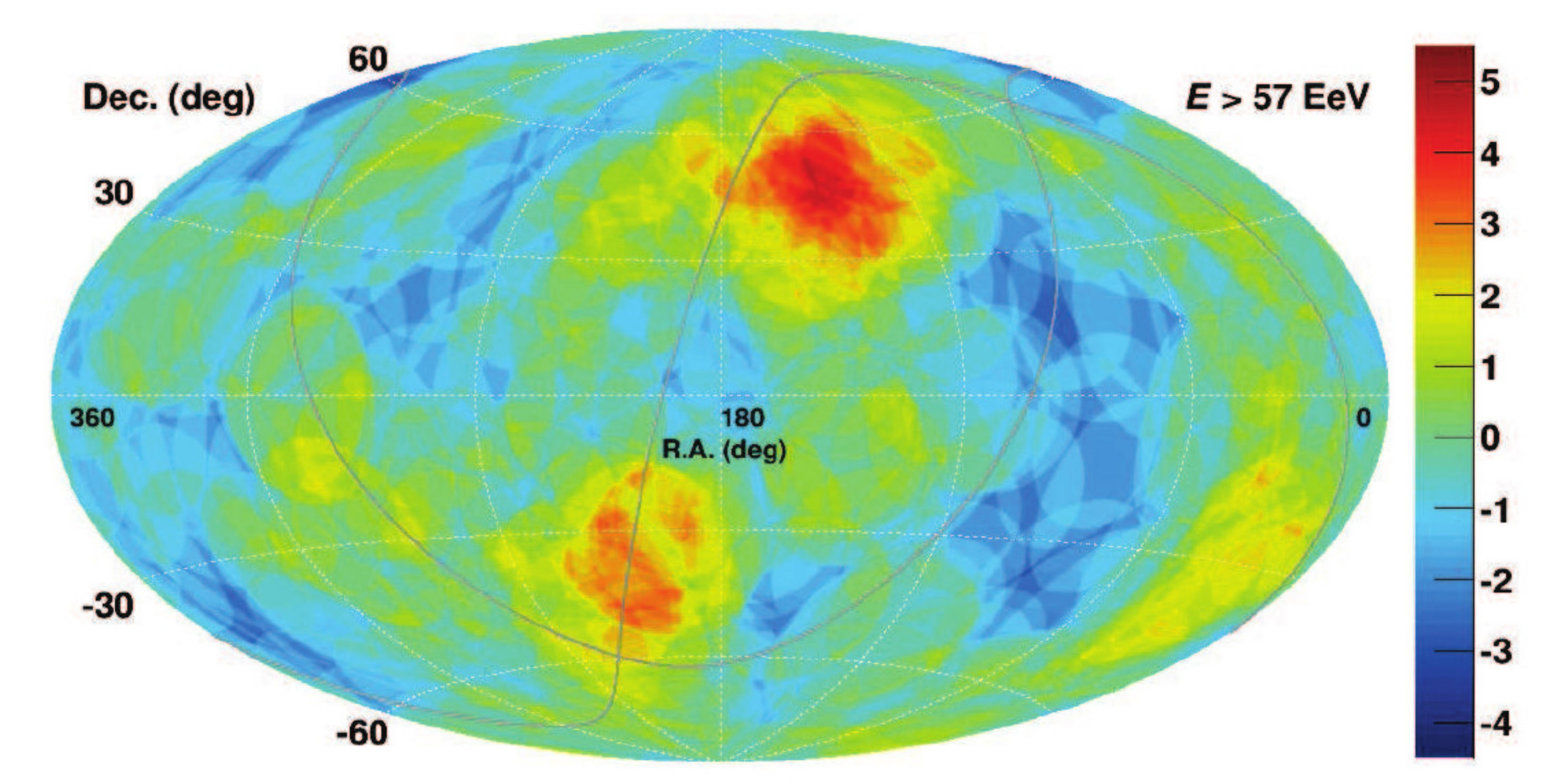}
  \caption{Sky map of UHECR events with $E>5.7\times 10^{19}$\,eV from TA and
    PAO; the arrival directions are smeared out on circles with radius
    $25^\circ$; from Ref.~\protect\cite{Matthews:2017waf}.}
\label{fig:TAhotspot}
\end{figure}

Combining the data of the AGASA, HiReS, Yakutsk, SUGAR, and other experiments
after rescaling their energies to a common scale, Ref.~\cite{Kachelriess:2005uf}
found in 2005 medium scale anisotropies with 20--25 degrees scale at
the $3\sigma$
confidence level. Anisotropies on such scales were seen later in the
experimental data from the next generation experiments
TA~\cite{Abu-Zayyad:2013vza,Abbasi:2014lda} and PAO~\cite{Aab:2018chp}.
In Fig.~\ref{fig:TAhotspot}, we show a sky map combining events with
$E>5.7\times 10^{19}$\,eV from these two experiments. The arrival directions
are smeared on circles with radius $25^\circ$ and the colour code indicates
the local significance of an over- or under-fluctuation.  The most
significant feature of this map is the ``TA hotspot'', which is centered
at ${\rm R.A.}\simeq 150^\circ$ and ${\rm dec}\simeq 40^\circ$.
This direction points towards the Ursa Major supercluster; note also that the
Supergalactic plane indicated by the thin line is close to the hot spot.
{ Morover, the starburst galaxy M82 coincides with the TA hot spot.}
The local significance of this hot spot is $5.2\sigma$, which is reduced to
$3.4\sigma$ taking into account that the spot could appear anywhere by random
fluctuations~\cite{Matthews:2017waf}. An additional ``warm spot''  can be
seen in the field-of-view of the PAO, which coincides with Cen~A.
Also the warm spot overlaps with the supergalactic plane.

An alternative method to search for UHECR sources are correlation
studies. { Several studies have shown that the distribution of the TA
  hotspot events is consistent with
the hypothesis of a single source, the nearby starburst galaxy
M82 being a promising candidate~\cite{Fang:2014uja,He:2014mqa}.}
The recent analysis~\cite{Aab:2018chp} of the PAO data showed
evidence for a correlation of the arrival directions of UHECRs with starburst
galaxies, i.e.\ galaxies which are characterized by exceptionally high
rates of star formation.  Specifically, for UHECR with observed energies
$E>39$\,EeV, a model which attributes 9.7\% of the UHECR flux to nearby
starburst galaxies
(and the remaining 90.3\% to an isotropic background) was found to be favoured,
with $4\sigma$ significance, over a completely isotropic test hypothesis.
About  90\% of the anisotropic flux was found to be associated to four nearby
starburst galaxies: NGC~4945, NGC~253, M82, and NGC~1068.
Alternatively, a correlation analysis with 17~bright nearby AGNs was
performed. Here, the warm spot is related to Cen~A and around 7\% of the
total flux is attributed to the selected AGNs. These claims were
reassessed in a new analysis of the TA data~\cite{Abbasi:2018tqo}, which gave
results consistent with both the PAO anisotropy and with isotropy, due to the
small number of events.

\subsection{Transition energy}

The question at which energy the transition from Galactic to extragalactic
CRs takes place is fundamental to our understanding of Galactic CR sources
and the requirements on their acceleration mechanisms as well as to the
determination of the nuclear composition and the injection spectrum of
extragalactic sources. In the past, the two most promising choices for the
transition energy were to associate it with one of the evident features of
the CR spectrum: The second knee around $E\simeq (1-5)\times10^{17}$\,eV or the
ankle at $E\simeq3\times 10^{18}$\,eV. The latter option offered a simple
explanation for the sharpness of the ankle as the cross-over between the end
of Galactic flux and the start of the extragalactic component. Moreover,
it allowed for an extragalactic injection spectrum  $Q(E)\propto 1/E^\beta$
with $\beta\approx 2$, i.e.\ close to the theoretical expectation for diffusive
shock acceleration. The main disadvantage of this suggestion is the enormous
pressure it puts on acceleration models for Galactic CR sources. Moreover,
this solution may lead to the following ``fine-tuning problem'': Since
the acceleration and diffusion of CRs is rigidity-independent, one expects
that the end of the Galactic CR spectrum shows a sequence of cutoffs at
$ZE_{\max}$, i.e.\ follows a Peter's cycle. If the ankle is identified with the
transition, it is natural to assume that the second knee corresponds to the
iron knee. Thus, in this interpretation, the second knee signals the end of
the Galactic iron flux from those sources which contribute the bulk of
Galactic CRs.
Therefore an additional Galactic population of CR sources would be required
to fill the gap between the second knee and the ankle. If this population is
unrelated to the standard population of Galactic CR sources, it is surprising
that the normalisation of the two fluxes is so close.

The challenge of models identifying the second knee as the transition to
extragalactic CRs is to find a physical mechanism which explains the ankle
as a consequence of either the propagation of extragalactic CRs or of 
interactions in their sources. The first successful model of this kind
explained the ankle by the dip  in the pair-production losses of protons
on cosmic microwave background (CMB) photons
$p+\gamma_{\rm CMB} \rightarrow p+e^++e^-$~\cite{Berezinsky:2005cq}.
This elegant possibility is however excluded by composition measurements.
Only recently, the authors of Ref.~\cite{Unger:2015laa} suggested as a
viable alternative a model which relies on photo-disintegration of CR nuclei
on background photons inside CR sources.
Note also that from a theoretical prespective, a successful model for the
transition has to address both the extragalactic and the Galactic
contributions to the measured fluxes of individual groups of elements,
$F_i^{\rm obs}(E)= F_i^{\rm exgal}(E)+F_i^{\rm gal}(E)$, since only the sum of
both is observed.

How can these two options for the transition energy, the second knee and the
ankle, be experimentally distinguished? It is natural to expect that the
nuclear composition of Galactic and extragalactic CRs should differ, because
of propagation effects and of the different nature of their sources.
In particular, 
the Galactic CR spectrum should become close to its end iron-dominated.
A similar behaviour is expected for the extragalactic flux, shifted however
to higher energies. Thus one expects the extragalactic composition at the
transition energy to be lighter than the Galactic one. Therefore the signature
of the transition in the composition is the disappearance of the (Galactic)
iron, and the increase of a light or intermediate  extragalactic component.
An additional powerful constraint comes from anisotropy measurements which
limit a light or intermediate Galactic component.

From the data presented in the previous section, the following conclusions
can be drawn: Using only the composition data, the limits on the iron fraction
from Fig.~\ref{fig:auger_composition2} imply that the Galactic contribution
to the observed CR spectrum has to die out before $7\times 10^{17}$\,eV. The
confidence in this conclusion is strengthened considerably, if one combines
the composition and anisotropy measurements: In Ref.~\cite{Giacinti:2011ww},
it was shown that a light (intermediate) Galactic CR flux leads to a dipole
of order $20\%$ (10\%), overshooting clearly the limits which are on the
percent level~\cite{Aab:2018mmi}. Thus the dominant light-intermediate
contribution to the CR flux measured by the PAO above $3\times 10^{17}$\,eV
has to be extragalactic. Finally, we recall that the smooth change of the
dipole phase at energies above $10^{17}$\,eV towards the Galactic anticentre
supports the suggestion that the transition from Galactic to extragalactic CRs
starts at $10^{17}$\,eV. Thus we conclude that the second knee marks the
transition  between Galactic and extragalactic CRs. Then it is natural
that the second knee is close to its upper end of the range
$(1-5)\times 10^{17}$\,eV of values
considered in the literature and, for definiteness, we choose
$E\simeq 5\times10^{17}$\,eV for the position of the second knee.

\subsection{Models for extragalactic CRs and the transition}
\label{modeltrans}

The first concrete model\footnote{For an early suggestion that the ankle is a propagation effect of UHECRs  see
  Ref.~\cite{1967PhLA...24..677H}} able to explain the ankle as a feature of the extragalactic
CR spectrum was the dip model~\cite{Berezinsky:2002nc,Berezinsky:2005cq}. The
main assumption of this model is that the extragalactic CR flux consists of
protons, with a maximal admixture of $\lsim 10\%$ of helium. Then the ankle
can be explained as a feature in the extragalactic CR spectrum imprinted
by pair-production losses of protons on CMB photons, cf.\ with the energy
loss rate of protons shown in Fig.~\ref{fig:UHECR_losses}.
This elegant possibility
has been however excluded by composition measurements, in particular of the
PAO. It is interesting to note that meanwhile even the non-observation of
cosmogenic neutrinos challenges this model~\cite{Heinze:2015hhp}.

Since  composition measurements, in particular of the fluctuations of the
shower maximum \X2, pointed to an increase of the mean mass number of CRs
with increasing energy, models including nuclei were proposed as alternative.
For instance, the models of
Refs.~\cite{Allard:2005ha,Allard:2005cx,Hooper:2006tn} used a  mixed
composition together with a power-law in rigidity and an exponential
cut-off, $Q(\R)=Q_i \R^{-\beta} \exp(-\R/\R_{\max})$, for the injection spectrum.
These models could reproduce \Xmax\ and \X2\ data, but lead to the ankle as
transition energy.
Therefore such models require an additional light extragalactic component
below the ankle. The spectral index of the injection spectrum of this
additional population should be steeper ($Q(\R)\propto \R^{-2.7}$)
than the one of the population responsible for the spectrum above the
ankle~\cite{Aloisio:2013hya}. Again, it is surprising that the normalisation
of these two contributions is so close, if these two populations are
unrelated.

The end of the proton component measured by KASCADE-Grande can be
extended smoothly to the one observed by the PAO by a power law
with slope $\alpha\sim 2.2$, cf.\ with Fig.~\ref{fig:secondary}.
Any model aiming to extend the extragalactic
flux below the ankle has to explain this proton component. In
Ref.~\cite{Giacinti:2015pya}, it was assumed
that this flux reflects the original injection spectrum of protons,
since the slope is consistent with the one expected from diffusive shock
acceleration. It was shown that star-forming galaxies cannot explain this
proton component, while BL Lacs/FR~I galaxies could both provide the
proton component and a dominant contribution to the observed neutrino
flux and the EGRB. However, this work did not address the fluxes of 
heavier nuclei required by the composition measurements.

\begin{figure}
  \hspace*{-0.5cm}
  \includegraphics[width=0.35\columnwidth,angle=0]{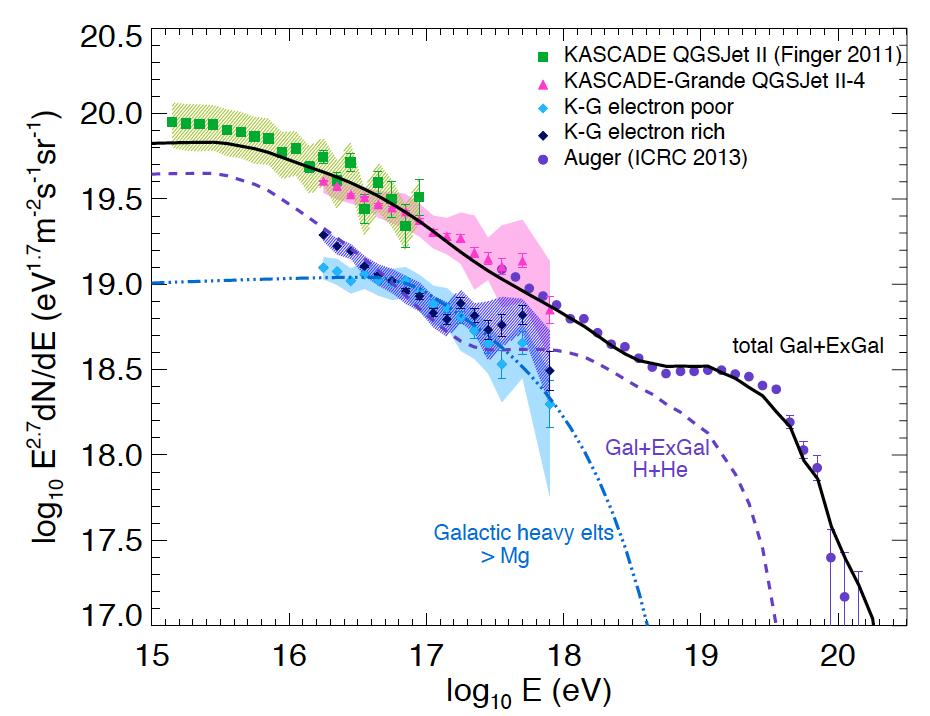}
  \includegraphics[width=0.65\columnwidth,angle=0]{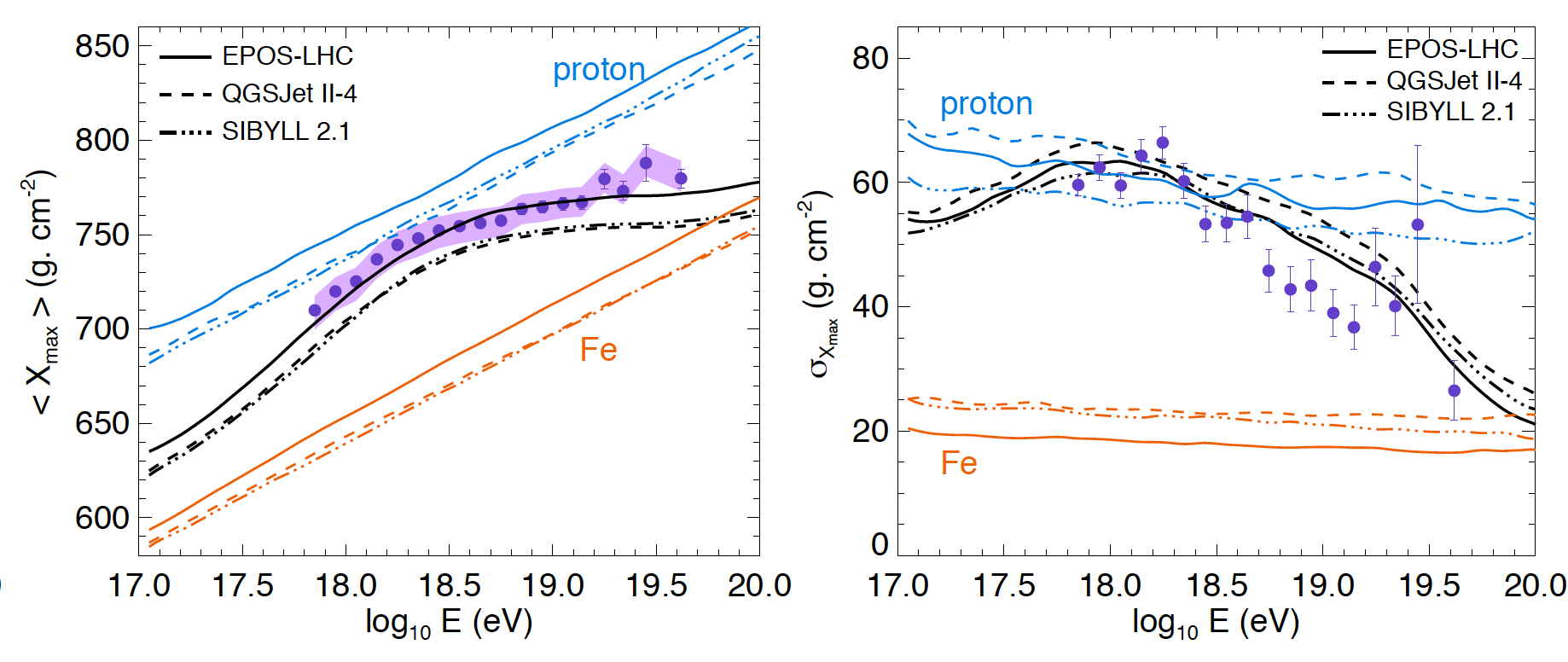}
  \caption{{\em Left panel:\/} Galactic and extraglactic contributions to the
    total CR flux in the GRB model of Ref.~\protect\cite{Globus:2015xga} compared to
    experimental data from PAO and KASCADE-Grande.
    {\em Right panel:\/} $X_{\max}$ and \X2 as function of energy in model of
    Ref.~\protect\cite{Globus:2015xga} compared to data from the  PAO.
    \label{GAP}}
\end{figure}

In an alternative scenario, the extragalactic proton component originates
from the photo-disintegration of heavier nuclei in photon fields present
in the source~\cite{Globus:2014fka,Unger:2015laa}. This  mechanism has been
employed in a specific model of UHECR acceleration by gamma-ray bursts (GRB)
in Refs.~\cite{Globus:2014fka,Globus:2015xga}.
In this model,  the photo-disintegration of low-energy nuclei leads to a
flattening 
of their spectra from $\beta\simeq 2.1-2.2$ to $\beta\simeq 1$.
Only  the proton spectrum follows the original acceleration spectrum
because of a decaying neutron component which escapes from the source,
cf.\ with Fig.~\ref{GAP}. While the original model resulted
in a transition at the ankle~\cite{Globus:2014fka}, choosing a stronger
redshift evolution $Q(z)\propto (1+z)^{m}$ with $m=3.5$ for GRBs increases the
extragalactic contribution below the ankle~\cite{Globus:2015xga}. 
Nevertheless, the required Galactic iron fraction is , e.g., 20\% at
$10^{18}$\,eV and thus relatively high compared to the determination from
Ref.~\cite{unger:2017cr}.

\begin{figure}
  \centering
  \includegraphics[width=0.99\columnwidth,angle=0]{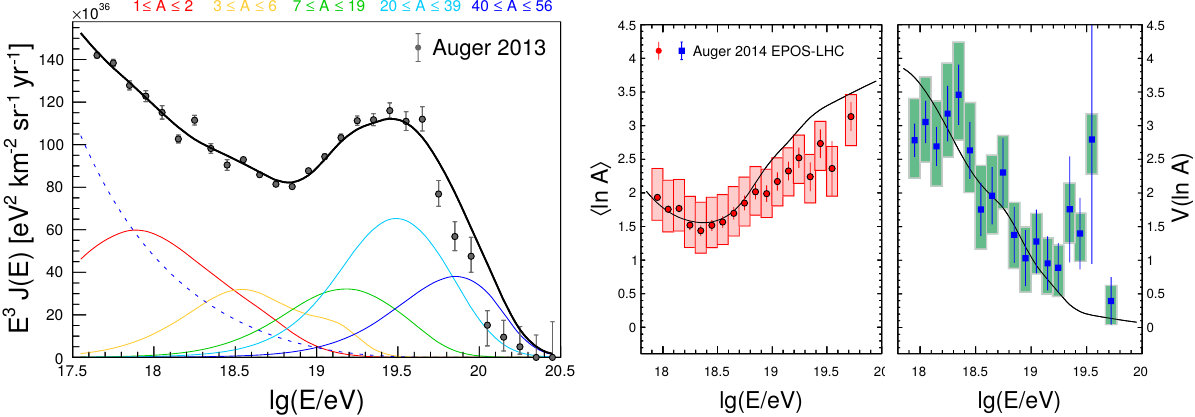}
  \caption{The energy spectrum, $\ln(A)$ and $\sigma(\ln(A))$ as functions
    of energy  for an injection composition following a Galactic mixture.
    The dotted line shows the assumed Galactic flux; adapted from
    Ref.~\protect\cite{Unger:2015laa}.
\label{UFA}}
\end{figure}

A generic calculation of the effect of photo-nuclear interactions in
astrophysical sources was performed in Ref.~\cite{Unger:2015laa}.
In their fiducial model a hard energy spectrum, $\beta=1$, was assumed and
the maximum proton energy and photon background energy were derived by
fitting the UHECR data. Within a variety of source evolutions and different
assumptions about the shape of the photon spectrum, the maximum energy of
protons and photon background were found to be typically
$E_{\max}^p=3\times 10^{18}$\,eV and $\eps_\gamma=0.1$\,eV, respectively.
A good fit to both the energy spectrum and the
composition data was obtained injecting a single nuclear species with
intermediate mass number like, e.g., silicon.
For a more natural mixed composition of the injected CRs, the transition
energy moves somewhat to higher energies, cf.\ with the left panel of
Fig.~\ref{UFA}. The agreement of the fit for $\ln(A)$ and the variance
$\sigma(\ln(A))$ shown in the other two panels is, taking into account
the large uncertainties of hadronic interactions in air showers, very good.

Both the GRB and AGN models of Refs.~\cite{Globus:2015xga,Unger:2015laa}
relied on interactions of nuclei with a photon background. As a result of the
threshold effect described in section~\ref{sec4second}, the resulting
neutrino fluxes
are therefore suppressed in the interesting TeV--PeV range and cannot
explain the observations of IceCube.
In contrast, models leading to large neutrino fluxes in the 0.1--1\,PeV 
energy range use typically $pp$ interactions and primaries with 10--100\,PeV
energies. Thus such models have no direct connection to the
sources of UHECRs.

A possible way of how a single source class can explain the extragalactic
CR flux, its nuclear  composition and the observed neutrino flux in IceCube
at the same time was suggested in Ref.~\cite{Kachelriess:2017tvs}. The model
presented there
assumes that UHECRs are accelerated in the core of (a subclass of) AGNs.
Subsequently, the CR nuclei diffuse first through a zone
dominated by photo-hadronic interactions, before they escape into a second
zone dominated by hadronic interactions with gas. In the first zone, the
energy-dependence of the photo-disintegration rates and the escape times
leads together with a rigidity-dependent cut-off to a rather small energy window
in which a single nuclear species is unsuppressed. The flat proton component
is generated again by escaping neutrons. In the second zone, on larger scales,
the escaping nuclei interact on gas and produce thereby a neutrino flux
which can give a substantial contribution to the flux observed by
IceCube. In an alternative scenario, the photon background  was assumed to
have negligible impact and only $Ap$ interactions were included.

\begin{figure*}
\hspace*{-0.4cm}    
\includegraphics[width=0.5\columnwidth,angle=0]{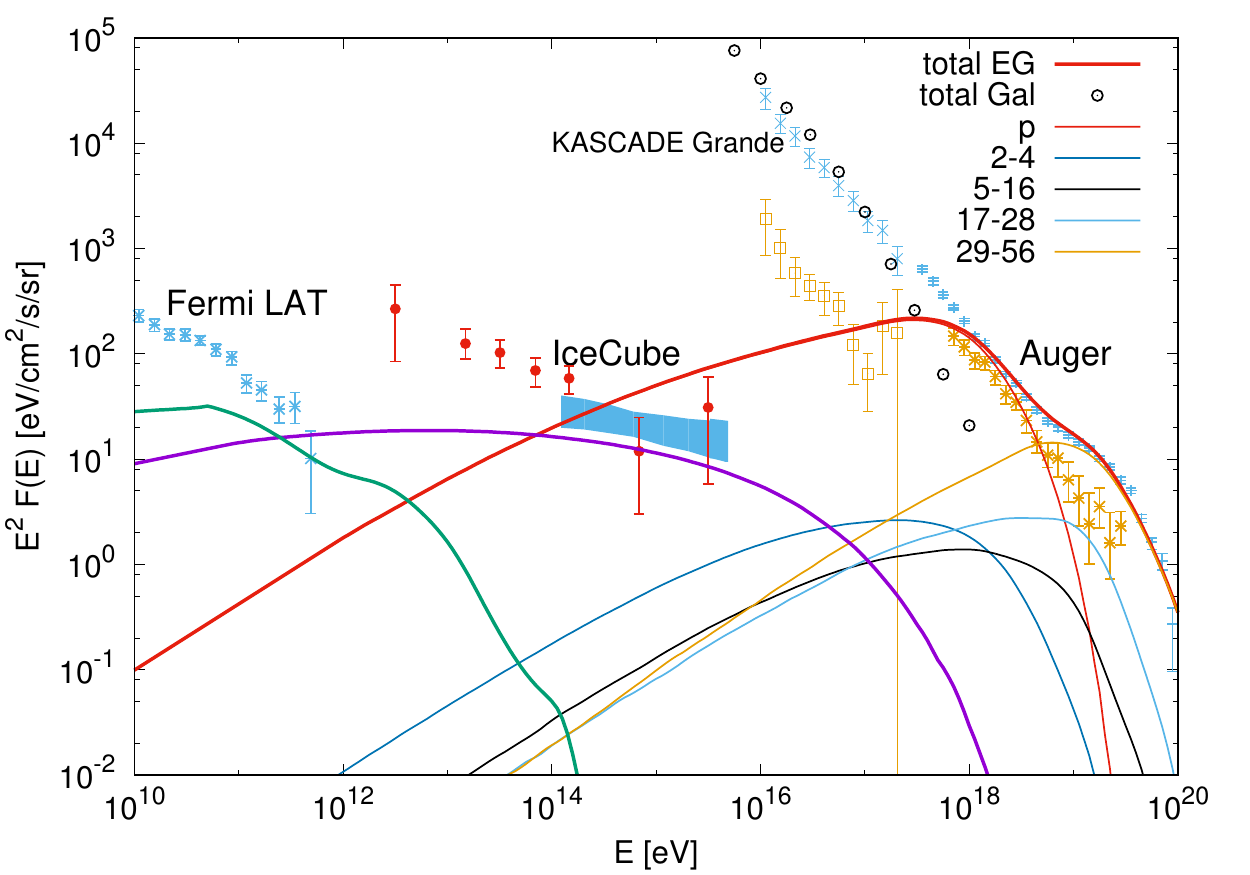}
\includegraphics[width=0.5\columnwidth,angle=0]{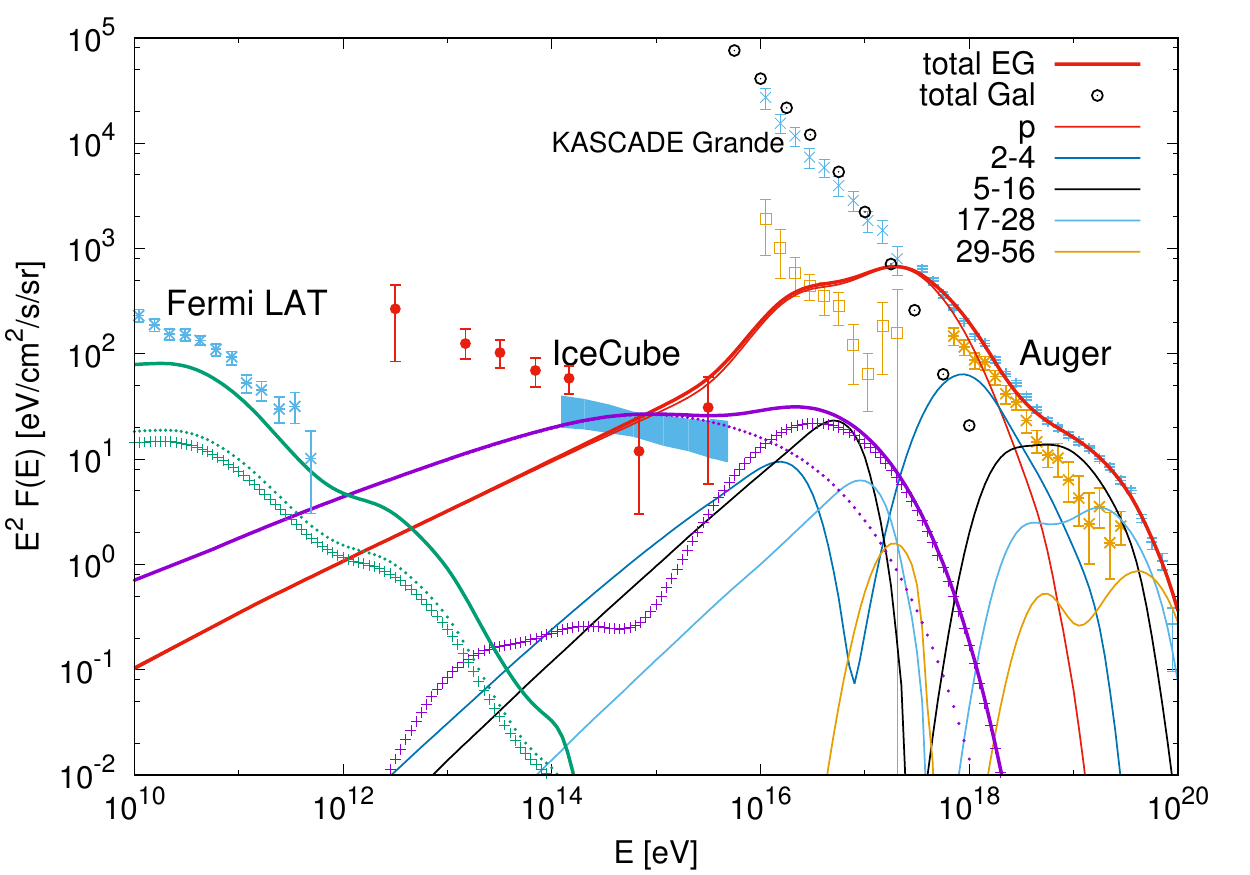}
\caption{Diffuse flux  of five elemental groups in the model of
  Ref.~\protect\cite{Kachelriess:2017tvs}.
\label{figbig}}
\end{figure*}

The diffuse fluxes of five elemental groups computed in this model are shown
on the left panel of Fig.~\ref{figbig}
for only $Ap$ interactions and on the right for $A\gamma$ and
$Ap$ interactions. In the case of only  hadronic interactions, 
$\beta=1.8$, $E_{\max}=3\times 10^{18}$\,eV and BL Lac evolution is used,
otherwise  $\beta=1.5$,  $E_{\max}=6\times 10^{18}$\,eV, $\tau^{p\gamma}=0.29$
and AGN evolution. The hadronic interaction depth is normalised as
$\tau^{pp}_0=0.035$ at $E=10^{19}$\,eV.
The diffuse fluxes are compared to experimental results for the proton
(orange error-bars) and the total flux from KASCADE, KASCADE-Grande
(light-blue error-bars)~\cite{Apel:2013uni} and
Auger (dark-blue error-bars)~\cite{Aab:2013ika,Aab:2014aea}, 
the EGRB from Fermi-LAT (light-blue error-bars)~\cite{Ackermann:2014usa}, and
the high-energy neutrino flux from IceCube (light-blue shaded
area)~\cite{Aartsen:2016xlq}.
In both cases, the total CR flux including the ankle feature is well-fitted.
Adding the Galactic CR flux determined in the escape
model~\cite{Giacinti:2014xya,Giacinti:2015hva} leads to a good description
of the total flux in the transition region and below.
Both cases lead to large neutrino fluxes, respecting at the same time however
the EGRB limit. For illustration,  the contribution of neutrinos and photons
from $A\gamma$ and $Ap$ interaction is shown on the right top separately by
crosses and dotted lines, respectively. 

\begin{figure*}
\hspace*{-0.4cm}    
\includegraphics[width=0.5\columnwidth,angle=0]{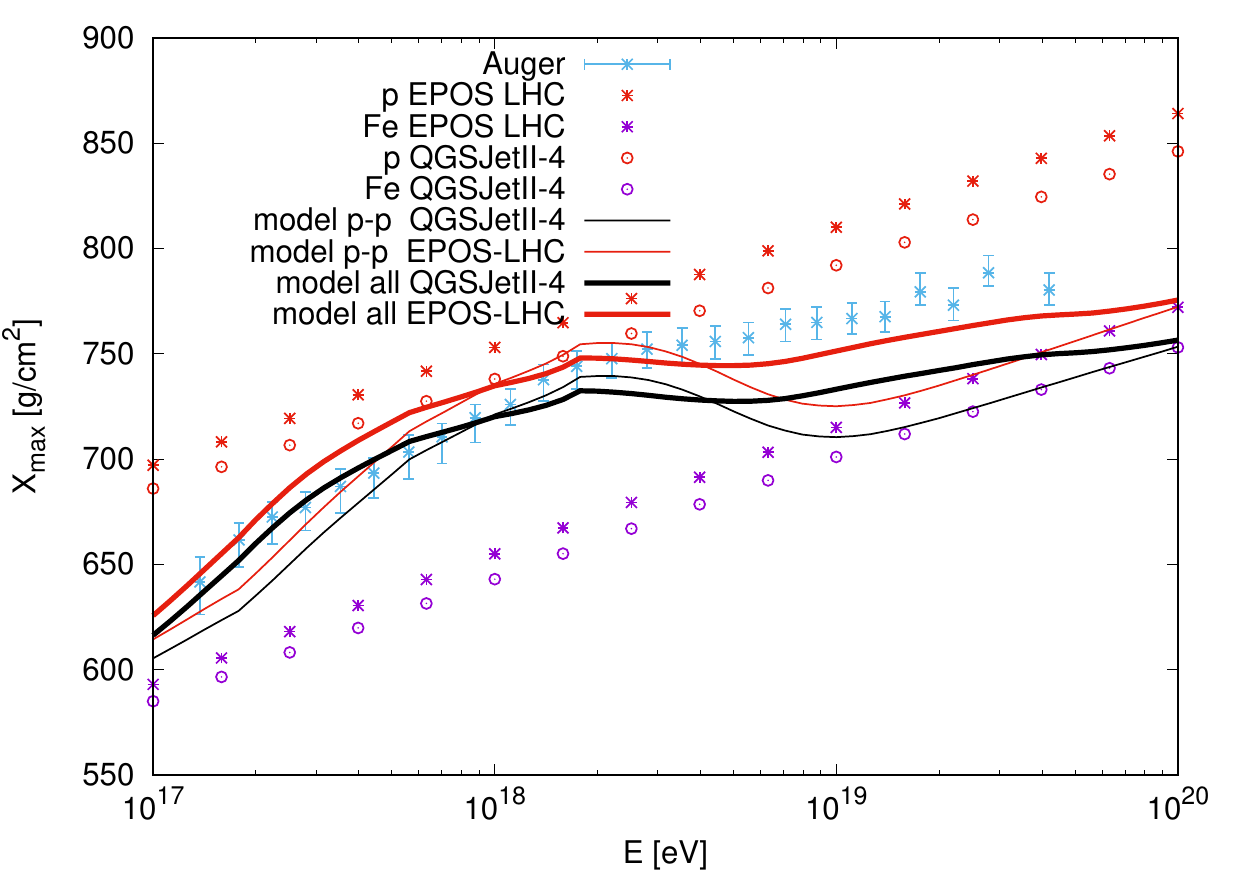}
\includegraphics[width=0.5\columnwidth,angle=0]{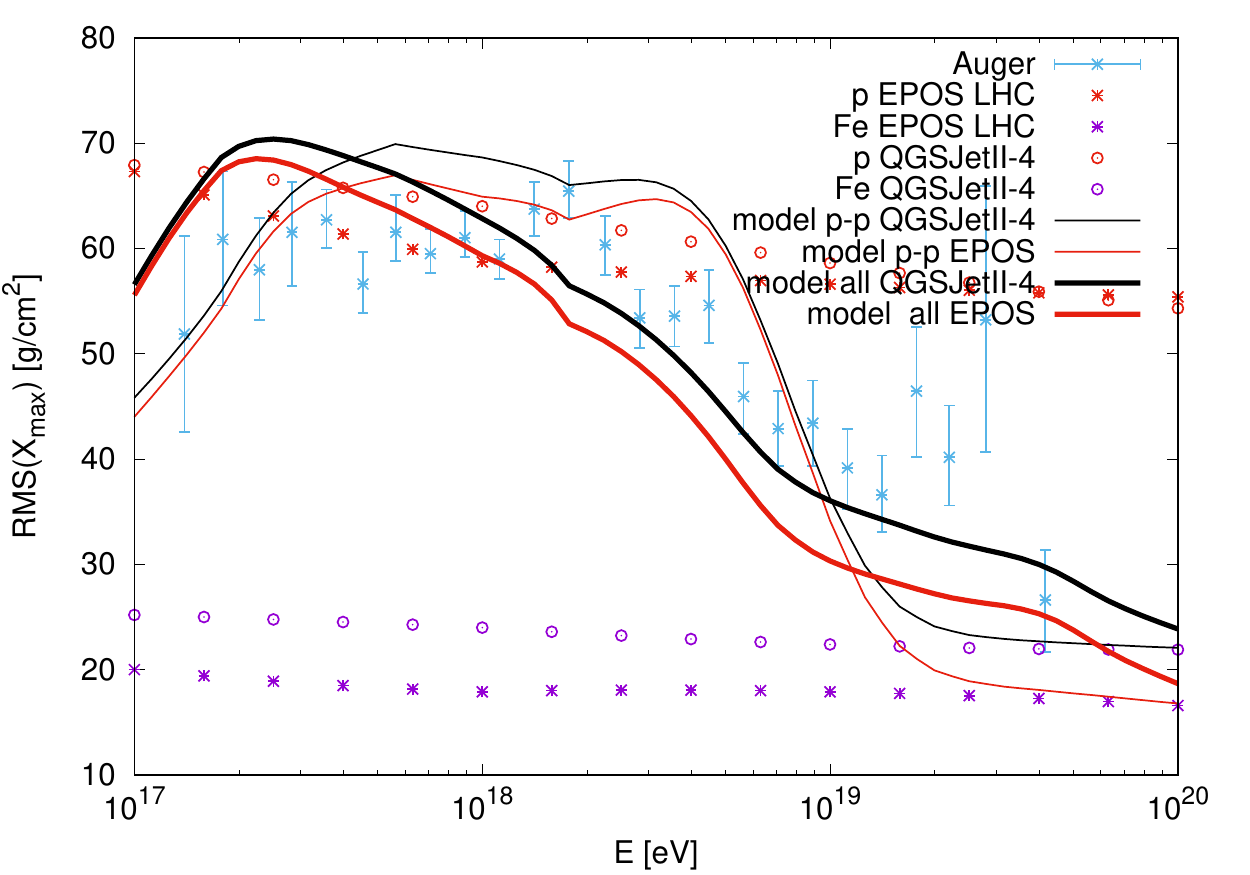}
\caption{The resulting \Xmax\ (middle) and \X2 (bottom) values in the model of
  Ref.~\protect\cite{Kachelriess:2017tvs}.
\label{figbig2}}
\end{figure*}

In Fig.~\ref{figbig2}, we compare the predictions of these two scenarios
for \Xmax and \X2 using the EPOS-LHC~\cite{Pierog:2013ria} and 
QGSJET-II-04~\cite{Ostapchenko:2010vb,Ostapchenko:2013pia} models
to data from Auger~\cite{Aab:2014kda}.  In the ``hadronic only'' scenario,
insisting to reproduce the ankle requires a relatively low cut-off energy,
and a small contribution of intermediate nuclei. This drives the composition
towards a two-component model, consisting mainly of protons and
iron. As a result, the  \Xmax data above $5\times 10^{18}$\,eV are not well
described. Since the spectra of intermediate CNO nuclei are cut off around
the ankle,  their contribution could be only increased if the proton flux
is reduced. But a  reduction of the proton flux would in turn reduce the
neutrino flux and the model will fall short of explaining the IceCube data.
In contrast, the scenario including photo-disintegration reproduces the
\Xmax and \X2 data well, taking into account the systematic uncertainties.

Finally, let us comment briefly on some other recent extragalactic CR
models which attempt at the same time to account for the diffuse neutrino
flux observed by IceCube.
The authors of Ref.~\cite{Fang:2017zjf} explain the IceCube neutrinos 
by CR interactions in the galaxy clusters surrounding UHECR sources. This
scheme may be considered as a concrete physical model of the
``only hadronic'' scenario presented above. The  CR flux in this model
has however no ankle feature and requires therefore a second
extragalactic component.
The same limitation applies to the  model of Ref.~\cite{Boncioli:2018lrv}
which suggests low-luminosity GRB as common source of UHECRs and
neutrinos.
The authors of Ref.~\cite{Supanitsky:2018jje} studied the central regions
of active galaxies as sources of UHECRs. They  followed the approach of
Ref.~\cite{Unger:2015laa} and concluded that low luminosity AGNs with
no source evolution are favoured as UHECR sources compared to models with
strong source evolution or large photon backgrounds in the source.

\section{Summary and outlook}
\label{conclusions}

The all-particle spectrum of CRs has been measured with good accuracy up to
the highest energies. Moreover, considerable progress has been made in the
last decade on the more challenging task of determining the nuclear
composition of the CR flux. At the highest energies, the composition results
from the PAO have led to a paradigm shift, making ``pure proton'' models for
the extragalactic CR flux { nonviable.} Similarly,
the results in the energy range up to the knee obtained by direct detection
experiments have challenged previously accepted models for Galactic CRs. In
particular, the increased precision of experiments like PAMELA,
CREAM and AMS-02 for the fluxes of individual nuclei as well as of electrons,
positrons and antiprotons have revealed several deviations from previous
expectations: The anomalies discovered include non-universal rigidity spectra
of CR nuclei with a break at 300\,GV, the hardening of the electron
spectrum around 50\,GeV and the positron excess.

A multitude of models for Galactic CRs has been developed which can explain
one or several of these features. In order to distinguish between them,
additional experimental data are necessary.
For instance, the confirmation of a power law-like positron spectrum with a
steepening by $\Delta\alpha=1$ at 300\,GeV would support the idea of a
2--3\,Myr old local source where the positrons are produced during
propagation~\cite{Kachelriess:2015oua}. The related suggestion that the
positrons
are produced close to a younger source would lead instead to a steeper break
in the positron spectrum at higher energy,
$E\sim 1$\,TeV~\cite{Fujita:2009wk,Kohri:2015mga}. In contrast to this
high-energy
suppression, the reacceleration model of Ref.~\cite{Blasi:2009hv} predicts
a rise of the positron flux with energy.
If the positrons are connected instead to 
nearby pulsars, several bumps related to the steep high-energy cutoff of the
nearest pulsars may become visible in the positron spectrum.
Another prediction of most pulsar models are similar fluxes of positrons and
electrons at high energies, $R_{e^+/(e^-+e^+)}\simeq 0.5$.
In contrast, antiproton and positron fluxes are unrelated in this model,
and the ratio $R_{e^+/\bar p}\simeq 2$ is a mere accident. Decreasing the
uncertainties in the predictions for antiproton production and testing
thereby the ratio $R_{\bar p/e^+}$ more precisely is therefore an additional
possibility to differentiate between hadronic and pulsar models for
the positron excess.

Another important handle to distinguish models are the secondary-to-primary
ratios of nuclei like B/C at high energies: While in the majority of models
these ratios should continue to decrease like $E^{-1/3}$ moving to higher
energies, some models predict a slower decrease as
$\propto E^{-0.15}$~\cite{Tomassetti:2015mha}, a
plateau~\cite{Kachelriess:2017yzq} or even
an increase~\cite{Mertsch:2009ph}. Extending the energy range of the B/C
measurements and adding additional ratios like Ti/Fe is therefore very
valuable.

Higher precision in the determination of the primary (and secondary) spectra
may reveal how granular the CR sea is. Since the number of CR sources
contributing to the locally measured CR flux depends strongly on how
anisotropic CR diffuse, this ``granularity'' provides an important
constraint on the CR  propagation. If the number of contributing sources
is reduced by a factor~100 as argued in Ref.~\cite{Giacinti:2017dgt},
only few sources may be responsible for the locally observed CR flux
above 200\,GeV. In this scenario, one expects also at lower energies a
reduced number of sources contributing which should manifest itself by
additional small breaks in the primary spectra. Similarly, the smooth B/C
ratio  as function of rigidity should dissolve into a series of small
plateaus in this scenario, if the experimental precision is increased.

At present, the  electron spectra determined by different experiments
show systematic discrepancies. A more precise, consistent determination
of the electron flux above 200\,GeV is required to constrain the properties
of local sources. The question at which energy
the spectrum contains cooling breaks, and if they agree or not with
those in the positron spectrum, is an important piece of information.
Combining the electron and positron measurements, a ratio $R_{e^+/(e^-+e^+)}$
much smaller than 0.5 would disfavour pulsar models.

The behaviour of the dipole anisotropy of CR nuclei as function of energy,
i.e.\ the phase flip by $180^\circ$  and the plateau in the amplitude,
suggests that the dipole is aligned with the local ordered magnetic field
and that CRs propagate anisotropically. A measurement of the dipole phase
of positrons provides therefore a test of both source and propagation models:
In naive diffusion models, it is assumed that the dipole is aligned with the
CR flux $\vec j\propto \vec\nabla n$. Thus the dipole should point to the
(flux weighted) location of the positron sources, which are known in the
case of pulsars like Geminga and PSR B0656+14.
In contrast, the dipole of positrons should agree with the one of CR nuclei,
if the positrons have an hadronic origin and are produced, e.g., in SNRs.
Going beyond the naive diffusion approach,  CRs with $R_{\rm L}\ll L_{\rm c}$
propagate anisotropically even in the absence of a regular magnetic field,
since the field modes with $k/(2\pi)\gg R_{\rm L}$ act locally as an ordered
field~\cite{Giacinti:2012ar}. Thus one should always expect that the dipole
is aligned with the local ordered magnetic field.
Finally, the relative size of the dipole amplitude of
positrons and nuclei can help to distinguish models.

The question how strong our nearby  environment in the Galaxy, in
particular the Local Bubble, influences local measurements has not attracted
much attention yet. The effect of the Local Bubble might be particularly
important for recent nearby sources as Vela and for the interpretation
of the CR dipole: The increased magnetic field in the bubble wall may
surppress at low-energies the flux from young, nearby sources and reduce
anisotropies by spreading CRs on the wall.
Additionally, the interactions of CRs in the  dense walls of the Local Bubble
and Loop~I have been suggested as sources of an extended
Galactic neutrino emission.

Another important requisite in advancing propagation models are  improved
models of the global GMF and an enhanced understanding of magnetic turbulence.
The advance of the SKA radio telescope array will dramatically increase the
number of RMs of
extragalactic sources from 42.000 to tens of millions, while pulsar RMs are
expected to increase from 1133 to 18.000~\cite{Haverkorn:2015wsa}. This
enormous increase of input data should in combination with more sophisticated
analysis methods~\cite{Boulanger:2018zrk} allow one to improve considerably
the reliability of models for the global GMF.
On the theoretical side, an
important open issue is the nature and importance of magnetic
turbulence. While it seems natural that at sufficiently low energies
self-generated turbulence becomes more important than the turbulence
injected by SNe and stellar winds, the transition energy between these
two regimes is unknown. Moreover, the theoretical understanding of
magnetic turbulence is rather incomplete~\cite{Brandenburg:2013vya}, and
a self-consistent description of the coupling at low energies between the
CR fluid, magnetic fields and the ISM is largely unexplored.

An important objective of new experiments like
LHAASO~\cite{DiSciascio:2016rgi} or the the extension of
IceTop~\cite{Haungs:2019ylq} is the measurement
of the mass composition in the knee region and above. In particular,
the resolution of the position of the proton knee---which is at present
discrepant between different experiments---is from a theoretical point
crucial in pinning down the maximal energy and the possible need for an
additional population of Galactic CR sources. Moreover, a better
knowledge of the mass composition close to the second  knee is needed to
constrain models for the extragalactic CR flux.

There has been considerable experimental progress at higher energies.
In particular, a cut-off in the energy spectrum of UHECRs has been firmly
established although its nature is yet unclear: Both the finite maximal
acceleration energy in the sources and GZK-like interactions of nuclei
on background photons may be the cause. 
The energy spectrum derived by PAO and TA agree well within their systematic
uncertainties, except for the highest energies where the differences may
be caused by the different field-of-view. The determination of the mass
composition of UHECRs has progressed by the much larger statistics of
fluorescence events available which allowed TA and PAO to go beyond the
analysis of the average \Xmax. In particular, using the full
information contained in the \Xmax distribution allowed the PAO collaboration
to fit the fraction of four elements in the CR flux. While these results have
to be taken with some caution, they suggest the presence of a Peters' cycle
above $2\times 10^{18}$\,eV. In the near future, the low-energy extensions
and upgrades of the TA~\cite{Ogio:2018hyq} and the PAO~\cite{Aab:2016vlz}
will allow one to test this scenario with increased accuracy.

While in the energy range $10^{16}-8\times 10^{18}$\,eV only upper
limits on the dipole anisotropy exist, the dipole vector with an
amplitude $6\%$ has been measured by the PAO using events with energy
$\geq 8\times 10^{18}$\,eV. Its direction points away from the Galactic
centre and is roughly consistent, taking into account deflections in the
GMF and EGMF, with an overdensity in the local galaxy distribution.
On smaller scales, the most prominent deviation from isotropy is the
TA hot spot. Additionally, the PAO claims evidence for correlations
of the UHECR arrival directions with specific types of CR sources,
as e.g.\ starburst galaxies. At present it is however unclear if these
correlations do not simply reflect the nearby large-scale structure.

Three pieces of evidence can be used to determine the energy at which the
transition from Galactic to extragalactic CRs takes place.
First, the behaviour of the dipole anisotropy suggests that the transition
starts at $10^{17}$\,eV. Second, the disappearance of the iron component
at $5\times 10^{17}$\,eV in the PAO composition data signals the end of
the Galactic CR spectrum. Finally,
combining the composition and anisotropy measurements, one can conclude
that the dominant light-intermediate contribution to the CR flux measured
above $7\times 10^{17}$\,eV  has to be extragalactic. Taking into account
these different pieces of evidence, we have concluded that
the transition from Galactic to extragalactic CRs happens at the second
knee, i.e.\ around $5\times 10^{17}$\,eV.

Using Occam's razor, one may dismiss the possibility that the ankle is the
cross-over of the fluxes of two independent extragalactic populations of
UHECR sources. Then the ankle has to be explained as a consequence of either
the propagation of extragalactic CRs or of interactions in their sources.
The dip model which relied on the first possibility is excluded by composition
measurements. A viable alternative uses photo-disintegration of CR nuclei on
background photons inside their sources, which might be either AGN cores or
GRBs. Successful models tend to have built-in unusual features as steep
injection spectra or a large contribution of injected  nuclei with
intermediate mass number which wait for a
physical motivation. Requiring additionally that these UHECR sources explain
the neutrino flux measured by IceCube poses another challenge.
Developing concrete models which reproduce all the experimental data which
have become available in the last years will be an important step towards
understanding the sources of UHECRs.

\section*{Acknowledgments}

\noindent
It is a pleasure to thank Gwenal Giacinti, Oleg Kalashev, Andrii Neronov, and
Sergey Ostapchenko for fruitful collaborations on topics related to this review.
We are grateful to Carmelo Evoli, Kazunori Kohri, Dmitry Podorozhniy, G\"unter
Sigl, Nicola Tomassetti and Michael Unger for providing updated figures and
data for this review. Last but not least we would like to thank Markus Ahlers,
Katia Ferri\`ere, Yutaka Fujita, Gwenael Giacinti, Kazunori Kohri, Paolo Lipari,
Igor Moskalenko, Sergey Ostapchenko, Peter Tinyakov, Nicola Tomassetti, and
Michael Unger for valuable comments on this article.


\itemsep -4pt 

{\small

\begin{thebibliography}{100}
\expandafter\ifx\csname url\endcsname\relax
  \def\url#1{\texttt{#1}}\fi
\expandafter\ifx\csname urlprefix\endcsname\relax\def\urlprefix{URL }\fi
\expandafter\ifx\csname href\endcsname\relax
  \def\href#1#2{#2} \def\path#1{#1}\fi

\bibitem{Coulomb:1785}
C.~A. de~Coulomb, {Troisi\`eme M\'emoire sur l'Electricit\'e et le
  Magn\'etiseme}, Histoire de l'Acad\'emie Royale des Sciences (1785) 612--638.

\bibitem{Hess:1912srp}
V.~F. Hess, {\"Uber Beobachtungen der durchdringenden Strahlung bei sieben
  Freiballonfahrten}, Phys. Z. 13 (1912) 1084--1091.

\bibitem{Clay27}
J.~Clay, {Penetrating Radiation}, Proc. Kon. Akademie (Amsterdam) 30~(9-10)
  (1927) 1115--1127.

\bibitem{Bothe29}
W.~Bothe, W.~Kolh{\"o}rster, {Das Wesen der H\"ohenstrahlung}, Z. Phys. 56
  (1929) 751--777.
\newblock \href {https://doi.org/10.1007/BF01340137}
  {\path{doi:10.1007/BF01340137}}.

\bibitem{Schein:1941}
M.~Schein, W.~P. Jesse, E.~O. Wollan, {The Nature of the Primary Cosmic
  Radiation and the Origin of the Mesotron}, Phys. Rev. 59 (1941) 615.
\newblock \href {https://doi.org/10.1103/PhysRev.59.615}
  {\path{doi:10.1103/PhysRev.59.615}}.

\bibitem{1938NW.....26Q.576K}
W.~{Kolh{\"o}rster}, I.~{Matthes}, E.~{Weber}, {Gekoppelte H{\"o}henstrahlen},
  Naturwissenschaften 26 (1938) 576--576.
\newblock \href {https://doi.org/10.1007/BF01773491}
  {\path{doi:10.1007/BF01773491}}.

\bibitem{Auger:1938}
P.~Auger, P.~Maze, T.~Grivet-Meyer, {Grandes gerbes cosmiques atmosph\'eriques
  contenant des corpuscules ultra-p\'en\'etrants}, Comptes Rendus de
  l'Acad\'emie des Sciences 206 (1938) 1721.

\bibitem{1959JETP...35....8K}
G.~V. Kulikov, G.~B. Khristiansen, {On the Size Spectrum of Extensive Air
  Showers}, J. Exp. Theor. Phys. 35 (1958) 8.

\bibitem{Linsley:1963km}
J.~Linsley, {Evidence for a primary cosmic-ray particle with energy
  $10^{20}$\,eV}, Phys. Rev. Lett. 10 (1963) 146--148.
\newblock \href {https://doi.org/10.1103/PhysRevLett.10.146}
  {\path{doi:10.1103/PhysRevLett.10.146}}.

\bibitem{1934PNAS...20..259B}
W.~{Baade}, F.~{Zwicky}, {Cosmic Rays from Super-novae}, Proceedings of the
  National Academy of Science 20 (1934) 259--263.
\newblock \href {https://doi.org/10.1073/pnas.20.5.259}
  {\path{doi:10.1073/pnas.20.5.259}}.

\bibitem{1949Natur.163..283H}
W.~A. {Hiltner}, {Polarization of Radiation from Distant Stars by the
  Interstellar Medium}, \nat 163 (1949) 283.
\newblock \href {https://doi.org/10.1038/163283a0}
  {\path{doi:10.1038/163283a0}}.

\bibitem{1949Sci...109..166H}
J.~S. {Hall}, {Observations of the Polarized Light from Stars}, Science 109
  (1949) 166--167.
\newblock \href {https://doi.org/10.1126/science.109.2825.166}
  {\path{doi:10.1126/science.109.2825.166}}.

\bibitem{Fermi:1949ee}
E.~Fermi, {On the Origin of the Cosmic Radiation}, Phys. Rev. 75 (1949)
  1169--1174.
\newblock \href {https://doi.org/10.1103/PhysRev.75.1169}
  {\path{doi:10.1103/PhysRev.75.1169}}.

\bibitem{Morrison:1954}
P.~Morrison, S.~Olbert, B.~Rossi, {The Origin of Cosmic Rays}, Phys. Rev.
  94~(2) (1953) 440--453.
\newblock \href {https://doi.org/10.1103/PhysRev.94.440}
  {\path{doi:10.1103/PhysRev.94.440}}.

\bibitem{1969ocr..book.....G}
V.~L. {Ginzburg}, S.~I. {Syrovatskii}, {The origin of cosmic rays}, New York:
  Gordon and Breach, 1969.

\bibitem{1977ICRC...11..132A}
W.~I. {Axford}, E.~{Leer}, G.~{Skadron}, {The acceleration of cosmic rays by
  shock waves}, International Cosmic Ray Conference 11 (1977) 132--137.

\bibitem{Blandford:1978ky}
R.~D. Blandford, J.~P. Ostriker, {Particle Acceleration by Astrophysical
  Shocks}, Astrophys. J. 221 (1978) L29--L32.
\newblock \href {https://doi.org/10.1086/182658} {\path{doi:10.1086/182658}}.

\bibitem{1977DoSSR.234.1306K}
G.~F. {Krymskii}, {A regular mechanism for the acceleration of charged
  particles on the front of a shock wave}, Akademiia Nauk SSSR Doklady 234
  (1977) 1306--1308.

\bibitem{1978MNRAS.182..147B}
A.~R. {Bell}, {The acceleration of cosmic rays in shock fronts. I}, \mnras 182
  (1978) 147--156.

\bibitem{Drury:1983zz}
L.~O. Drury, {An introduction to the theory of diffusive shock acceleration of
  energetic particles in tenuous plasmas}, Rept. Prog. Phys. 46 (1983)
  973--1027.
\newblock \href {https://doi.org/10.1088/0034-4885/46/8/002}
  {\path{doi:10.1088/0034-4885/46/8/002}}.

\bibitem{1984ARA&A..22..425H}
A.~M. {Hillas}, {The Origin of Ultra-High-Energy Cosmic Rays}, \araa 22 (1984)
  425--444.
\newblock \href {https://doi.org/10.1146/annurev.aa.22.090184.002233}
  {\path{doi:10.1146/annurev.aa.22.090184.002233}}.

\bibitem{1987ApJ...313..842J}
J.~R. {Jokipii}, {Rate of energy gain and maximum energy in diffusive shock
  acceleration}, \apj 313 (1987) 842--846.
\newblock \href {https://doi.org/10.1086/165022} {\path{doi:10.1086/165022}}.

\bibitem{Lagage:1983zz}
P.~O. Lagage, C.~J. Cesarsky, {The maximum energy of cosmic rays accelerated by
  supernova shocks}, Astron. Astrophys. 125 (1983) 249--257.

\bibitem{2001MNRAS.321..433B}
A.~R. {Bell}, S.~G. {Lucek}, {Cosmic ray acceleration to very high energy
  through the non-linear amplification by cosmic rays of the seed magnetic
  field}, \mnras 321 (2001) 433--438.
\newblock \href {https://doi.org/10.1046/j.1365-8711.2001.04063.x}
  {\path{doi:10.1046/j.1365-8711.2001.04063.x}}.

\bibitem{2004MNRAS.353..550B}
A.~R. {Bell}, {Turbulent amplification of magnetic field and diffusive shock
  acceleration of cosmic rays}, \mnras 353 (2004) 550--558.
\newblock \href {https://doi.org/10.1111/j.1365-2966.2004.08097.x}
  {\path{doi:10.1111/j.1365-2966.2004.08097.x}}.

\bibitem{Berezhko:2003ej}
E.~G. Berezhko, L.~T. Ksenofontov, H.~J. V{\"o}lk, {Confirmation of strong
  magnetic field amplification and nuclear cosmic ray acceleration in SN~1006},
  Astron. Astrophys. 412 (2003) L11--L14.
\newblock \href {http://arxiv.org/abs/astro-ph/0310862}
  {\path{arXiv:astro-ph/0310862}}, \href
  {https://doi.org/10.1051/0004-6361:20031667}
  {\path{doi:10.1051/0004-6361:20031667}}.

\bibitem{Vink:2002yx}
J.~Vink, J.~M. Laming, {On the magnetic fields and particle acceleration in Cas
  A}, Astrophys. J. 584 (2003) 758--769.
\newblock \href {http://arxiv.org/abs/astro-ph/0210669}
  {\path{arXiv:astro-ph/0210669}}, \href {https://doi.org/10.1086/345832}
  {\path{doi:10.1086/345832}}.

\bibitem{Bamba:2004zr}
A.~Bamba, R.~Yamazaki, T.~Yoshida, T.~Terasawa, K.~Koyama, {A Spatial and
  spectral study of nonthermal filaments in historical supernova remnants:
  Observational results with Chandra}, Astrophys. J. 621 (2005) 793--802.
\newblock \href {http://arxiv.org/abs/astro-ph/0411326}
  {\path{arXiv:astro-ph/0411326}}, \href {https://doi.org/10.1086/427620}
  {\path{doi:10.1086/427620}}.

\bibitem{Kachelriess:2008ze}
M.~Kachelrie\ss, {Lecture notes on high energy cosmic rays}\href
  {http://arxiv.org/abs/0801.4376} {\path{arXiv:0801.4376}}.

\bibitem{Ptitsyna:2008zs}
K.~V. Ptitsyna, S.~V. Troitsky, {Physical conditions in potential sources of
  ultra-high-energy cosmic rays. I. Updated Hillas plot and radiation-loss
  constraints}, Phys. Usp. 53 (2010) 691--701.
\newblock \href {http://arxiv.org/abs/0808.0367} {\path{arXiv:0808.0367}},
  \href {https://doi.org/10.3367/UFNe.0180.201007c.0723}
  {\path{doi:10.3367/UFNe.0180.201007c.0723}}.

\bibitem{Spurio:2018knn}
M.~Spurio, {Probes of Multimessenger Astrophysics}, Astron. Astrophys. Lib.
  9783319968544 (2018) pp.1--591.
\newblock \href {https://doi.org/10.1007/978-3-319-96854-4}
  {\path{doi:10.1007/978-3-319-96854-4}}.

\bibitem{GBM:2017lvd}
B.~P. Abbott, et~al., {Multi-messenger Observations of a Binary Neutron Star
  Merger}, Astrophys. J. 848~(2) (2017) L12.
\newblock \href {http://arxiv.org/abs/1710.05833} {\path{arXiv:1710.05833}},
  \href {https://doi.org/10.3847/2041-8213/aa91c9}
  {\path{doi:10.3847/2041-8213/aa91c9}}.

\bibitem{Ginzburg:1990sk}
V.~S. Berezinsky, S.~V. Bulanov, V.~A. Dogiel, V.~S. Ptuskin, {Astrophysics of
  cosmic rays}, Amsterdam, Netherlands: North-Holland, 1990.

\bibitem{Schlickeiser:2002pg}
R.~Schlickeiser, {Cosmic ray astrophysics}, Berlin: Springer, 2002.

\bibitem{Gaisser:2016uoy}
T.~K. Gaisser, R.~Engel, E.~Resconi, {Cosmic Rays and Particle Physics},
  Cambridge University Press, 2016.

\bibitem{Potgieter:2013pdj}
M.~Potgieter, {Solar Modulation of Cosmic Rays}, Living Rev. Solar Phys. 10
  (2013) 3.
\newblock \href {http://arxiv.org/abs/1306.4421} {\path{arXiv:1306.4421}},
  \href {https://doi.org/10.12942/lrsp-2013-3}
  {\path{doi:10.12942/lrsp-2013-3}}.

\bibitem{Marcowith:2016vzl}
A.~Marcowith, et~al., {The microphysics of collisionless shock waves}, Rept.
  Prog. Phys. 79 (2016) 046901.
\newblock \href {http://arxiv.org/abs/1604.00318} {\path{arXiv:1604.00318}},
  \href {https://doi.org/10.1088/0034-4885/79/4/046901}
  {\path{doi:10.1088/0034-4885/79/4/046901}}.

\bibitem{Bykov:2018wrt}
A.~M. Bykov, D.~C. Ellison, A.~Marcowith, S.~M. Osipov, {Cosmic ray production
  in supernovae}, Space Sci. Rev. 214~(1) (2018) 41.
\newblock \href {http://arxiv.org/abs/1801.08890} {\path{arXiv:1801.08890}},
  \href {https://doi.org/10.1007/s11214-018-0479-4}
  {\path{doi:10.1007/s11214-018-0479-4}}.

\bibitem{Strong:2007nh}
A.~W. Strong, I.~V. Moskalenko, V.~S. Ptuskin, {Cosmic-ray propagation and
  interactions in the Galaxy}, Ann. Rev. Nucl. Part. Sci. 57 (2007) 285--327.
\newblock \href {http://arxiv.org/abs/astro-ph/0701517}
  {\path{arXiv:astro-ph/0701517}}, \href
  {https://doi.org/10.1146/annurev.nucl.57.090506.123011}
  {\path{doi:10.1146/annurev.nucl.57.090506.123011}}.

\bibitem{DeAngelis:2018lra}
A.~De~Angelis, M.~Mallamaci, {Gamma-Ray Astrophysics}, Eur. Phys. J. Plus 133
  (2018) 324.
\newblock \href {http://arxiv.org/abs/1805.05642} {\path{arXiv:1805.05642}},
  \href {https://doi.org/10.1140/epjp/i2018-12181-0}
  {\path{doi:10.1140/epjp/i2018-12181-0}}.

\bibitem{vanEldik:2015qla}
C.~van Eldik, {Gamma rays from the Galactic Centre region: a review},
  Astropart. Phys. 71 (2015) 45--70.
\newblock \href {http://arxiv.org/abs/1505.06055} {\path{arXiv:1505.06055}},
  \href {https://doi.org/10.1016/j.astropartphys.2015.05.002}
  {\path{doi:10.1016/j.astropartphys.2015.05.002}}.

\bibitem{Madejski:2016oqg}
G.~G. Madejski, M.~Sikora, {Gamma-Ray Observations of Active Galactic Nuclei},
  Ann. Rev. Astron. Astrophys. 54 (2016) 725--760.
\newblock \href {https://doi.org/10.1146/annurev-astro-081913-040044}
  {\path{doi:10.1146/annurev-astro-081913-040044}}.

\bibitem{Mollerach:2017idb}
S.~Mollerach, E.~Roulet, {Progress in high-energy cosmic ray physics}, Prog.
  Part. Nucl. Phys. 98 (2018) 85--118.
\newblock \href {http://arxiv.org/abs/1710.11155} {\path{arXiv:1710.11155}},
  \href {https://doi.org/10.1016/j.ppnp.2017.10.002}
  {\path{doi:10.1016/j.ppnp.2017.10.002}}.

\bibitem{Anchordoqui:2018qom}
L.~A. Anchordoqui, {Ultra-High-Energy Cosmic Rays}\href
  {http://arxiv.org/abs/1807.09645} {\path{arXiv:1807.09645}}.

\bibitem{Dorman:2014nka}
I.~V. Dorman, L.~I. Dorman, {How cosmic rays were discovered and why they
  received this misnomer}, Adv. Space Res. 53 (2014) 1388--1404.
\newblock \href {https://doi.org/10.1016/j.asr.2013.04.022}
  {\path{doi:10.1016/j.asr.2013.04.022}}.

\bibitem{Dorman:2014mka}
L.~I. Dorman, I.~V. Dorman, {The beginning of cosmic ray astrophysics}, Adv.
  Space Res. 53 (2014) 1379--1387.
\newblock \href {https://doi.org/10.1016/j.asr.2013.11.046}
  {\path{doi:10.1016/j.asr.2013.11.046}}.

\bibitem{DeAngelis:2010pj}
A.~De~Angelis, N.~Giglietto, S.~Stramaglia, {Domenico Pacini, the forgotten
  pioneer of the discovery of cosmic rays}\href
  {http://arxiv.org/abs/1002.2888} {\path{arXiv:1002.2888}}.

\bibitem{Kampert:2012qen}
K.-H. Kampert, A.~A. Watson, {Development of Ultra High-Energy Cosmic Ray
  Research}, in: B.~Falkenburg, W.~Rhode (Eds.), From Ultra Rays to
  Astroparticles: A Historical Introduction to Astroparticle Physics, 2012, pp.
  103--141.
\newblock \href {https://doi.org/10.1007/978-94-007-5422-5_5}
  {\path{doi:10.1007/978-94-007-5422-5_5}}.

\bibitem{Kachelriess:2008bk}
M.~Kachelrie\ss, {The rise and fall of top-down models as main UHECR sources},
  in: {Proceedings, 20th Rencontres de Blois on Challenges in Particle
  Astrophysics: Blois, France, May 18-23, 2008}, 2008, pp. 215--224.
\newblock \href {http://arxiv.org/abs/0810.3017} {\path{arXiv:0810.3017}}.

\bibitem{Aab:2016agp}
A.~Aab, et~al., {Search for photons with energies above 10$^{18}$ eV using the
  hybrid detector of the Pierre Auger Observatory}, JCAP 1704~(04) (2017) 009.
\newblock \href {http://arxiv.org/abs/1612.01517} {\path{arXiv:1612.01517}},
  \href {https://doi.org/10.1088/1475-7516/2017/04/009}
  {\path{doi:10.1088/1475-7516/2017/04/009}}.

\bibitem{Bhattacharya:2019ucd}
A.~Bhattacharya, A.~Esmaili, S.~Palomares-Ruiz, I.~Sarcevic, {Update on
  decaying and annihilating heavy dark matter with the 6-year IceCube HESE
  data}\href {http://arxiv.org/abs/1903.12623} {\path{arXiv:1903.12623}}.

\bibitem{Gaskins:2016cha}
J.~M. Gaskins, {A review of indirect searches for particle dark matter},
  Contemp. Phys. 57~(4) (2016) 496--525.
\newblock \href {http://arxiv.org/abs/1604.00014} {\path{arXiv:1604.00014}},
  \href {https://doi.org/10.1080/00107514.2016.1175160}
  {\path{doi:10.1080/00107514.2016.1175160}}.

\bibitem{2003ApJ...591.1220L}
K.~{Lodders}, {Solar System Abundances and Condensation Temperatures of the
  Elements}, \apj 591 (2003) 1220--1247.
\newblock \href {https://doi.org/10.1086/375492} {\path{doi:10.1086/375492}}.

\bibitem{2002ApJ...564..244W}
J.~Z. {Wang}, E.~S. {Seo}, K.~{Anraku}, M.~{Fujikawa}, M.~{Imori}, T.~{Maeno},
  N.~{Matsui}, H.~{Matsunaga}, M.~{Motoki}, S.~{Orito}, T.~{Saeki},
  T.~{Sanuki}, I.~{Ueda}, K.~{Yoshimura}, Y.~{Makida}, J.~{Suzuki},
  K.~{Tanaka}, A.~{Yamamoto}, T.~{Yoshida}, T.~{Mitsui}, H.~{Matsumoto},
  M.~{Nozaki}, M.~{Sasaki}, J.~{Mitchell}, A.~{Moiseev}, J.~{Ormes},
  R.~{Streitmatter}, J.~{Nishimura}, Y.~{Yajima}, T.~{Yamagami}, {Measurement
  of Cosmic-Ray Hydrogen and Helium and Their Isotopic Composition with the
  BESS Experiment}, \apj 564 (2002) 244--259.
\newblock \href {https://doi.org/10.1086/324140} {\path{doi:10.1086/324140}}.

\bibitem{deNolfo:2006qj}
G.~A. de~Nolfo, et~al., {Observations of the Li, Be, and B isotopes and
  constraints on cosmic-ray propagation}, Adv. Space Res. 38 (2006) 1558--1564.
\newblock \href {http://arxiv.org/abs/astro-ph/0611301}
  {\path{arXiv:astro-ph/0611301}}, \href
  {https://doi.org/10.1016/j.asr.2006.09.008}
  {\path{doi:10.1016/j.asr.2006.09.008}}.

\bibitem{2009ApJ...698.1666G}
J.~S. {George}, K.~A. {Lave}, M.~E. {Wiedenbeck}, W.~R. {Binns}, A.~C.
  {Cummings}, A.~J. {Davis}, G.~A. {de Nolfo}, P.~L. {Hink}, M.~H. {Israel},
  R.~A. {Leske}, R.~A. {Mewaldt}, L.~M. {Scott}, E.~C. {Stone}, T.~T. {von
  Rosenvinge}, N.~E. {Yanasak}, {Elemental Composition and Energy Spectra of
  Galactic Cosmic Rays During Solar Cycle 23}, \apj 698 (2009) 1666--1681.
\newblock \href {https://doi.org/10.1088/0004-637X/698/2/1666}
  {\path{doi:10.1088/0004-637X/698/2/1666}}.

\bibitem{Brandenburg:2013vya}
A.~Brandenburg, A.~Lazarian, {Astrophysical hydromagnetic turbulence}, Space
  Sci. Rev. 178 (2013) 163--200.
\newblock \href {http://arxiv.org/abs/1307.5496} {\path{arXiv:1307.5496}},
  \href {https://doi.org/10.1007/s11214-013-0009-3}
  {\path{doi:10.1007/s11214-013-0009-3}}.

\bibitem{1941DoSSR..30..301K}
A.~{Kolmogorov}, {The Local Structure of Turbulence in Incompressible Viscous
  Fluid for Very Large Reynolds' Numbers}, Akademiia Nauk SSSR Doklady 30
  (1941) 301--305.

\bibitem{1964SvA.....7..566I}
P.~S. {Iroshnikov}, {Turbulence of a Conducting Fluid in a Strong Magnetic
  Field}, \sovast 7 (1964) 566.

\bibitem{1965PhFl....8.1385K}
R.~H. {Kraichnan}, {Inertial-Range Spectrum of Hydromagnetic Turbulence},
  Physics of Fluids 8 (1965) 1385--1387.
\newblock \href {https://doi.org/10.1063/1.1761412}
  {\path{doi:10.1063/1.1761412}}.

\bibitem{Shukurov:2017jxf}
A.~Shukurov, A.~Snodin, A.~Seta, P.~Bushby, T.~Wood, {Cosmic Rays in
  Intermittent Magnetic Fields}, Astrophys. J. 839~(1) (2017) L16.
\newblock \href {http://arxiv.org/abs/1702.06193} {\path{arXiv:1702.06193}},
  \href {https://doi.org/10.3847/2041-8213/aa6aa6}
  {\path{doi:10.3847/2041-8213/aa6aa6}}.

\bibitem{Armstrong:1995zc}
J.~W. Armstrong, B.~J. Rickett, S.~R. Spangler, {Electron density power
  spectrum in the local interstellar medium}, Astrophys. J. 443 (1995)
  209--221.
\newblock \href {https://doi.org/10.1086/175515} {\path{doi:10.1086/175515}}.

\bibitem{2015ApJ...804L..31B}
L.~F. {Burlaga}, V.~{Florinski}, N.~F. {Ness}, {In Situ Observations of
  Magnetic Turbulence in the Local Interstellar Medium}, \apj 804 (2015) L31.
\newblock \href {https://doi.org/10.1088/2041-8205/804/2/L31}
  {\path{doi:10.1088/2041-8205/804/2/L31}}.

\bibitem{1996ApJ...458..194M}
A.~H. {Minter}, S.~R. {Spangler}, {Observation of Turbulent Fluctuations in the
  Interstellar Plasma Density and Magnetic Field on Spatial Scales of 0.01 to
  100 Parsecs}, \apj 458 (1996) 194.
\newblock \href {https://doi.org/10.1086/176803} {\path{doi:10.1086/176803}}.

\bibitem{Haverkorn:2008tb}
M.~Haverkorn, J.~C. Brown, B.~M. Gaensler, N.~M. McClure-Griffiths, {The outer
  scale of turbulence in the magneto-ionized Galactic interstellar medium},
  Astrophys. J. 680 (2008) 362.
\newblock \href {http://arxiv.org/abs/0802.2740} {\path{arXiv:0802.2740}},
  \href {https://doi.org/10.1086/587165} {\path{doi:10.1086/587165}}.

\bibitem{Iacobelli:2013fqa}
M.~Iacobelli, et~al., {Studying Galactic interstellar turbulence through
  fluctuations in synchrotron emission: First LOFAR Galactic foreground
  detection}, Astron. Astrophys. 558 (2013) A72.
\newblock \href {http://arxiv.org/abs/1308.2804} {\path{arXiv:1308.2804}},
  \href {https://doi.org/10.1051/0004-6361/201322013}
  {\path{doi:10.1051/0004-6361/201322013}}.

\bibitem{Beck:2000dc}
R.~Beck, {Galactic and extragalactic magnetic fields}, Space Sci. Rev. 99
  (2001) 243--260.
\newblock \href {http://arxiv.org/abs/astro-ph/0012402}
  {\path{arXiv:astro-ph/0012402}}, \href
  {https://doi.org/10.1023/A:1013805401252}
  {\path{doi:10.1023/A:1013805401252}}.

\bibitem{Haverkorn:2014jka}
M.~Haverkorn, {Magnetic Fields in the Milky Way}, in: A.~{Lazarian}, E.~M. {de
  Gouveia Dal Pino}, C.~{Melioli} (Eds.), Magnetic Fields in Diffuse Media,
  Vol. 407 of Astrophysics and Space Science Library, 2015, p. 483.
\newblock \href {http://arxiv.org/abs/1406.0283} {\path{arXiv:1406.0283}},
  \href {https://doi.org/10.1007/978-3-662-44625-6_17}
  {\path{doi:10.1007/978-3-662-44625-6_17}}.

\bibitem{Boulanger:2018zrk}
F.~Boulanger, et~al., {IMAGINE: A comprehensive view of the interstellar
  medium, Galactic magnetic fields and cosmic rays}, JCAP 1808~(08) (2018) 049.
\newblock \href {http://arxiv.org/abs/1805.02496} {\path{arXiv:1805.02496}},
  \href {https://doi.org/10.1088/1475-7516/2018/08/049}
  {\path{doi:10.1088/1475-7516/2018/08/049}}.

\bibitem{Pshirkov:2011um}
M.~S. Pshirkov, P.~G. Tinyakov, P.~P. Kronberg, K.~J. Newton-McGee, {Deriving
  global structure of the Galactic Magnetic Field from Faraday Rotation
  Measures of extragalactic sources}, Astrophys. J. 738 (2011) 192.
\newblock \href {http://arxiv.org/abs/1103.0814} {\path{arXiv:1103.0814}},
  \href {https://doi.org/10.1088/0004-637X/738/2/192}
  {\path{doi:10.1088/0004-637X/738/2/192}}.

\bibitem{Jansson:2012pc}
R.~Jansson, G.~R. Farrar, {A New Model of the Galactic Magnetic Field},
  Astrophys.J. 757 (2012) 14.
\newblock \href {http://arxiv.org/abs/1204.3662} {\path{arXiv:1204.3662}},
  \href {https://doi.org/10.1088/0004-637X/757/1/14}
  {\path{doi:10.1088/0004-637X/757/1/14}}.

\bibitem{Jaffe:2013yi}
T.~R. Jaffe, K.~M. Ferriere, A.~J. Banday, A.~W. Strong, E.~Orlando, J.~F.
  Macias-Perez, L.~Fauvet, C.~Combet, E.~Falgarone, {Comparing Polarised
  Synchrotron and Thermal Dust Emission in the Galactic Plane}, Mon. Not. Roy.
  Astron. Soc. 431 (2013) 683.
\newblock \href {http://arxiv.org/abs/1302.0143} {\path{arXiv:1302.0143}},
  \href {https://doi.org/10.1093/mnras/stt200}
  {\path{doi:10.1093/mnras/stt200}}.

\bibitem{Goldreich:1994zz}
P.~Goldreich, S.~Sridhar, {Toward a theory of interstellar turbulence. 2.
  Strong Alfvenic turbulence}, Astrophys. J. 438 (1995) 763--775.
\newblock \href {https://doi.org/10.1086/175121} {\path{doi:10.1086/175121}}.

\bibitem{Cho:2002qi}
J.~Cho, A.~Lazarian, {Compressible sub-alfvenic MHD turbulence in low-Beta
  plasmas}, Phys. Rev. Lett. 88 (2002) 245001.
\newblock \href {http://arxiv.org/abs/astro-ph/0205282}
  {\path{arXiv:astro-ph/0205282}}, \href
  {https://doi.org/10.1103/PhysRevLett.88.245001}
  {\path{doi:10.1103/PhysRevLett.88.245001}}.

\bibitem{vanMarle:2012bs}
A.~J. van Marle, Z.~Meliani, A.~Marcowith, {A hydrodynamical model of the
  circumstellar bubble created by two massive stars}, Astron. Astrophys. 541
  (2012) L8.
\newblock \href {http://arxiv.org/abs/1204.2078} {\path{arXiv:1204.2078}},
  \href {https://doi.org/10.1051/0004-6361/201219180}
  {\path{doi:10.1051/0004-6361/201219180}}.

\bibitem{Krause:2013qar}
M.~Krause, C.~Charbonnel, T.~Decressin, G.~Meynet, N.~Prantzos, {Superbubble
  dynamics in globular cluster infancy II. Consequences for secondary star
  formation in the context of self-enrichment via fast rotating massive stars},
  Astron. Astrophys. 552 (2013) A121.
\newblock \href {http://arxiv.org/abs/1302.2494} {\path{arXiv:1302.2494}},
  \href {https://doi.org/10.1051/0004-6361/201220694}
  {\path{doi:10.1051/0004-6361/201220694}}.

\bibitem{vanMarle:2015oca}
A.~J. van Marle, Z.~Meliani, A.~Marcowith, {Shape and evolution of wind-blown
  bubbles of massive stars: on the effect of the interstellar magnetic field},
  Astron. Astrophys. 584 (2015) A49.
\newblock \href {http://arxiv.org/abs/1509.00192} {\path{arXiv:1509.00192}},
  \href {https://doi.org/10.1051/0004-6361/201425230}
  {\path{doi:10.1051/0004-6361/201425230}}.

\bibitem{2003A&A...411..447L}
R.~{Lallement}, B.~Y. {Welsh}, J.~L. {Vergely}, F.~{Crifo}, D.~{Sfeir}, {3D
  mapping of the dense interstellar gas around the Local Bubble}, \aap 411
  (2003) 447--464.
\newblock \href {https://doi.org/10.1051/0004-6361:20031214}
  {\path{doi:10.1051/0004-6361:20031214}}.

\bibitem{2019arXiv190107692M}
I.~{Medan}, B.-G. {Andersson}, {Magnetic Field Strengths and Grain Alignment
  Variations in the Local Bubble Wall}, arXiv e-prints (2019)
  arXiv:1901.07692\href {http://arxiv.org/abs/1901.07692}
  {\path{arXiv:1901.07692}}.

\bibitem{2019arXiv190105971L}
R.~H. {Leike}, T.~A. {En{\ss}lin}, {Charting nearby dust clouds using Gaia data
  only}, arXiv e-prints (2019) arXiv:1901.05971\href
  {http://arxiv.org/abs/1901.05971} {\path{arXiv:1901.05971}}.

\bibitem{Breitschwerdt:1999sv}
D.~Breitschwerdt, S.~Komossa, {Galactic fountains and galactic winds},
  Astrophys. Space Sci. 272 (2000) 3--13.
\newblock \href {http://arxiv.org/abs/astro-ph/9908003}
  {\path{arXiv:astro-ph/9908003}}, \href
  {https://doi.org/10.1023/A:1002661516435}
  {\path{doi:10.1023/A:1002661516435}}.

\bibitem{Schulreich:2017dyt}
M.~M. Schulreich, D.~Breitschwerdt, J.~Feige, C.~Dettbarn, {Numerical studies
  on the link between radioisotopic signatures on Earth and the formation of
  the Local Bubble - I. 60Fe transport to the solar system by turbulent mixing
  of ejecta from nearby supernovae into a locally homogeneous interstellar
  medium}, Astron. Astrophys. 604 (2017) A81.
\newblock \href {http://arxiv.org/abs/1704.08221} {\path{arXiv:1704.08221}},
  \href {https://doi.org/10.1051/0004-6361/201629837}
  {\path{doi:10.1051/0004-6361/201629837}}.

\bibitem{Wolleben:2010se}
M.~Wolleben, A.~Fletcher, T.~L. Landecker, E.~Carretti, J.~M. Dickey, B.~M.
  Gaensler, M.~Haverkorn, N.~McClure-Griffiths, W.~Reich, A.~R. Taylor,
  {Antisymmetry in the Faraday Rotation Sky Caused by a Nearby Magnetized
  Bubble}, Astrophys. J. 724 (2010) L48.
\newblock \href {http://arxiv.org/abs/1011.0341} {\path{arXiv:1011.0341}},
  \href {https://doi.org/10.1088/2041-8205/724/1/L48}
  {\path{doi:10.1088/2041-8205/724/1/L48}}.

\bibitem{2006ApJ...640L..51A}
B.-G. {Andersson}, S.~B. {Potter}, {The Magnetic Field Strength in the Wall of
  the Local Bubble toward $l, b\sim 300^\circ, 0^\circ$}, \apjl 640 (2006)
  L51--L54.
\newblock \href {https://doi.org/10.1086/503199} {\path{doi:10.1086/503199}}.

\bibitem{Mao:2010zr}
S.~A. Mao, B.~M. Gaensler, M.~Haverkorn, E.~G. Zweibel, G.~J. Madsen, N.~M.
  McClure-Griffiths, A.~Shukurov, P.~P. Kronberg, {A Survey of Extragalactic
  Faraday Rotation at High Galactic Latitude: The Vertical Magnetic Field of
  the Milky Way towards the Galactic Poles}, Astrophys. J. 714 (2010)
  1170--1186.
\newblock \href {http://arxiv.org/abs/1003.4519} {\path{arXiv:1003.4519}},
  \href {https://doi.org/10.1088/0004-637X/714/2/1170}
  {\path{doi:10.1088/0004-637X/714/2/1170}}.

\bibitem{Opher:2007ce}
M.~Opher, E.~C. Stone, T.~I. Gombosi, {The Orientation of the Local
  Interstellar Magnetic Field}, Science 316 (2007) 875--878.
\newblock \href {http://arxiv.org/abs/0705.1841} {\path{arXiv:0705.1841}},
  \href {https://doi.org/10.1126/science.1139480}
  {\path{doi:10.1126/science.1139480}}.

\bibitem{Wood:2007br}
B.~E. Wood, V.~V. Izmodenov, J.~L. Linsky, D.~Alexashov, {Dependence of
  Heliospheric Lyman-alpha Absorption on the Interstellar Magnetic Field},
  Astrophys. J. 659 (2007) 1784--1791.
\newblock \href {http://arxiv.org/abs/astro-ph/0701274}
  {\path{arXiv:astro-ph/0701274}}, \href {https://doi.org/10.1086/512482}
  {\path{doi:10.1086/512482}}.

\bibitem{2018A&A...611L...5A}
M.~I.~R. {Alves}, F.~{Boulanger}, K.~{Ferri{\`e}re}, L.~{Montier}, {The Local
  Bubble: a magnetic veil to our Galaxy}, \aap 611 (2018) L5.
\newblock \href {http://arxiv.org/abs/1803.05251} {\path{arXiv:1803.05251}},
  \href {https://doi.org/10.1051/0004-6361/201832637}
  {\path{doi:10.1051/0004-6361/201832637}}.

\bibitem{1998ApJ...493..715S}
S.~L. {Snowden}, R.~{Egger}, D.~P. {Finkbeiner}, M.~J. {Freyberg}, P.~P.
  {Plucinsky}, {Progress on Establishing the Spatial Distribution of Material
  Responsible for the 1 4 keV Soft X-Ray Diffuse Background Local and Halo
  Components}, \apj 493 (1998) 715--729.
\newblock \href {https://doi.org/10.1086/305135} {\path{doi:10.1086/305135}}.

\bibitem{2004A&A...422..391L}
R.~{Lallement}, {On the contribution of charge-exchange induced X-ray emission
  in the ISM and ICM}, \aap 422 (2004) 391--400.
\newblock \href {https://doi.org/10.1051/0004-6361:20035625}
  {\path{doi:10.1051/0004-6361:20035625}}.

\bibitem{2001Ap&SS.276..163B}
D.~{Breitschwerdt}, {Modeling the Local Interstellar Medium}, \apss 276 (2001)
  163--176.

\bibitem{Casse:2001be}
F.~Casse, M.~Lemoine, G.~Pelletier, {Transport of cosmic rays in chaotic
  magnetic fields}, Phys. Rev. D65 (2002) 023002.
\newblock \href {http://arxiv.org/abs/astro-ph/0109223}
  {\path{arXiv:astro-ph/0109223}}, \href
  {https://doi.org/10.1103/PhysRevD.65.023002}
  {\path{doi:10.1103/PhysRevD.65.023002}}.

\bibitem{Parizot:2004wh}
E.~Parizot, {GZK horizon and magnetic fields}, Nucl. Phys. Proc. Suppl. 136
  (2004) 169--178.
\newblock \href {http://arxiv.org/abs/astro-ph/0409191}
  {\path{arXiv:astro-ph/0409191}}, \href
  {https://doi.org/10.1016/j.nuclphysbps.2004.10.034}
  {\path{doi:10.1016/j.nuclphysbps.2004.10.034}}.

\bibitem{Giacinti:2012ar}
G.~Giacinti, M.~Kachelrie{\ss}, D.~V. Semikoz, {Filamentary Diffusion of Cosmic
  Rays on Small Scales}, Phys. Rev. Lett. 108 (2012) 261101.
\newblock \href {http://arxiv.org/abs/1204.1271} {\path{arXiv:1204.1271}},
  \href {https://doi.org/10.1103/PhysRevLett.108.261101}
  {\path{doi:10.1103/PhysRevLett.108.261101}}.

\bibitem{1991A&A...245...79B}
D.~{Breitschwerdt}, J.~F. {McKenzie}, H.~J. {Voelk}, {Galactic winds. I -
  Cosmic ray and wave-driven winds from the Galaxy}, \aap 245 (1991) 79--98.

\bibitem{Recchia:2017coe}
S.~Recchia, P.~Blasi, G.~Morlino, {Cosmic ray driven winds in the Galactic
  environment and the cosmic ray spectrum}, Mon. Not. Roy. Astron. Soc. 470~(1)
  (2017) 865--881.
\newblock \href {http://arxiv.org/abs/1703.04490} {\path{arXiv:1703.04490}},
  \href {https://doi.org/10.1093/mnras/stx1214}
  {\path{doi:10.1093/mnras/stx1214}}.

\bibitem{2005ARA&A..43..337C}
D.~P. {Cox}, {The Three-Phase Interstellar Medium Revisited}, \araa 43 (2005)
  337--385.
\newblock \href {https://doi.org/10.1146/annurev.astro.43.072103.150615}
  {\path{doi:10.1146/annurev.astro.43.072103.150615}}.

\bibitem{2016ApJ...816L..19G}
P.~{Girichidis}, T.~{Naab}, S.~{Walch}, M.~{Hanasz}, M.-M. {Mac Low}, J.~P.
  {Ostriker}, A.~{Gatto}, T.~{Peters}, R.~{W{\"u}nsch}, S.~C.~O. {Glover},
  R.~S. {Klessen}, P.~C. {Clark}, C.~{Baczynski}, {Launching Cosmic-Ray-driven
  Outflows from the Magnetized Interstellar Medium}, \apjl 816 (2016) L19.
\newblock \href {http://arxiv.org/abs/1509.07247} {\path{arXiv:1509.07247}},
  \href {https://doi.org/10.3847/2041-8205/816/2/L19}
  {\path{doi:10.3847/2041-8205/816/2/L19}}.

\bibitem{Pfrommer:2016mvy}
C.~Pfrommer, R.~Pakmor, K.~Schaal, C.~M. Simpson, V.~Springel, {Simulating
  cosmic ray physics on a moving mesh}, Mon. Not. Roy. Astron. Soc. 465~(4)
  (2017) 4500--4529.
\newblock \href {http://arxiv.org/abs/1604.07399} {\path{arXiv:1604.07399}},
  \href {https://doi.org/10.1093/mnras/stw2941}
  {\path{doi:10.1093/mnras/stw2941}}.

\bibitem{Thomas:2018xnj}
T.~Thomas, C.~Pfrommer, {Cosmic-ray hydrodynamics: Alfv\'en-wave regulated
  transport of cosmic rays}\href {http://arxiv.org/abs/1805.11092}
  {\path{arXiv:1805.11092}}.

\bibitem{Evoli:2018nmb}
C.~Evoli, P.~Blasi, G.~Morlino, R.~Aloisio, {Origin of the Cosmic Ray Galactic
  Halo Driven by Advected Turbulence and Self-Generated Waves}, Phys. Rev.
  Lett. 121~(2) (2018) 021102.
\newblock \href {http://arxiv.org/abs/1806.04153} {\path{arXiv:1806.04153}},
  \href {https://doi.org/10.1103/PhysRevLett.121.021102}
  {\path{doi:10.1103/PhysRevLett.121.021102}}.

\bibitem{Ptuskin:2005ax}
V.~S. Ptuskin, I.~V. Moskalenko, F.~C. Jones, A.~W. Strong, V.~N. Zirakashvili,
  {Dissipation of magnetohydrodynamic waves on energetic particles: impact on
  interstellar turbulence and cosmic ray transport}, Astrophys. J. 642 (2006)
  902--916.
\newblock \href {http://arxiv.org/abs/astro-ph/0510335}
  {\path{arXiv:astro-ph/0510335}}, \href {https://doi.org/10.1086/501117}
  {\path{doi:10.1086/501117}}.

\bibitem{2008JGRA..11311106W}
W.~R. {Webber}, P.~R. {Higbie}, {Limits on the interstellar cosmic ray electron
  spectrum below $\sim 1$--2\,GeV derived from the galactic polar radio
  spectrum and constrained by new Voyager 1 measurements}, Journal of
  Geophysical Research (Space Physics) 113~(A12) (2008) A11106.
\newblock \href {https://doi.org/10.1029/2008JA013386}
  {\path{doi:10.1029/2008JA013386}}.

\bibitem{Blasi:2012yr}
P.~Blasi, E.~Amato, P.~D. Serpico, {Spectral breaks as a signature of cosmic
  ray induced turbulence in the Galaxy}, Phys. Rev. Lett. 109 (2012) 061101.
\newblock \href {http://arxiv.org/abs/1207.3706} {\path{arXiv:1207.3706}},
  \href {https://doi.org/10.1103/PhysRevLett.109.061101}
  {\path{doi:10.1103/PhysRevLett.109.061101}}.

\bibitem{Aloisio:2013tda}
R.~Aloisio, P.~Blasi, {Propagation of galactic cosmic rays in the presence of
  self-generated turbulence}, JCAP 1307 (2013) 001.
\newblock \href {http://arxiv.org/abs/1306.2018} {\path{arXiv:1306.2018}},
  \href {https://doi.org/10.1088/1475-7516/2013/07/001}
  {\path{doi:10.1088/1475-7516/2013/07/001}}.

\bibitem{Aloisio:2015rsa}
R.~Aloisio, P.~Blasi, P.~Serpico, {Nonlinear cosmic ray Galactic transport in
  the light of AMS-02 and Voyager data}, Astron. Astrophys. 583 (2015) A95.
\newblock \href {http://arxiv.org/abs/1507.00594} {\path{arXiv:1507.00594}},
  \href {https://doi.org/10.1051/0004-6361/201526877}
  {\path{doi:10.1051/0004-6361/201526877}}.

\bibitem{Schure:2013yga}
K.~M. Schure, A.~R. Bell, {From cosmic ray source to the Galactic pool}, Mon.
  Not. Roy. Astron. Soc. 437~(3) (2014) 2802--2805.
\newblock \href {http://arxiv.org/abs/1310.7027} {\path{arXiv:1310.7027}},
  \href {https://doi.org/10.1093/mnras/stt2089}
  {\path{doi:10.1093/mnras/stt2089}}.

\bibitem{Cardillo:2015zda}
M.~Cardillo, E.~Amato, P.~Blasi, {On the cosmic ray spectrum from type II
  Supernovae expanding in their red giant presupernova wind}, Astropart. Phys.
  69 (2015) 1--10.
\newblock \href {http://arxiv.org/abs/1503.03001} {\path{arXiv:1503.03001}},
  \href {https://doi.org/10.1016/j.astropartphys.2015.03.002}
  {\path{doi:10.1016/j.astropartphys.2015.03.002}}.

\bibitem{Tatischeff:2018cyp}
V.~Tatischeff, S.~Gabici, {Particle acceleration by supernova shocks and
  spallogenic nucleosynthesis of light elements}, Ann. Rev. Nucl. Part. Sci. 68
  (2018) 377--404.
\newblock \href {http://arxiv.org/abs/1803.01794} {\path{arXiv:1803.01794}},
  \href {https://doi.org/10.1146/annurev-nucl-101917-021151}
  {\path{doi:10.1146/annurev-nucl-101917-021151}}.

\bibitem{Green:2015isa}
D.~A. Green, {Constraints on the distribution of supernova remnants with
  Galactocentric radius}, Mon. Not. Roy. Astron. Soc. 454~(2) (2015)
  1517--1524.
\newblock \href {http://arxiv.org/abs/1508.02931} {\path{arXiv:1508.02931}},
  \href {https://doi.org/10.1093/mnras/stv1885}
  {\path{doi:10.1093/mnras/stv1885}}.

\bibitem{Carlson:2016iis}
E.~Carlson, T.~Linden, S.~Profumo, {Improved Cosmic-Ray Injection Models and
  the Galactic Center Gamma-Ray Excess}, Phys. Rev. D94~(6) (2016) 063504.
\newblock \href {http://arxiv.org/abs/1603.06584} {\path{arXiv:1603.06584}},
  \href {https://doi.org/10.1103/PhysRevD.94.063504}
  {\path{doi:10.1103/PhysRevD.94.063504}}.

\bibitem{Drury:2016ubm}
L.~O. Drury, A.~W. Strong, {Power requirements for cosmic ray propagation
  models involving diffusive reacceleration; estimates and implications for the
  damping of interstellar turbulence}, Astron. Astrophys. 597 (2017) A117.
\newblock \href {http://arxiv.org/abs/1608.04227} {\path{arXiv:1608.04227}},
  \href {https://doi.org/10.1051/0004-6361/201629526}
  {\path{doi:10.1051/0004-6361/201629526}}.

\bibitem{1979ApJ...229..747J}
F.~C. {Jones}, {The dynamical halo and the variation of cosmic-ray path length
  with energy}, \apj 229 (1979) 747--752.
\newblock \href {https://doi.org/10.1086/157010} {\path{doi:10.1086/157010}}.

\bibitem{2002ApJ...565..280M}
I.~V. {Moskalenko}, A.~W. {Strong}, J.~F. {Ormes}, M.~S. {Potgieter},
  {Secondary Antiprotons and Propagation of Cosmic Rays in the Galaxy and
  Heliosphere}, \apj 565 (2002) 280--296.
\newblock \href {http://arxiv.org/abs/astro-ph/0106567}
  {\path{arXiv:astro-ph/0106567}}, \href {https://doi.org/10.1086/324402}
  {\path{doi:10.1086/324402}}.

\bibitem{2002A&A...394.1039M}
D.~{Maurin}, R.~{Taillet}, F.~{Donato}, {New results on source and diffusion
  spectral features of Galactic cosmic rays: I B/C ratio}, \aap 394 (2002)
  1039--1056.
\newblock \href {http://arxiv.org/abs/astro-ph/0206286}
  {\path{arXiv:astro-ph/0206286}}, \href
  {https://doi.org/10.1051/0004-6361:20021176}
  {\path{doi:10.1051/0004-6361:20021176}}.

\bibitem{2011MNRAS.414.2446M}
P.~J. {McMillan}, {Mass models of the Milky Way}, \mnras 414 (2011) 2446--2457.
\newblock \href {http://arxiv.org/abs/1102.4340} {\path{arXiv:1102.4340}},
  \href {https://doi.org/10.1111/j.1365-2966.2011.18564.x}
  {\path{doi:10.1111/j.1365-2966.2011.18564.x}}.

\bibitem{2016A&A...594A.116H}
{HI4PI Collaboration}, N.~{Ben Bekhti}, L.~{Fl{\"o}er}, R.~{Keller}, J.~{Kerp},
  D.~{Lenz}, B.~{Winkel}, J.~{Bailin}, M.~R. {Calabretta}, L.~{Dedes}, H.~A.
  {Ford}, B.~K. {Gibson}, U.~{Haud}, S.~{Janowiecki}, P.~M.~W. {Kalberla},
  F.~J. {Lockman}, N.~M. {McClure- Griffiths}, T.~{Murphy}, H.~{Nakanishi},
  D.~J. {Pisano}, L.~{Staveley-Smith}, {HI4PI: A full-sky H I survey based on
  EBHIS and GASS}, \aap 594 (2016) A116.
\newblock \href {http://arxiv.org/abs/1610.06175} {\path{arXiv:1610.06175}},
  \href {https://doi.org/10.1051/0004-6361/201629178}
  {\path{doi:10.1051/0004-6361/201629178}}.

\bibitem{Genolini:2018ekk}
Y.~Genolini, D.~Maurin, I.~V. Moskalenko, M.~Unger, {Current status and desired
  precision of the isotopic production cross sections relevant to astrophysics
  of cosmic rays: Li, Be, B, C, and N}, Phys. Rev. C98~(3) (2018) 034611.
\newblock \href {http://arxiv.org/abs/1803.04686} {\path{arXiv:1803.04686}},
  \href {https://doi.org/10.1103/PhysRevC.98.034611}
  {\path{doi:10.1103/PhysRevC.98.034611}}.

\bibitem{Aduszkiewicz:2287004}
A.~Aduszkiewicz, \href{https://cds.cern.ch/record/2287004}{{Feasibility Study
  for the Measurement of Nuclear Fragmentation Cross Sections with NA61/SHINE
  at the CERN SPS}}, Tech. Rep. CERN-SPSC-2017-035. SPSC-P-330-ADD-9, CERN,
  Geneva (Oct 2017).
\newline\urlprefix\url{https://cds.cern.ch/record/2287004}

\bibitem{Aduszkiewicz:2309890}
A.~Aduszkiewicz, \href{https://cds.cern.ch/record/2309890}{{Study of
  Hadron-Nucleus and Nucleus-Nucleus Collisions at the CERN SPS: Early Post-LS2
  Measurements and Future Plans}}, Tech. Rep. CERN-SPSC-2018-008.
  SPSC-P-330-ADD-10, CERN, Geneva (Mar 2018).
\newline\urlprefix\url{https://cds.cern.ch/record/2309890}

\bibitem{Tripathi:1996}
R.~K. Tripathi, F.~A. Cucinotta, J.~W. Wilson, {Accurate universal
  parametrization of absorption cross-sections}, Nucl. Instrum. Meth. B117
  (1996) 347--349.
\newblock \href {https://doi.org/10.1016/0168-583X(96)00331-X}
  {\path{doi:10.1016/0168-583X(96)00331-X}}.

\bibitem{2003ApJS..144..153W}
W.~R. {Webber}, A.~{Soutoul}, J.~C. {Kish}, J.~M. {Rockstroh}, {Updated Formula
  for Calculating Partial Cross Sections for Nuclear Reactions of Nuclei with
  $Z\leq 28$ and $E>150$\,MeV/Nucleon in Hydrogen Targets}, \apjs 144 (2003)
  153--167.
\newblock \href {https://doi.org/10.1086/344051} {\path{doi:10.1086/344051}}.

\bibitem{Ostapchenko:2010vb}
S.~Ostapchenko, {Monte Carlo treatment of hadronic interactions in enhanced
  Pomeron scheme: I. QGSJET-II model}, Phys. Rev. D83 (2011) 014018.
\newblock \href {http://arxiv.org/abs/1010.1869} {\path{arXiv:1010.1869}},
  \href {https://doi.org/10.1103/PhysRevD.83.014018}
  {\path{doi:10.1103/PhysRevD.83.014018}}.

\bibitem{Ostapchenko:2013pia}
S.~Ostapchenko, {QGSJET-II: physics, recent improvements, and results for air
  showers}, EPJ Web Conf. 52 (2013) 02001.
\newblock \href {https://doi.org/10.1051/epjconf/20125202001}
  {\path{doi:10.1051/epjconf/20125202001}}.

\bibitem{1998A&A...337..859P}
V.~S. {Ptuskin}, A.~{Soutoul}, {Decaying cosmic ray nuclei in the local
  interstellar medium}, \aap 337 (1998) 859--865.

\bibitem{Aguilar:2015ooa}
M.~Aguilar, et~al., {Precision Measurement of the Proton Flux in Primary Cosmic
  Rays from Rigidity 1 GV to 1.8 TV with the Alpha Magnetic Spectrometer on the
  International Space Station}, Phys. Rev. Lett. 114 (2015) 171103.
\newblock \href {https://doi.org/10.1103/PhysRevLett.114.171103}
  {\path{doi:10.1103/PhysRevLett.114.171103}}.

\bibitem{Aguilar:2016vqr}
M.~Aguilar, et~al., {Precision Measurement of the Boron to Carbon Flux Ratio in
  Cosmic Rays from 1.9 GV to 2.6 TV with the Alpha Magnetic Spectrometer on the
  International Space Station}, Phys. Rev. Lett. 117~(23) (2016) 231102.
\newblock \href {https://doi.org/10.1103/PhysRevLett.117.231102}
  {\path{doi:10.1103/PhysRevLett.117.231102}}.

\bibitem{Ackermann:2013wqa}
M.~Ackermann, et~al., {Detection of the Characteristic Pion-Decay Signature in
  Supernova Remnants}, Science 339 (2013) 807.
\newblock \href {http://arxiv.org/abs/1302.3307} {\path{arXiv:1302.3307}},
  \href {https://doi.org/10.1126/science.1231160}
  {\path{doi:10.1126/science.1231160}}.

\bibitem{Honda:2004yz}
M.~Honda, T.~Kajita, K.~Kasahara, S.~Midorikawa, {A New calculation of the
  atmospheric neutrino flux in a 3-dimensional scheme}, Phys. Rev. D70 (2004)
  043008.
\newblock \href {http://arxiv.org/abs/astro-ph/0404457}
  {\path{arXiv:astro-ph/0404457}}, \href
  {https://doi.org/10.1103/PhysRevD.70.043008}
  {\path{doi:10.1103/PhysRevD.70.043008}}.

\bibitem{Kachelriess:2014mga}
M.~Kachelrie{\ss}, I.~V. Moskalenko, S.~S. Ostapchenko, {Nuclear enhancement of
  the photon yield in cosmic ray interactions}, Astrophys. J. 789 (2014) 136.
\newblock \href {http://arxiv.org/abs/1406.0035} {\path{arXiv:1406.0035}},
  \href {https://doi.org/10.1088/0004-637X/789/2/136}
  {\path{doi:10.1088/0004-637X/789/2/136}}.

\bibitem{Kachelriess:2019}
M.~Kachelrie{\ss}, I.~V. Moskalenko, S.~S. Ostapchenko, {AAfrag: Interpolation
  routines for Monte Carlo results on secondary production in proton-proton,
  proton-nucleus and nucleus-nucleus interactions}\href
  {http://arxiv.org/abs/1904.05129} {\path{arXiv:1904.05129}}.

\bibitem{Kachelriess:2015wpa}
M.~Kachelrie{\ss}, I.~V. Moskalenko, S.~S. Ostapchenko, {New calculation of
  antiproton production by cosmic ray protons and nuclei}, Astrophys. J.
  803~(2) (2015) 54.
\newblock \href {http://arxiv.org/abs/1502.04158} {\path{arXiv:1502.04158}},
  \href {https://doi.org/10.1088/0004-637X/803/2/54}
  {\path{doi:10.1088/0004-637X/803/2/54}}.

\bibitem{diMauro:2014zea}
M.~di~Mauro, F.~Donato, A.~Goudelis, P.~D. Serpico, {New evaluation of the
  antiproton production cross section for cosmic ray studies}, Phys. Rev.
  D90~(8) (2014) 085017, [Erratum: Phys. Rev.D98,no.4,049901(2018)].
\newblock \href {http://arxiv.org/abs/1408.0288} {\path{arXiv:1408.0288}},
  \href {https://doi.org/10.1103/PhysRevD.98.049901,
  10.1103/PhysRevD.90.085017} {\path{doi:10.1103/PhysRevD.98.049901,
  10.1103/PhysRevD.90.085017}}.

\bibitem{Kappl:2015bqa}
R.~Kappl, A.~Reinert, M.~W. Winkler, {AMS-02 Antiprotons Reloaded}, JCAP
  1510~(10) (2015) 034.
\newblock \href {http://arxiv.org/abs/1506.04145} {\path{arXiv:1506.04145}},
  \href {https://doi.org/10.1088/1475-7516/2015/10/034}
  {\path{doi:10.1088/1475-7516/2015/10/034}}.

\bibitem{Winkler:2017xor}
M.~W. Winkler, {Cosmic Ray Antiprotons at High Energies}, JCAP 1702~(02) (2017)
  048.
\newblock \href {http://arxiv.org/abs/1701.04866} {\path{arXiv:1701.04866}},
  \href {https://doi.org/10.1088/1475-7516/2017/02/048}
  {\path{doi:10.1088/1475-7516/2017/02/048}}.

\bibitem{Parizot:2004em}
E.~Parizot, A.~Marcowith, E.~van~der Swaluw, A.~M. Bykov, V.~Tatischeff,
  {Superbubbles and energetic particles in the galaxy. 1. Collective effects of
  particle acceleration}, Astron. Astrophys. 424 (2004) 747--760.
\newblock \href {http://arxiv.org/abs/astro-ph/0405531}
  {\path{arXiv:astro-ph/0405531}}, \href
  {https://doi.org/10.1051/0004-6361:20041269}
  {\path{doi:10.1051/0004-6361:20041269}}.

\bibitem{Parizot:2000ts}
E.~Parizot, {Superbubbles and the galactic evolution of Li, Be and B}, Astron.
  Astrophys. 362 (2000) 786.
\newblock \href {http://arxiv.org/abs/astro-ph/0006099}
  {\path{arXiv:astro-ph/0006099}}.

\bibitem{1996ApJ...466..457D}
M.~A. {Duvernois}, M.~{Garcia-Munoz}, K.~R. {Pyle}, J.~A. {Simpson}, M.~R.
  {Thayer}, {The Isotopic Composition of Galactic Cosmic-Ray Elements from
  Carbon to Silicon: The Combined Release and Radiation Effects Satellite
  Investigation}, \apj 466 (1996) 457.
\newblock \href {https://doi.org/10.1086/177524} {\path{doi:10.1086/177524}}.

\bibitem{2011ApJ...729L..13O}
Y.~{Ohira}, K.~{Ioka}, {Cosmic-ray Helium Hardening}, \apjl 729 (2011) L13.
\newblock \href {https://doi.org/10.1088/2041-8205/729/1/L13}
  {\path{doi:10.1088/2041-8205/729/1/L13}}.

\bibitem{Ohira:2015ega}
Y.~Ohira, N.~Kawanaka, K.~Ioka, {Cosmic-ray hardenings in light of AMS-02
  data}, Phys. Rev. D93~(8) (2016) 083001.
\newblock \href {http://arxiv.org/abs/1506.01196} {\path{arXiv:1506.01196}},
  \href {https://doi.org/10.1103/PhysRevD.93.083001}
  {\path{doi:10.1103/PhysRevD.93.083001}}.

\bibitem{Biermann:2018clk}
P.~L. Biermann, et~al., {Supernova explosions of massive stars and cosmic
  rays}, Adv. Space Res. 62 (2018) 2773--2816.
\newblock \href {http://arxiv.org/abs/1803.10752} {\path{arXiv:1803.10752}},
  \href {https://doi.org/10.1016/j.asr.2018.03.028}
  {\path{doi:10.1016/j.asr.2018.03.028}}.

\bibitem{Soderberg:2009ps}
A.~M. Soderberg, et~al., {Discovery of a Relativistic Supernova Without a
  Gamma-ray Trigger}, Nature 463 (2010) 513.
\newblock \href {http://arxiv.org/abs/0908.2817} {\path{arXiv:0908.2817}},
  \href {https://doi.org/10.1038/nature08714} {\path{doi:10.1038/nature08714}}.

\bibitem{Neronov:2017syf}
A.~Neronov, {Supernova Origin of Cosmic Rays from a Gamma-Ray Signal in the
  Constellation III Region of the Large Magellanic Cloud}, Phys. Rev. Lett.
  119~(19) (2017) 191102.
\newblock \href {http://arxiv.org/abs/1711.02734} {\path{arXiv:1711.02734}},
  \href {https://doi.org/10.1103/PhysRevLett.119.191102}
  {\path{doi:10.1103/PhysRevLett.119.191102}}.

\bibitem{Zatsepin:2006ci}
V.~I. Zatsepin, N.~V. Sokolskaya, {Three component model of cosmic ray spectra
  from 100\,GeV up to 100\,PeV}, Astron. Astrophys. 458 (2006) 1--5.
\newblock \href {http://arxiv.org/abs/astro-ph/0601475}
  {\path{arXiv:astro-ph/0601475}}, \href
  {https://doi.org/10.1051/0004-6361:20065108}
  {\path{doi:10.1051/0004-6361:20065108}}.

\bibitem{Wick:2003ex}
S.~D. Wick, C.~D. Dermer, A.~Atoyan, {High-energy cosmic rays from gamma-ray
  bursts}, Astropart. Phys. 21 (2004) 125--148.
\newblock \href {http://arxiv.org/abs/astro-ph/0310667}
  {\path{arXiv:astro-ph/0310667}}, \href
  {https://doi.org/10.1016/j.astropartphys.2003.12.008}
  {\path{doi:10.1016/j.astropartphys.2003.12.008}}.

\bibitem{1993ApJ...418..386L}
A.~{Levinson}, D.~{Eichler}, {Baryon Purity in Cosmological Gamma-Ray Bursts as
  a Manifestation of Event Horizons}, \apj 418 (1993) 386.
\newblock \href {https://doi.org/10.1086/173397} {\path{doi:10.1086/173397}}.

\bibitem{Eichler:2016mut}
D.~Eichler, N.~Globus, R.~Kumar, E.~Gavish, {Ultrahigh Energy Cosmic Rays: a
  Galactic Origin?}, Astrophys. J. 821~(2) (2016) L24.
\newblock \href {http://arxiv.org/abs/1604.05721} {\path{arXiv:1604.05721}},
  \href {https://doi.org/10.3847/2041-8205/821/2/L24}
  {\path{doi:10.3847/2041-8205/821/2/L24}}.

\bibitem{Miller:2016chr}
M.~J. Miller, J.~N. Bregman, {The Interaction of the Fermi Bubbles With the
  Milky Ways hot gas Halo}, Astrophys. J. 829~(1) (2016) 9.
\newblock \href {http://arxiv.org/abs/1607.04906} {\path{arXiv:1607.04906}},
  \href {https://doi.org/10.3847/0004-637X/829/1/9}
  {\path{doi:10.3847/0004-637X/829/1/9}}.

\bibitem{Jaupart:2018eev}
E.~Jaupart, E.~Parizot, D.~Allard, {Contribution of the Galactic center to the
  local cosmic-ray flux}, Astron. Astrophys. 619 (2018) A12.
\newblock \href {http://arxiv.org/abs/1808.02322} {\path{arXiv:1808.02322}},
  \href {https://doi.org/10.1051/0004-6361/201833683}
  {\path{doi:10.1051/0004-6361/201833683}}.

\bibitem{Giacinti:2017dgt}
G.~Giacinti, M.~Kachelrie{\ss}, D.~V. Semikoz, {Reconciling cosmic ray
  diffusion with Galactic magnetic field models}, JCAP 1807~(07) (2018) 051.
\newblock \href {http://arxiv.org/abs/1710.08205} {\path{arXiv:1710.08205}},
  \href {https://doi.org/10.1088/1475-7516/2018/07/051}
  {\path{doi:10.1088/1475-7516/2018/07/051}}.

\bibitem{Johannesson:2016rlh}
G.~J{\'o}hannesson, et~al., {Bayesian analysis of cosmic-ray propagation:
  evidence against homogeneous diffusion}, Astrophys. J. 824~(1) (2016) 16.
\newblock \href {http://arxiv.org/abs/1602.02243} {\path{arXiv:1602.02243}},
  \href {https://doi.org/10.3847/0004-637X/824/1/16}
  {\path{doi:10.3847/0004-637X/824/1/16}}.

\bibitem{Evoli:2008dv}
C.~Evoli, D.~Gaggero, D.~Grasso, L.~Maccione, {Cosmic-Ray Nuclei, Antiprotons
  and Gamma-rays in the Galaxy: a New Diffusion Model}, JCAP 0810 (2008) 018,
  [Erratum: JCAP1604,no.04,E01(2016)].
\newblock \href {http://arxiv.org/abs/0807.4730} {\path{arXiv:0807.4730}},
  \href {https://doi.org/10.1088/1475-7516/2008/10/018,
  10.1088/1475-7516/2016/04/E01} {\path{doi:10.1088/1475-7516/2008/10/018,
  10.1088/1475-7516/2016/04/E01}}.

\bibitem{2014ApJ...785..129K}
R.~{Kumar}, D.~{Eichler}, {Large-scale Anisotropy of TeV-band Cosmic Rays},
  \apj 785 (2014) 129.
\newblock \href {https://doi.org/10.1088/0004-637X/785/2/129}
  {\path{doi:10.1088/0004-637X/785/2/129}}.

\bibitem{Bouyahiaoui:2018lew}
M.~Bouyahiaoui, M.~Kachelrie\ss, D.~V. Semikoz, {Vela as the Source of Galactic
  Cosmic Rays above 100 TeV}, JCAP 1901 (2019) 046.
\newblock \href {http://arxiv.org/abs/1812.03522} {\path{arXiv:1812.03522}},
  \href {https://doi.org/10.1088/1475-7516/2019/01/046}
  {\path{doi:10.1088/1475-7516/2019/01/046}}.

\bibitem{Jansson:2012rt}
R.~Jansson, G.~R. Farrar, {The Galactic Magnetic Field}, Astrophys.J. 761
  (2012) L11.
\newblock \href {http://arxiv.org/abs/1210.7820} {\path{arXiv:1210.7820}},
  \href {https://doi.org/10.1088/2041-8205/761/1/L11}
  {\path{doi:10.1088/2041-8205/761/1/L11}}.

\bibitem{Giacinti:2014xya}
G.~Giacinti, M.~Kachelrie{\ss}, D.~V. Semikoz, {Explaining the Spectra of
  Cosmic Ray Groups above the Knee by Escape from the Galaxy}, Phys. Rev.
  D90~(4) (2014) 041302.
\newblock \href {http://arxiv.org/abs/1403.3380} {\path{arXiv:1403.3380}},
  \href {https://doi.org/10.1103/PhysRevD.90.041302}
  {\path{doi:10.1103/PhysRevD.90.041302}}.

\bibitem{Donato:2001eq}
F.~Donato, D.~Maurin, R.~Taillet, {Beta-radioactive cosmic rays in a diffusion
  model: test for a local bubble?}, Astron. Astrophys. 381 (2002) 539--559.
\newblock \href {http://arxiv.org/abs/astro-ph/0108079}
  {\path{arXiv:astro-ph/0108079}}, \href
  {https://doi.org/10.1051/0004-6361:20011447}
  {\path{doi:10.1051/0004-6361:20011447}}.

\bibitem{Moskalenko:2002yx}
I.~V. Moskalenko, A.~W. Strong, S.~G. Mashnik, J.~F. Ormes, {Challenging cosmic
  ray propagation with antiprotons. Evidence for a fresh nuclei component?},
  Astrophys. J. 586 (2003) 1050--1066.
\newblock \href {http://arxiv.org/abs/astro-ph/0210480}
  {\path{arXiv:astro-ph/0210480}}, \href {https://doi.org/10.1086/367697}
  {\path{doi:10.1086/367697}}.

\bibitem{2005ICRC....3..157S}
R.~E. {Streitmatter}, F.~C. {Jones}, {Superbubbles and Local Cosmic Rays},
  International Cosmic Ray Conference 3 (2005) 157.

\bibitem{An17}
K.~J. Andersen, \href{http://hdl.handle.net/11250/2456366}{{Charged Particle
  Trajectories in the Local Superbubble}}, Master's thesis, NTNU Trondheim,
  available at http://hdl.handle.net/11250/2456366 (2016).
\newline\urlprefix\url{http://hdl.handle.net/11250/2456366}

\bibitem{Andersen:2017yyg}
K.~J. Andersen, M.~Kachelrie{\ss}, D.~V. Semikoz, {High-energy Neutrinos from
  Galactic Superbubbles}, Astrophys. J. 861~(2) (2018) L19.
\newblock \href {http://arxiv.org/abs/1712.03153} {\path{arXiv:1712.03153}},
  \href {https://doi.org/10.3847/2041-8213/aacefd}
  {\path{doi:10.3847/2041-8213/aacefd}}.

\bibitem{2016ApJ...818L..18Z}
E.~J. {Zirnstein}, J.~{Heerikhuisen}, H.~O. {Funsten}, G.~{Livadiotis}, D.~J.
  {McComas}, N.~V. {Pogorelov}, {Local Interstellar Magnetic Field Determined
  from the Interstellar Boundary Explorer Ribbon}, \apjl 818 (2016) L18.
\newblock \href {https://doi.org/10.3847/2041-8205/818/1/L18}
  {\path{doi:10.3847/2041-8205/818/1/L18}}.

\bibitem{Aartsen:2018ppz}
A.~U. Abeysekara, et~al., {All-Sky Measurement of the Anisotropy of Cosmic Rays
  at 10 TeV and Mapping of the Local Interstellar Magnetic Field}, Astrophys.
  J. 871 (2019) 97.
\newblock \href {http://arxiv.org/abs/1812.05682} {\path{arXiv:1812.05682}},
  \href {https://doi.org/10.3847/1538-4357/aaf5cc}
  {\path{doi:10.3847/1538-4357/aaf5cc}}.

\bibitem{Giacinti:2016tld}
G.~Giacinti, J.~G. Kirk, {Large-Scale Cosmic-Ray Anisotropy as a Probe of
  Interstellar Turbulence}, Astrophys. J. 835~(2) (2017) 258.
\newblock \href {http://arxiv.org/abs/1610.06134} {\path{arXiv:1610.06134}},
  \href {https://doi.org/10.3847/1538-4357/835/2/258}
  {\path{doi:10.3847/1538-4357/835/2/258}}.

\bibitem{Ahlers:2016rox}
M.~Ahlers, P.~Mertsch, {Origin of Small-Scale Anisotropies in Galactic Cosmic
  Rays}, Prog. Part. Nucl. Phys. 94 (2017) 184--216.
\newblock \href {http://arxiv.org/abs/1612.01873} {\path{arXiv:1612.01873}},
  \href {https://doi.org/10.1016/j.ppnp.2017.01.004}
  {\path{doi:10.1016/j.ppnp.2017.01.004}}.

\bibitem{Deligny:2018blo}
O.~Deligny, {Measurements and implications of cosmic ray anisotropies from TeV
  to trans-EeV energies}, Astropart. Phys. 104 (2019) 13--41.
\newblock \href {http://arxiv.org/abs/1808.03940} {\path{arXiv:1808.03940}},
  \href {https://doi.org/10.1016/j.astropartphys.2018.08.005}
  {\path{doi:10.1016/j.astropartphys.2018.08.005}}.

\bibitem{Bartoli:2015ysa}
B.~Bartoli, et~al., {ARGO-YBJ Observation of the Large-scale Cosmic Ray
  Anisotropy During the Solar Minimum Between Cycles 23 and 24}, Astrophys. J.
  809~(1) (2015) 90.
\newblock \href {https://doi.org/10.1088/0004-637X/809/1/90}
  {\path{doi:10.1088/0004-637X/809/1/90}}.

\bibitem{Amenomori:2005dy}
M.~Amenomori, et~al., {Large-scale sidereal anisotropy of Galactic cosmic-ray
  intensity observed by the Tibet air shower array}, Astrophys. J. 626 (2005)
  L29--L32.
\newblock \href {http://arxiv.org/abs/astro-ph/0505114}
  {\path{arXiv:astro-ph/0505114}}, \href {https://doi.org/10.1086/431582}
  {\path{doi:10.1086/431582}}.

\bibitem{Amenomori:2017jbv}
M.~Amenomori, {Northern sky Galactic Cosmic Ray anisotropy between 10-1000 TeV
  with the Tibet Air Shower Array}, Astrophys. J. 836~(2) (2017) 153.
\newblock \href {http://arxiv.org/abs/1701.07144} {\path{arXiv:1701.07144}},
  \href {https://doi.org/10.3847/1538-4357/836/2/153}
  {\path{doi:10.3847/1538-4357/836/2/153}}.

\bibitem{Abeysekara:2018qho}
A.~U. Abeysekara, et~al., {Observation of Anisotropy of TeV Cosmic Rays with
  Two Years of HAWC}, Astrophys. J. 865~(1) (2018) 57.
\newblock \href {http://arxiv.org/abs/1805.01847} {\path{arXiv:1805.01847}},
  \href {https://doi.org/10.3847/1538-4357/aad90c}
  {\path{doi:10.3847/1538-4357/aad90c}}.

\bibitem{Aartsen:2016ivj}
M.~G. Aartsen, et~al., {Anisotropy in Cosmic-ray Arrival Directions in the
  Southern Hemisphere Based on six Years of Data From the Icecube Detector},
  Astrophys. J. 826~(2) (2016) 220.
\newblock \href {http://arxiv.org/abs/1603.01227} {\path{arXiv:1603.01227}},
  \href {https://doi.org/10.3847/0004-637X/826/2/220}
  {\path{doi:10.3847/0004-637X/826/2/220}}.

\bibitem{TheHAWC:2017uyf}
J.~C.~D. Velez, et~al., {Combined Analysis of Cosmic-Ray Anisotropy with
  IceCube and HAWC}, PoS ICRC2017 (2018) 539.
\newblock \href {http://arxiv.org/abs/1708.03005} {\path{arXiv:1708.03005}},
  \href {https://doi.org/10.22323/1.301.0539} {\path{doi:10.22323/1.301.0539}}.

\bibitem{Chiavassa:2015jbg}
A.~Chiavassa, et~al., {A study of the first harmonic of the large scale
  anisotropies with the KASCADE-Grande experiment}, PoS ICRC2015 (2016) 281.
\newblock \href {https://doi.org/10.22323/1.236.0281}
  {\path{doi:10.22323/1.236.0281}}.

\bibitem{Aab:2018mmi}
A.~Aab, et~al., {Large-scale cosmic-ray anisotropies above 4 EeV measured by
  the Pierre Auger Observatory}, Astrophys. J. 868~(1) (2018) 4.
\newblock \href {http://arxiv.org/abs/1808.03579} {\path{arXiv:1808.03579}},
  \href {https://doi.org/10.3847/1538-4357/aae689}
  {\path{doi:10.3847/1538-4357/aae689}}.

\bibitem{Hillas:2005cs}
A.~M. Hillas, {Can diffusive shock acceleration in supernova remnants account
  for high-energy galactic cosmic rays?}, J. Phys. G31 (2005) R95--R131.
\newblock \href {https://doi.org/10.1088/0954-3899/31/5/R02}
  {\path{doi:10.1088/0954-3899/31/5/R02}}.

\bibitem{Blasi:2011fi}
P.~Blasi, E.~Amato, {Diffusive propagation of cosmic rays from supernova
  remnants in the Galaxy. I: spectrum and chemical composition}, JCAP 1201
  (2012) 010.
\newblock \href {http://arxiv.org/abs/1105.4521} {\path{arXiv:1105.4521}},
  \href {https://doi.org/10.1088/1475-7516/2012/01/010}
  {\path{doi:10.1088/1475-7516/2012/01/010}}.

\bibitem{Aguilar:2017hno}
M.~Aguilar, et~al., {Observation of the Identical Rigidity Dependence of He, C,
  and O Cosmic Rays at High Rigidities by the Alpha Magnetic Spectrometer on
  the International Space Station}, Phys. Rev. Lett. 119~(25) (2017) 251101.
\newblock \href {https://doi.org/10.1103/PhysRevLett.119.251101}
  {\path{doi:10.1103/PhysRevLett.119.251101}}.

\bibitem{Yoon:2017qjx}
Y.~S. Yoon, et~al., {Proton and Helium Spectra from the CREAM-III Flight},
  Astrophys. J. 839~(1) (2017) 5.
\newblock \href {http://arxiv.org/abs/1704.02512} {\path{arXiv:1704.02512}},
  \href {https://doi.org/10.3847/1538-4357/aa68e4}
  {\path{doi:10.3847/1538-4357/aa68e4}}.

\bibitem{Gorbunov:2018stf}
N.~Gorbunov, et~al., {Energy spectra of abundant cosmic-ray nuclei in the
  NUCLEON experiment}\href {http://arxiv.org/abs/1809.05333}
  {\path{arXiv:1809.05333}}.

\bibitem{Atkin:2019xxt}
E.~V. Atkin, et~al., {Energy Spectra of Cosmic-Ray Protons and Nuclei Measured
  in the NUCLEON Experiment Using a New Method}, Astron. Rep. 63~(1) (2019)
  66--78.
\newblock \href {https://doi.org/10.1134/S1063772919010013}
  {\path{doi:10.1134/S1063772919010013}}.

\bibitem{Ahn:2010gv}
H.~S. Ahn, et~al., {Discrepant hardening observed in cosmic-ray elemental
  spectra}, Astrophys. J. 714 (2010) L89--L93.
\newblock \href {http://arxiv.org/abs/1004.1123} {\path{arXiv:1004.1123}},
  \href {https://doi.org/10.1088/2041-8205/714/1/L89}
  {\path{doi:10.1088/2041-8205/714/1/L89}}.

\bibitem{Adriani:2011cu}
O.~Adriani, et~al., {PAMELA Measurements of Cosmic-ray Proton and Helium
  Spectra}, Science 332 (2011) 69--72.
\newblock \href {http://arxiv.org/abs/1103.4055} {\path{arXiv:1103.4055}},
  \href {https://doi.org/10.1126/science.1199172}
  {\path{doi:10.1126/science.1199172}}.

\bibitem{Ackermann:2014ula}
M.~Ackermann, et~al., {Inferred Cosmic-Ray Spectrum from Fermi Large Area
  Telescope $\gamma$-Ray Observations of Earths Limb}, Phys. Rev. Lett. 112
  (2014) 151103.
\newblock \href {http://arxiv.org/abs/1403.5372} {\path{arXiv:1403.5372}},
  \href {https://doi.org/10.1103/PhysRevLett.112.151103}
  {\path{doi:10.1103/PhysRevLett.112.151103}}.

\bibitem{Aguilar:2015ctt}
M.~Aguilar, et~al., {Precision Measurement of the Helium Flux in Primary Cosmic
  Rays of Rigidities 1.9 GV to 3 TV with the Alpha Magnetic Spectrometer on the
  International Space Station}, Phys. Rev. Lett. 115~(21) (2015) 211101.
\newblock \href {https://doi.org/10.1103/PhysRevLett.115.211101}
  {\path{doi:10.1103/PhysRevLett.115.211101}}.

\bibitem{Atkin:2018wsp}
E.~Atkin, et~al., {New Universal Cosmic-Ray Knee near a Magnetic Rigidity of 10
  TV with the NUCLEON Space Observatory}, JETP Lett. 108~(1) (2018) 5--12.
\newblock \href {http://arxiv.org/abs/1805.07119} {\path{arXiv:1805.07119}},
  \href {https://doi.org/10.1134/S0021364018130015}
  {\path{doi:10.1134/S0021364018130015}}.

\bibitem{Neronov:2017lqd}
A.~Neronov, D.~Malyshev, D.~V. Semikoz, {Cosmic ray spectrum in the local
  Galaxy}, Astron. Astrophys. 606 (2017) A22.
\newblock \href {http://arxiv.org/abs/1705.02200} {\path{arXiv:1705.02200}},
  \href {https://doi.org/10.1051/0004-6361/201731149}
  {\path{doi:10.1051/0004-6361/201731149}}.

\bibitem{Neronov:2011wi}
A.~Neronov, D.~V. Semikoz, A.~M. Taylor, {Low-energy break in the spectrum of
  Galactic cosmic rays}, Phys. Rev. Lett. 108 (2012) 051105.
\newblock \href {http://arxiv.org/abs/1112.5541} {\path{arXiv:1112.5541}},
  \href {https://doi.org/10.1103/PhysRevLett.108.051105}
  {\path{doi:10.1103/PhysRevLett.108.051105}}.

\bibitem{Kachelriess:2012fz}
M.~Kachelrie\ss, S.~Ostapchenko, {Deriving the cosmic ray spectrum from
  gamma-ray observations}, Phys. Rev. D86 (2012) 043004.
\newblock \href {http://arxiv.org/abs/1206.4705} {\path{arXiv:1206.4705}},
  \href {https://doi.org/10.1103/PhysRevD.86.043004}
  {\path{doi:10.1103/PhysRevD.86.043004}}.

\bibitem{Dermer:2012bz}
C.~D. Dermer, {Diffuse Galactic Gamma Rays from Shock-Accelerated Cosmic Rays},
  Phys. Rev. Lett. 109 (2012) 091101.
\newblock \href {http://arxiv.org/abs/1206.2899} {\path{arXiv:1206.2899}},
  \href {https://doi.org/10.1103/PhysRevLett.109.091101}
  {\path{doi:10.1103/PhysRevLett.109.091101}}.

\bibitem{Neronov:2015vua}
A.~Neronov, D.~Malyshev, {Hard spectrum of cosmic rays in the Disks of Milky
  Way and Large Magellanic Cloud}\href {http://arxiv.org/abs/1505.07601}
  {\path{arXiv:1505.07601}}.

\bibitem{Yang:2016jda}
R.~Yang, F.~Aharonian, C.~Evoli, {Radial distribution of the diffuse gamma-ray
  emissivity in the Galactic disk}, Phys. Rev. D93~(12) (2016) 123007.
\newblock \href {http://arxiv.org/abs/1602.04710} {\path{arXiv:1602.04710}},
  \href {https://doi.org/10.1103/PhysRevD.93.123007}
  {\path{doi:10.1103/PhysRevD.93.123007}}.

\bibitem{Aharonian:2018rob}
F.~Aharonian, G.~Peron, R.~Yang, S.~Casanova, R.~Zanin, {Probing the "Sea" of
  Galactic Cosmic Rays with Fermi-LAT}\href {http://arxiv.org/abs/1811.12118}
  {\path{arXiv:1811.12118}}.

\bibitem{Aguilar:2014mma}
M.~Aguilar, et~al., {Electron and Positron Fluxes in Primary Cosmic Rays
  Measured with the Alpha Magnetic Spectrometer on the International Space
  Station}, Phys. Rev. Lett. 113 (2014) 121102.
\newblock \href {https://doi.org/10.1103/PhysRevLett.113.121102}
  {\path{doi:10.1103/PhysRevLett.113.121102}}.

\bibitem{Abdollahi:2017nat}
S.~Abdollahi, et~al., {Cosmic-ray electron-positron spectrum from 7 GeV to 2
  TeV with the Fermi Large Area Telescope}, Phys. Rev. D95~(8) (2017) 082007.
\newblock \href {http://arxiv.org/abs/1704.07195} {\path{arXiv:1704.07195}},
  \href {https://doi.org/10.1103/PhysRevD.95.082007}
  {\path{doi:10.1103/PhysRevD.95.082007}}.

\bibitem{Abdalla:2017brm}
H.~Abdalla, et~al., {Contributions of the High Energy Stereoscopic System
  (H.E.S.S.) to the 35th International Cosmic Ray Conference (ICRC), Busan,
  Korea}, 2017.
\newblock \href {http://arxiv.org/abs/1709.06442} {\path{arXiv:1709.06442}}.

\bibitem{Ambrosi:2017wek}
G.~Ambrosi, et~al., {Direct detection of a break in the teraelectronvolt
  cosmic-ray spectrum of electrons and positrons}, Nature 552 (2017) 63--66.
\newblock \href {http://arxiv.org/abs/1711.10981} {\path{arXiv:1711.10981}},
  \href {https://doi.org/10.1038/nature24475} {\path{doi:10.1038/nature24475}}.

\bibitem{Adriani:2018ktz}
O.~Adriani, et~al., {Extended Measurement of the Cosmic-Ray Electron and
  Positron Spectrum from 11 GeV to 4.8 TeV with the Calorimetric Electron
  Telescope on the International Space Station}, Phys. Rev. Lett. 120~(26)
  (2018) 261102.
\newblock \href {http://arxiv.org/abs/1806.09728} {\path{arXiv:1806.09728}},
  \href {https://doi.org/10.1103/PhysRevLett.120.261102}
  {\path{doi:10.1103/PhysRevLett.120.261102}}.

\bibitem{Lipari:2018usj}
P.~Lipari, {The spectral shapes of the fluxes of electrons and positrons and
  the average residence time of cosmic rays in the Galaxy}, Phys. Rev. D99~(4)
  (2019) 043005.
\newblock \href {http://arxiv.org/abs/1810.03195} {\path{arXiv:1810.03195}},
  \href {https://doi.org/10.1103/PhysRevD.99.043005}
  {\path{doi:10.1103/PhysRevD.99.043005}}.

\bibitem{Ahn:2008my}
H.~S. Ahn, et~al., {Measurements of cosmic-ray secondary nuclei at high
  energies with the first flight of the CREAM balloon-borne experiment},
  Astropart. Phys. 30 (2008) 133--141.
\newblock \href {http://arxiv.org/abs/0808.1718} {\path{arXiv:0808.1718}},
  \href {https://doi.org/10.1016/j.astropartphys.2008.07.010}
  {\path{doi:10.1016/j.astropartphys.2008.07.010}}.

\bibitem{2018arXiv180909665G}
V.~Grebenyuk, et~al., {Secondary cosmic rays in the NUCLEON space
  experiment}\href {http://arxiv.org/abs/1809.09665} {\path{arXiv:1809.09665}}.

\bibitem{2016Natur.532...69W}
A.~{Wallner}, J.~{Feige}, N.~{Kinoshita}, M.~{Paul}, L.~K. {Fifield},
  R.~{Golser}, M.~{Honda}, U.~{Linnemann}, H.~{Matsuzaki}, S.~{Merchel},
  G.~{Rugel}, S.~G. {Tims}, P.~{Steier}, T.~{Yamagata}, S.~R. {Winkler},
  {Recent near-Earth supernovae probed by global deposition of interstellar
  radioactive $^{60}$Fe}, \nat 532 (2016) 69--72.
\newblock \href {https://doi.org/10.1038/nature17196}
  {\path{doi:10.1038/nature17196}}.

\bibitem{2001ApJ...563..768Y}
N.~E. {Yanasak}, M.~E. {Wiedenbeck}, R.~A. {Mewaldt}, A.~J. {Davis}, A.~C.
  {Cummings}, J.~S. {George}, R.~A. {Leske}, E.~C. {Stone}, E.~R. {Christian},
  T.~T. {von Rosenvinge}, W.~R. {Binns}, P.~L. {Hink}, M.~H. {Israel},
  {Measurement of the Secondary Radionuclides $^{10}$Be, $^{26}$Al, $^{36}$Cl,
  $^{54}$Mn, and $^{14}$C and Implications for the Galactic Cosmic-Ray Age},
  \apj 563 (2001) 768--792.
\newblock \href {https://doi.org/10.1086/323842} {\path{doi:10.1086/323842}}.

\bibitem{Ellis:1995qb}
J.~R. Ellis, B.~D. Fields, D.~N. Schramm, {Geological isotope anomalies as
  signatures of nearby supernovae}, Astrophys. J. 470 (1996) 1227--1236.
\newblock \href {http://arxiv.org/abs/astro-ph/9605128}
  {\path{arXiv:astro-ph/9605128}}, \href {https://doi.org/10.1086/177945}
  {\path{doi:10.1086/177945}}.

\bibitem{Rugel:2009zz}
G.~Rugel, T.~Faestermann, K.~Knie, G.~Korschinek, M.~Poutivtsev, D.~Schumann,
  N.~Kivel, I.~Gunther-Leopold, R.~Weinreich, M.~Wohlmuther, {New Measurement
  of the Fe-60 Half-Life}, Phys. Rev. Lett. 103 (2009) 072502.
\newblock \href {https://doi.org/10.1103/PhysRevLett.103.072502}
  {\path{doi:10.1103/PhysRevLett.103.072502}}.

\bibitem{Knie:1999zz}
K.~Knie, G.~Korschinek, T.~Faestermann, C.~Wallner, J.~Scholten, et~al.,
  {Indication for Supernova Produced Fe-60 Activity on Earth}, Phys.Rev.Lett.
  83 (1999) 18--21.
\newblock \href {https://doi.org/10.1103/PhysRevLett.83.18}
  {\path{doi:10.1103/PhysRevLett.83.18}}.

\bibitem{Benitez:2002jt}
N.~Benitez, J.~Maiz-Apellaniz, M.~Canelles, {Evidence for nearby supernova
  explosions}, Phys. Rev. Lett. 88 (2002) 081101.
\newblock \href {http://arxiv.org/abs/astro-ph/0201018}
  {\path{arXiv:astro-ph/0201018}}, \href
  {https://doi.org/10.1103/PhysRevLett.88.081101}
  {\path{doi:10.1103/PhysRevLett.88.081101}}.

\bibitem{Fitoussi:2007ef}
C.~Fitoussi, et~al., {Search for supernova-produced Fe-60 in a marine
  sediment}, Phys. Rev. Lett. 101 (2008) 121101.
\newblock \href {http://arxiv.org/abs/0709.4197} {\path{arXiv:0709.4197}},
  \href {https://doi.org/10.1103/PhysRevLett.101.121101}
  {\path{doi:10.1103/PhysRevLett.101.121101}}.

\bibitem{2016PhRvL.116o1104F}
L.~{Fimiani}, D.~L. {Cook}, T.~{Faestermann}, J.~M. {G{\'o}mez-Guzm{\'a}n},
  K.~{Hain}, G.~{Herzog}, K.~{Knie}, G.~{Korschinek}, P.~{Ludwig}, J.~{Park},
  R.~C. {Reedy}, G.~{Rugel}, {Interstellar $^{60}$Fe on the Surface of the
  Moon}, Phys. Rev. Lett. 116~(15) (2016) 151104.
\newblock \href {https://doi.org/10.1103/PhysRevLett.116.151104}
  {\path{doi:10.1103/PhysRevLett.116.151104}}.

\bibitem{2016Natur.532...73B}
D.~{Breitschwerdt}, J.~{Feige}, M.~M. {Schulreich}, M.~A.~D. {Avillez},
  C.~{Dettbarn}, B.~{Fuchs}, {The locations of recent supernovae near the Sun
  from modelling $^{60}$Fe transport}, \nat 532 (2016) 73--76.
\newblock \href {https://doi.org/10.1038/nature17424}
  {\path{doi:10.1038/nature17424}}.

\bibitem{Fry:2014yqa}
B.~J. Fry, B.~D. Fields, J.~R. Ellis, {Astrophysical Shrapnel: Discriminating
  Among Near-Earth Stellar Explosion Sources of Live Radioactive Isotopes},
  Astrophys. J. 800~(1) (2015) 71.
\newblock \href {http://arxiv.org/abs/1405.4310} {\path{arXiv:1405.4310}},
  \href {https://doi.org/10.1088/0004-637X/800/1/71}
  {\path{doi:10.1088/0004-637X/800/1/71}}.

\bibitem{Feige:2012be}
J.~Feige, A.~Wallner, S.~R. Winkler, S.~Merchel, L.~K. Fifield, G.~Korschinek,
  G.~Rugel, D.~Breitschwerdt, {The Search for Supernova-produced Radionuclides
  in Terrestrial Deep-sea Archives}, Publ. Astron. Soc. Austral. 29 (2012) 109.
\newblock \href {http://arxiv.org/abs/1204.4320} {\path{arXiv:1204.4320}},
  \href {https://doi.org/10.1071/AS11070} {\path{doi:10.1071/AS11070}}.

\bibitem{Fields:2019jtj}
B.~D. Fields, et~al., {Near-Earth Supernova Explosions: Evidence, Implications,
  and Opportunities}\href {http://arxiv.org/abs/1903.04589}
  {\path{arXiv:1903.04589}}.

\bibitem{Aguilar:2019owu}
M.~Aguilar, et~al., {Towards Understanding the Origin of Cosmic-Ray Positrons},
  Phys. Rev. Lett. 122~(4) (2019) 041102.
\newblock \href {https://doi.org/10.1103/PhysRevLett.122.041102}
  {\path{doi:10.1103/PhysRevLett.122.041102}}.

\bibitem{Aguilar:2016kjl}
M.~Aguilar, et~al., {Antiproton Flux, Antiproton-to-Proton Flux Ratio, and
  Properties of Elementary Particle Fluxes in Primary Cosmic Rays Measured with
  the Alpha Magnetic Spectrometer on the International Space Station}, Phys.
  Rev. Lett. 117~(9) (2016) 091103.
\newblock \href {https://doi.org/10.1103/PhysRevLett.117.091103}
  {\path{doi:10.1103/PhysRevLett.117.091103}}.

\bibitem{Kachelriess:2015oua}
M.~Kachelrie{\ss}, A.~Neronov, D.~V. Semikoz, {Signatures of a two million year
  old supernova in the spectra of cosmic ray protons, antiprotons and
  positrons}, Phys. Rev. Lett. 115~(18) (2015) 181103.
\newblock \href {http://arxiv.org/abs/1504.06472} {\path{arXiv:1504.06472}},
  \href {https://doi.org/10.1103/PhysRevLett.115.181103}
  {\path{doi:10.1103/PhysRevLett.115.181103}}.

\bibitem{Lipari:2016vqk}
P.~Lipari, {Interpretation of the cosmic ray positron and antiproton fluxes},
  Phys. Rev. D95~(6) (2017) 063009.
\newblock \href {http://arxiv.org/abs/1608.02018} {\path{arXiv:1608.02018}},
  \href {https://doi.org/10.1103/PhysRevD.95.063009}
  {\path{doi:10.1103/PhysRevD.95.063009}}.

\bibitem{Weng:2018}
Z.~{Weng}, {Distinctive Properties of Cosmic Electrons and Positrons Measured
  by AMS on ISS}, talk on ICHEP 2018.

\bibitem{Giesen:2015ufa}
G.~Giesen, M.~Boudaud, Y.~Ganolini, V.~Poulin, M.~Cirelli, P.~Salati, P.~D.
  Serpico, {AMS-02 antiprotons, at last! Secondary astrophysical component and
  immediate implications for Dark Matter}, JCAP 1509~(09) (2015) 023.
\newblock \href {http://arxiv.org/abs/1504.04276} {\path{arXiv:1504.04276}},
  \href {https://doi.org/10.1088/1475-7516/2015/09/023}
  {\path{doi:10.1088/1475-7516/2015/09/023}}.

\bibitem{Boschini:2017fxq}
M.~J. Boschini, et~al., {Solution of heliospheric propagation: unveiling the
  local interstellar spectra of cosmic ray species}, Astrophys. J. 840~(2)
  (2017) 115.
\newblock \href {http://arxiv.org/abs/1704.06337} {\path{arXiv:1704.06337}},
  \href {https://doi.org/10.3847/1538-4357/aa6e4f}
  {\path{doi:10.3847/1538-4357/aa6e4f}}.

\bibitem{1990ApJ...361..162J}
F.~C. {Jones}, {The generalized diffusion-convection equation}, \apj 361 (1990)
  162--172.
\newblock \href {https://doi.org/10.1086/169179} {\path{doi:10.1086/169179}}.

\bibitem{Savchenko:2015dha}
V.~Savchenko, M.~Kachelrie{\ss}, D.~V. Semikoz, {Imprint of a 2 Million Year
  old Source on the Cosmic-ray Anisotropy}, Astrophys. J. 809~(2) (2015) L23.
\newblock \href {http://arxiv.org/abs/1505.02720} {\path{arXiv:1505.02720}},
  \href {https://doi.org/10.1088/2041-8205/809/2/L23}
  {\path{doi:10.1088/2041-8205/809/2/L23}}.

\bibitem{Giacinti:2015hva}
G.~Giacinti, M.~Kachelrie{\ss}, D.~V. Semikoz, {Escape model for Galactic
  cosmic rays and an early extragalactic transition}, Phys. Rev. D91~(8) (2015)
  083009.
\newblock \href {http://arxiv.org/abs/1502.01608} {\path{arXiv:1502.01608}},
  \href {https://doi.org/10.1103/PhysRevD.91.083009}
  {\path{doi:10.1103/PhysRevD.91.083009}}.

\bibitem{Kachelriess:2017yzq}
M.~Kachelrie{\ss}, A.~Neronov, D.~V. Semikoz, {Cosmic ray signatures of a 2-3
  Myr old local supernova}, Phys. Rev. D97~(6) (2018) 063011.
\newblock \href {http://arxiv.org/abs/1710.02321} {\path{arXiv:1710.02321}},
  \href {https://doi.org/10.1103/PhysRevD.97.063011}
  {\path{doi:10.1103/PhysRevD.97.063011}}.

\bibitem{2013APh....50...33S}
L.~G. {Sveshnikova}, O.~N. {Strelnikova}, V.~S. {Ptuskin}, {Spectrum and
  anisotropy of cosmic rays at TeV-PeV-energies and contribution of nearby
  sources}, Astroparticle Physics 50 (2013) 33--46.
\newblock \href {http://arxiv.org/abs/1301.2028} {\path{arXiv:1301.2028}},
  \href {https://doi.org/10.1016/j.astropartphys.2013.08.007}
  {\path{doi:10.1016/j.astropartphys.2013.08.007}}.

\bibitem{Ahlers:2016njd}
M.~Ahlers, {Deciphering the Dipole Anisotropy of Galactic Cosmic Rays}, Phys.
  Rev. Lett. 117~(15) (2016) 151103.
\newblock \href {http://arxiv.org/abs/1605.06446} {\path{arXiv:1605.06446}},
  \href {https://doi.org/10.1103/PhysRevLett.117.151103}
  {\path{doi:10.1103/PhysRevLett.117.151103}}.

\bibitem{Mertsch:2014cua}
P.~Mertsch, S.~Funk, {Solution to the cosmic ray anisotropy problem}, Phys.
  Rev. Lett. 114~(2) (2015) 021101.
\newblock \href {http://arxiv.org/abs/1408.3630} {\path{arXiv:1408.3630}},
  \href {https://doi.org/10.1103/PhysRevLett.114.021101}
  {\path{doi:10.1103/PhysRevLett.114.021101}}.

\bibitem{Blum:2017iwq}
K.~Blum, R.~Sato, E.~Waxman, {Cosmic-ray Antimatter}\href
  {http://arxiv.org/abs/1709.06507} {\path{arXiv:1709.06507}}.

\bibitem{Kirk:2007tn}
J.~G. Kirk, Y.~Lyubarsky, J.~Petri, {The theory of pulsar winds and
  nebulae}[Astrophys. Space Sci. Libr.357,421(2009)].
\newblock \href {http://arxiv.org/abs/astro-ph/0703116}
  {\path{arXiv:astro-ph/0703116}}, \href
  {https://doi.org/10.1007/978-3-540-76965-1_16}
  {\path{doi:10.1007/978-3-540-76965-1_16}}.

\bibitem{1987ICRC....2...92H}
A.~K. {Harding}, R.~{Ramaty}, {The Pulsar Contribution to Galactic Cosmic Ray
  Positrons}, International Cosmic Ray Conference 2 (1987) 92.

\bibitem{1989ApJ...342..807B}
A.~{Boulares}, {The nature of the cosmic-ray electron spectrum, and supernova
  remnant contributions}, \apj 342 (1989) 807--813.
\newblock \href {https://doi.org/10.1086/167637} {\path{doi:10.1086/167637}}.

\bibitem{Hooper:2008kg}
D.~Hooper, P.~Blasi, P.~D. Serpico, {Pulsars as the Sources of High Energy
  Cosmic Ray Positrons}, JCAP 0901 (2009) 025.
\newblock \href {http://arxiv.org/abs/0810.1527} {\path{arXiv:0810.1527}},
  \href {https://doi.org/10.1088/1475-7516/2009/01/025}
  {\path{doi:10.1088/1475-7516/2009/01/025}}.

\bibitem{Grasso:2009ma}
D.~Grasso, et~al., {On possible interpretations of the high energy
  electron-positron spectrum measured by the Fermi Large Area Telescope},
  Astropart. Phys. 32 (2009) 140--151.
\newblock \href {http://arxiv.org/abs/0905.0636} {\path{arXiv:0905.0636}},
  \href {https://doi.org/10.1016/j.astropartphys.2009.07.003}
  {\path{doi:10.1016/j.astropartphys.2009.07.003}}.

\bibitem{Abeysekara:2017old}
A.~U. Abeysekara, et~al., {Extended gamma-ray sources around pulsars constrain
  the origin of the positron flux at Earth}, Science 358~(6365) (2017)
  911--914.
\newblock \href {http://arxiv.org/abs/1711.06223} {\path{arXiv:1711.06223}},
  \href {https://doi.org/10.1126/science.aan4880}
  {\path{doi:10.1126/science.aan4880}}.

\bibitem{Lopez-Coto:2017pbk}
R.~L\'opez-Coto, G.~Giacinti, {Constraining the properties of the magnetic
  turbulence in the Geminga region using HAWC $\gamma$-ray data}, Mon. Not.
  Roy. Astron. Soc. 479 (2018) 4526.
\newblock \href {http://arxiv.org/abs/1712.04373} {\path{arXiv:1712.04373}},
  \href {https://doi.org/10.1093/mnras/sty1821}
  {\path{doi:10.1093/mnras/sty1821}}.

\bibitem{DAngelo:2015cfw}
M.~D'Angelo, P.~Blasi, E.~Amato, {Grammage of cosmic rays around Galactic
  supernova remnants}, Phys. Rev. D94~(8) (2016) 083003.
\newblock \href {http://arxiv.org/abs/1512.05000} {\path{arXiv:1512.05000}},
  \href {https://doi.org/10.1103/PhysRevD.94.083003}
  {\path{doi:10.1103/PhysRevD.94.083003}}.

\bibitem{Nava:2016szf}
L.~Nava, S.~Gabici, A.~Marcowith, G.~Morlino, V.~S. Ptuskin, {Non-linear
  diffusion of cosmic rays escaping from supernova remnants I. The effect of
  neutrals}, Mon. Not. Roy. Astron. Soc. 461~(4) (2016) 3552--3562.
\newblock \href {http://arxiv.org/abs/1606.06902} {\path{arXiv:1606.06902}},
  \href {https://doi.org/10.1093/mnras/stw1592}
  {\path{doi:10.1093/mnras/stw1592}}.

\bibitem{Hooper:2017gtd}
D.~Hooper, I.~Cholis, T.~Linden, K.~Fang, {HAWC Observations Strongly Favor
  Pulsar Interpretations of the Cosmic-Ray Positron Excess}, Phys. Rev.
  D96~(10) (2017) 103013.
\newblock \href {http://arxiv.org/abs/1702.08436} {\path{arXiv:1702.08436}},
  \href {https://doi.org/10.1103/PhysRevD.96.103013}
  {\path{doi:10.1103/PhysRevD.96.103013}}.

\bibitem{Fang:2018qco}
K.~Fang, X.-J. Bi, P.-F. Yin, Q.~Yuan, {Two-zone diffusion of electrons and
  positrons from Geminga explains the positron anomaly}, Astrophys. J. 863~(1)
  (2018) 30.
\newblock \href {http://arxiv.org/abs/1803.02640} {\path{arXiv:1803.02640}},
  \href {https://doi.org/10.3847/1538-4357/aad092}
  {\path{doi:10.3847/1538-4357/aad092}}.

\bibitem{Profumo:2018fmz}
S.~Profumo, J.~Reynoso-Cordova, N.~Kaaz, M.~Silverman, {Lessons from HAWC
  pulsar wind nebulae observations: The diffusion constant is not a constant;
  pulsars remain the likeliest sources of the anomalous positron fraction;
  cosmic rays are trapped for long periods of time in pockets of inefficient
  diffusion}, Phys. Rev. D97~(12) (2018) 123008.
\newblock \href {http://arxiv.org/abs/1803.09731} {\path{arXiv:1803.09731}},
  \href {https://doi.org/10.1103/PhysRevD.97.123008}
  {\path{doi:10.1103/PhysRevD.97.123008}}.

\bibitem{Shao-Qiang:2018zla}
S.-Q. Xi, R.-Y. Liu, Z.-Q. Huang, K.~Fang, H.~Yan, X.-Y. Wang, {GeV
  observations of the extended pulsar wind nebulae challenge the pulsar
  interpretations of the cosmic-ray positron excess}\href
  {http://arxiv.org/abs/1810.10928} {\path{arXiv:1810.10928}}.

\bibitem{DiMauro:2019yvh}
M.~Di~Mauro, S.~Manconi, F.~Donato, {Detection of a $\gamma$-ray halo around
  Geminga with the Fermi-LAT and implications for the positron flux}\href
  {http://arxiv.org/abs/1903.05647} {\path{arXiv:1903.05647}}.

\bibitem{Cholis:2018izy}
I.~Cholis, T.~Karwal, M.~Kamionkowski, {Studying the Milky Way pulsar
  population with cosmic-ray leptons}, Phys. Rev. D98~(6) (2018) 063008.
\newblock \href {http://arxiv.org/abs/1807.05230} {\path{arXiv:1807.05230}},
  \href {https://doi.org/10.1103/PhysRevD.98.063008}
  {\path{doi:10.1103/PhysRevD.98.063008}}.

\bibitem{Venter:2015gga}
C.~Venter, A.~Kopp, A.~K. Harding, P.~L. Gonthier, I.~B{\"u}sching, {Cosmic-ray
  positrons from millisecond pulsars}, Astrophys. J. 807~(2) (2015) 130.
\newblock \href {http://arxiv.org/abs/1506.01211} {\path{arXiv:1506.01211}},
  \href {https://doi.org/10.1088/0004-637X/807/2/130}
  {\path{doi:10.1088/0004-637X/807/2/130}}.

\bibitem{Lopez-Coto:2018ksn}
R.~L\'opez-Coto, R.~D. Parsons, J.~A. Hinton, G.~Giacinti, {Undiscovered Pulsar
  in the Local Bubble as an Explanation of the Local High Energy Cosmic Ray
  All-Electron Spectrum}, Phys. Rev. Lett. 121~(25) (2018) 251106.
\newblock \href {http://arxiv.org/abs/1811.04123} {\path{arXiv:1811.04123}},
  \href {https://doi.org/10.1103/PhysRevLett.121.251106}
  {\path{doi:10.1103/PhysRevLett.121.251106}}.

\bibitem{Blasi:2010de}
P.~Blasi, E.~Amato, {Positrons from pulsar winds}, in: {Proceedings, 1st
  Session of the Sant Cugat Forum on Astrophysics: High-Energy Emission from
  Pulsars and their Systems: Sant Cugat, Catalonia, Spain, April 12-16, 2010},
  2011, pp. 623--641.
\newblock \href {http://arxiv.org/abs/1007.4745} {\path{arXiv:1007.4745}},
  \href {https://doi.org/10.1007/978-3-642-17251-9_50}
  {\path{doi:10.1007/978-3-642-17251-9_50}}.

\bibitem{Bykov:2017xpo}
A.~M. Bykov, E.~Amato, A.~E. Petrov, A.~M. Krassilchtchikov, K.~P. Levenfish,
  {Pulsar wind nebulae with bow shocks: non-thermal radiation and cosmic ray
  leptons}, Space Sci. Rev. 207~(1-4) (2017) 235--290.
\newblock \href {http://arxiv.org/abs/1705.00950} {\path{arXiv:1705.00950}},
  \href {https://doi.org/10.1007/s11214-017-0371-7}
  {\path{doi:10.1007/s11214-017-0371-7}}.

\bibitem{Blasi:2009hv}
P.~Blasi, {The origin of the positron excess in cosmic rays}, Phys. Rev. Lett.
  103 (2009) 051104.
\newblock \href {http://arxiv.org/abs/0903.2794} {\path{arXiv:0903.2794}},
  \href {https://doi.org/10.1103/PhysRevLett.103.051104}
  {\path{doi:10.1103/PhysRevLett.103.051104}}.

\bibitem{Blasi:2009bd}
P.~Blasi, P.~D. Serpico, {High-energy antiprotons from old supernova remnants},
  Phys. Rev. Lett. 103 (2009) 081103.
\newblock \href {http://arxiv.org/abs/0904.0871} {\path{arXiv:0904.0871}},
  \href {https://doi.org/10.1103/PhysRevLett.103.081103}
  {\path{doi:10.1103/PhysRevLett.103.081103}}.

\bibitem{Mertsch:2009ph}
P.~Mertsch, S.~Sarkar, {Testing astrophysical models for the PAMELA positron
  excess with cosmic ray nuclei}, Phys. Rev. Lett. 103 (2009) 081104.
\newblock \href {http://arxiv.org/abs/0905.3152} {\path{arXiv:0905.3152}},
  \href {https://doi.org/10.1103/PhysRevLett.103.081104}
  {\path{doi:10.1103/PhysRevLett.103.081104}}.

\bibitem{Mertsch:2014poa}
P.~Mertsch, S.~Sarkar, {AMS-02 data confront acceleration of cosmic ray
  secondaries in nearby sources}, Phys. Rev. D90 (2014) 061301.
\newblock \href {http://arxiv.org/abs/1402.0855} {\path{arXiv:1402.0855}},
  \href {https://doi.org/10.1103/PhysRevD.90.061301}
  {\path{doi:10.1103/PhysRevD.90.061301}}.

\bibitem{Kachelriess:2011qv}
M.~Kachelrie\ss, S.~Ostapchenko, R.~Tom{\`a}s, {Antimatter production in
  supernova remnants}, Astrophys. J. 733 (2011) 119.
\newblock \href {http://arxiv.org/abs/1103.5765} {\path{arXiv:1103.5765}},
  \href {https://doi.org/10.1088/0004-637X/733/2/119}
  {\path{doi:10.1088/0004-637X/733/2/119}}.

\bibitem{Kachelriess:2012ag}
M.~Kachelrie\ss, S.~Ostapchenko, {B/C ratio and the PAMELA positron excess},
  Phys. Rev. D87~(4) (2013) 047301.
\newblock \href {http://arxiv.org/abs/1211.1033} {\path{arXiv:1211.1033}},
  \href {https://doi.org/10.1103/PhysRevD.87.047301}
  {\path{doi:10.1103/PhysRevD.87.047301}}.

\bibitem{Fujita:2009wk}
Y.~Fujita, K.~Kohri, R.~Yamazaki, K.~Ioka, {Is the PAMELA anomaly caused by the
  supernova explosions near the Earth?}, Phys. Rev. D80 (2009) 063003.
\newblock \href {http://arxiv.org/abs/0903.5298} {\path{arXiv:0903.5298}},
  \href {https://doi.org/10.1103/PhysRevD.80.063003}
  {\path{doi:10.1103/PhysRevD.80.063003}}.

\bibitem{Yamazaki:2006uf}
R.~Yamazaki, K.~Kohri, A.~Bamba, T.~Yoshida, T.~Tsuribe, F.~Takahara, {TeV
  gamma-rays from old supernova remnants}, Mon. Not. Roy. Astron. Soc. 371
  (2006) 1975--1982.
\newblock \href {http://arxiv.org/abs/astro-ph/0601704}
  {\path{arXiv:astro-ph/0601704}}, \href
  {https://doi.org/10.1111/j.1365-2966.2006.10832.x}
  {\path{doi:10.1111/j.1365-2966.2006.10832.x}}.

\bibitem{Kohri:2015mga}
K.~Kohri, K.~Ioka, Y.~Fujita, R.~Yamazaki, {Can we explain AMS-02 antiproton
  and positron excesses simultaneously by nearby supernovae without pulsars or
  dark matter?}, PTEP 2016~(2) (2016) 021E01.
\newblock \href {http://arxiv.org/abs/1505.01236} {\path{arXiv:1505.01236}},
  \href {https://doi.org/10.1093/ptep/ptv193} {\path{doi:10.1093/ptep/ptv193}}.

\bibitem{Tomassetti:2015mha}
N.~Tomassetti, {Cosmic-ray protons, nuclei, electrons, and antiparticles under
  a two-halo scenario of diffusive propagation}, Phys. Rev. D92~(8) (2015)
  081301.
\newblock \href {http://arxiv.org/abs/1509.05775} {\path{arXiv:1509.05775}},
  \href {https://doi.org/10.1103/PhysRevD.92.081301}
  {\path{doi:10.1103/PhysRevD.92.081301}}.

\bibitem{Gaggero:2014xla}
D.~Gaggero, A.~Urbano, M.~Valli, P.~Ullio, {Gamma-ray sky points to radial
  gradients in cosmic-ray transport}, Phys. Rev. D91~(8) (2015) 083012.
\newblock \href {http://arxiv.org/abs/1411.7623} {\path{arXiv:1411.7623}},
  \href {https://doi.org/10.1103/PhysRevD.91.083012}
  {\path{doi:10.1103/PhysRevD.91.083012}}.

\bibitem{Gaggero:2015xza}
D.~Gaggero, D.~Grasso, A.~Marinelli, A.~Urbano, M.~Valli, {The gamma-ray and
  neutrino sky: A consistent picture of Fermi-LAT, Milagro, and IceCube
  results}, Astrophys. J. 815~(2) (2015) L25.
\newblock \href {http://arxiv.org/abs/1504.00227} {\path{arXiv:1504.00227}},
  \href {https://doi.org/10.1088/2041-8205/815/2/L25}
  {\path{doi:10.1088/2041-8205/815/2/L25}}.

\bibitem{Tomassetti:2017umm}
N.~Tomassetti, J.~Feng, A.~Oliva, {Fresh Insights on Cosmic-ray Propagation
  from the New AMS Data}, Astron. Astrophys. Suppl. Ser. 1 (2017) 35.
\newblock \href {http://arxiv.org/abs/1712.03176} {\path{arXiv:1712.03176}},
  \href {https://doi.org/10.3847/2515-5172/aa9f11}
  {\path{doi:10.3847/2515-5172/aa9f11}}.

\bibitem{Tomassetti:2017izg}
N.~Tomassetti, A.~Oliva, {Production and acceleration of antinuclei in
  supernova shockwaves}, Astrophys. J. Lett. 844 (2017) L26, [Astrophys.
  J.844,no.2,L26(2017)].
\newblock \href {http://arxiv.org/abs/1707.06915} {\path{arXiv:1707.06915}},
  \href {https://doi.org/10.3847/2041-8213/aa80da}
  {\path{doi:10.3847/2041-8213/aa80da}}.

\bibitem{1995ApJ...452..912M}
J.~A. {Miller}, D.~A. {Roberts}, {Stochastic Proton Acceleration by Cascading
  Alfven Waves in Impulsive Solar Flares}, \apj 452 (1995) 912.
\newblock \href {https://doi.org/10.1086/176359} {\path{doi:10.1086/176359}}.

\bibitem{1975MNRAS.173..255S}
J.~{Skilling}, {Cosmic ray streaming. III - Self-consistent solutions}, \mnras
  173 (1975) 255--269.
\newblock \href {https://doi.org/10.1093/mnras/173.2.255}
  {\path{doi:10.1093/mnras/173.2.255}}.

\bibitem{2013PhPl...20e5501Z}
E.~G. {Zweibel}, {The microphysics and macrophysics of cosmic rays}, Physics of
  Plasmas 20~(5) (2013) 055501.
\newblock \href {https://doi.org/10.1063/1.4807033}
  {\path{doi:10.1063/1.4807033}}.

\bibitem{Thoudam:2011aa}
S.~Thoudam, J.~R. H{\"o}randel, {Nearby supernova remnants and the cosmic-ray
  spectral hardening at high energies}, Mon. Not. Roy. Astron. Soc. 421 (2012)
  1209.
\newblock \href {http://arxiv.org/abs/1112.3020} {\path{arXiv:1112.3020}},
  \href {https://doi.org/10.1111/j.1365-2966.2011.20385.x}
  {\path{doi:10.1111/j.1365-2966.2011.20385.x}}.

\bibitem{Thoudam:2013iia}
S.~Thoudam, J.~R. H{\"o}randel, {Cosmic-ray spectral anomaly at GeV-TeV
  energies as due to re-accel\-eration by weak shocks in the Galaxy}, in:
  {Proceedings, 33rd International Cosmic Ray Conference (ICRC2013): Rio de
  Janeiro, Brazil, July 2-9, 2013}, 2013, p. 1022.
\newblock \href {http://arxiv.org/abs/1308.1357} {\path{arXiv:1308.1357}}.

\bibitem{Bernard:2012pia}
G.~Bernard, T.~Delahaye, Y.-Y. Keum, W.~Liu, P.~Salati, R.~Taillet, {No More
  Anomaly in the TeV Cosmic Ray Proton and Helium Spectra}, Astron. Astrophys.
  555 (2013) A48.
\newblock \href {http://arxiv.org/abs/1207.4670} {\path{arXiv:1207.4670}},
  \href {https://doi.org/10.1051/0004-6361/201321202}
  {\path{doi:10.1051/0004-6361/201321202}}.

\bibitem{Tomassetti:2015cva}
N.~Tomassetti, F.~Donato, {The Connection Between the Positron Fraction Anomaly
  and the Spectral Features in Galactic Cosmic-Ray Hadrons}, Astrophys. J.
  803~(2) (2015) L15.
\newblock \href {http://arxiv.org/abs/1502.06150} {\path{arXiv:1502.06150}},
  \href {https://doi.org/10.1088/2041-8205/803/2/L15}
  {\path{doi:10.1088/2041-8205/803/2/L15}}.

\bibitem{Yang:2018nhs}
R.-z. Yang, F.~Aharonian, {Interpretation of the "Excess" of Antiparticles
  within the Standard Cosmic Ray Paradigm with a Slight Modification}\href
  {http://arxiv.org/abs/1812.04364} {\path{arXiv:1812.04364}}.

\bibitem{Yoon:2011cr}
Y.~S. Yoon, et~al., {Cosmic-ray Proton and Helium Spectra from the First CREAM
  Flight}, Astrophys. J. 728~(122) (2011) 8.
\newblock \href {http://arxiv.org/abs/1602.04710} {\path{arXiv:1602.04710}},
  \href {https://doi.org/10.1103/PhysRevD.93.123007}
  {\path{doi:10.1103/PhysRevD.93.123007}}.

\bibitem{Apel:2011mi}
W.~D. Apel, et~al., {Kneelike structure in the spectrum of the heavy component
  of cosmic rays observed with KASCADE-Grande}, Phys. Rev. Lett. 107 (2011)
  171104.
\newblock \href {http://arxiv.org/abs/1107.5885} {\path{arXiv:1107.5885}},
  \href {https://doi.org/10.1103/PhysRevLett.107.171104}
  {\path{doi:10.1103/PhysRevLett.107.171104}}.

\bibitem{ATIC_spec}
A.~D. Panov, et~al., {Energy Spectra of Abundant Nuclei of Primary Cosmic Rays
  from the Data of ATIC-2 Experiment: Final Results}, Bull. Russ. Acad. Sci.
  Phys. 73~(5) (2009) 564--567, [Izv. Ross. Akad. Nauk Ser. Fiz.73,602(2009)].
\newblock \href {http://arxiv.org/abs/1101.3246} {\path{arXiv:1101.3246}},
  \href {https://doi.org/10.3103/S1062873809050098}
  {\path{doi:10.3103/S1062873809050098}}.

\bibitem{TUNKA_spec}
V.~V. Prosin, et~al., {Primary CR energy spectrum and mass composition by the
  data of Tunka-133 array}, EPJ Web Conf. 99 (2015) 04002.
\newblock \href {https://doi.org/10.1051/epjconf/20159904002}
  {\path{doi:10.1051/epjconf/20159904002}}.

\bibitem{TUNKA_HS_spec}
M.~Tluczykont, et~al., {TAIGA-HiSCORE: results from the first two operation
  seasons}, PoS ICRC2017 (2018) 759.
\newblock \href {https://doi.org/10.22323/1.301.0759}
  {\path{doi:10.22323/1.301.0759}}.

\bibitem{IceTop_spec}
K.~Rawlins, {Cosmic ray spectrum and composition from three years of IceTop and
  IceCube}, J. Phys. Conf. Ser. 718~(5) (2016) 052033.
\newblock \href {https://doi.org/10.1088/1742-6596/718/5/052033}
  {\path{doi:10.1088/1742-6596/718/5/052033}}.

\bibitem{Abbasi:2018xsn}
R.~U. Abbasi, et~al., {The Cosmic-Ray Energy Spectrum between 2 PeV and 2 EeV
  Observed with the TALE detector in monocular mode}, Astrophys. J. 865~(1)
  (2018) 74.
\newblock \href {http://arxiv.org/abs/1803.01288} {\path{arXiv:1803.01288}},
  \href {https://doi.org/10.3847/1538-4357/aada05}
  {\path{doi:10.3847/1538-4357/aada05}}.

\bibitem{PAO_spec}
A.~Aab, et~al., {Highlights from the Pierre Auger Observatory}, PoS ICRC2017
  (2018) 1102, [35,1102(2017)].
\newblock \href {http://arxiv.org/abs/1710.09478} {\path{arXiv:1710.09478}},
  \href {https://doi.org/10.22323/1.301.1102} {\path{doi:10.22323/1.301.1102}}.

\bibitem{TA_spec}
R.~U. Abbasi, et~al., {The energy spectrum of cosmic rays above 10$^{17.2}$ eV
  measured by the fluorescence detectors of the Telescope Array experiment in
  seven years}, Astropart. Phys. 80 (2016) 131--140.
\newblock \href {http://arxiv.org/abs/1511.07510} {\path{arXiv:1511.07510}},
  \href {https://doi.org/10.1016/j.astropartphys.2016.04.002}
  {\path{doi:10.1016/j.astropartphys.2016.04.002}}.

\bibitem{Nagano:1991jz}
M.~Nagano, M.~Teshima, Y.~Matsubara, H.~Y. Dai, T.~Hara, N.~Hayashida,
  M.~Honda, H.~Ohoka, S.~Yoshida, {Energy spectrum of primary cosmic rays above
  $10^{17}$\,eV determined from the extensive air shower experiment at Akeno},
  J. Phys. G18 (1992) 423--442.
\newblock \href {https://doi.org/10.1088/0954-3899/18/2/022}
  {\path{doi:10.1088/0954-3899/18/2/022}}.

\bibitem{Bluemer:2009zf}
J.~Bl{\"u}mer, R.~Engel, J.~R. H{\"o}randel, {Cosmic Rays from the Knee to the
  Highest Energies}, Prog. Part. Nucl. Phys. 63 (2009) 293--338.
\newblock \href {http://arxiv.org/abs/0904.0725} {\path{arXiv:0904.0725}},
  \href {https://doi.org/10.1016/j.ppnp.2009.05.002}
  {\path{doi:10.1016/j.ppnp.2009.05.002}}.

\bibitem{Apel:2013uni}
W.~D. Apel, et~al., {KASCADE-Grande measurements of energy spectra for
  elemental groups of cosmic rays}, Astropart. Phys. 47 (2013) 54--66.
\newblock \href {http://arxiv.org/abs/1306.6283} {\path{arXiv:1306.6283}},
  \href {https://doi.org/10.1016/j.astropartphys.2013.06.004}
  {\path{doi:10.1016/j.astropartphys.2013.06.004}}.

\bibitem{Aglietta:2004np}
M.~Aglietta, et~al., {The cosmic ray primary composition in the 'knee' region
  through the EAS electromagnetic and muon measurements at EAS-TOP}, Astropart.
  Phys. 21 (2004) 583--596.
\newblock \href {https://doi.org/10.1016/j.astropartphys.2004.04.005}
  {\path{doi:10.1016/j.astropartphys.2004.04.005}}.

\bibitem{Antoni:2005wq}
T.~Antoni, et~al., {KASCADE measurements of energy spectra for elemental groups
  of cosmic rays: Results and open problems}, Astropart. Phys. 24 (2005) 1--25.
\newblock \href {http://arxiv.org/abs/astro-ph/0505413}
  {\path{arXiv:astro-ph/0505413}}, \href
  {https://doi.org/10.1016/j.astropartphys.2005.04.001}
  {\path{doi:10.1016/j.astropartphys.2005.04.001}}.

\bibitem{IceCube:2012vv}
R.~Abbasi, et~al., {Cosmic Ray Composition and Energy Spectrum from 1-30 PeV
  Using the 40-String Configuration of IceTop and IceCube}, Astropart. Phys. 42
  (2013) 15--32.
\newblock \href {http://arxiv.org/abs/1207.3455} {\path{arXiv:1207.3455}},
  \href {https://doi.org/10.1016/j.astropartphys.2012.11.003}
  {\path{doi:10.1016/j.astropartphys.2012.11.003}}.

\bibitem{Bartoli:2014eua}
B.~Bartoli, et~al., {Energy Spectrum of Cosmic Protons and Helium Nuclei by a
  Hybrid Measurement at 4300 m a.s.l}, Chin. Phys. C38 (2014) 045001.
\newblock \href {http://arxiv.org/abs/1401.6987} {\path{arXiv:1401.6987}},
  \href {https://doi.org/10.1088/1674-1137/38/4/045001}
  {\path{doi:10.1088/1674-1137/38/4/045001}}.

\bibitem{Bartoli:2015fhw}
B.~Bartoli, et~al., {Cosmic ray proton plus helium energy spectrum measured by
  the ARGO-YBJ experiment in the energy range 3-300 TeV}, Phys. Rev. D91~(11)
  (2015) 112017.
\newblock \href {http://arxiv.org/abs/1503.07136} {\path{arXiv:1503.07136}},
  \href {https://doi.org/10.1103/PhysRevD.91.112017}
  {\path{doi:10.1103/PhysRevD.91.112017}}.

\bibitem{Amenomori:2011zza}
M.~Amenomori, et~al., {Cosmic-ray energy spectrum around the knee obtained by
  the Tibet experiment and future prospects}, Adv. Space Res. 47 (2011)
  629--639.
\newblock \href {https://doi.org/10.1016/j.asr.2010.08.029}
  {\path{doi:10.1016/j.asr.2010.08.029}}.

\bibitem{Montini:2016osn}
P.~Montini, {Cosmic ray physics with ARGO-YBJ}, Nucl. Part. Phys. Proc. 279-281
  (2016) 7--14.
\newblock \href {http://arxiv.org/abs/1608.01251} {\path{arXiv:1608.01251}},
  \href {https://doi.org/10.1016/j.nuclphysbps.2016.10.003}
  {\path{doi:10.1016/j.nuclphysbps.2016.10.003}}.

\bibitem{dEnterria:2011twh}
D.~d'Enterria, R.~Engel, T.~Pierog, S.~Ostapchenko, K.~Werner, {Constraints
  from the first LHC data on hadronic event generators for ultra-high energy
  cosmic-ray physics}, Astropart. Phys. 35 (2011) 98--113.
\newblock \href {http://arxiv.org/abs/1101.5596} {\path{arXiv:1101.5596}},
  \href {https://doi.org/10.1016/j.astropartphys.2011.05.002}
  {\path{doi:10.1016/j.astropartphys.2011.05.002}}.

\bibitem{Stanev:1993tx}
T.~Stanev, P.~L. Biermann, T.~K. Gaisser, {Cosmic rays. 4. The Spectrum and
  chemical composition above $10^4$\,GeV}, Astron. Astrophys. 274 (1993) 902.
\newblock \href {http://arxiv.org/abs/astro-ph/9303006}
  {\path{arXiv:astro-ph/9303006}}.

\bibitem{Kobayakawa:2000nq}
K.~Kobayakawa, Y.~Sato, T.~Samura, {Acceleration of particles by oblique shocks
  and cosmic ray spectra around the knee region}, Phys. Rev. D66 (2002) 083004.
\newblock \href {http://arxiv.org/abs/astro-ph/0008209}
  {\path{arXiv:astro-ph/0008209}}, \href
  {https://doi.org/10.1103/PhysRevD.66.083004}
  {\path{doi:10.1103/PhysRevD.66.083004}}.

\bibitem{Drury:2003fd}
L.~O. Drury, E.~van~der Swaluw, O.~Carroll, {Particle acceleration in supernova
  remnants, the Bell - Lucek hypothesis and the cosmic ray knee}\href
  {http://arxiv.org/abs/astro-ph/0309820} {\path{arXiv:astro-ph/0309820}}.

\bibitem{1993A&A...268..726P}
V.~S. {Ptuskin}, S.~I. {Rogovaya}, V.~N. {Zirakashvili}, L.~G. {Chuvilgin},
  G.~B. {Khristiansen}, E.~G. {Klepach}, G.~V. {Kulikov}, {Diffusion and drift
  of very high energy cosmic rays in galactic magnetic fields}, \aap 268 (1993)
  726--735.

\bibitem{Candia:2002we}
J.~Candia, E.~Roulet, L.~N. Epele, {Turbulent diffusion and drift in galactic
  magnetic fields and the explanation of the knee in the cosmic ray spectrum},
  JHEP 12 (2002) 033.
\newblock \href {http://arxiv.org/abs/astro-ph/0206336}
  {\path{arXiv:astro-ph/0206336}}, \href
  {https://doi.org/10.1088/1126-6708/2002/12/033}
  {\path{doi:10.1088/1126-6708/2002/12/033}}.

\bibitem{Candia:2003dk}
J.~Candia, S.~Mollerach, E.~Roulet, {Cosmic ray spectrum and anisotropies from
  the knee to the second knee}, JCAP 0305 (2003) 003.
\newblock \href {http://arxiv.org/abs/astro-ph/0302082}
  {\path{arXiv:astro-ph/0302082}}, \href
  {https://doi.org/10.1088/1475-7516/2003/05/003}
  {\path{doi:10.1088/1475-7516/2003/05/003}}.

\bibitem{1971CoASP...3..155S}
S.~I. {Syrovatskii}, {Cosmic Rays of Ultra-High Energy}, Comments on
  Astrophysics and Space Physics 3 (1971) 155.

\bibitem{Peters61}
B.~Peters, {Primary Cosmic Radiation and Extensive Air Showers}, Nuovo Cim. 22
  (1961) 800.

\bibitem{Zatsepin62}
G.~Zatsepin, N.~Gorunov, L.~Dedenko, lzv.\ Akad. Nauk USSR Ser. Fiz. 26 (1962)
  685.

\bibitem{Gaisser:2013bla}
T.~K. Gaisser, T.~Stanev, S.~Tilav, {Cosmic Ray Energy Spectrum from
  Measurements of Air Showers}, Front. Phys.(Beijing) 8 (2013) 748--758.
\newblock \href {http://arxiv.org/abs/1303.3565} {\path{arXiv:1303.3565}},
  \href {https://doi.org/10.1007/s11467-013-0319-7}
  {\path{doi:10.1007/s11467-013-0319-7}}.

\bibitem{Thoudam:2016syr}
S.~Thoudam, J.~P. Rachen, A.~van Vliet, A.~Achterberg, S.~Buitink, H.~Falcke,
  J.~R. H{\"o}randel, {Cosmic-ray energy spectrum and composition up to the
  ankle: the case for a second Galactic component}, Astron. Astrophys. 595
  (2016) A33.
\newblock \href {http://arxiv.org/abs/1605.03111} {\path{arXiv:1605.03111}},
  \href {https://doi.org/10.1051/0004-6361/201628894}
  {\path{doi:10.1051/0004-6361/201628894}}.

\bibitem{Giacinti:2011ww}
G.~Giacinti, M.~Kachelrie{\ss}, D.~V. Semikoz, G.~Sigl, {Cosmic Ray Anisotropy
  as Signature for the Transition from Galactic to Extragalactic Cosmic Rays},
  JCAP 1207 (2012) 031.
\newblock \href {http://arxiv.org/abs/1112.5599} {\path{arXiv:1112.5599}},
  \href {https://doi.org/10.1088/1475-7516/2012/07/031}
  {\path{doi:10.1088/1475-7516/2012/07/031}}.

\bibitem{Protheroe:2004rt}
R.~J. Protheroe, {Effect of energy losses and interactions during diffusive
  shock acceleration: Applications to SNR, AGN and UHE cosmic rays}, Astropart.
  Phys. 21 (2004) 415--431.
\newblock \href {http://arxiv.org/abs/astro-ph/0401523}
  {\path{arXiv:astro-ph/0401523}}, \href
  {https://doi.org/10.1016/j.astropartphys.2004.02.004}
  {\path{doi:10.1016/j.astropartphys.2004.02.004}}.

\bibitem{Biermann:1993wy}
P.~L. Biermann, {Cosmic rays. 1. The Cosmic ray spectrum between $10^4$\,GeV
  and $3\times 10^9$\,GeV}, Astron. Astrophys. 271 (1993) 649.
\newblock \href {http://arxiv.org/abs/astro-ph/9301008}
  {\path{arXiv:astro-ph/9301008}}.

\bibitem{Erlykin:1997bs}
A.~D. Erlykin, A.~W. Wolfendale, {A single source of cosmic rays in the range
  $10^{15}$\,eV to $10^{16}$\,eV}, J. Phys. G23 (1997) 979--989.
\newblock \href {https://doi.org/10.1088/0954-3899/23/8/012}
  {\path{doi:10.1088/0954-3899/23/8/012}}.

\bibitem{Erlykin:2000jm}
A.~D. Erlykin, A.~W. Wolfendale, {Models for the origin of the knee in the
  cosmic ray spectrum}, Adv. Space Res. 27 (2001) 803--812.
\newblock \href {http://arxiv.org/abs/astro-ph/0011057}
  {\path{arXiv:astro-ph/0011057}}, \href
  {https://doi.org/10.1016/S0273-1177(01)00125-9}
  {\path{doi:10.1016/S0273-1177(01)00125-9}}.

\bibitem{Erlykin:2015hha}
A.~D. Erlykin, A.~W. Wolfendale, {Vela as the source of cosmic rays responsible
  for the formation of the knee in their energy spectrum}, Bull. Russ. Acad.
  Sci. Phys. 79~(3) (2015) 308--310, [Izv. Ross. Akad. Nauk Ser.
  Fiz.79,no.3,340(2015)].
\newblock \href {https://doi.org/10.3103/S1062873815030181}
  {\path{doi:10.3103/S1062873815030181}}.

\bibitem{Batista:2016yrx}
R.~Alves~Batista, A.~Dundovic, M.~Erdmann, K.-H. Kampert, D.~Kuempel,
  G.~Muller, G.~Sigl, A.~van Vliet, D.~Walz, T.~Winchen, {CRPropa 3 - a Public
  Astrophysical Simulation Framework for Propagating Extraterrestrial
  Ultra-High Energy Particles}, JCAP 1605~(05) (2016) 038.
\newblock \href {http://arxiv.org/abs/1603.07142} {\path{arXiv:1603.07142}},
  \href {https://doi.org/10.1088/1475-7516/2016/05/038}
  {\path{doi:10.1088/1475-7516/2016/05/038}}.

\bibitem{Sigl:2017wfx}
G.~Sigl, {Astroparticle Physics: Theory and Phenomenology}, Vol.~1 of Atlantis
  Studies in Astroparticle Physics and Cosmology, Atlantis Press, 2017.
\newblock \href {https://doi.org/10.2991/978-94-6239-243-4}
  {\path{doi:10.2991/978-94-6239-243-4}}.

\bibitem{Greisen:1966jv}
K.~Greisen, {End to the cosmic ray spectrum?}, Phys. Rev. Lett. 16 (1966)
  748--750.
\newblock \href {https://doi.org/10.1103/PhysRevLett.16.748}
  {\path{doi:10.1103/PhysRevLett.16.748}}.

\bibitem{Zatsepin:1966jv}
G.~T. Zatsepin, V.~A. Kuzmin, {Upper limit of the spectrum of cosmic rays},
  JETP Lett. 4 (1966) 78--80, [Pisma Zh. Eksp. Teor. Fiz.4,114(1966)].

\bibitem{Hill:1983mk}
C.~T. Hill, D.~N. Schramm, {The Ultrahigh-Energy Cosmic Ray Spectrum}, Phys.
  Rev. D31 (1985) 564.
\newblock \href {https://doi.org/10.1103/PhysRevD.31.564}
  {\path{doi:10.1103/PhysRevD.31.564}}.

\bibitem{Berezinsky:1988wi}
V.~S. Berezinsky, S.~I. Grigor'eva, {A Bump in the ultrahigh-energy cosmic ray
  spectrum}, Astron. Astrophys. 199 (1988) 1--12.

\bibitem{Stanev:2000fb}
T.~Stanev, R.~Engel, A.~Mucke, R.~J. Protheroe, J.~P. Rachen, {Propagation of
  ultrahigh-energy protons in the nearby universe}, Phys. Rev. D62 (2000)
  093005.
\newblock \href {http://arxiv.org/abs/astro-ph/0003484}
  {\path{arXiv:astro-ph/0003484}}, \href
  {https://doi.org/10.1103/PhysRevD.62.093005}
  {\path{doi:10.1103/PhysRevD.62.093005}}.

\bibitem{Berezinsky:2002nc}
V.~Berezinsky, A.~Z. Gazizov, S.~I. Grigorieva, {On astrophysical solution to
  ultrahigh-energy cosmic rays}, Phys. Rev. D74 (2006) 043005.
\newblock \href {http://arxiv.org/abs/hep-ph/0204357}
  {\path{arXiv:hep-ph/0204357}}, \href
  {https://doi.org/10.1103/PhysRevD.74.043005}
  {\path{doi:10.1103/PhysRevD.74.043005}}.

\bibitem{Abbasi:2007sv}
R.~U. Abbasi, et~al., {First observation of the Greisen-Zatsepin-Kuzmin
  suppression}, Phys. Rev. Lett. 100 (2008) 101101.
\newblock \href {http://arxiv.org/abs/astro-ph/0703099}
  {\path{arXiv:astro-ph/0703099}}, \href
  {https://doi.org/10.1103/PhysRevLett.100.101101}
  {\path{doi:10.1103/PhysRevLett.100.101101}}.

\bibitem{Berezinsky:2006mk}
V.~Berezinsky, A.~Gazizov, M.~Kachelrie\ss, {Second dip as a signature of
  ultrahigh energy proton interactions with cosmic microwave background
  radiation}, Phys. Rev. Lett. 97 (2006) 231101.
\newblock \href {http://arxiv.org/abs/astro-ph/0612247}
  {\path{arXiv:astro-ph/0612247}}, \href
  {https://doi.org/10.1103/PhysRevLett.92.231101}
  {\path{doi:10.1103/PhysRevLett.92.231101}}.

\bibitem{Ivanov:2017juh}
D.~Ivanov, {Report of the Telescope Array - Pierre Auger Observatory Working
  Group on Energy Spectrum}, PoS ICRC2017 (2018) 498.
\newblock \href {https://doi.org/10.22323/1.301.0498}
  {\path{doi:10.22323/1.301.0498}}.

\bibitem{Yushkov:2018}
A.~Yushkov, {Report of the Auger-TA Working Group on the Composition of
  UHECRs}, presentation at UHECR 2018.

\bibitem{Abbasi:2018nun}
R.~U. Abbasi, et~al., {Depth of Ultra High Energy Cosmic Ray Induced Air Shower
  Maxima Measured by the Telescope Array Black Rock and Long Ridge FADC
  Fluorescence Detectors and Surface Array in Hybrid Mode}, Astrophys. J.
  858~(2) (2018) 76.
\newblock \href {http://arxiv.org/abs/1801.09784} {\path{arXiv:1801.09784}},
  \href {https://doi.org/10.3847/1538-4357/aabad7}
  {\path{doi:10.3847/1538-4357/aabad7}}.

\bibitem{Aab:2017cgk}
A.~Aab, et~al., {Inferences on mass composition and tests of hadronic
  interactions from 0.3 to 100 EeV using the water-Cherenkov detectors of the
  Pierre Auger Observatory}, Phys. Rev. D96~(12) (2017) 122003.
\newblock \href {http://arxiv.org/abs/1710.07249} {\path{arXiv:1710.07249}},
  \href {https://doi.org/10.1103/PhysRevD.96.122003}
  {\path{doi:10.1103/PhysRevD.96.122003}}.

\bibitem{Bellido:2017cgf}
J.~Bellido, {Depth of maximum of air-shower profiles at the Pierre Auger
  Observatory: Measurements above $10^{17.2}$ eV and Composition Implications},
  PoS ICRC2017 (2018) 506.
\newblock \href {https://doi.org/10.22323/1.301.0506}
  {\path{doi:10.22323/1.301.0506}}.

\bibitem{Bergmann:2006yz}
T.~Bergmann, R.~Engel, D.~Heck, N.~N. Kalmykov, S.~Ostapchenko, T.~Pierog,
  T.~Thouw, K.~Werner, {One-dimensional Hybrid Approach to Extensive Air Shower
  Simulation}, Astropart. Phys. 26 (2007) 420--432.
\newblock \href {http://arxiv.org/abs/astro-ph/0606564}
  {\path{arXiv:astro-ph/0606564}}, \href
  {https://doi.org/10.1016/j.astropartphys.2006.08.005}
  {\path{doi:10.1016/j.astropartphys.2006.08.005}}.

\bibitem{AlvesBatista:2019tlv}
R.~Alves~Batista, et~al., {Open Questions in Cosmic-Ray Research at Ultrahigh
  Energies}\href {http://arxiv.org/abs/1903.06714} {\path{arXiv:1903.06714}}.

\bibitem{Pierog:2013ria}
T.~Pierog, I.~Karpenko, J.~M. Katzy, E.~Yatsenko, K.~Werner, {EPOS LHC: Test of
  collective hadronization with data measured at the CERN Large Hadron
  Collider}, Phys. Rev. C92~(3) (2015) 034906.
\newblock \href {http://arxiv.org/abs/1306.0121} {\path{arXiv:1306.0121}},
  \href {https://doi.org/10.1103/PhysRevC.92.034906}
  {\path{doi:10.1103/PhysRevC.92.034906}}.

\bibitem{Riehn:2017mfm}
F.~Riehn, H.~P. Dembinski, R.~Engel, A.~Fedynitch, T.~K. Gaisser, T.~Stanev,
  {The hadronic interaction model SIBYLL 2.3c and Feynman scaling}, PoS
  ICRC2017 (2018) 301, [35,301(2017)].
\newblock \href {http://arxiv.org/abs/1709.07227} {\path{arXiv:1709.07227}},
  \href {https://doi.org/10.22323/1.301.0301} {\path{doi:10.22323/1.301.0301}}.

\bibitem{unger:2017cr}
M.~Unger, {Highlights from the Pierre Auger Observatory}, PoS ICRC2017 (2018)
  1102.
\newblock \href {http://arxiv.org/abs/1710.09478} {\path{arXiv:1710.09478}},
  \href {https://doi.org/10.22323/1.301.1102} {\path{doi:10.22323/1.301.1102}}.

\bibitem{deSouza:2017wgx}
V.~de~Souza, {Testing the agreement between the $X_\mathrm{max}$ distributions
  measured by the Pierre Auger and Telescope Array Observatories}, PoS ICRC2017
  (2018) 522.
\newblock \href {https://doi.org/10.22323/1.301.0522}
  {\path{doi:10.22323/1.301.0522}}.

\bibitem{Dembinski:2019uta}
H.~P. Dembinski, et~al., {Report on Tests and Measurements of Hadronic
  Interaction Properties with Air Showers}, 2019.
\newblock \href {http://arxiv.org/abs/1902.08124} {\path{arXiv:1902.08124}}.

\bibitem{Kachelriess:2017tvs}
M.~Kachelrie\ss, O.~Kalashev, S.~Ostapchenko, D.~V. Semikoz, {Minimal model for
  extragalactic cosmic rays and neutrinos}, Phys. Rev. D96~(8) (2017) 083006.
\newblock \href {http://arxiv.org/abs/1704.06893} {\path{arXiv:1704.06893}},
  \href {https://doi.org/10.1103/PhysRevD.96.083006}
  {\path{doi:10.1103/PhysRevD.96.083006}}.

\bibitem{strong74}
A.~W. {Strong}, J.~{Wdowczyk}, A.~W. {Wolfendale}, {The gamma-ray background: a
  consequence of metagalactic cosmic ray origin?}, Journal of Physics A
  Mathematical General 7 (1974) 120--134.

\bibitem{Berezinsky:1975zz}
V.~S. Berezinsky, A.~{\relax Yu}. Smirnov, {Cosmic neutrinos of ultra-high
  energies and detection possibility}, Astrophys. Space Sci. 32 (1975)
  461--482.
\newblock \href {https://doi.org/10.1007/BF00643157}
  {\path{doi:10.1007/BF00643157}}.

\bibitem{Kalashev:2007sn}
O.~E. Kalashev, D.~V. Semikoz, G.~Sigl, {Ultra-High Energy Cosmic Rays and the
  GeV-TeV Diffuse Gamma-Ray Flux}, Phys. Rev. D79 (2009) 063005.
\newblock \href {http://arxiv.org/abs/0704.2463} {\path{arXiv:0704.2463}},
  \href {https://doi.org/10.1103/PhysRevD.79.063005}
  {\path{doi:10.1103/PhysRevD.79.063005}}.

\bibitem{Berezinsky:2010xa}
V.~Berezinsky, A.~Gazizov, M.~Kachelrie{\ss}, S.~Ostapchenko, {Restricting
  UHECRs and cosmogenic neutrinos with Fermi-LAT}, Phys. Lett. B695 (2011)
  13--18.
\newblock \href {http://arxiv.org/abs/1003.1496} {\path{arXiv:1003.1496}},
  \href {https://doi.org/10.1016/j.physletb.2010.11.019}
  {\path{doi:10.1016/j.physletb.2010.11.019}}.

\bibitem{Ackermann:2014usa}
M.~Ackermann, et~al., {The spectrum of isotropic diffuse gamma-ray emission
  between 100 MeV and 820 GeV}, Astrophys. J. 799 (2015) 86.
\newblock \href {http://arxiv.org/abs/1410.3696} {\path{arXiv:1410.3696}},
  \href {https://doi.org/10.1088/0004-637X/799/1/86}
  {\path{doi:10.1088/0004-637X/799/1/86}}.

\bibitem{Neronov:2011kg}
A.~Neronov, D.~V. Semikoz, {Extragalactic Very-High-Energy gamma-ray
  background}, Astrophys. J. 757 (2012) 61.
\newblock \href {http://arxiv.org/abs/1103.3484} {\path{arXiv:1103.3484}},
  \href {https://doi.org/10.1088/0004-637X/757/1/61}
  {\path{doi:10.1088/0004-637X/757/1/61}}.

\bibitem{DiMauro:2013zfa}
M.~Di~Mauro, F.~Donato, G.~Lamanna, D.~A. Sanchez, P.~D. Serpico, {Diffuse
  $\gamma$-ray emission from unresolved BL Lac objects}, Astrophys. J. 786
  (2014) 129.
\newblock \href {http://arxiv.org/abs/1311.5708} {\path{arXiv:1311.5708}},
  \href {https://doi.org/10.1088/0004-637X/786/2/129}
  {\path{doi:10.1088/0004-637X/786/2/129}}.

\bibitem{TheFermi-LAT:2015ykq}
M.~Ackermann, et~al., {Resolving the Extragalactic $\gamma$-Ray Background
  above 50 GeV with the Fermi Large Area Telescope}, Phys. Rev. Lett. 116~(15)
  (2016) 151105.
\newblock \href {http://arxiv.org/abs/1511.00693} {\path{arXiv:1511.00693}},
  \href {https://doi.org/10.1103/PhysRevLett.116.151105}
  {\path{doi:10.1103/PhysRevLett.116.151105}}.

\bibitem{Kachelriess:2011bi}
M.~Kachelrie\ss, S.~Ostapchenko, R.~Tom{\`a}s, {ELMAG: A Monte Carlo simulation
  of electromagnetic cascades on the extragalactic background light and in
  magnetic fields}, Comput. Phys. Commun. 183 (2012) 1036--1043.
\newblock \href {http://arxiv.org/abs/1106.5508} {\path{arXiv:1106.5508}},
  \href {https://doi.org/10.1016/j.cpc.2011.12.025}
  {\path{doi:10.1016/j.cpc.2011.12.025}}.

\bibitem{Aartsen:2016xlq}
M.~G. Aartsen, et~al., {Observation and Characterization of a Cosmic Muon
  Neutrino Flux from the Northern Hemisphere using six years of IceCube data},
  Astrophys. J. 833~(1) (2016) 3.
\newblock \href {http://arxiv.org/abs/1607.08006} {\path{arXiv:1607.08006}},
  \href {https://doi.org/10.3847/0004-637X/833/1/3}
  {\path{doi:10.3847/0004-637X/833/1/3}}.

\bibitem{Aartsen:2017mau}
M.~G. Aartsen, et~al., {The IceCube Neutrino Observatory - Contributions to
  ICRC 2017 Part II: Properties of the Atmospheric and Astrophysical Neutrino
  Flux}\href {http://arxiv.org/abs/1710.01191} {\path{arXiv:1710.01191}}.

\bibitem{Murase:2013rfa}
K.~Murase, M.~Ahlers, B.~C. Lacki, {Testing the Hadronuclear Origin of PeV
  Neutrinos Observed with IceCube}, Phys. Rev. D88~(12) (2013) 121301.
\newblock \href {http://arxiv.org/abs/1306.3417} {\path{arXiv:1306.3417}},
  \href {https://doi.org/10.1103/PhysRevD.88.121301}
  {\path{doi:10.1103/PhysRevD.88.121301}}.

\bibitem{Murase:2010cu}
K.~Murase, T.~A. Thompson, B.~C. Lacki, J.~F. Beacom, {New Class of High-Energy
  Transients from Crashes of Supernova Ejecta with Massive Circumstellar
  Material Shells}, Phys. Rev. D84 (2011) 043003.
\newblock \href {http://arxiv.org/abs/1012.2834} {\path{arXiv:1012.2834}},
  \href {https://doi.org/10.1103/PhysRevD.84.043003}
  {\path{doi:10.1103/PhysRevD.84.043003}}.

\bibitem{Katz:2011zx}
B.~Katz, N.~Sapir, E.~Waxman, {X-rays, gamma-rays and neutrinos from
  collisionless shocks in supernova wind breakouts}\href
  {http://arxiv.org/abs/1106.1898} {\path{arXiv:1106.1898}}.

\bibitem{Zirakashvili:2015mua}
V.~N. Zirakashvili, V.~S. Ptuskin, {Type IIn supernovae as sources of high
  energy astrophysical neutrinos}, Astropart. Phys. 78 (2016) 28--34.
\newblock \href {http://arxiv.org/abs/1510.08387} {\path{arXiv:1510.08387}},
  \href {https://doi.org/10.1016/j.astropartphys.2016.02.004}
  {\path{doi:10.1016/j.astropartphys.2016.02.004}}.

\bibitem{Taylor:2014hya}
A.~M. Taylor, S.~Gabici, F.~Aharonian, {Galactic halo origin of the neutrinos
  detected by IceCube}, Phys. Rev. D89~(10) (2014) 103003.
\newblock \href {http://arxiv.org/abs/1403.3206} {\path{arXiv:1403.3206}},
  \href {https://doi.org/10.1103/PhysRevD.89.103003}
  {\path{doi:10.1103/PhysRevD.89.103003}}.

\bibitem{Blasi:2019obb}
P.~Blasi, E.~Amato, {Escape of cosmic rays from the Galaxy and effects on the
  circumgalactic medium}, Phys. Rev. Lett. 122~(5) (2019) 051101.
\newblock \href {http://arxiv.org/abs/1901.03609} {\path{arXiv:1901.03609}},
  \href {https://doi.org/10.1103/PhysRevLett.122.051101}
  {\path{doi:10.1103/PhysRevLett.122.051101}}.

\bibitem{Ahlers:2013xia}
M.~Ahlers, K.~Murase, {Probing the Galactic Origin of the IceCube Excess with
  Gamma-Rays}, Phys. Rev. D90~(2) (2014) 023010.
\newblock \href {http://arxiv.org/abs/1309.4077} {\path{arXiv:1309.4077}},
  \href {https://doi.org/10.1103/PhysRevD.90.023010}
  {\path{doi:10.1103/PhysRevD.90.023010}}.

\bibitem{Neronov:2018ibl}
A.~Neronov, M.~Kachelrie{\ss}, D.~V. Semikoz, {Multimessenger gamma-ray
  counterpart of the IceCube neutrino signal}, Phys. Rev. D98~(2) (2018)
  023004.
\newblock \href {http://arxiv.org/abs/1802.09983} {\path{arXiv:1802.09983}},
  \href {https://doi.org/10.1103/PhysRevD.98.023004}
  {\path{doi:10.1103/PhysRevD.98.023004}}.

\bibitem{Dzhappuev:2017vfy}
D.~D. Dzhappuev, et~al., {The Carpet-3 air shower array to search for diffuse
  gamma rays with energy $E_\gamma>100$\,TeV}, J. Phys. Conf. Ser. 934~(1)
  (2017) 012022.
\newblock \href {https://doi.org/10.1088/1742-6596/934/1/012022}
  {\path{doi:10.1088/1742-6596/934/1/012022}}.

\bibitem{Murase:2016gly}
K.~Murase, E.~Waxman, {Constraining High-Energy Cosmic Neutrino Sources:
  Implications and Prospects}, Phys. Rev. D94~(10) (2016) 103006.
\newblock \href {http://arxiv.org/abs/1607.01601} {\path{arXiv:1607.01601}},
  \href {https://doi.org/10.1103/PhysRevD.94.103006}
  {\path{doi:10.1103/PhysRevD.94.103006}}.

\bibitem{IceCube:2018dnn}
M.~G. Aartsen, et~al., {Multimessenger observations of a flaring blazar
  coincident with high-energy neutrino IceCube-170922A}, Science 361~(6398)
  (2018) eaat1378.
\newblock \href {http://arxiv.org/abs/1807.08816} {\path{arXiv:1807.08816}},
  \href {https://doi.org/10.1126/science.aat1378}
  {\path{doi:10.1126/science.aat1378}}.

\bibitem{Neronov:2018wuo}
A.~Neronov, D.~V. Semikoz, {Self-consistent model of extragalactic neutrino
  flux from evolving blazar population}\href {http://arxiv.org/abs/1811.06356}
  {\path{arXiv:1811.06356}}.

\bibitem{Kachelriess:2006aq}
M.~Kachelrie\ss, P.~D. Serpico, {The Compton-Getting effect on ultra-high
  energy cosmic rays of cosmological origin}, Phys. Lett. B640 (2006) 225--229.
\newblock \href {http://arxiv.org/abs/astro-ph/0605462}
  {\path{arXiv:astro-ph/0605462}}, \href
  {https://doi.org/10.1016/j.physletb.2006.08.006}
  {\path{doi:10.1016/j.physletb.2006.08.006}}.

\bibitem{Waxman:1996zn}
E.~Waxman, J.~Miralda-Escude, {Images of bursting sources of high-energy cosmic
  rays. 1. Effects of magnetic fields}, Astrophys. J. 472 (1996) L89--L92.
\newblock \href {http://arxiv.org/abs/astro-ph/9607059}
  {\path{arXiv:astro-ph/9607059}}, \href {https://doi.org/10.1086/310367}
  {\path{doi:10.1086/310367}}.

\bibitem{MiraldaEscude:1996kf}
J.~Miralda-Escude, E.~Waxman, {Signatures of the origin of high-energy cosmic
  rays in cosmological gamma-ray bursts}, Astrophys. J. 462 (1996) L59--L62.
\newblock \href {http://arxiv.org/abs/astro-ph/9601012}
  {\path{arXiv:astro-ph/9601012}}, \href {https://doi.org/10.1086/310042}
  {\path{doi:10.1086/310042}}.

\bibitem{Berezinsky:2005fa}
V.~Berezinsky, A.~Z. Gazizov, {Diffusion of cosmic rays in expanding universe},
  Astrophys. J. 643 (2006) 8--13.
\newblock \href {http://arxiv.org/abs/astro-ph/0512090}
  {\path{arXiv:astro-ph/0512090}}, \href {https://doi.org/10.1086/502626}
  {\path{doi:10.1086/502626}}.

\bibitem{Giacinti:2011uj}
G.~Giacinti, M.~Kachelrie{\ss}, D.~V. Semikoz, G.~Sigl, {Ultrahigh Energy
  Nuclei in the Turbulent Galactic Magnetic Field}, Astropart. Phys. 35 (2011)
  192--200.
\newblock \href {http://arxiv.org/abs/1104.1141} {\path{arXiv:1104.1141}},
  \href {https://doi.org/10.1016/j.astropartphys.2011.07.006}
  {\path{doi:10.1016/j.astropartphys.2011.07.006}}.

\bibitem{Sigl:2004yk}
G.~Sigl, F.~Miniati, T.~A. Ensslin, {Ultrahigh energy cosmic ray probes of
  large scale structure and magnetic fields}, Phys. Rev. D70 (2004) 043007.
\newblock \href {http://arxiv.org/abs/astro-ph/0401084}
  {\path{arXiv:astro-ph/0401084}}, \href
  {https://doi.org/10.1103/PhysRevD.70.043007}
  {\path{doi:10.1103/PhysRevD.70.043007}}.

\bibitem{Dolag:2004kp}
K.~Dolag, D.~Grasso, V.~Springel, I.~Tkachev, {Constrained simulations of the
  magnetic field in the local Universe and the propagation of UHECRs}, JCAP
  0501 (2005) 009.
\newblock \href {http://arxiv.org/abs/astro-ph/0410419}
  {\path{arXiv:astro-ph/0410419}}, \href
  {https://doi.org/10.1088/1475-7516/2005/01/009}
  {\path{doi:10.1088/1475-7516/2005/01/009}}.

\bibitem{Bruggen:2005ti}
M.~Bruggen, M.~Ruszkowski, A.~Simionescu, M.~Hoeft, C.~Dalla~Vecchia,
  {Simulations of magnetic fields in filaments}, Astrophys. J. 631 (2005)
  L21--L24.
\newblock \href {http://arxiv.org/abs/astro-ph/0508231}
  {\path{arXiv:astro-ph/0508231}}, \href {https://doi.org/10.1086/497004}
  {\path{doi:10.1086/497004}}.

\bibitem{Hackstein:2016pwa}
S.~Hackstein, F.~Vazza, M.~Br{\"u}ggen, G.~Sigl, A.~Dundovic, {Propagation of
  ultrahigh energy cosmic rays in extragalactic magnetic fields: a view from
  cosmological simulations}, Mon. Not. Roy. Astron. Soc. 462~(4) (2016)
  3660--3671.
\newblock \href {http://arxiv.org/abs/1607.08872} {\path{arXiv:1607.08872}},
  \href {https://doi.org/10.1093/mnras/stw1903}
  {\path{doi:10.1093/mnras/stw1903}}.

\bibitem{Pshirkov:2015tua}
M.~S. Pshirkov, P.~G. Tinyakov, F.~R. Urban, {New limits on extragalactic
  magnetic fields from rotation measures}, Phys. Rev. Lett. 116~(19) (2016)
  191302.
\newblock \href {http://arxiv.org/abs/1504.06546} {\path{arXiv:1504.06546}},
  \href {https://doi.org/10.1103/PhysRevLett.116.191302}
  {\path{doi:10.1103/PhysRevLett.116.191302}}.

\bibitem{Aab:2017tyv}
A.~Aab, et~al., {Observation of a Large-scale Anisotropy in the Arrival
  Directions of Cosmic Rays above $8 \times 10^{18}$ eV}, Science 357~(6537)
  (2017) 1266--1270.
\newblock \href {http://arxiv.org/abs/1709.07321} {\path{arXiv:1709.07321}},
  \href {https://doi.org/10.1126/science.aan4338}
  {\path{doi:10.1126/science.aan4338}}.

\bibitem{Erdogdu:2005wi}
P.~Erdogdu, et~al., {The Dipole anisotropy of the 2 Micron All-Sky Redshift
  Survey}, Mon. Not. Roy. Astron. Soc. 368 (2006) 1515--1526.
\newblock \href {http://arxiv.org/abs/astro-ph/0507166}
  {\path{arXiv:astro-ph/0507166}}, \href
  {https://doi.org/10.1111/j.1365-2966.2006.10243.x}
  {\path{doi:10.1111/j.1365-2966.2006.10243.x}}.

\bibitem{Abreu:2012ybu}
P.~Abreu, et~al., {Constraints on the origin of cosmic rays above $10^{18}$ eV
  from large scale anisotropy searches in data of the Pierre Auger
  Observatory}, Astrophys. J. 762 (2012) L13.
\newblock \href {http://arxiv.org/abs/1212.3083} {\path{arXiv:1212.3083}},
  \href {https://doi.org/10.1088/2041-8205/762/1/L13}
  {\path{doi:10.1088/2041-8205/762/1/L13}}.

\bibitem{Kumar:2013jaa}
R.~Kumar, D.~Eichler, {The isotropy problem of Sub-ankle Ultra-high energy
  cosmic rays}, Astrophys. J. 781~(1) (2014) 47.
\newblock \href {http://arxiv.org/abs/1311.1208} {\path{arXiv:1311.1208}},
  \href {https://doi.org/10.1088/0004-637X/781/1/47}
  {\path{doi:10.1088/0004-637X/781/1/47}}.

\bibitem{Matthews:2017waf}
J.~Matthews, {Highlights from the Telescope Array Experiment}, PoS ICRC2017
  (2018) 1096.
\newblock \href {https://doi.org/10.22323/1.301.1096}
  {\path{doi:10.22323/1.301.1096}}.

\bibitem{Kachelriess:2005uf}
M.~Kachelrie{\ss}, D.~V. Semikoz, {Clustering of ultrahigh energy cosmic ray
  arrival directions on medium scales}, Astropart. Phys. 26 (2006) 10--15.
\newblock \href {http://arxiv.org/abs/astro-ph/0512498}
  {\path{arXiv:astro-ph/0512498}}, \href
  {https://doi.org/10.1016/j.astropartphys.2006.04.006}
  {\path{doi:10.1016/j.astropartphys.2006.04.006}}.

\bibitem{Abu-Zayyad:2013vza}
T.~Abu-Zayyad, et~al., {Correlations of the Arrival Directions of Ultra-high
  Energy Cosmic Rays with Extragalactic Objects as Observed by the Telescope
  Array Experiment}, Astrophys. J. 777 (2013) 88.
\newblock \href {http://arxiv.org/abs/1306.5808} {\path{arXiv:1306.5808}},
  \href {https://doi.org/10.1088/0004-637X/777/2/88}
  {\path{doi:10.1088/0004-637X/777/2/88}}.

\bibitem{Abbasi:2014lda}
R.~U. Abbasi, et~al., {Indications of Intermediate-Scale Anisotropy of Cosmic
  Rays with Energy Greater Than 57 EeV in the Northern Sky Measured with the
  Surface Detector of the Telescope Array Experiment}, Astrophys. J. 790 (2014)
  L21.
\newblock \href {http://arxiv.org/abs/1404.5890} {\path{arXiv:1404.5890}},
  \href {https://doi.org/10.1088/2041-8205/790/2/L21}
  {\path{doi:10.1088/2041-8205/790/2/L21}}.

\bibitem{Aab:2018chp}
A.~Aab, et~al., {An Indication of anisotropy in arrival directions of
  ultra-high-energy cosmic rays through comparison to the flux pattern of
  extragalactic gamma-ray sources}, Astrophys. J. 853~(2) (2018) L29.
\newblock \href {http://arxiv.org/abs/1801.06160} {\path{arXiv:1801.06160}},
  \href {https://doi.org/10.3847/2041-8213/aaa66d}
  {\path{doi:10.3847/2041-8213/aaa66d}}.

\bibitem{Fang:2014uja}
K.~Fang, T.~Fujii, T.~Linden, A.~V. Olinto, {Is the Ultra-High Energy
  Cosmic-Ray Excess Observed by the Telescope Array Correlated with IceCube
  Neutrinos?}, Astrophys. J. 794~(2) (2014) 126.
\newblock \href {http://arxiv.org/abs/1404.6237} {\path{arXiv:1404.6237}},
  \href {https://doi.org/10.1088/0004-637X/794/2/126}
  {\path{doi:10.1088/0004-637X/794/2/126}}.

\bibitem{He:2014mqa}
H.-N. He, A.~Kusenko, S.~Nagataki, B.-B. Zhang, R.-Z. Yang, Y.-Z. Fan, {Monte
  Carlo Bayesian search for the plausible source of the Telescope Array
  hotspot}, Phys. Rev. D93 (2016) 043011.
\newblock \href {http://arxiv.org/abs/1411.5273} {\path{arXiv:1411.5273}},
  \href {https://doi.org/10.1103/PhysRevD.93.043011}
  {\path{doi:10.1103/PhysRevD.93.043011}}.

\bibitem{Abbasi:2018tqo}
R.~U. Abbasi, et~al., {Testing a Reported Correlation between Arrival
  Directions of Ultra-high-energy Cosmic Rays and a Flux Pattern from nearby
  Starburst Galaxies using Telescope Array Data}, Astrophys. J. 867~(2) (2018)
  L27.
\newblock \href {http://arxiv.org/abs/1809.01573} {\path{arXiv:1809.01573}},
  \href {https://doi.org/10.3847/2041-8213/aaebf9}
  {\path{doi:10.3847/2041-8213/aaebf9}}.

\bibitem{Berezinsky:2005cq}
V.~Berezinsky, A.~Z. Gazizov, S.~I. Grigorieva, {Dip in UHECR spectrum as
  signature of proton interaction with CMB}, Phys. Lett. B612 (2005) 147--153.
\newblock \href {http://arxiv.org/abs/astro-ph/0502550}
  {\path{arXiv:astro-ph/0502550}}, \href
  {https://doi.org/10.1016/j.physletb.2005.02.058}
  {\path{doi:10.1016/j.physletb.2005.02.058}}.

\bibitem{Unger:2015laa}
M.~Unger, G.~R. Farrar, L.~A. Anchordoqui, {Origin of the ankle in the
  ultrahigh energy cosmic ray spectrum, and of the extragalactic protons below
  it}, Phys. Rev. D92~(12) (2015) 123001.
\newblock \href {http://arxiv.org/abs/1505.02153} {\path{arXiv:1505.02153}},
  \href {https://doi.org/10.1103/PhysRevD.92.123001}
  {\path{doi:10.1103/PhysRevD.92.123001}}.

\bibitem{1967PhLA...24..677H}
A.~M. {Hillas}, {The energy spectrum of cosmic rays in an evolving universe},
  Physics Letters A 24 (1967) 677--678.
\newblock \href {https://doi.org/10.1016/0375-9601(67)91023-7}
  {\path{doi:10.1016/0375-9601(67)91023-7}}.

\bibitem{Heinze:2015hhp}
J.~Heinze, D.~Boncioli, M.~Bustamante, W.~Winter, {Cosmogenic Neutrinos
  Challenge the Cosmic Ray Proton Dip Model}, Astrophys. J. 825~(2) (2016) 122.
\newblock \href {http://arxiv.org/abs/1512.05988} {\path{arXiv:1512.05988}},
  \href {https://doi.org/10.3847/0004-637X/825/2/122}
  {\path{doi:10.3847/0004-637X/825/2/122}}.

\bibitem{Allard:2005ha}
D.~Allard, E.~Parizot, E.~Khan, S.~Goriely, A.~V. Olinto, {UHE nuclei
  propagation and the interpretation of the ankle in the cosmic-ray spectrum},
  Astron. Astrophys. 443 (2005) L29--L32.
\newblock \href {http://arxiv.org/abs/astro-ph/0505566}
  {\path{arXiv:astro-ph/0505566}}, \href
  {https://doi.org/10.1051/0004-6361:200500199}
  {\path{doi:10.1051/0004-6361:200500199}}.

\bibitem{Allard:2005cx}
D.~Allard, E.~Parizot, A.~V. Olinto, {On the transition from galactic to
  extragalactic cosmic-rays: spectral and composition features from two
  opposite scenarios}, Astropart. Phys. 27 (2007) 61--75.
\newblock \href {http://arxiv.org/abs/astro-ph/0512345}
  {\path{arXiv:astro-ph/0512345}}, \href
  {https://doi.org/10.1016/j.astropartphys.2006.09.006}
  {\path{doi:10.1016/j.astropartphys.2006.09.006}}.

\bibitem{Hooper:2006tn}
D.~Hooper, S.~Sarkar, A.~M. Taylor, {The intergalactic propagation of ultrahigh
  energy cosmic ray nuclei}, Astropart. Phys. 27 (2007) 199--212.
\newblock \href {http://arxiv.org/abs/astro-ph/0608085}
  {\path{arXiv:astro-ph/0608085}}, \href
  {https://doi.org/10.1016/j.astropartphys.2006.10.008}
  {\path{doi:10.1016/j.astropartphys.2006.10.008}}.

\bibitem{Aloisio:2013hya}
R.~Aloisio, V.~Berezinsky, P.~Blasi, {Ultra high energy cosmic rays:
  implications of Auger data for source spectra and chemical composition}, JCAP
  1410~(10) (2014) 020.
\newblock \href {http://arxiv.org/abs/1312.7459} {\path{arXiv:1312.7459}},
  \href {https://doi.org/10.1088/1475-7516/2014/10/020}
  {\path{doi:10.1088/1475-7516/2014/10/020}}.

\bibitem{Giacinti:2015pya}
G.~Giacinti, M.~Kachelrie{\ss}, O.~Kalashev, A.~Neronov, D.~V. Semikoz,
  {Unified model for cosmic rays above 10$^{17}$ eV and the diffuse gamma-ray
  and neutrino backgrounds}, Phys. Rev. D92~(8) (2015) 083016.
\newblock \href {http://arxiv.org/abs/1507.07534} {\path{arXiv:1507.07534}},
  \href {https://doi.org/10.1103/PhysRevD.92.083016}
  {\path{doi:10.1103/PhysRevD.92.083016}}.

\bibitem{Globus:2015xga}
N.~Globus, D.~Allard, E.~Parizot, {A complete model of the cosmic ray spectrum
  and composition across the Galactic to extragalactic transition}, Phys. Rev.
  D92~(2) (2015) 021302.
\newblock \href {http://arxiv.org/abs/1505.01377} {\path{arXiv:1505.01377}},
  \href {https://doi.org/10.1103/PhysRevD.92.021302}
  {\path{doi:10.1103/PhysRevD.92.021302}}.

\bibitem{Globus:2014fka}
N.~Globus, D.~Allard, R.~Mochkovitch, E.~Parizot, {UHECR acceleration at GRB
  internal shocks}, Mon. Not. Roy. Astron. Soc. 451~(1) (2015) 751--790.
\newblock \href {http://arxiv.org/abs/1409.1271} {\path{arXiv:1409.1271}},
  \href {https://doi.org/10.1093/mnras/stv893}
  {\path{doi:10.1093/mnras/stv893}}.

\bibitem{Aab:2013ika}
A.~Letessier-Selvon, et~al., {Highlights from the Pierre Auger Observatory},
  Braz. J. Phys. 44 (2014) 560--570.
\newblock \href {http://arxiv.org/abs/1310.4620} {\path{arXiv:1310.4620}},
  \href {https://doi.org/10.1007/s13538-014-0218-6}
  {\path{doi:10.1007/s13538-014-0218-6}}.

\bibitem{Aab:2014aea}
A.~Aab, et~al., {Depth of maximum of air-shower profiles at the Pierre Auger
  Observatory. II. Composition implications}, Phys. Rev. D90~(12) (2014)
  122006.
\newblock \href {http://arxiv.org/abs/1409.5083} {\path{arXiv:1409.5083}},
  \href {https://doi.org/10.1103/PhysRevD.90.122006}
  {\path{doi:10.1103/PhysRevD.90.122006}}.

\bibitem{Aab:2014kda}
A.~Aab, et~al., {Depth of maximum of air-shower profiles at the Pierre Auger
  Observatory. I. Measurements at energies above $10^{17.8}$ eV}, Phys. Rev.
  D90~(12) (2014) 122005.
\newblock \href {http://arxiv.org/abs/1409.4809} {\path{arXiv:1409.4809}},
  \href {https://doi.org/10.1103/PhysRevD.90.122005}
  {\path{doi:10.1103/PhysRevD.90.122005}}.

\bibitem{Fang:2017zjf}
K.~Fang, K.~Murase, {Linking High-Energy Cosmic Particles by Black Hole Jets
  Embedded in Large-Scale Structures}, Nature Phys. 14 (2018) 396.
\newblock \href {http://arxiv.org/abs/1704.00015} {\path{arXiv:1704.00015}},
  \href {https://doi.org/10.1038/s41567-017-0025-4}
  {\path{doi:10.1038/s41567-017-0025-4}}.

\bibitem{Boncioli:2018lrv}
D.~Boncioli, D.~Biehl, W.~Winter, {On the common origin of cosmic rays across
  the ankle and diffuse neutrinos at the highest energies from low-luminosity
  Gamma-Ray Bursts}, Astrophys. J. 872~(1) (2019) 110.
\newblock \href {http://arxiv.org/abs/1808.07481} {\path{arXiv:1808.07481}},
  \href {https://doi.org/10.3847/1538-4357/aafda7}
  {\path{doi:10.3847/1538-4357/aafda7}}.

\bibitem{Supanitsky:2018jje}
A.~D. Supanitsky, A.~Cobos, A.~Etchegoyen, {Origin of the light cosmic ray
  component below the ankle}, Phys. Rev. D98~(10) (2018) 103016.
\newblock \href {http://arxiv.org/abs/1810.12367} {\path{arXiv:1810.12367}},
  \href {https://doi.org/10.1103/PhysRevD.98.103016}
  {\path{doi:10.1103/PhysRevD.98.103016}}.

\bibitem{Haverkorn:2015wsa}
M.~Haverkorn, et~al., {Measuring magnetism in the Milky Way with the Square
  Kilometre Array}, PoS AASKA14 (2015) 096.
\newblock \href {https://doi.org/10.22323/1.215.0096}
  {\path{doi:10.22323/1.215.0096}}.

\bibitem{DiSciascio:2016rgi}
G.~Di~Sciascio, {The LHAASO experiment: from Gamma-Ray Astronomy to Cosmic
  Rays}, Nucl. Part. Phys. Proc. 279-281 (2016) 166--173.
\newblock \href {http://arxiv.org/abs/1602.07600} {\path{arXiv:1602.07600}},
  \href {https://doi.org/10.1016/j.nuclphysbps.2016.10.024}
  {\path{doi:10.1016/j.nuclphysbps.2016.10.024}}.

\bibitem{Haungs:2019ylq}
A.~Haungs, {A Scintillator and Radio Enhancement of the IceCube Surface
  Detector Array}, in: {Ultra High Energy Cosmic Rays (UHECR 2018) Paris,
  France, October 8-12, 2018}, 2019.
\newblock \href {http://arxiv.org/abs/1903.04117} {\path{arXiv:1903.04117}}.

\bibitem{Ogio:2018hyq}
S.~Ogio, {Telescope Array Low energy Extension: TALE}, JPS Conf. Proc. 19
  (2018) 011026.
\newblock \href {https://doi.org/10.7566/JPSCP.19.011026}
  {\path{doi:10.7566/JPSCP.19.011026}}.

\bibitem{Aab:2016vlz}
A.~Aab, et~al., {The Pierre Auger Observatory Upgrade - Preliminary Design
  Report}\href {http://arxiv.org/abs/1604.03637} {\path{arXiv:1604.03637}}.

\end{thebibliography}

}

\bibliographystyle{h-elsevier}

\end{document}